\documentclass[preprint,prd,onecolumn]{revtex4}%
\usepackage{amsfonts}
\usepackage{amsmath}
\usepackage{amssymb}
\usepackage{graphicx}%
\numberwithin{equation}{section}

\numberwithin{figure}{section}

\begin{document}
\title{Two Extended New Approaches to Vacuum, Matter \& Fields}
\author{Alex Kaivarainen}
\affiliation{University of Turku, Department of Physics,Vesilinnantie 5, FIN-20014, Turku, Finland,\\ H2o@karelia.ru  www.karelia.ru/\~\ alexk}
\author{Bo Lehnert}
\affiliation{Alfven Laboratory, Royal Institute of Technology, Stockholm, S-10044, Sweden \\ bo.lehnert@alfvenlab.kth.se}
\begin{abstract}
There are unsolved problems in physics leading to difficulties with Maxwell's
equations that are not removed by and not directly associated with quantum
mechanics. Recently a number of extended and modified theories have therefore
been put forward. For a manifold of these the concepts of the vacuum state
have played a fundamental r\^{o}le. The two theories by the authors \ (AK and
BL) are related to this state. The first theory consists of a revised
formulation of electromagnetics with an electric vacuum charge density(BL),
and the second is a unified theory of Bivacuum, matter and fields(AK).

In the first theory by BL a space-charge current density arises from the
condition of Lorentz invariance, in addition to the conventional displacement
current. This leads both to steady electromagnetic states and to new wave
types. The steady states provide models for the leptons, having an internal
structure associated with vortex motion around the axis of symmetry at the
velocity of light. For the electron the radial forces are kept in balance such
as to prevent it from \textquotedblleft exploding\textquotedblright\ under the
action of its net self-charge. The electron model has the deduced character of
a point-charge-like state with a very small radius, in which the infinite
self-energy problem is removed, and a more surveyable alternative to
renormalization is implemented. The electron charge is deduced from the
theory, and it deviates by only one percent from its experimental value. Two
individual photon models are further investigated. Both consist of
axisymmetric configurations having the capacity of a boson particle of limited
spatial extent and nonzero angular momentum. One model leads to the geometry
of needle-shaped radiation. These photon models behave simultaneously as
particles and as waves.\medskip

In the second theory by AK a Bivacuum model is established as a consequence of
a new interpretation and development of Dirac theory, pointing to equal
probability of positive and negative energy. Bivacuum is considered as the
infinite dynamic superfluid matrix of virtual dipoles, named Bivacuum fermions
($\mathbf{BVF}^{\updownarrow}$)$^{i}$ and Bivacuum bosons ($\mathbf{BVB}^{\pm
})^{i}$, formed by correlated torus ($\mathbf{V}^{+})$ and antitorus
($\mathbf{V}^{-})$ (donuts), as a collective excitations of subquantum
particles and antiparticles of opposite energy, charge and magnetic moments,
separated by an energy gap. The spatial and energetic parameters of the torus
and antitorus of primordial Bivacuum, i.e. in the absence of matter and field
influence, correspond to electrons, muons and tauons ($i=e,\mu,\tau)$. The
positive and negative Virtual Pressure Waves ($\mathbf{VPW}^{\pm})$ and
Virtual Spin Waves ($\mathbf{VirSW}^{S=\pm1/2})$ are the result of emission
and absorption of positive and negative Virtual Clouds ($\mathbf{VC}^{\pm}),$
resulting from transitions of \textbf{V}$^{+}$ and \textbf{V}$^{-}$ between
different states of excitation. The formation of real sub-elementary fermions
and their fusion to stable triplets of elementary fermions, corresponding to
the electron and proton rest mass and charge origination, become possible at a
certain symmetry shift between the \textbf{V}$^{+}\Updownarrow~$%
\textbf{V}$^{-}$ parameters. This shift appears at the velocity of angular
rotation of pairs of $\left[  \mathbf{BVF}^{\uparrow}\mathbf{\bowtie
BVF}^{\downarrow}\right]  $ around a common axis, being determined by the
Golden Mean condition: (\textbf{v/c)}$^{2}=\phi=0.618.$ The photon in such an
approach is a result of fusion (annihilation) of two triplets of particle and
antiparticle: $\left(  electron+positron\right)  $ or $\left(
proton+antiproton\right)  $. It represents a sextet of sub-elementary fermions
and antifermions with axial structural symmetry.

New expressions for total, potential and kinetic energies of de Broglie waves
of elementary particles, extending the relativity formalism, were obtained.
The dynamic mechanism of [corpuscle (C) $\rightleftharpoons$ wave (W)] duality
is proposed. It involves the modulation of the internal (hidden) quantum beats
frequency between the asymmetric 'actual' (torus) and 'complementary'
(antitorus) states of sub-elementary fermions or antifermions by the external
- empirical de Broglie wave frequency of the whole particles. It is
demonstrated, that the different kind of Bivacuum matrix excitations,
accompanied $\left[  Corpuscle~\leftrightharpoons~Wave\right]  $ pulsations of
sub-elementary fermions in triplets of fermions and their fact rotation are
responsible for electric, magnetic and gravitational fields origination. The
zero-point vibrations of particle and evaluated zero-point velocity of these
vibrations are also a result of $\left[  recoil\leftrightharpoons
antirecoil\right]  $\ effects, accompanied $\left[
\mathbf{C\leftrightharpoons W}\right]  $ pulsations. The electromagnetic
field, is a result of Corpuscle - Wave pulsation of photon and its fast
rotation with angle velocity ($\omega_{rot}$), equal to pulsation frequency.
The clockwise or anticlockwise direction of photon rotation as respect to
direction of its propagation, corresponds to its two polarization.

The Maxwell displacement current and the additional currents, increasing the
refraction index of Bivacuum, are also the consequences of the Bivacuum dipole
(\textbf{BVF}$^{\Updownarrow}$ and \textbf{BVB}$^{\pm}$) vibrations and
transitions, induced by an interaction with matter and fields. The nonzero
contributions of the rest mass of photons and neutrinos in this perturbed
Bivacuum follow from an elevated refraction index.

It is demonstrated that the dimensionless 'pace of time' ($\mathbf{dt/t=-dT}%
_{k}\mathbf{/T}_{k}$) and time itself (\textbf{t)} for each closed system are
determined by the change of this system's kinetic energy. They are positive if
the particle dynamics of the system is slowing down, and negative in the
opposite case. The concept of Virtual Replica (VR) or virtual hologram of any
material object is developed. The \textbf{VR }or virtual hologram\textbf{ }is
a result of interference of Virtual Pressure Waves ($\mathbf{VPW}_{q}^{+}$ and
$\mathbf{VPW}_{q}^{-})$ and Virtual Spin waves ($\mathbf{VirSW}_{\mathbf{q}%
}^{\mathbf{S=\pm1/2}}$) of the Bivacuum, representing \emph{\textquotedblright
reference waves\textquotedblright} with \emph{\textquotedblright object
waves\textquotedblright: } $\mathbf{VPW}_{\mathbf{m}}^{\pm}$ \ and
$\mathbf{VirSW}_{\mathbf{m}}^{\pm1/2}$, modulated by de Broglie waves of
molecules$.$

A possible Mechanism of Quantum entanglement between remote elementary
particles via Virtual Guides of spin, momentum and energy ($\mathbf{VirG}%
_{\mathbf{S,M,E}})$ is proposed. The consequences of the Unified theory (UT)
of Bivacuum, matter and fields are in good agreement with known experimental
data. The theory of AK makes an attempt to explain and unify the electric,
magnetic and gravitation fields in the framework of the proposed mechanism of
corpuscle-wave pulsations of elementary fermions, accompanied by different
kinds of their dynamic interaction with the Bivacuum superfluid matrix.

The theories of BL and AK reinforce each other, by having several points in
common. There are the similarities with the Dirac theory, symmetry with
respect to the concepts of matter and antimatter, the non-appearance of
magnetic monopoles, a space-charge vacuum current in addition to the
displacement current, models of the leptons having an internal vortex-like
structure and force balance, photon models with a small but nonzero rest mass
and a slightly reduced velocity of propagation, and particle-wave properties,
explaining self-interference even of single electrons or photons.\medskip

\textbf{PACS indexing codes:} 03; 04; 11\medskip
\end{abstract}
\maketitle
\section*{CONTENTS\bigskip}

\begin{quote}
ABSTRACT

\textbf{I. Introduction}

\textbf{II. Electromagnetics with a Vacuum Charge Density (by Bo Lehnert)}

A. Basis of the Theory

\emph{A.1 The Extended Field Equations}

\emph{A.2 Comparison with the Dirac Theory}

\emph{A.3 The Quantization Procedure}

\emph{A.4 Momentum and Energy}

B. New Aspects and Fundamental Applications

\emph{B.1 Steady Phenomena}

\emph{B.2 Time-dependent Phenomena}

C. General Features of Steady Axisymmetric States

\emph{C.1 The Generating Function}

\emph{C.2 Quantum Conditions}

D. A Model of the Electron

\emph{D.1 Generating Function and Integrated Field Quantities}

\emph{D.2 The Magnetic Flux}

\emph{D.3 Quantum Conditions of the Electron}

\emph{D.4 Comparison with the Conventional Renormalization Procedure}

\emph{D.5 Variational Analysis of the Integrated Charge}

\emph{D.6 Force Balance of the Electron Model}

E. Axisymmetric Wave Modes and Their Photon Models

\emph{E.1 Elementary Normal Modes}

\emph{E.2 Wave Packets}

\emph{E.3 Spatially Integrated Field Quantities}

\emph{E.4 Features of Present Photon Model}

F. Summary of Obtained Main Results in Section II\medskip

\textbf{III. Unified Theory of Bivacuum, \ Particles Duality, Fields and Time
(by A.Kaivarainen)}

\emph{A.1. Properties of Bivacuum dipoles - Bivacuum fermions and Bivacuum
bosons}

\emph{A.2 \ The basic (carrying)Virtual Pressure Waves (VPW}$^{\pm}$\emph{)
and }

\emph{Virtual spin waves (VirSW}$^{\pm1/2})$\emph{ of Bivacuum}

\emph{A.3 \ Virtual Bose condensation (VirBC), as a base of Bivacuum
nonlocality}

B. Basic postulates of Unified Theory and their consequences

\emph{B.1 The relation between the external and internal parameters of
Bivacuum fermions \& quantum roots of Golden mean. The rest mass and charge
origination}

C. \ Fusion of triplets of elementary particles from sub-elementary fermions
and antifermions at Golden mean conditions
\end{quote}

\begin{quotation}
\emph{C.1 \ Correlation between our new model of hadrons and conventional
quark model of }

\emph{protons, neutrons and mesons}

D. \ Total, potential and kinetic energies of elementary de Broglie waves

E.\ The dynamic mechanism of corpuscle-wave duality

F. The nature of electrostatic, magnetic and gravitational potentials, based
on Unified theory\ \ 

\emph{F.1 \ Electromagnetic dipole radiation, as a consequence of charge
oscillation}

\emph{F.2 }\ \emph{The different kind of Bivacuum dipoles perturbation,
induced by dynamics of elementary particles }

\emph{F.3 } \emph{The new approach to quantum gravity and antigravity. The
equality of gravitational and inertial mass of particles }

\emph{F.4.\ Possible nature of neutrino and antineutrino}

\emph{F.5 The Bivacuum dipoles symmetry shift and linear and circular
ordering, }

\emph{as a background of electrostatic and magnetic fields origination}

\emph{F.6 \ Interpretation of Maxwell displacement current, based on Bivacuum
model}

\emph{F.7 \ New kind of current in secondary Bivacuum, additional to
displacement one,}

\emph{increasing the refraction index of Bivacuum }

\emph{F.8 The mechanism, increasing the refraction index of Bivacuum}

G. The Principle of least action, as a consequence of Bivacuum basic Virtual
Pressure Waves (VPW$_{q=1}^{\pm}$) resonance interaction with particles.
Possible source of energy for overunity devices.

\emph{G.1 The new approach to problem of Time }

H. Virtual Replica (VR) of matter in Bivacuum

\emph{H.1 Bivacuum perturbations, induced by the oscillation of the total
energy of de Broglie waves, }

\emph{accompanied their thermal vibrations and recoil }$\rightleftharpoons$
\emph{antirecoil effects}

\emph{H.2 \ Modulation of basic Virtual Pressure Waves (}$VPW_{q}^{\pm}%
)$\emph{\ and Virtual Spin Waves (}$VirSW_{q}^{\pm1/2})$\emph{\ of Bivacuum by
translational and librational dynamics (de Broglie waves) of the object
molecules, correspondingly}

I. \ Possible Mechanism of Quantum entanglement between remote elementary
particles via Virtual Guides of spin, momentum and energy
(VirG$_{\mathbf{S,M,E}})$

\emph{I.1 }\ \emph{The role of tuning} \emph{force (}$\mathbf{F}%
_{\mathbf{VPW}^{\pm}})$\emph{ of virtual pressure waves }$\mathbf{VPW}%
_{q}^{\pm}$\emph{ of Bivacuum in entanglement}

J. \ Experimental data, confirming Unified theory (UT)

\emph{J.1 \ Radiation of accelerating charges, magnetic moment of the
electron}

\emph{J.2 \ The double turn (720}$^{0}),$\emph{\ as a condition of the
fermions spin state reversibility}

\emph{J.3 Michelson-Morley experiment, as the evidence for the Virtual Replica
of the Earth}

\emph{J.4 Interaction of particles with their Virtual Replicas, as a
background of two slit experiments explanation }

K. The main Conclusions of Unified Theory (section III)
\end{quotation}

\begin{quote}
\textbf{IV. Common Points and Main Features of the two Present Theories}

\textbf{V. General Conclusions}
\end{quote}

\begin{quotation}
\bigskip
\end{quotation}

\section{Introduction}

Conventional field theory based on quantum mechanics and Maxwell's equations
has been very successful in its applications to numerous problems in physics,
and has sometimes manifested itself in an extremely good agreement with
experimental results. Nevertheless there are certain areas within which these
joint theories do not seem to provide fully adequate descriptions of physical
reality. Thus, as already pointed out by Feynman\cite{Feyn64}, there are
unsolved problems leading to difficulties with Maxwell's equations that are
not removed by and not directly associated with quantum mechanics. Physics is
therefore a far from completed field of research. Consequently, a number of
extended and modified theories have recently been put forward, and for a
manifold of these the concepts of the vacuum state have played a fundamental
r\^{o}le. It would lead too far to present a detailed description of all these
approaches within this review. The latter is thus \ limited to two theories by
the authors (AK and BL) being related to the properties of the vacuum state.
The first theory consists of a revised formulation of electromagnetics with a
vacuum charge density(BL), and the second is a unified theory of Bivacuum,
matter, fields and time (AK).

Among the present shortcomings of conventional theory, a number of examples
can be given here. Turning first to charged leptons in the capacity of
elementary particles such as the electron, there are the following problems:

\begin{itemize}
\item In conventional theory these particles are treated strictly as
mathematical points, with no internal structure. This is unsatisfactory from
the physical point of view.

\item According to Maxwell's equations the electron cannot be kept in
equilibrium by its own static forces. It tends to \textquotedblleft
explode\textquotedblright\ under the action of its self-charge, and has
therefore to be kept in balance by some forces of nonelectromagnetic
character, as stated by Jackson\cite{Jack62} among others. This explanation in
terms of unspecified forces is incomplete.

\item The infinite self-energy of the point charge is a serious problem. It
has so far been solved in terms of a renormalization procedure, where the
difference between two \textquotedblleft infinities\textquotedblright\ gives a
finite result. However, a more satisfactory procedure from the physical point
of view has been called for by Ryder\cite{Ryde96}.

\item There is so far no explanation by conventional theory why the free
elementary electronic charge has a discrete minimum value, as found from experiments.
\end{itemize}

When further considering the photons and their wave phenomena, there are
additional problems as follows:

\begin{itemize}
\item As pointed out by Heitler\cite{Heit54} among others, the photon in the
form of a plane conventional electromagnetic wave does not possess any angular
momentum (spin). For a nonzero spin, derivatives are therefore needed in the
transverse direction of a photon configuration which has the capacity of a
boson particle. In the case of conventional electromagnetic theory, however,
such transverse derivatives lead to divergent solutions in space, as already
found by Thomson\cite{Thom36} and being further discussed by several
investigators. Such solutions become physically unacceptable. An even more
serious complication is due to the fact that there does not arise any spin of
an axisymmetric wave deduced from Maxwell's equations for which there is a
vanishing electric field divergence.

\item There are no explicit solutions from conventional theory which become
consistent with needle-like radiation. Such radiation has been suggested by
Einstein, and it becomes associated with the photoelectric effect where a
photon \textquotedblleft knocks out\textquotedblright\ an atomic electron, as
well as with the observed dot-shaped marks on a screen in two-slit experiments.

\item The particle and wave nature of the photon remain in conventional theory
as two separate concepts. In two-slit experiments the individual photon
behaves on the other hand in terms of both concepts at the same time, as
demonstrated by Tsuchiya et. al.\cite{Tsuc85}.
\end{itemize}

These problems have been tackled in terms of an extended Lorentz invariant
theory described in Section II of this review and based on the concept of a
nonzero electric charge density and electric field divergence in the vacuum
(Lehnert
\cite{Lehn86,Lehn95,Lehn98,Lehn01,Lehn02a,Lehn02b,Lehn03,Lehn04a,Lehn04b}).
This gives rise to a space-charge current density in the vacuum state, in
addition to the displacement current, and it leads both to steady
electromagnetic states and to new types of wave phenomena.

The steady states provide models of the leptons, such as the electron and the
neutrino. These states possess an internal structure which comes out from the
theory, thereby being associated with a vortex motion around the axis of
symmetry at the velocity of light, and with related electric and magnetic
field components. This vortex-like structure is not hypothetical, but it comes
out from the solutions of the extended basic equations. There is an
\textquotedblleft electromagnetic confinement\textquotedblright\ by which the
radial forces are kept in balance, also in the case of the electron which is
then prevented from \textquotedblleft exploding\textquotedblright\ under the
action of its net self-charge. The electron model has further the deduced
character of a point-charge-like state with a very small radius, in which the
infinite self-energy problem is removed, and a physically more surveyable
alternative to the renormalization process is being implemented. The deduced
elementary electronic charge deviates by only one percent from its
experimental \ value, and there are proposed mechanisms for the possible
removal of this small deviation.

An application of the theory to photon physics further leads to several
options of photon models. One resulting axisymmetric configuration has the
capacity of a boson particle of limited spatial extent and a nonzero angular
momentum. Another option leads to the geometry \ of a strongly limited
transverse dimension, in the form of needle-shaped radiation. Thereby the
elaborated photon models behave simultaneously as particles and as waves.

The Dirac equation points to equal probability of positive and negative
energy\cite{Dira47}. In the Dirac vacuum its realm of negative energy is
saturated by an infinite number of electrons. However, it was assumed that
these electrons, following the Pauli principle, have not any gravitational or
viscosity effects. Positrons in the Dirac model represent the
\textquotedblleft holes\textquotedblright, originated as a result of the
electron jumps in the realm of positive energy, over the energetic gap
$\Delta=2m_{0}c^{2}$. Currently it becomes clear that the Dirac model of the
vacuum is not general enough to explain all known experimental data, e.g. the
boson emergency. The model of the Bivacuum presented in Section III of this
review appears to be more advanced. However, it uses the same starting point
of equal probability of positive and negative energy, being confined in each
of the Bivacuum dipoles. These virtual dipoles are correlated pairs of torus
and antitorus, representing, in turn, the collective vortical excitations of
subquantum particles and antiparticles and separated by an energy gap.

The theory of AK explains the absence of magnetic monopoles, through the
absence of a symmetry shift between the internal opposite actual quantities
$|\mathbf{\mu}_{+}|$ and the complementary $|\mathbf{\mu}_{-}|$ \ magnetic
moments of the virtual torus and antitorus which form Bivacuum fermions. This
is in contrast to the mass and charge symmetry shifts, providing the rest mass
and charge origination.

In a book by Bohm and Hiley\cite{Bohm93} the electron is considered as a
particle with well-defined position and momentum which are, however, under the
influence of a special wave(quantum potential). In accordance with these
authors an elementary particle is a \emph{sequence of incoming and outgoing
waves}, which are very close to each other. But the particle itself does not
have a wave nature. The interference pattern in double-slit experiments is
after Bohm the result of a periodically \textquotedblleft
bunched\textquotedblright\ character of a quantum potential. There is a basic
difference in this respect with a model of duality in Section III, where it is
assumed that the wave and corpuscle phases are realized alternatively at a
high Compton frequency. This frequency is modulated by the empirical de
Broglie wave frequency of particles, thereby making possible their
interference with their own 'virtual replica', formed by Bivacuum Virtual
Pressure Waves(VPW$^{\pm}$). In 1950 Wheeler and Misner\cite{Whee68} published
\textquotedblleft Geometrodynamics\textquotedblright, a new description of
space-time properties based on topology. They supposed that elementary
particles and antiparticles, their spins, and positive and negative charges
can be presented as interconnected black and white holes. The connecting tube
exists in another space-time than the holes themselves and the process of
energy transmission looks as instantaneous. In conventional space-time the two
ends of the tube, termed \textquotedblleft wormholes\textquotedblright, can be
a vast distance apart. This is one of the possible explanations of quantum
non-locality. The \textquotedblleft wormhole\textquotedblright\ idea has
common features with our model of entanglement between coherent remote
elementary particles via Virtual Guides of spin, momentum, and
energy(VirG$_{\text{SME}}$) - microtubules, formed by pairs of torus and antitorus.

Sidhart\cite{Sidh98,Sidh99} considered the elementary particle as a
\emph{relativistic vortex} of Compton radius, from which he recovered its mass
and quantized spin. A particle was pictured as a fluid vortex, steadily
circulating at the light velocity along a 2D ring or spherical 3D shell with
radius $L=\hbar/2mc$. Inside such a vortex the notions of negative energy,
superluminal velocities and nonlocality are acceptable without contradiction
with conventional theory. This view has some common features with the earlier
models of extended electromagnetic steady states
\cite{Lehn86,Lehn95,Lehn98,Lehn01,Lehn02a,Lehn02b,Lehn03,Lehn04a,Lehn04b}
described in Section II, and with the models of elementary and sub-elementary
particles and antiparticles\cite{Kaiv04a,Kaiv05a,Kaiv05b} described in Section
III. If the measurements are averaged over time $t\sim mc^{2}/\hbar$ and over
space $L\sim\hbar/mc$, the imaginary part of the particle position disappears,
and we are back in usual physics\cite{Sidh98}.

Aspden\cite{Aspd03} introduced in his aether theory a basic unit, named Quon,
as a pair of virtual muons of opposite charges, i.e. a muon and an antimuon.
This idea is close to the model of Bivacuum dipoles consisting of a torus and
an antitorus of opposite energy/mass, charge, and magnetic moment with Compton
radii of the electron, muon and tauon\cite{Kaiv04a,Kaiv05a,Kaiv05b}. In the
book entitled \textquotedblleft The physics of creation\textquotedblright,
Aspden used the Thomson model of the electron, as a sphere with the radius
$a$, charge $e$, and energy $2e^{2}/3a=mc^{2}$. This model strongly differs
from that presented in the unified theory of Section III.

Barut and Bracken\cite{Baru81} considered the \emph{zitterbewegung}, a rapidly
oscillating imaginary part of particle position leading from the Dirac
theory\cite{Dira47}, as a harmonic oscillator in the Compton wavelength region
of the particle. Also Einstein\cite{Eins71,Eins82,Stac89} and
Schr\"{o}dinger\cite{Schr80} considered the oscillation of the electron at the
frequency $\nu=m_{0}c^{2}/h$ and the amplitude $\zeta_{\max}=\hbar/(2m_{0}c).$
It was demonstrated by Schr\"{o}dinger that the position of the free electron
can be presented as $x=\overline{x}+\zeta$ where $\overline{x}$ characterizes
the average position of the free electron, and $\zeta$ is its instantaneous
position related to the oscillations. Hestness\cite{Hest90} further proposed
that the zitterbewegung arises from self-interaction, resulting from
wave-particle duality. These ideas are close to our explanation of zero-point
oscillations of elementary particles, as a consequence of their recoil and
antirecoil vibrations, induced by the particles corpuscle-wave pulsations.

Puthoff\cite{Puth91} developed the idea of \textquotedblleft vacuum
engineering\textquotedblright, using the hypothesis of a polarizable vacuum.
The electric permittivity $\varepsilon_{0}$ and the magnetic permeability
$\mu_{0}$ are interrelated in symmetric vacuum, as $\varepsilon_{0}\mu
_{0}=1/c^{2}$. It is conceived that a change of the vacuum refraction index
$n=c/\mathbf{v}=\varepsilon^{1/2}$ in gravitational or electric fields is
accompanied by a variation of a number of space-time parameters.
Fock\cite{Fock64} earlier explained the bending of a light beam, induced by
gravitation near massive bodies, also by a change of the refraction in the
vacuum, i.e. in another way than that of the general theory of relativity.
However, the mechanism of this vacuum polarization and its related refraction
index remains obscure.

The transformation of a neutron after scattering on a neutrino to a proton and
electron in accordance with the electro-weak (EW) theory developed by
Glashov\cite{Glas61}, Weinberg\cite{Wein67}, and Salam\cite{Sala68}, is
mediated by a negative W$^{-}$ particle. The reverse reaction in EW theory is
mediated by a positive massless W$^{+}$ boson. Scattering of the electron on a
neutrino, not being accompanied by charge transfer, is mediated by a third
massless neutral boson Z$^{0}$. For the explanation in EW theory of
spontaneous symmetry violation of intermediate vector bosons, such as charged
W$^{\pm}$ and neutral Z$^{0}$ with spin 1, the Higgs field was introduced and
accompanied by a big mass in the origination of these exchange particles. The
EW theory also needs the quantum of the Higgs field, namely Higgs bosons with
big mass, zero charge and integer spin. The experimental discovery of the
heavy W$^{+}$ and Z$^{0}$ particles in 1983 was considered as a confirmation
of the EW theory. However, the Higgs field and Higgs bosons are still not
found. The Unified theory of Section III proposes another explanation of spin,
mass and charge origination, related to the symmetry shift of Bivacuum dipoles.

Thomson, Heaviside and Searl supposed that mass is an electrical phenomenon
\cite{Aspd03}. In the theory of Rueda and Haisch \cite{Rued01} it was proposed
that the inertia is a reaction force, originating in a course of dynamic
interaction between the electromagnetic zero-point field (ZPF) of the vacuum
and the charges of elementary particles. However, it is not clear from their
theory how the charges themselves are being created. In EW theory, mentioned
above, two parameters of W$^{\pm}$ particles, such as charge and mass, are
considered to be independent. The theory of AK tries to unify these parameters
with the Bivacuum dipole symmetry shift, induced by rotation of their pairs
around a common axis, fusion to triplets and external translational motion of
triplets \cite{Kaiv04a,Kaiv05a,Kaiv05b,Kaiv04b}. This theory also makes an
attempt to explain and unify the electric, magnetic and gravitation fields in
the framework of the proposed mechanism of correlated corpuscle - wave
pulsations of sub-elementary fermions in triplets, accompanied by different
kinds of Bivacuum superfluid matrix perturbation. The internal time (t) and
its pace (dt/t) of the each closed system of particles are parameters,
characterizing both - the average velocity and acceleration of these
particles. Such new approach to time problem differs from the conventional
one, following from relativistic theory, however, they are compatible.

The conventional quantum field theory do not explain a lot of fundamental
phenomena, like corpuscle-wave duality, origination of mass, charge, etc. This
approach face a serious problem of singularity and related energy divergence.
Our Unified theory goes from Bivacuum dipoles to particles and then to fields
and contains the following main stages:

\textbf{1.} The Bivacuum dipoles symmetry shift, induced by their external
translational-rotational motion, turning them to sub-elementary fermions at
Golden mean conditions $\mathbf{(v/c)}^{2}=0.618$; \ \textbf{2.} Origination
of the rest mass, charge and fusion of triplets of sub-elementary fermions,
representing elementary fermions. This process, like the previous one can be
considered, as a kind of self-organization in Bivacuum; \textbf{3.} The
correlated $\mathbf{Corpuscle\rightleftharpoons Wave}$ pulsation of
sub-elementary particles, composing elementary particles; \textbf{4.}
Different kind of Bivacuum medium perturbations, induced by
$\mathbf{C\rightleftharpoons W}$ pulsation of unpaired and paired
sub-elementary fermions, responsible for excitation of the electric, magnetic
and gravitational potentials. \medskip

\section{Electromagnetics with a Vacuum Charge Density}

The vacuum is not merely an empty space. Its zero-point energy is related to
electromagnetic vacuum fluctuations. Moreover, the observed electron-positron
pair formation from an energetic photon indicates that electric charges can be
created out of an electrically neutral vacuum state. It can therefore be
conceived that a current density related to an electric space charge density
and a nonzero electric field divergence can arise in the vacuum, in addition
to the displacement current by Maxwell. As a consequence, the theory of this
Section consists of a modified form extended beyond Maxwell's equations in the
vacuum state, and having the latter equations as a special case. This approach
is thus based on two mutually independent hypotheses:

\begin{itemize}
\item A nonzero charge density and a related electric field divergence can
exist in the vacuum state. This should in principle not become less
conceivable than the nonzero curl of the magnetic field strength for a
conventional electromagnetic wave in the vacuum. All these concepts can be
regarded as intrinsic properties of the electromagnetic field.

\item The resulting extended (revised) form of the field equations should
remain Lorentz invariant. Physical experience supports such a statement, as
long as there are no observations getting into conflict with it.
\end{itemize}

This section presents a condensed review of the theory, and for its details
reference is made to earlier
publications\cite{Lehn86,Lehn95,Lehn98,Lehn01,Lehn02a,Lehn02b,Lehn03,Lehn04a,Lehn04b}%
.

\subsection{Basis of the Theory}

The basic concepts of the theory are first presented here, with resulting
field equations, main features, and the related quantization procedure.

\subsubsection{The Extended Field Equations}

In presence of a charge density and a related current density in the vacuum,
the extended field equations are now written in the four-dimensional form of a
Proca-Type equation
\begin{align}
\square A_{\mu}  &  \equiv\left(  \frac{1}{c^{2}}\frac{\partial^{2}}{\partial
t^{2}}-\nabla^{2}\right)  A_{\mu}=\mu_{0}J_{\mu}\\
\mu &  =1,2,3,4\nonumber
\end{align}
given in SI units In this equation
\begin{equation}
A_{\mu}=\left(  \mathbf{A},i\phi/c\right)
\end{equation}
where $\mathbf{A}$ and $\mathbf{\phi}$ are the magnetic vector potential and
the electrostatic potential in three-space, and
\begin{equation}
J_{\mu}=\left(  \mathbf{j},ic\bar{\rho}\right)
\end{equation}
is the additional four-current density which includes the corresponding
three-space current density $\mathbf{j}$ and the electric charge density
$\mathbf{\bar{\rho}}$.

Maxwell's equations in the vacuum are recovered as a special case in which the
current density $\mathit{J}_{\mu}$ disappears.

The current density (2.3) appearing in the right-hand side of the Proca-type
equation (2.1) is here required to transform as a four-vector. This implies
that the square of $J_{\mu}$ should become invariant to a transition from one
inertial frame K to another such frame $\operatorname{K}^{\prime}$. Then
Eq.(2.3) yields
\begin{equation}
j^{2}-c^{2}\bar{\rho}^{2}=j^{\prime}{^{2}}-{c^{2}}\bar{\rho}^{\prime}{^{2}%
}=const
\end{equation}
In addition, the corresponding three-space current density $\mathbf{j}$ is
required to exist only when there is also an electric charge density$\bar
{\rho}$ associated with the nonzero electric field divergence. This implies
that the constant in Eq.(2.4) vanishes. Consequently the three-space current
density becomes
\begin{equation}
\mathbf{j}=\bar{\rho}\mathbf{C}=\varepsilon_{0}(\nabla\cdot\mathbf{E}%
)\mathbf{C}\qquad\mathbf{C}^{2}=c^{2}%
\end{equation}
where $\mathbf{C}$ is a velocity vector having a modulus equal to the velocity
$c$ of light. The final form of the four-current density is then given by
\begin{equation}
J_{\mu}=(\mathbf{j},ic\bar{\rho})=\varepsilon_{0}(\nabla\cdot\mathbf{E}%
)(\mathbf{C},ic)
\end{equation}

In analogy with the direction to be determined for the current density in
conventional theory, the unit vector $\mathbf{C}/c$ will depend on the
geometry of the particular configuration to be considered. This unit vector
can in principle have components which vary from point to point in space,
whereas the modulus of $\mathbf{C}$ remains constant according to the
invariance condition (2.5). Both curl $\mathbf{C}$ and div $\mathbf{C}$ can
also differ from zero, but in all cases treated here we restrict ourselves to
\begin{equation}
\nabla\cdot\mathbf{C}=0
\end{equation}
and to a time-independent velocity vector $\mathbf{C}$.

In the three-dimensional representation the extended equations of the vacuum
state now become
\begin{align}
\nabla\times\mathbf{B}/\mu_{0}  &  =\varepsilon_{0}(\nabla\cdot\mathbf{E}%
)\mathbf{C}+\varepsilon_{0}\partial\mathbf{E}/\partial t\\[0.2cm]
\nabla\times\mathbf{E}  &  =-\partial\mathbf{B}/\partial t\\[0.2cm]
\mathbf{B}  &  =\nabla\times\mathbf{A}\qquad\nabla\cdot\mathbf{B}=0\\[0.2cm]
\mathbf{E}  &  =-\nabla\phi-\partial\mathbf{A}/\partial t\\[0.2cm]
\nabla\cdot\mathbf{E}  &  =\bar{\rho}/\varepsilon_{0}%
\end{align}
where the first term of the right-hand member of Eq.(2.8) and Eq.(2.12) are
the new parts being introduced by the present theory. The nonzero electric
field divergence introduces an asymmetry in the appearance of the electric and
magnetic fields in these equations.

The equations (2.8)--(2.12) are gauge invariant \cite{Lehn01,Lehn04b}, but not
those which explicitly include a particle mass \cite{Ryde96}.

The presence in Eqs.(2.12) and (2.8) of the dielectric constant $\varepsilon
_{0}$ and the magnetic permeability $\mu_{0}$ of the conventional vacuum state
may require further explanation. Liquid and solid matter consist of atoms and
molecules which often behave as electric and magnetic dipoles. Such matter can
thus become electrically polarized and magnetized when external electric and
magnetic fields are being imposed. Then the constants $\varepsilon_{0}$ and
$\mu_{0}$ have to be replaced by modified values $\varepsilon$ and $\mu$,
which take these polarization effects into account. One consequence of this is
that a conventional plane electromagnetic wave would propagate in such matter
at a velocity $(1/\varepsilon\mu)^{1/2}$ being smaller than the velocity
$c=(1/\varepsilon_{0}\mu_{0})^{1/2}$ in empty vacuum space. In the present
extended theory on the vacuum state, however, no electrically polarized and
magnetized atoms or molecules are present. There are only electromagnetic
fluctuations due to the zero-point field. In other words, the vacuum state is
here conceived as a background of empty space upon which are superimposed zero
point electromagnetic wave-like fluctuations, as well as phenomena of a more
regular character which could take the form of the wave phenomena and field
configurations to be treated in the present theory. In this respect Eq.(2.12)
becomes identical with that used in the conventional theory on hot plasmas
which contain freely moving charged particles of both polarities, and where
electric charge separation and a resulting electric field can arise, such as
in the case of plasma oscillations.

The question may also be raised why only $\nabla\cdot\mathbf{E}$ and not
$\nabla\cdot\mathbf{B}$ should be permitted to become nonzero in an extended
approach. It has then to be noticed that the nonzero electric field divergence
has its support in the observed pair formation and the vacuum fluctuations,
whereas a nonzero magnetic field divergence is so far a possible but not
proved theoretical supposition. With Dirac\cite{Dira78} we shall therefore
leave the magnetic monopole concept as an open question, but not include it at
the present stage in the theory.

As shown later, this theory will lead to a small but nonzero photon rest mass.
This raises a further question about the gauge invariance. According to
earlier investigations it was concluded that the gauge invariance does not
require the photon rest mass to be zero\cite{Lehn98,Feld63}. In fact, the
field equations (2.8)-(2.9) can easily be seen to become gauge invariant. A
new gauge with a magnetic vector potential $\mathbf{A}^{\prime}$ and an
electrostatic potential $\mathbf{\phi}^{\prime}$ is defined by
\begin{equation}
\mathbf{A}^{\prime}=\mathbf{A}+\bigtriangledown\psi\qquad\phi^{\prime}%
=\phi-\partial\psi/\partial t
\end{equation}
When being inserted into Eqs.(2.10) and (2.11), these gauge relations lead to
field strengths $\mathbf{E}^{\prime}=\mathbf{E}$ and $\mathbf{B}^{\prime
}=\mathbf{B}$ which thus become invariant. The field equations (2.8) and (2.9)
only contain the field strengths and therefore become gauge invariant.

\subsubsection{Comparison with the Dirac Theory}

In the theory of the electron by Dirac\cite{Dira28} the relativistic wave
function $\boldsymbol{\psi}$ has four components in spin space. The
three-space charge and current densities become\cite{Mors53}
\begin{equation}
\bar{\rho}=e\bar{\boldsymbol{\psi}}\boldsymbol{\psi}%
\end{equation}
and
\begin{equation}
\mathbf{j}=ce(\bar{\boldsymbol{\psi}}\boldsymbol{\alpha}_{i}\boldsymbol{\psi
})\qquad i=1,2,3
\end{equation}
where $\boldsymbol{\alpha}_{i}$ are the Dirac matrices of the three spatial
directions, and the expression for $\bar{\boldsymbol{\psi}}$ is obtained from
that of $\boldsymbol{\psi}$ by replacing the included unit vectors and
functions by their complex conjugate values. These equations are then seen to
be similar to the corresponding relation (2.5) for the current density in the
present theory where $|\mathbf{C}|=c$. The angular momentum in the Dirac
theory emerges from the spin matrices $\boldsymbol{\alpha}_{i}$, whereas the
relation $\mathbf{C}^{2}=c^{2}$ leads to the two spin directions of the same
momentum when the present theory is applied to an axially symmetric
configuration such as that in a model of the electron.

It should further be observed that the present theory does not have to include
the concepts of a net electric charge and a particle rest mass from the
beginning. Such concepts will first come out from spatial integration of the
electric charge density and the electromagnetic energy density. When relating
this approach to the Dirac theory, wave functions have thus to be considered
which only represent states without an initially included net charge and rest
mass. With $u$ as an arbitrary function and $U$ as a constant, this yields a
charge density\cite{Leig59}
\begin{equation}
\bar{\rho}=2e\bar{U}U\bar{u}u
\end{equation}
and the corresponding current density components
\begin{equation}
j_{z}=\pm c\bar{\rho};\ j_{x}=0\ \operatorname{and}\ j_{y}=0
\end{equation}
where a bar over $U$ and $u$ indicates the complex conjugate value. Other
analogous forms can be chosen where instead $j_{y}=\pm c\bar{\rho}$ or
$j_{x}=\pm c\bar{\rho}$.

This result indicates that there is a connection between the present theory
and that by Dirac. But the former theory sometimes applies to a larger class
of phenomena, in the capacity of both "bound" and of "free" states. Thereby
the elementary charge is a given quantity in Dirac's theory, whereas the total
net charge of the present approach is deduced from the field equations in a
steady axisymmetric state, as shown later.

\subsubsection{The Quantization Procedure}

The relevant quantum conditions to be imposed are those on the angular
momentum, the magnetic moment, and on the magnetic flux of the steady or
time-dependent configuration to be analyzed. The rigorous and most complete
way to proceed is then to quantize the field equations already from the outset.

As a first step, however, a simplification will instead be made here, by first
determining the general solutions of the basic field equations, and then
imposing the quantum conditions afterwards. This is at least justified by the
fact that the quantized electrodynamic equations become equivalent to the
field equations (2.1) in which the potentials $A_{\mu}$ and currents $J_{\mu}$
are replaced by their expectation values, as shown by Heitler\cite{Heit54}. In
this connection Gersten\cite{Gers87,Gers99} states that Maxwell's equations
should be used as a guideline for proper interpretations of quantum theories.
This should also apply to the inhomogeneous equations (2.1) being used here.
The present theory may therefore not be too far from the truth, by
representing the most probable states in a first approximation to a rigorous
quantum-theoretical approach. Thus the theory could be designated as a type of
extended quantum electrodynamics (\textquotedblleft EQED\textquotedblright).

\subsubsection{Momentum and Energy}

In analogy with conventional electromagnetic theory, the basic equations (2.8)
and (2.9) can be subject to vector and scalar multiplication to result in new
but still identical forms. This leads to the momentum equation
\begin{equation}
\nabla\cdot^{2}\mathbf{S=\bar{\rho}(E+C\times B)}+\varepsilon_{0}%
\ \frac{\partial}{\partial t}(\mathbf{E\times B)}%
\end{equation}
where $^{2}\mathbf{S}$ is the electromagnetic stress tensor\cite{Stra41}, and
where
\begin{equation}
\mathbf{f}=\bar{\rho}(\mathbf{E}+\mathbf{C}\times\mathbf{B})
\end{equation}
is the total local volume force. The corresponding integral form is
\begin{equation}
\int{^{2}\mathbf{S\cdot n}}\ dS=\mathbf{F}_{e}+\mathbf{F}_{m}+\frac{\partial
}{\partial t}\int\mathbf{g}\ dV
\end{equation}
Here $dS$ and $dV$ are surface and volume elements, $\mathbf{n}$ is the
surface normal,
\begin{align}
\mathbf{F}_{e}  &  =\int\bar{\rho}\mathbf{E}\ dV\\
\mathbf{F}_{m}  &  =\int\bar{\rho}\mathbf{C}\times\mathbf{B}\ dV\nonumber
\end{align}
are the integrated electric and magnetic volume forces, and
\begin{equation}
\mathbf{g}=\varepsilon_{0}\mathbf{E}\times\mathbf{B}=\frac{1}{c^{2}}\mathbf{S}%
\end{equation}
can be interpreted as an electromagnetic momentum density, with $\mathbf{S}$
denoting the Poynting vector. The latter can be conceived to represent the
magnitude and direction of the energy flow in space.

One further obtains the energy equation
\begin{equation}
-\nabla\cdot\mathbf{S}=-\left(  \frac{1}{\mu_{0}}\right)  \nabla
\cdot(\mathbf{E}\times\mathbf{B})=
\bar{\rho}\mathbf{E}\cdot\mathbf{C}+\frac{1}{2}\varepsilon_{0}\frac{\partial
}{\partial t}(\mathbf{E}^{2}+c^{2}\mathbf{B}^{2})
\end{equation}
with its integral form
\begin{equation}
\int\mathbf{S}\cdot\mathbf{n}\ dS+\int\bar{\rho}\mathbf{E}\cdot\mathbf{C}%
\ dV=\\
-\frac{1}{2}\varepsilon_{0}\int\frac{\partial}{\partial t}(\mathbf{E}%
^{2}+c^{2}\mathbf{B}^{2})\ dV
\end{equation}

Using the basic equations (2.8)--(2.12) and some vector identities,
expressions can be obtained for the energy densities, as given by
\begin{align}
w_{f}  &  =\frac{1}{2}\left(  \varepsilon_{0}\mathbf{E}^{2}+\mathbf{B}^{2}%
/\mu_{0}\right) \\
w_{s}  &  =\frac{1}{2}\left(  \bar{\rho}\phi+\mathbf{j}\cdot\mathbf{A}\right)
=\frac{1}{2}\bar{\rho}\left(  \phi+\mathbf{C}\cdot\mathbf{A}\right)
\end{align}
Here $w_{f}$ can be interpreted as a \textquotedblleft field energy
density\textquotedblright\ being associated with the field strengths
$\mathbf{E}$ and $\mathbf{B}$, and $w_{s}$ as a \textquotedblleft source
energy density\textquotedblright\ being associated with the sources $\bar
{\rho}$ and $\mathbf{j}$ of the electromagnetic field. Conditions for the
integral forms of $w_{f}$ and $w_{s}$ to become equal or to be different have
been investigated, as shown in detail elsewhere\cite{Lehn01}.

\subsection{New Aspects and Fundamental Applications}

The introduction of an additional degree of freedom in the form of a nonzero
electric field divergence, and the resulting transition from a homogeneous
d'Alembert equation to an inhomogeneous Proca-type equation, lead to a number
of new aspects and applications. These include the new class of steady states,
and an extended class of time-dependent states with new wave phenomena. Within
these classes there are special applications to models of the electron,
neutrino, and photon, as well as to string-shaped geometry, and to total
reflection of dissipative plane waves. This condensed review is limited to
models of the electron and photon.

The analysis which follows mainly concerns axially symmetric systems, but
asymmetric configurations can in principle also be treated by the theory.

\subsubsection{Steady Phenomena}

In a time-independent case the space-charge current density makes possible the
existence of steady electromagnetic states which are absent within the frame
of Maxwell's equations. From relations (2.7)-(2.12) the basic steady-state
equations
\begin{equation}
c^{2}\nabla\times\nabla\times\mathbf{A}=-\mathbf{C}(\nabla^{2}\phi)=(\bar
{\rho}/\epsilon_{0})\mathbf{C}%
\end{equation}
are then obtained. On the basis of the electromagnetic forces given by
Eqs.(2.18) and (2.21), the corresponding steady electromagnetic equilibrium of
the electron will later be discussed.

Axisymmetric states are of special interest and will be treated in detail in
the following subsections. These states can roughly be pictured as a result of
``self-confined'' circulating electromagnetic radiation, on which relevant
quantum conditions are being imposed. Among these ``bound'' states two
subclasses are of special interest:

\begin{itemize}
\item Particle-shaped states where the geometrical configuration is
boun\-d\-ed both in the axial and in the radial directions. There are states
both with a nonzero as well as with a zero net electric charge to be
investigated. The corresponding solutions and models may have some bearing on
and contribute to the understanding of such truly elementary particles as the leptons.

\item String shaped states where the geometrical configuration is uniform in
the axial direction and becomes piled up near the axis of symmetry. These
equilibria can in an analogous manner reproduce several desirable features of
the earlier proposed hadron string model.
\end{itemize}

\subsubsection{Time-dependent Phenomena}

The basic equations for time-dependent states combine to
\begin{equation}
(\frac{\partial^{2}}{\partial t^{2}}-c^{2}\nabla^{2})\mathbf{E}+(c^{2}%
\nabla+\mathbf{C}\frac{\partial}{\partial t})(\nabla\cdot\mathbf{E})=0
\end{equation}
for the electric field. From the corresponding solution the magnetic field
$\mathbf{B}$ can be obtained by means of Eq.(2.9). A divergence operation on
Eq.(2.8) further yields
\begin{equation}
(\frac{\partial}{\partial t}+\mathbf{C}\cdot\nabla)(\nabla\cdot\mathbf{E})=0
\end{equation}
in combination with Eq.(2.7). In some cases this equation becomes useful to
the analysis, but it does not introduce more information than that already
contained in Eq.(2.28).

Three limiting cases can be identified on the basis of Eq.(2.28):

\begin{itemize}
\item When $\nabla\cdot\mathbf{E}=0$ and $\nabla\times\mathbf{E}\neq0$ the
result is a conventional \emph{transverse} electromagnetic wave, henceforth
denoted as an \textquotedblleft EM wave\textquotedblright.

\item When $\nabla\cdot\mathbf{E}\neq0$ and $\nabla\times\mathbf{E}=0$ a
purely \emph{longitudinal} electric space-charge wave arises, here being
denoted as an \textquotedblleft S wave\textquotedblright.

\item When both $\nabla\cdot\mathbf{E}\neq0$ and $\nabla\times\mathbf{E}\neq0$
a hybrid \emph{nontransverse} electromagnetic space-charge wave appears, here
denoted as an \textquotedblleft EMS wave\textquotedblright.
\end{itemize}

In a general case these various modes can become superimposed, also the EMS
modes with different velocity vectors $\mathbf{C}$. That the basic equations
can give rise both to modes with a nonvanishing and a vanishing electric field
divergence is not less conceivable than the analogous property of the
conventional basic equations governing a magnetized plasma. The frame of the
latter accommodates both longitudinal electrostatic waves and transverse
Alfv\'{e}n waves. These are, like the EM, S, and EMS waves, separate modes
even if they originate from the same basic formalism.

The applications which follow are related to the EMS waves. The conditions
under which the S wave may exist are not clear at this stage.

\subsection{General Features of Steady Axisymmetric States}

The analysis is now restricted to particle-shaped axisymmetric states. For
these a frame $(r,\theta,\varphi)$ of spherical coordinates is introduced
where all relevant quantities are independent of the angle $\varphi$. The
analysis is further restricted to a current density $\mathbf{j}=(0,0,C\bar
{\rho})$ and a magnetic vector potential $\mathbf{A}=(0,0,A)$. Here $C=\pm c$
represents the two possible spin directions. The basic equations (27) then
take the form
\begin{equation}
\frac{(r_{0}\rho)^{2}\bar{\rho}}{\varepsilon_{0}}=D\phi=[D+(\sin\theta
)^{-2}](CA)
\end{equation}
where the dimensionless radial coordinate
\begin{equation}
\rho=r/r_{0}%
\end{equation}
has been introduced, $r_{0}$ is a characteristic radial dimension, and the
operator $D$ is given by
\begin{align}
D  &  =D_{\rho}+D_{\theta}\\
D_{\rho}  &  =-\frac{\partial}{\partial\rho}(\rho^{2}\frac{\partial}%
{\partial\rho})\nonumber\\
D_{\theta}  &  =-\frac{\partial^{2}}{\partial\theta^{2}}-\frac{\cos\theta
}{\sin\theta}\frac{\partial}{\partial\theta}\nonumber
\end{align}

\subsubsection{The Generating Function}

The general solution of Eqs.(2.30) in particle-shaped geometry can now be
obtained in terms of a \emph{generating function}
\begin{equation}
F(r,\theta)=CA-\phi=G_{0}.G(\rho,\theta)
\end{equation}
where $G_{0}$ stands for a characteristic amplitude and $G$ for a normalized
dimensionless part. This yields the solution
\begin{align}
CA  &  =-(\sin\theta)^{2}DF\\
\phi &  =-[1+(\sin\theta)^{2}D]F\\
\bar{\rho}  &  =-(\frac{\varepsilon_{0}}{r_{0}^{2}\rho^{2}})D[1+(\sin
\theta)^{2}D]F
\end{align}
which is easily seen by direct insertion to satisfy Eqs.(2.30). Starting from
an arbitrary function $F$, the corresponding spatial distributions of the
potentials $CA$ and $\phi$ and of the space-charge density $\bar{\rho}$ are
thus generated, i.e. those which simultaneously satisfy the set (2.30) of
equations. If one would instead start from a given distribution of only one of
the field quantities $CA$, $\phi$, or $\bar{\rho}$, this would not provide a
simple solution such as that obtained from the set (2.33)--(2.36).

The quantities (2.34)--(2.36) are uniquely determined by the generating
function. The latter can be chosen such that all field quantities decrease
rapidly to zero at large radial distances from the origin. In this way the
charge and current density, as well as all associated features of an
electromagnetic field configuration, can be confined to a limited region of space.

In the analysis of particle-shaped states the functions
\begin{align}
f(\rho,\theta)  &  =-(\sin\theta)D[1+(\sin\theta)^{2}D]G\\
g(\rho,\theta)  &  =-[1+2(\sin\theta)^{2}D]G
\end{align}
will now be introduced. Using expressions (2.34)--(2.36), (2.26), (2.37) and
(2.38), integrated field quantities can be obtained which represent a net
electric charge $q_{0}$, magnetic moment $M_{0}$, mass $m_{0}$, and angular
momentum $s_{0}$. The magnetic moment is obtained from the local contributions
provided by the current density (2.5). The mass and the angular momentum are
deduced from the local contributions of $w_{s}/c^{2}$ being given by the
source energy density (2.26) and the energy relation by Einstein. The current
density (2.5) then behaves as a convection current being common to all
contributions from the charge density. The corresponding mass flow originates
from the velocity vector $\mathbf{C}$ which has the same direction for
positive as for negative charge elements. The integrated field quantities
finally become

\bigskip%
\begin{align}
q_{0}   =2\pi\varepsilon_{0}r_{0}G_{0}J_{q}\qquad
& I_{q}   =f\\
M_{0}   =\pi\varepsilon_{0}Cr_{0}^{2}G_{0}J_{M}\qquad
& I_{M}   =\rho(\sin\theta)f\\
m_{0}   =\pi(\varepsilon_{0}/c^{2})r_{0}G_{0}^{2}J_{m}\qquad
& I_{m}   =fg\\
s_{0}   =\pi(\varepsilon_{0}C/c^{2})r_{0}^{2}G_{0}^{2}J_{s}\qquad
&I_{s}   =\rho(\sin\theta)fg
\end{align}
These relations include normalized integrals defined by
\begin{equation}
J_{k}   =\int_{\rho_{k}}^{\infty}\!\!\int_{0}^{\pi}I_{k}\ d\rho d\theta\qquad
k   =q,M,m,s
\end{equation}
where $\rho_{k}$ are small radii of circles centered around the origin
$\rho=0$ when $G$ is divergent there, and $\rho_{k}=0$ when $G$ is convergent
at $\rho=0$. The case $\rho_{k}\neq0$ will later be treated in detail. So far
the integrals (2.39)--(2.43) are not uniquely determined but depend on the
distribution of the function $G$ in space. They will first become fully
determined when further conditions are being imposed, such as those of a
quantization. The general forms (2.39)--(2.43) thus provide the integrated
quantities with a certain degree of flexibility, being related to the extra
degree of freedom which has been introduced through the nonzero electric field divergence.

At this point a further step is taken by imposing the restriction of a
separable generating function
\begin{equation}
G(\rho,\theta)=R(\rho).T(\theta)
\end{equation}
The integrands of the normalized form (2.43) then become
\begin{equation}
I_{q}=\tau_{0}R+\tau_{1}(D_{\rho}R)+\tau_{2}D_{\rho}(D_{\rho}R)
\end{equation}%
\begin{equation}
I_{M}=\rho(\sin\theta)I_{q}%
\end{equation}%
\begin{multline}
I_{m}=\tau_{0}\tau_{3}R^{2}+(\tau_{0}\tau_{4}+\tau_{1}\tau_{3})R(D_{\rho}R)+
\tau_{1}\tau_{4}(D_{\rho}R)^{2}+\\
+\tau_{2}\tau_{3}RD_{\rho}(D_{\rho}R)+
\tau_{2}\tau_{4}(D_{\rho}R)[D_{\rho}(D_{\rho}R)]
\end{multline}%
\begin{equation}
I_{s}=\rho(\sin\theta)I_{m}%
\end{equation}
where
\begin{equation}
\tau_{0}=-(\sin\theta)(D_{\theta}T)-(\sin\theta)D_{\theta}[(\sin^{2}\theta)(D_{\theta}T)]
\end{equation}%
\begin{equation}
\tau_{1}=-(\sin\theta)T-(\sin\theta)D_{\theta}[(\sin^{2}\theta)T]-\sin^{3}\theta(D_{\theta}T)
\end{equation}%
\begin{align}
\tau_{2}  &  =-(\sin^{3}\theta)T\\
\tau_{3}  &  =-T-2(\sin^{2}\theta)(D_{\theta}T)\\
\tau_{4}  &  =-2(\sin^{2}\theta)T
\end{align}

Among the possible forms to be adopted for the radial function $R(\rho)$ and
the polar function $T(\theta)$, we will here consider the following cases:

\begin{itemize}
\item The radial part $R$ can become convergent or divergent at the origin
$\rho=0$ but it always goes strongly towards zero at large distances from it.

\item The polar part $T$ is always finite and has finite derivatives. It can
become symmetric or antisymmetric with respect to the ``equatorial plane''
(midplane) defined by $\theta=\pi/2$.
\end{itemize}

We first consider the radial part $R$ and the corresponding normalized
integrals $J_{q}$ and $J_{M}$ of the integrated charge and magnetic moment.
After a number of partial integrations this yields the following conclusions:

\begin{itemize}
\item Convergent radial functions $R$ lead to a class of electrically neutral
particle states where $q_{0}$ and $M_{0}$ both vanish, whereas $m_{0}$ and
$s_{0}$ are nonzero.

\item For the present particle-shaped geometry to result in electrically
charged states, the \emph{divergence} of the radial function becomes a
necessary but not sufficient condition. This leads to the subsequent question
whether the corresponding integrals (2.43) would then be able to form the
basis of a steady state having \emph{finite} and nonzero values of all the
integrated field quantities (2.39)--(2.42). This will later be shown to become possible.
\end{itemize}

We then turn to the polar part $T$ of the generating function. The top-bottom
symmetry properties of the polar function $T$ are now considered with respect
to the equatorial plane. The integrals (2.43) which are evaluated in the
interval $0\leq\theta\leq\pi$ become nonzero for symmetric integrands $I_{k}$,
but vanish for antisymmetric ones. These symmetry properties are evident from
expressions (2.45)--(2.53). Thus, the product of two symmetric or of two
antisymmetric functions becomes symmetric, whereas a product of a symmetric
function with an antisymmetric one becomes antisymmetric. The symmetry or
antisymmetry of $T$ leads to a corresponding symmetry or antisymmetry of
$D_{\theta}T$, $D_{\theta}\big[(\sin\theta)^{2}T\big]$, and $D_{\theta
}\big[(\sin\theta)^{2}(D_{\theta}T)\big]$. Therefore all functions
(2.49)--(2.53) are either symmetric or antisymmetric in the same way as $T$.
The symmetry rules with respect to $T$ thus lead to two general conclusions:

\begin{itemize}
\item The charge and the magnetic moment vanish for antisymmetric forms of
$T$, irrespective of the form of $R$.

\item The mass and the angular momentum generally differ from zero, both when
$T$ is symmetric and when it is antisymmetric.
\end{itemize}

To sum up, it is thus found that a model of the electron can be based on $R$
being divergent at the origin and $T$ having top-bottom symmetry with respect
to the equatorial plane, whereas a model of the neutrino can be based on any
of the remaining alternatives being considered here.

Finally a few words should be said about matter and antimatter particle
models. With $C=\pm c$ and $G_{0}=\pm|G_{0}|$ it is seen from
Eqs.(2.39)--(2.53) that there are pairs of solutions having the same geometry
of their local distributions, but having opposite signs of the corresponding
integrated field quantities in places where $G_{0}$ and $C$ appear linearly.
This could become the basis for models of charged and neutral particles of
matter as well as of antimatter. A minor difference between matter and
antimatter will then not be included in the models.

\subsubsection{Quantum Conditions}

The angular momentum condition to be imposed on the models of the electron in
the capacity of a fermion particle, as well as of the neutrino, is combined
with Eq.(2.42) to result in
\begin{equation}
s_{0}=\pi(\varepsilon_{0}C/c^{2})r_{0}^{2}G_{0}^{2}J_{s}=\pm h/4\pi
\end{equation}
This condition becomes compatible with the two signs of $C=\pm c$, as obtained
from the Lorentz invariance.

In particular, for a charged particle such as the electron, muon, tauon or
their antiparticles, Eqs.(2.39) and (2.42) combine to
\begin{align}
q^{\ast}  &  \equiv|q_{0}/e|=(f_{0}J_{q}^{2}/2J_{s})^{1/2}\\
f_{0}  &  =2\varepsilon_{0}ch/e^{2}\nonumber
\end{align}
where $q^{\ast}$ is a dimensionless charge being normalized with respect to
the experimentally determined electronic charge \textquotedblleft%
$e$\textquotedblright, and $f_{0}\approx137.036$ is the inverted value of the
fine-structure constant.

According to Dirac\cite{Dira28}, Schwinger\cite{Schw49} and
Feynman\cite{Feyn90} the quantum condition on the magnetic moment $M_{0}$ of a
charged particle such as the electron becomes
\begin{equation}
M_{0}m_{0}/q_{0}s_{0}=1+\delta_{M}\qquad\delta_{M}=1/2\pi f_{0}%
\end{equation}
which shows excellent agreement with experiments. Here the unity term of the
right-hand member is due to Dirac who obtained the correct Land\'{e} factor by
considering the electron to be situated in an imposed external magnetic field.
This leads to a magnetic moment being twice as large as that expected from an
elementary proportionality relation between the electron spin and the magnetic
moment\cite{Ryde96}. Further, in Eq.(2.56) the term $\delta_{M}$ is a small
quantum mechanical correction due to Schwinger and Feynman.

Conditions (2.54) and (2.56) can also be made plausible by elementary physical
arguments based on the present picture of a particle-shaped state of
\textquotedblleft self-confined\textquotedblright\ radiation. In the latter
picture there is a circulation of radiation at the velocity of light around
the axis of symmetry. If this circulation takes place at the average
characteristic radial distance $r_{0}$, the corresponding frequency of
revolution would become $\nu\cong c/2\pi r_{0}$. In combination with the
energy relations $W=m_{0}c^{2}$ and $W=h\nu$ by Einstein and Planck, one then
obtains
\begin{equation}
r_{0}m_{0}\cong\frac{h}{2\pi c}%
\end{equation}
where the radius $r_{0}$ has the same form as the Compton radius. We shall
later return to this concept, at the end of Section D.1. The result (2.57)
should, however, only be taken as a crude physical indication where the radius
in Eq.(2.57) comes out to be far too large as compared to observations. With
the total current $q_{0}\nu$, the magnetic moment further becomes%

\begin{equation}
M_{0}=\pi r_{0}^{2}q_{0}\nu=\frac{1}{2}q_{0}cr_{0}=\frac{q_{0}h}{4\pi m_{0}}%
\end{equation}
which agrees wih conditions (2.54) and (2.56) when $\delta_{M}$ is neglected.
Thereby the angular momentum becomes $|s_{0}|=r_{0}cm_{0}/2=h/4\pi$ when its
inwardly peaked spatial distribution corresponds to an equivalent radius
$r_{0}/2$.

The unequal integrands of the magnetic moment and the angular momentum in
Eqs.(2.40) and (2.42) could also be reconcilable with a proportionality
relation between $s_{0}$ and $M_{0}$ which leads to the correct Land\'{e}
factor. Another possibility of obtaining its correct value is to modify
Eq.(2.57) to represent the first \textquotedblleft
subharmonic\textquotedblright\ given by a path length $4\pi r_{0}$ around the
axis. In an axisymmetric case this would, however, become a rather far-fetched
proposal. In any case, when using the correct value of this factor, the
electronic charge deduced later in Section D turns out to be close to its
experimental value. This would not be the case for a Land\'{e} factor being
half as large.

Finally, in a charged particle-shaped state with a nonzero magnetic moment,
the electric current distribution will generate a total magnetic flux
$\Gamma_{tot}$ and a corresponding total magnetic energy. In such a state the
quantized value of the angular momentum further depends on the type of
configuration being considered. It thus becomes $|s_{0}|=h/4\pi$ for a
fermion, but $|s_{0}|=h/2\pi$ for a boson. We now consider the electron to be
a system also having a quantized charge $q_{0}$. As in a number of other
physical systems, the flux should then become quantized as well, and be given
by the two quantized concepts $s_{0}$ and $q_{0}$, in a relation having the
dimension of magnetic flux. This leads to the quantum
condition\cite{Lehn01,Lehn02b}
\begin{equation}
\Gamma_{tot}=|s_{0}/q_{0}|
\end{equation}

\subsection{A Model of the Electron}

A model of the electron will now be elaborated which in principle also applies
to the muon, tauon and to corresponding antiparticles\cite{Cohe98}. According
to the previous analysis in Section C.1, a charged particle-shaped state only
becomes possible by means of a generating function having a radial part $R$
which is divergent at the origin, and a polar part $T$ with top-bottom
symmetry. In its turn, the divergence of $R$ leads to the question how to
obtain finite and nonzero integrated field quantities (2.39)--(2.43). The
following analysis will show this to become possible, by shrinking the
characteristic radius $r_{0}$ to the very small values of a \textquotedblleft
point-charge-like\textquotedblright\ state. It does on the other hand not
imply that $r_{0}$ has to become strictly equal to zero, which would end up
into the unphysical situation of a structureless point.

\subsubsection{Generating Function and Integrated Field Quantities}

The generating function to be considered has the parts%

\begin{equation}
R=\rho^{-\gamma}e^{-\rho}\qquad\gamma>0
\end{equation}

\begin{multline}
T=1+\sum_{v=1}^{n}\{a_{2v-1}\sin[(2v-1)\theta]+a_{2v}\cos(2v\theta)\}=\\
=1+a_{1}\sin\theta+a_{2}\cos2\theta+a_{3}\sin3\theta+\ldots
\end{multline}
Concerning the radial part (2.60), it may at a first glance appear to be
somewhat special and artificial. Under general conditions one could thus have
introduced a negative power series of $\rho$ instead of the single term
$\rho^{-\gamma}$. However, for a limited number of terms in such a series,
that with the largest negative power will in any case dominate near the
origin. Moreover, due to the following analysis the same series has to contain
one term only, with a locked special value\cite{Lehn02b} of the radial
parameter $\gamma$. The exponential factor in expression (2.60) has been
included to secure the convergence of any moment with $R$ at large distances
from the origin. This factor will not appear in the end results of the analysis.

The polar part (2.61) represents a general form of axisymmetric geometry
having top-bottom symmetry with respect to the equatorial plane. Here it
should be noticed that all sine and cosine terms can be rewritten into a
corresponding power series of $\sin\theta$, as well as the entire resulting series.

The radial form given by Eq.(2.60) can now be inserted into the integrands
(2.45)--(2.48). Since the radial integrals (2.43) are extended to very low
limits $\rho_{k}$, they become dominated by contributions from terms of the
strongest negative power. Keeping these contributions only, the integrands
reduce to
\begin{equation}
I_{k}=I_{k\rho}I_{k\theta}\qquad k=q,M,m,s
\end{equation}
where
\begin{align}
I_{q\rho}  &  =R\qquad I_{M\rho}=\rho R\\
I_{m\rho}  &  =R^{2}\qquad I_{s\rho}=\rho R^{2}\nonumber
\end{align}
and
\begin{equation}
I_{q\theta}=\tau_{0}-\gamma(\gamma-1)\tau_{1}+\gamma^{2}(\gamma-1)^{2}\tau_{2}%
\end{equation}%
\begin{equation}
I_{M\theta}=(\sin\theta)I_{q\theta}%
\end{equation}%
\begin{multline}
I_{m\theta}=\tau_{0}\tau_{3}-\gamma(\gamma-1)(\tau_{0}\tau_{4}+\tau_{1}%
\tau_{3})+\\
+\gamma^{2}(\gamma-1)^{2}(\tau_{1}\tau_{4}+\tau_{2}\tau_{3})-
\gamma^{3}(\gamma-1)^{3}\tau_{2}\tau_{4}%
\end{multline}%
\begin{equation}
I_{s\theta}=(\sin\theta)I_{m\theta}%
\end{equation}
$I_{M\theta}=(\sin\theta)I_{q\theta}$The corresponding integrals (2.43) then
become
\begin{equation}
J_{k}=J_{k\rho}J_{k\theta}%
\end{equation}
where
\begin{align}
J_{q\rho}  &  =\frac{1}{\gamma-1}\rho_{q}^{-(\gamma-1)}\\
J_{M\rho}  &  =\frac{1}{\gamma-2}\rho_{M}^{-(\gamma-2)}\nonumber
\end{align}%
\begin{align}
J_{m\rho}  &  =\frac{1}{2\gamma-1}\rho_{m}^{-(2\gamma-1)}\\
J_{s\rho}  &  =\frac{1}{2(\gamma-1)}\rho_{s}^{-2(\gamma-1)}\nonumber
\end{align}
and
\begin{equation}
J_{k\theta}=\int_{0}^{\pi}I_{k\theta}\ d\theta
\end{equation}
In expression (2.69) for the magnetic moment a point-charge-like behavior
excludes the range $\gamma<2$.

The divergences of the so far undetermined radial expressions (2.69)--(2.70)
appear in the integrals of Eqs.(2.39)--(2.43) when the lower limits $\rho_{k}$
approach zero. To outbalance these, one has to introduce a shrinking
characteristic radius defined by
\begin{equation}
r_{0}=c_{0}\varepsilon\qquad c_{0}>0\qquad0<\varepsilon\ll1
\end{equation}
Here $c_{0}$ is a positive constant having the dimension of length, and
$\varepsilon$ is a dimensionless characteristic radius which has the role of a
decreasing smallness parameter. Replacing $\varepsilon$ by any power of
$\varepsilon$ would not lead to more generality. Combination of Eqs.(2.69),
(2.70), and (2.72) with Eqs.(2.39)--(2.43) gives the result
\begin{equation}
q_{0}=2\pi\varepsilon_{0}c_{0}G_{0}[J_{q\theta}/(\gamma-1)](\varepsilon
/\rho_{q}^{\gamma-1})
\end{equation}%
\begin{equation}
M_{0}m_{0}=\pi^{2}(\varepsilon_{0}^{2}C/c^{2})c_{0}^{3}G_{0}^{3}\cdot
\lbrack J_{M\theta}J_{m\theta}/(\gamma-2)(2\gamma-1)]\cdot
(\varepsilon^{3}/\rho_{M}^{\gamma-2}\rho_{m}^{2\gamma-1})
\end{equation}%
\begin{equation}
s_{0}=\pi(\varepsilon_{0}C/c^{2})c_{0}^{2}G_{0}^{2}\cdot
\lbrack J_{s\theta}/2(\gamma-1)](\varepsilon/\rho_{s}^{\gamma-1})^{2}%
\end{equation}
The reason for introducing the compound quantity $M_{0}m_{0}$ in Eq.(2.74) is
that this quantity appears as a single entity of all finally obtained
relations in the present theory.

For all four quantities $q_{0}$, $M_{0}$, $m_{0}$, $s_{0}$ to become finite
when $\varepsilon$ approaches zero, we must satisfy the
conditions\cite{Lehn86}
\begin{align}
\rho_{q}  &  =\varepsilon^{1/(\gamma-1)}\qquad\rho_{M}=\varepsilon
^{2/(\gamma-2)}\\
\rho_{m}  &  =\varepsilon^{1/(2\gamma-1)}\qquad\rho_{s}=\varepsilon
^{1/(\gamma-1)}\nonumber
\end{align}
whereas the condition for $M_{0}m_{0}$ to be finite is thus
\begin{equation}
\rho_{M}^{\gamma-2}\rho_{m}^{2\gamma-1}=\varepsilon^{3}%
\end{equation}

Now we only require the configuration represented by $q_{0}$, $s_{0}$ and the
compound quantity $M_{0}m_{0}$ of Eqs.(2.73)--(2.75) to scale in such a way
that the geometry is preserved by becoming independent of $\rho_{k}$ and
$\varepsilon$, within the entire range of small $\varepsilon$. Such a uniform
scaling implies that
\begin{equation}
\rho_{q}=\rho_{M}=\rho_{m}=\rho_{s}=\varepsilon
\end{equation}
and that the radial parameter $\gamma$ has to approach the value $2$ from
above, i.e.
\begin{align}
\gamma(\gamma-1)  &  =2+\delta\qquad0\leq\delta\ll1\\
\gamma &  \approx2+\delta/3\nonumber
\end{align}
These relations thus hold for condition (2.77) where $M_{0}m_{0}$ is treated
as one entity, but not for the two intermediate conditions (2.76) where
$M_{0}$ and $m_{0}$ are treated separately. For both these options $M_{0}%
m_{0}$ becomes finite in any case.%

$>$%
>From the earlier obtained results of partial integration it is further seen
that the contribution from $\tau_{0}$ in Eq.(2.64) vanishes as well as
$J_{M\theta}$ when $\gamma=2$. In the limit (2.79) the integrands
(2.64)--(2.67) can then be replaced by
\begin{align}
I_{q\theta}  &  =-2\tau_{1}+4\tau_{2}\\
I_{M\theta}/\delta &  =(\sin\theta)(-\tau_{1}+4\tau_{2})
\end{align}%
\begin{equation}
I_{m\theta}=\tau_{0}\tau_{3}-2(\tau_{0}\tau_{4}+\tau_{1}\tau_{3})+
4(\tau_{1}\tau_{4}+\tau_{2}\tau_{3})-8\tau_{2}\tau_{4}%
\end{equation}%
\begin{equation}
I_{s\theta}=(\sin\theta)I_{m\theta}%
\end{equation}
With the integrals (2.71) and the definitions
\begin{align}
A_{q}  &  \equiv J_{q\theta}\qquad A_{M}\equiv J_{M\theta}/\delta\\
A_{m}  &  \equiv J_{m\theta}\qquad A_{s}\equiv J_{s\theta}\nonumber
\end{align}
the integrated quantities (2.73)--(2.75) take the final form
\begin{align}
q_{0}=  &  2\pi\varepsilon_{0}c_{0}G_{0}A_{q}\\
M_{0}m_{0}=  &  \pi^{2}(\varepsilon_{0}^{2}C/c^{2})c_{0}^{3}G_{0}^{3}%
A_{M}A_{m}\\
s_{0}=  &  (1/2)\pi(\varepsilon_{0}C/c^{2})c_{0}^{2}G_{0}^{2}A_{s}%
\end{align}
for small $\delta$.

Through combination of expressions (2.41), (2.43), (2.58), (2.62), (2.70),
(2.72), (2.84), and (2.87), and due to the fact that the ratio $A_{m}/A_{s}$
is close to unity, it is readily seen that the radial constant $c_{0}$ becomes
nearly equal to the Compton wavelength $h/m_{0}c$ divided by $6\pi$.

\subsubsection{The Magnetic Flux}

According to Eq.(2.34) the magnetic flux function becomes
\begin{equation}
\Gamma=2\pi r(\sin\theta)A=
-2\pi r_{0}(G_{0}/c)\rho(\sin\theta)^{3}DG
\end{equation}
which vanishes at $\theta=(0,\pi)$. Making use of Eqs.(2.32), (2.60), and
(2.72) the flux becomes%
\begin{equation}
\Gamma=2\pi(c_{0}G_{0}/C)\sin^{3}\theta\{[\gamma(\gamma-1)+
2(\gamma-1)\rho+\rho^{2}]T-D_{\theta}T\}(\varepsilon/\rho^{\gamma-1})e^{-\rho}%
\end{equation}
This relation shows that the flux increases strongly as $\rho$ decreases
towards small values, in accordance with a point-charge-like behavior. To
obtain a nonzero and finite magnetic flux when $\gamma$ approaches the value
$2$ from above, one has to choose a corresponding dimensionless lower radius
limit $\rho_{\Gamma}=\varepsilon$, in analogy with expressions (2.72) and
(2.78). There is then a magnetic flux which intersects the equatorial plane.
It is counted from the point $\rho=\rho_{\Gamma}=\varepsilon$ and outwards, as
given by
\begin{equation}
\Gamma_{0}=-\Gamma(\rho=\varepsilon,\theta=\pi/2)=
2\pi(c_{0}G_{0}\varepsilon^{-\delta/3}/C)A_{\Gamma}%
\end{equation}
where
\begin{equation}
A_{\Gamma}=[D_{\theta}T-2T]_{\theta=\pi/2}%
\end{equation}
for $\gamma=2$. The flux (2.90) can be regarded as being generated by a
configuration of thin current loops. These have almost all their currents
located to a spherical surface with the radius $\rho=\varepsilon$, and the
corresponding magnetic field lines cut the equatorial plane at right angles.
The described field pattern is based on a separable generating function and
applies to this near-field region.

It has to be observed that the flux (2.90) is not necessarily the total flux
which is generated by the current system as a whole. There are cases in which
magnetic islands are formed above and below the equatorial plane, and where
these islands possess an isolated circulating extra flux which does not
intersect the same plane. The total flux $\Gamma_{tot}$ then consists of the
\textquotedblleft main flux\textquotedblright\ $-\Gamma_{0}$ of Eq.(2.90),
plus the extra \textquotedblleft island flux\textquotedblright\ $\Gamma_{i}$,
which can be deduced from the function (2.89), as will be shown later. This
type of field geometry turns out to prevail in the analysis which follows.

In the analysis of the contribution from the magnetic islands we introduce the
normalized flux function defined by Eq.(2.89) in the upper half-plane of the
sphere $\rho=\varepsilon\ll1$. It becomes
\begin{equation}
\Psi\equiv\Gamma(\rho=\varepsilon,\theta)/2\pi(c_{0}G_{0}/C)=
\sin^{3}\theta(D_{\theta}T-2T)
\end{equation}
In the range of increasing $\theta$, the radial magnetic field component
$B_{\rho}$ then vanishes along the lines $\theta=\theta_{1}$ and
$\theta=\theta_{2}$, whereas the component $B_{\theta}$ vanishes along the
lines $\theta=\theta_{3}$ and $\theta=\theta_{4}$. Here $\theta_{3}$ is
situated between $\theta_{1}$ and $\theta_{2}$, and $\theta_{4}$ between
$\theta_{2}$ and $\pi/2$. Consequently, the increase of $\theta$ from the axis
at $\theta=0$ first leads to an increasing flux $\Psi$ up to a maximum at the
angle $\theta=\theta_{1}$. Then there follows an interval $\theta_{1}%
<\theta<\theta_{2}$ of decreasing flux, down to a minimum at $\theta
=\theta_{2}$. In the range $\theta_{2}\leq\theta\leq\pi/2$ there is again an
increasing flux, up to the value
\begin{equation}
\Psi_{0}=\Psi(\pi/2)=A_{\Gamma}%
\end{equation}
which is equal to the main flux.

We further introduce the parts of the integrated magnetic flux defined by
\begin{align}
\Psi_{1}  &  =\Psi(\theta_{1})-\Psi(0)=\Psi(\theta_{1})\\
\Psi_{2}  &  =\Psi(\pi/2)-\Psi(\theta_{2})\nonumber
\end{align}
and associated with the ranges $0<\theta<\theta_{1}$ and $\theta_{2}%
<\theta<\pi/2$. The sum of these parts includes the main flux $\Psi_{0}$, plus
an outward directed flux from one magnetic island. The contribution from the
latter becomes $\Psi_{1}+\Psi_{2}-\Psi_{0}$. The total flux which includes the
main flux and that from two islands is then given by
\begin{equation}
\Psi_{tot}=2(\Psi_{1}+\Psi_{2}-\Psi_{0})+\Psi_{0}%
\end{equation}
which also can be written as
\begin{align}
\Psi_{tot}  &  =f_{\Gamma f}\Psi_{0}\\
f_{\Gamma f}  &  =[2(\Psi_{1}+\Psi_{2})-\Psi_{0}]/\Psi_{0}\nonumber
\end{align}
where $f_{\Gamma f}>1$ is a corresponding flux factor.

\subsubsection{Quantum Conditions of the Electron}

The analysis is now at a stage where the relevant quantum conditions can be
imposed. For the angular momentum (2.54) and its associated charge relation
(2.55) the result becomes
\begin{equation}
q^{\ast}=(f_{0}A_{q}^{2}/A_{s})^{1/2}%
\end{equation}
according to Eqs.(2.68)--(2.71), (2.78) and (2.84) in the limit $\gamma=2$.

For the magnetic moment condition (2.56) reduces to
\begin{equation}
A_{M}A_{m}/A_{q}A_{s}=1+\delta_{M}%
\end{equation}
when applying Eqs.(2.85)--(2.87).

Magnetic flux quantization is expressed by condition (2.59). Combination of
Eqs.(2.93), (2.73), (2.75) and (2.84) then yields
\begin{equation}
8\pi f_{\Gamma q}A_{\Gamma}A_{q}=A_{s}%
\end{equation}
where $f_{\Gamma q}$ is the flux factor being \emph{required} by the
corresponding quantum condition. This factor should not be confused with the
factor $f_{\Gamma f}$ of Eq.(2.96) which \emph{results} from the magnetic
field geometry. Only when one arrives at a self-consistent solution will these
two factors become equal to the common flux factor
\begin{equation}
f_{\Gamma}=f_{\Gamma f}=f_{\Gamma q}%
\end{equation}

\subsubsection{Comparison with the Conventional Renormalization Procedure}

The previous analysis has shown that well-defined convergent and nonzero
integrated physical quantities can be obtained in a point-charge-like steady
state, thereby avoiding the problem of an infinite self-energy. Attention may
here be called to Ryder\cite{Ryde96} who has stressed that, despite the
success of the conventional renormalization procedure, a more physically
satisfactory way is needed concerning the infinite self-energy problem.
Possibly the present theory could provide such an alternative, by tackling
this problem in a more surveyable manner. The finite result due to a
difference between two \textquotedblleft infinities\textquotedblright\ in
renormalization theory, i.e. by adding extra counter-terms to the Lagrangian,
is then replaced by a finite result obtained from the product of an
\textquotedblleft infinity\textquotedblright\ with a \textquotedblleft
zero\textquotedblright, as determined by the combination of the present
divergent integrands with a shrinking characteristic radius.

As expressed by Eq.(2.72), the latter concept also has an impact on the
question of Lorentz invariance of the electron radius. In the limit
$\varepsilon=0$ of a vanishing radius $r_{0}$ corresponding to a structureless
mass point, the deductions of this chapter will thus in a formal way satisfy
the requirement of such an invariance. At the same time the analytic
continuation of the obtained solutions can also be applied to the physically
relevant case of a very small but nonzero radius of a configuration having an
internal structure.

\subsubsection{Variational Analysis of the Integrated Charge}

The elementary electronic charge has so far been considered as an independent
and fundamental physical constant of nature, being determined by measurements
only\cite{Cohe98}. However, since it appears to represent the smallest quantum
of free electric charge, the question can be raised whether there is a more
profound reason for such a minimum charge to exist, possibly in terms of a
quantized variational analysis.

The present theory can provide the basis for such an analysis. In a first
attempt one would use a conventional procedure including Lagrange multipliers
in searching for an extremum of the normalized charge $q^{\ast}$ of Eq.(2.97),
under the subsidiary quantum conditions (2.98)--(2.100) and (2.96). The
available variables are then the amplitudes $a_{2\nu-1}$ and $a_{2\nu}$ of the
polar function (2.61). Such an analysis has unfortunately been found to become
quite complicated\cite{Lehn02b}, partly due to its nonlinear character and a
high degree of the resulting equations. But there is even a more serious
difficulty which upsets such a conventional procedure. The latter namely
applies only when there are well-defined and localized points of an extremum,
in the form of a maximum or minimum or a saddle-point in parameter space, but
not when such single points are replaced by a flat plateau which has the
effect of an infinite number of extremum points being distributed over the
same space.

The plateau behavior is in fact what occurs here\cite{Lehn02b}, and an
alternative approach therefore has to be elaborated. The analysis then
proceeds by successively including an increasing number of amplitudes
$(a_{1},a_{2},a_{3},\ldots)$ which are being \textquotedblleft
swept\textquotedblright\ (scanned) across their entire range of variation. The
flux factor (2.100) has at the same time to be determined in a self-consistent
way through an iteration process. For each iteration the lowest occurring
value of $q^{\ast}$ can then be determined. Thereby both conditions (2.98) and
(2.99) and the flux factors of Eqs.(2.96) and (2.99) include variable
parameters. Thus the flux factor $f_{\Gamma f}$ of Eq.(2.99) does not become
constant but varies with the amplitudes of the polar function (2.61) when
there are magnetic islands which contribute to the magnetic flux. At a first
sight this appears also to result in a complicated and work-consuming process.
However, the flat plateau behavior is found to simplify the corresponding
iteration scheme and the physical interpretation of its results.

In the numerical analysis which follows, solutions with two real roots have
always been found. Of these roots only that resulting in the lowest value of
$q^{*}$ will be treated in detail in the following parts of this chapter.

Here the analysis will be illustrated by an example where four amplitudes
$(a_{1},a_{2},a_{3},a_{4})$ are included in the polar function (2.61). The
normalized charge $q^{\ast}$ can then be plotted as a function of the two
amplitudes $a_{3}$ and $a_{4}$ which are swept across their entire range of
variation, but for a fixed flux factor $f_{\Gamma}=f_{\Gamma q}=1.82$. The
result is shown in Fig.2.1 for the lowest of the two roots obtained for
$q^{\ast}$. In fact, the corresponding result of a self-consistent analysis
which takes Eq.(2.100) into account, leads only to small modifications which
hardly become visible on the scale of Fig.2.1. From the figure is seen that
there is a steep barrier in its upper part, from which $q^{\ast}$ drops down
to a flat plateau being close to the level $q^{\ast}=1$.

\begin{figure}
\begin{center}
\includegraphics[width=0.8\textwidth]{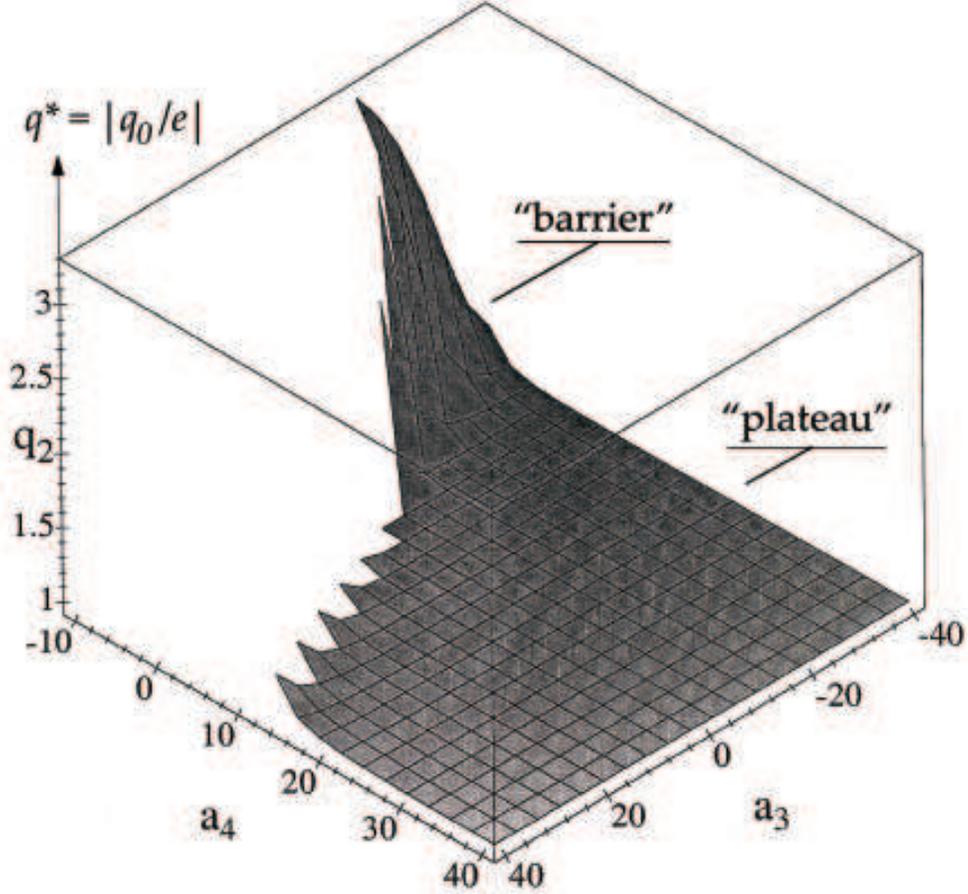}
\caption{The normalized electron charge $q^{\ast}\equiv|q_{0}/e|$ as a
function of the two amplitudes $a_{3}$ and $a_{4}$, for solutions satisfying
the subsidiary quantum conditions for a fixed flux factor $f_{\Gamma
}=f_{\Gamma q}=1.82$, and being based on a polar function $T$ having four
amplitudes $(a_{1},a_{2},a_{3},a_{4})$. The profile of $q^{\ast}$ consists of
a steep \textquotedblleft barrier\textquotedblright\ in the upper part of the
figure, and a flat \textquotedblleft plateau\textquotedblright\ in the lower
part. The plateau is close to the level $q^{\ast}=1$. The figure only
demonstrates the ranges of the real solutions of the first (lowest) root. The
deviations of this profile from that obtained for the self-consistent
solutions which obey condition (2.100) are hardly visible on the scale of the
figure. }
\end{center}
\end{figure}

When further increasing the number of amplitudes, the results were still found
to become plateau-like. The detailed shape of the plateau was seen to become
slightly \textquotedblleft warped\textquotedblright, thereby locally being
partly below and partly above the level $q^{\ast}=1$, i.e. in a range
${0.97\leq q^{\ast}\leq1.03}$.

The flat plateau behavior of $q^{\ast}$ at an increasing number of included
amplitudes can be understood from the fact that the expression (2.97) for the
normalized charge solely depends on the profile shape. It therefore becomes a
slow function of the higher \textquotedblleft multipole\textquotedblright%
\ terms in the expansion (2.61).

\subsubsection{Force Balance of the Electron Model}

The fundamental description of a charged particle in conventional theory is
deficient in several respects. Thus, an equilibrium cannot be maintained by
the classical electrostatic forces, but has been assumed to require forces of
a nonelectromagnetic character to be present\cite{Jack62,Heit54,Stra41}. In
other words, the electron would otherwise \textquotedblleft
explode\textquotedblright\ under the action of its self-charge. Here it will
be shown that a steady electromagnetic equilibrium can under certain
conditions be established by the present extended theory.

We now turn to the momentum balance of the present axisymmetric electron model
in a frame $(r,\theta,\varphi)$ of spherical coordinates. The integral form of
the equivalent forces is in a steady state given by Eqs.(2.20) and (2.21), on
the form
\begin{equation}
\mathbf{F}=\mathbf{F}_{e}+\mathbf{F}_{m}=\int\bar{\rho}(\mathbf{E}%
+\mathbf{C}\times\mathbf{B})\ dV
\end{equation}
When first considering the polar direction represented by the angle $\theta$,
it is readily seen that the corresponding integrated force $F_{\theta}$
vanishes due to the axial symmetry.

Turning then to the radial direction, the earlier obtained results of
Eqs.(2.30)--(2.36) and (2.44) are applied to obtain the radial force
component\cite{Lehn04a}
\begin{equation}
F_{r}=-2\pi\varepsilon_{0}G_{0}^{2}\int\!\!\int[DG+D(s^{2}DG)]\cdot
\Big[\frac{\partial G}{\partial\rho}-\frac{1}{\rho}s^{2}DG\Big]\rho
^{2}s\ d\rho d\theta
\end{equation}
where $s=\sin\theta$. For the point-charge-like model of Eqs.(2.60)--(2.61)
with $G=RT$, $\gamma\longrightarrow2$ and $R\longrightarrow1/\rho^{2}$ at
small $\rho$, we then have
\begin{equation}
\rho^{2}DG=D_{\theta}T-2T
\end{equation}
to be inserted into the integrand of expression (2.102). The latter can then
be represented by the form
\begin{equation}
F_{r}=I_{+}-I_{-}%
\end{equation}
where $I_{+}$ and $I_{-}$ are the positive and negative contributions to the
radial force $F_{r}$.

Consequently, there will arise an integrated radial force balance in the form
of electromagnetic confinement when $I_{+}=I_{-}$. This applies to very small
but nonzero values of the characteristic radius (2.72), i.e. when not
proceeding all the way to the structureless point case where $r_{0}=0$.

The integrals (2.162) and (2.164) are now applied to the plateau region. It is
then found that the ratio $I_{+}/I_{-}$ decreases from $I_{+}/I_{-}=1.27$ at
$q^{\ast}=0.98$ to $I_{+}/I_{-}=0.37$ at $q^{\ast}=1.01$, thus passing the
equilibrium point $I_{+}/I_{-}=1$ at $q^{\ast}\cong0.988$. The remaining
degrees of freedom being available in the parameter ranges of the plateau have
then been used up by the condition of a radially balanced equilibrium.

To sum up, the variational analysis has resulted in a plateau region at the
perimeter of which the normalized charge $q^{*}$ varies in the range
$0.97<q^{*}<1.03$, and where the lowest possible value $q^{*}\cong0.97$ is
obtained. However, this latter value does not satisfy the requirement of a
radial force balance. Such a balance is on the other hand realized within the
plateau region, but at the value $q^{*}\cong0.99$, i.e. where the deduced
charge only deviates by about one percent from the experimental value of the
elementary charge.

The reason for this deviation of the deduced charge from the experimentally
determined value is not clear at this stage, but a quantum mechanical
correction may offer one possibility for its removal. A successful
modification of the magnetic moment has earlier been made by Schwinger and
Feynman, as given by Eq.(2.56). Without getting into the details of the
advanced and work-consuming analysis which leads to this correction, it may
here be proposed that also the magnetic flux and its related quantum condition
would then have to be modified into the form\cite{Lehn04b}
\begin{align}
\tilde{f}_{\Gamma q}  &  =(\bar{A}_{s}/8\pi\bar{A}_{\Gamma}\bar{A}%
_{q})(1+\delta_{\Gamma})\\
\delta_{\Gamma}  &  =c_{\Gamma}/f_{0}\nonumber
\end{align}
where $c_{\Gamma}$ is a so far unspecified constant. For this proposal to lead
to an agreement with the experimental value $q^{\ast}=1$, a modified
self-consistency relation
\begin{equation}
\tilde{f}_{\Gamma q}=\bar{f}_{\Gamma f}%
\end{equation}
would have to be satisfied, and the constant $c_{\Gamma}$ would then have to
be of the order of unity.

At the extremely small dimensions of the present point-charge-like model, it
should become justified to consider the integrated force balance instead of
its localized counterpart. The quantum mechanical wave nature is then expected
to have a smoothing and overlapping effect on the local variations within the
configuration. In addition to this, a requirement of the local electrodynamic
force of Eq.(2.19) to vanish identically, would result in the unacceptable
situation of an overdetermined system of basic equations.

The obtained deviation of $q^{*}$ by about one percent only from the value
$q^{*}=1$ can, in itself, be interpreted as an experimental support of the
present theory. This is particularly the case as the deduced result has been
obtained from two independent aspects, namely the minimization of the charge
by a variational analysis, and the determination of the charge from the
requirement of a radial balance of forces.

\subsection{Axisymmetric Wave Modes and Their Photon Models}

During several decades a number of investigators have discussed the nature of
light and photon physics, not only in relation to the propagation of plane
wave fronts but also to axisymmetric wave packets, the concept of a photon
rest mass, the existence of a magnetic field component in the direction of
propagation, and to an associated angular momentum. The aim is here to
elaborate a model of the individual photon in the capacity of a boson particle
with an angular momentum(spin), propagating with preserved geometry and
limited extensions in a defined direction of space. This leads to the concept
of cylindrical waves and wave packets.

\subsubsection{Elementary Normal Modes}

A cylindrical frame $(r,\varphi,z)$ is introduced where $\varphi$ is an
ignorable coordinate. In this frame the velocity vector is assumed to have the
form
\begin{equation}
\mathbf{C}=c(0,\cos\alpha,\sin\alpha)
\end{equation}
with a constant angle $\alpha$. We further define the operators
\begin{align}
D_{1}  &  =\frac{\partial^{2}}{\partial r^{2}}+\frac{1}{r}\frac{\partial
}{\partial r}+\frac{\partial^{2}}{\partial z^{2}}-\frac{1}{c^{2}}%
\frac{\partial^{2}}{\partial t^{2}}\\
D_{2}  &  =\frac{\partial}{\partial t}+c(\sin\alpha)\frac{\partial}{\partial
z}%
\end{align}
The basic equations (2.28) and (2.29) are then represented by the system
\begin{align}
\Big(  &  D_{1}-\frac{1}{r^{2}}\Big)E_{r}=\frac{\partial}{\partial r}%
(\nabla\cdot\mathbf{E})\\
\Big(  &  D_{1}-\frac{1}{r^{2}}\Big)E_{\varphi}=\frac{1}{c}(\cos\alpha
)\frac{\partial}{\partial t}(\nabla\cdot\mathbf{E})\\
&  D_{1}E_{z}=\Big[\frac{\partial}{\partial z}+\frac{1}{c}(\sin\alpha
)\frac{\partial}{\partial t}\Big](\nabla\cdot\mathbf{E})
\end{align}
and
\begin{equation}
D_{2}(\nabla\cdot\mathbf{E})=0
\end{equation}
Three of the four relations (2.110)--(2.113) thereby form a complete system of
equations, because Eq.(2.113) originates from Eq.(2.29). By further defining
the operator
\begin{equation}
D_{3}=\frac{\partial^{2}}{\partial z^{2}}-\frac{1}{c^{2}}\frac{\partial^{2}%
}{\partial t^{2}}%
\end{equation}
Eq. (2.110) reduces to
\begin{equation}
D_{3}E_{r}=\frac{\partial^{2}E_{z}}{\partial r\partial z}%
\end{equation}
Combination with Eq.(2.111) then yields
\begin{equation}
D_{3}\Big(D_{1}-\frac{1}{r^{2}}\Big)E_{\varphi}=
\frac{1}{c}(\cos\alpha)\frac{\partial^{2}}{\partial z\partial t}D_{1}E_{z}%
\end{equation}

Normal modes depending on $z$ and $t$ as $\operatorname{exp}[i(-\omega
t+kz))]$ are now considered, first in the conventional case where $\nabla
\cdot\mathbf{E}=0$, and then for the EMS mode where $\nabla\cdot\mathbf{E}%
\neq0$. In fact, there are a number of choices with respect to the form
(2.107) as represented by $\pm\cos\alpha$ and $\pm\sin\alpha$ which satisfy
the condition of Eq.(2.5), thereby corresponding to the two directions along
and around the axis of symmetry.

In the case of a vanishing electric field divergence, Eq.(2.110)--(2.112)
reduce to those of an axisymmetric EM mode. Then
\begin{align}
\Big[\bar{D}_{\rho}-\Big(\frac{1}{\rho^{2}}\Big)\Big](E_{r},E_{\varphi})  &
=0\\
\bar{D}_{\rho}E_{z}  &  =0\nonumber
\end{align}
where
\begin{equation}
\bar{D}_{\rho}=\frac{\partial^{2}}{\partial\rho^{2}}+\frac{1}{\rho}%
\frac{\partial}{\partial\rho}%
\end{equation}
The solutions of Eqs.(2.117) become
\begin{align}
E_{r}  &  =k_{r1}\rho+k_{r2}/\rho\\
E_{\varphi}  &  =k_{\varphi1}\rho+k_{\varphi2}/\rho\nonumber\\
E_{z}  &  =k_{z1}\ln\rho+k_{z2}\nonumber
\end{align}
where $k_{r1},k_{r2},k_{\varphi1},k_{\varphi2},k_{z1}$ and $k_{z2}$ are
constants. The magnetic field is obtained from Eq.(2.9).

A similar divergence of the field at the axis $\rho=0$ and at large $\rho$ was
already realized by Thomson\cite{Thom36} and further discussed by
Heitler\cite{Heit54}, as well as by Hunter and Wadlinger\cite{Hunt89}. It
leaves the problem with the radial dependence unresolved. Thus the field does
not converge within the entire vacuum space, and cannot be made to vanish at
large radial distances as long as $k_{r1},k_{\varphi1}$ and $k_{z1}$ differ
from zero. All parts of the solutions (2.119) further result in integrals of
the energy density which become divergent when being extended all over the
vacuum space. An additional and even more severe point of concern is due to
the electric field divergence of Eq.(2.119) which becomes
\begin{equation}
\nabla\cdot\mathbf{E}=2k_{r1}+ik(k_{z1}\ln\rho+k_{z2})
\end{equation}
and has to vanish identically for all $(\rho,k)$ in the case of conventional
electromagnetic waves. Thereby the last part of Eq.(2.120) requires the
parallel electric field component $E_{z}$ to vanish identically, but then it
also becomes necessary for the constant $k_{r1}$ in Eq.(2.120) to disappear.
There are further analogous solutions for the magnetic field, by which also
the axial component $B_{z}$ vanishes in the conventional case. In is turn,
this leads to a Poynting vector (2.22) having only a component in the
direction of propagation, thus resulting in a vanishing angular momentum
(spin) with respect to the axial direction.

Returning to the axisymmetric EMS mode of Eqs.(2.107)--(2.116), a dispersion
relation
\begin{equation}
\omega=kv\qquad v=c(\sin\alpha)
\end{equation}
is obtained from Eq.(2.113), having the phase and group velocities $v$. Here
we limit ourselves to positive values of $\cos\alpha$ and $\sin\alpha$. With
the dispersion relation (2.121), Eq.(2.116) takes the form%
\begin{equation}
\Big[\frac{\partial^{2}}{\partial r^{2}}+\frac{1}{r}\frac{\partial}{\partial
r}-\frac{1}{r^{2}}-k^{2}(\cos\alpha)^{2}\Big]E_{\varphi}=
-(\operatorname{tg}\alpha)\Big[\frac{\partial^{2}}{\partial r^{2}}+\frac{1}%
{r}\frac{\partial^{2}}{\partial r}-
k^{2}(\cos\alpha)^{2}\Big]E_{z}%
\end{equation}
We can now introduce a \emph{generating function}
\begin{align}
G_{0}\cdot G  &  =E_{z}+(\cot\alpha)E_{\varphi}\\
G  &  =R(\rho)e^{i(-\omega t+kz)}\nonumber
\end{align}
where $G_{0}$ stands for the amplitude, $G$ for a normalized form,
$\rho=r/r_{0}$ is again a normalized coordinate with respect to the
characteristic radius $r_{0}$ of the configuration, and $R(\rho)$ is a
dimensionless function of $\rho$. With the operator of Eq.(2.118), the
definition
\begin{equation}
D=\bar{D}_{\rho}-\theta^{2}(\cos\alpha)^{2}\qquad\theta=kr_{0}%
\end{equation}
and using Eqs.(2.121), (2.115), and (2.9), insertion into Eq.(2.122) results
in
\begin{equation}
E_{r}=-iG_{o}[\theta(\cos\alpha)^{2}]^{-1}\cdot
\frac{\partial}{\partial\rho}[(1-\rho^{2}D)G]=-\frac{1}{r_{0}}\frac
{\partial\phi}{\partial\rho}+i\omega A_{r}%
\end{equation}%
\begin{align}
E_{\varphi}  &  =G_{0}(\operatorname{tg}\alpha)\rho^{2}DG=i\omega A_{\varphi
}\\
E_{z}  &  =G_{0}(1-\rho^{2}D)G=-ik\phi+i\omega A_{z}%
\end{align}
and
\begin{equation}
B_{r}=-G_{0}[c(\cos\alpha)]^{-1}\cdot
\rho^{2}DG=-ikA_{\varphi}%
\end{equation}%
\begin{equation}
B_{\varphi}=-iG_{0}(\sin\alpha)[\theta c(\cos\alpha)^{2}]^{-1}\cdot
\frac{\partial}{\partial\rho}[(1-\rho^{2}D)G]=ikA_{r}-\frac{1}{r_{0}}%
\frac{\partial A_{z}}{\partial\rho}%
\end{equation}%
\begin{equation}
B_{z}=-iG_{0}[\theta c(\cos\alpha)]^{-1}\Big(\frac{\partial}{\partial\rho
}+\frac{1}{\rho}\Big)\cdot
(\rho^{2}DG)=\frac{1}{r_{0}}\frac{1}{\rho}\frac{\partial}{\partial\rho}(\rho
A_{\varphi})
\end{equation}
This result makes possible a choice of generating functions and corresponding
modes which are physically relevant within the entire vacuum space, and which
become consistent with the imposed quantum conditions. The solutions
(2.125)--(2.127) are reconfirmed by insertion into the basic equations. The
result implies that the three equations (2.110)--(2.112) do not become
independent of each other.

Due to the dispersion relation (2.121) the EMS mode propagates at phase and
group velocities which are smaller than $c$. Not to get into conflict with the
experimental observations by Michelson and Moreley, the condition
\begin{equation}
0<\cos\alpha\ll1
\end{equation}
has then to be taken into account. Since the velocity $v$ of Eq.(2.121)
becomes slightly less than $c$, the condition (2.131) will be related to a
very small but nonzero rest mass of the photon, as shown later in more detail.

With the condition (2.131) the present mode has a dominant radially polarized
electric field component (2.125), and an associated magnetic field component
(2.129). The field configuration of Eqs.(2.125)--(2.130) thus differs in
several respects from elliptically, circularly and linearly polarized plane waves.

The components (2.125)--(2.130) are reconcilable with the field configuration
by Evans and Vigier\cite{Evan94,Evan96}, in the sense that all components are
nonzero and form a helical structure which also includes an axial magnetic
field component in the direction of propagation. The present result is,
however, not identical with that by Evans and Vigier.

So far the deductions have been performed with respect to the laboratory frame
K. In the present case where the phase and group velocities $v<c$, a
physically relevant rest frame K$^{\prime}$ can be defined, but not in the
conventional case where there is no rest mass, $\cos\alpha=0$, and $v=c$. The
details of a transformation to the rest frame are demonstrated
elsewhere\cite{Lehn01}.

\subsubsection{Wave Packets}

>From a spectrum of normal modes with different wave numbers, a wave packet
solution is now being formed which has finite radial and axial extensions and
a narrow line width in wavelength space. Such a line width is consistent with
experimental observations. We are free to rewrite the amplitude factor of the
generating function (2.123) as
\begin{equation}
G_{0}=g_{0}(\cos\alpha)^{2}%
\end{equation}
The normal modes are superimposed to result in a wave packet having the
amplitude
\begin{equation}
A_{k}=\Big(\frac{k}{k_{0}^{2}}\Big)e^{-z_{0}^{2}(k-k_{0})^{2}}%
\end{equation}
in the interval $dk$ and being centered around the wave number $k_{0}$.
Further $2z_{0}$ represents the effective axial length of the packet.

The physically relevant restriction to a narrow line width implies that the
amplitude (2.133) drops to $1/e$ of its maximum value for a small deviation of
$k$ from the maximum at $k=k_{0}$. With $\Delta k=k-k_{0}=1/z_{0}$ this
implies that $\Delta k/k_{0}=1/k_{0}z_{0}=\lambda_{0}/2\pi z_{0}\ll1$, where
$\lambda_{0}=2\pi/k_{0}$ is the average wave length of the packet. With
$\bar{z}=z-vt$ and $v=c(\sin\alpha)$, and after working out the integration
with respect to $k$, the spectral averages of the packet field components
become
\begin{equation}
\bar{E}_{r}=-iE_{0}[R_{5}+(\theta_{0}^{\prime})^{2}R_{2}]
\end{equation}%
\begin{equation}
\bar{E}_{\varphi}=E_{0}\theta_{0}(\sin\alpha)(\cos\alpha)\cdot
\lbrack R_{3}-(\theta_{0}^{\prime})^{2}R_{1}]
\end{equation}%
\begin{equation}
\bar{E}_{z}=E_{0}\theta_{0}(\cos\alpha)^{2}[R_{4}+(\theta_{0}^{\prime}%
)^{2}R_{1}]
\end{equation}%
\begin{align}
\bar{B}_{r}  &  =-\Big(\frac{1}{c}\Big)(\sin\alpha)^{-1}E_{\varphi}\\
\bar{B}_{\varphi}  &  =\Big(\frac{1}{c}\Big)(\sin\alpha)E_{r}\\
\bar{B}_{z}  &  =-i\Big(\frac{1}{c}\Big)E_{0}(\cos\alpha)[R_{8}-(\theta
_{0}^{\prime})^{2}R_{7}]
\end{align}
where $\theta_{0}=k_{0}r_{0}$, $\theta_{0}^{\prime}=\theta_{0}(\cos\alpha)$,
\begin{equation}
E_{0}=E_{0}(\bar{z})=\Big(\frac{g_{0}}{k_{0}r_{0}}\Big)\Big(\frac{\sqrt{\pi}%
}{k_{0}z_{0}}\Big)
\cdot\operatorname{exp}\Big[-\Big(\frac{\bar{z}}{2z_{0}}\Big)^{2}+ik_{0}%
\bar{z}\Big]
\end{equation}
and
\begin{equation}
R_{1}   =\rho^{2}R\qquad R_{2}=\frac{d}{d\rho}(\rho^{2}R) \qquad
R_{3}   =\rho^{2}\bar{D}_{\rho}R
\end{equation}%
\begin{equation}
R_{4}   =(1-\rho^{2}\bar{D}_{\rho})R \qquad
R_{5}   =\frac{d}{d\rho}[(1-\rho^{2}\bar{D}_{\rho})R]
\end{equation}%
\begin{equation}
R_{6}   =\bar{D}_{\rho}[(1-\rho^{2}\bar{D}_{\rho})R] \qquad
R_{7}   =\Big(\frac{d}{d\rho}+\frac{1}{\rho}\Big)(\rho^{2}R)
\end{equation}
with the operator $\bar{D}_{\rho}$ given by Eq.(2.118).

It should be noticed that expressions (2.137)--(2.139) are approximate, on
account of the restriction to a narrow line width. Therefore the condition
$\nabla\cdot\bar{\mathbf{B}}=0$ is only satisfied approximately. In the limit
of zero line width we would on the other hand be back to the exact form
(2.128)--(2.130) for an elementary normal mode where $\nabla\cdot\mathbf{B}=0$.

For the wave packet fields (2.134)--(2.139) the components $(\bar{E}_{\varphi
},\bar{E}_{z},\bar{B}_{r})$ are in phase with the generating function (2.123),
whereas the components $(\bar{E}_{r},\bar{B}_{\varphi},\bar{B}_{z})$ are
ninety degrees out of phase with it. In the analysis which follows we choose a
generating function which is symmetric with respect to the axial centre
$\bar{z}=0$ of the moving wave packet. Thus
\begin{equation}
G=R(\rho)\cos k\bar{z}%
\end{equation}
when the real parts of the forms (2.123) and (2.140) are adopted. Then $G$ and
$(\bar{E}_{\varphi},\bar{E}_{z},\bar{B_{r}})$ are symmetric and $(\bar{E}%
_{r},\bar{B}_{\varphi},\bar{B}_{z})$ are antisymmetric functions of $\bar{z}$
with respect to the centre $\bar{z}=0$.

\subsubsection{Spatially Integrated Field Quantities}

When elaborating a model for the individual photon, integrals have to be
formed to obtain net values of the electric charge $q$, magnetic moment $M$,
the total mass $m$ which represents the total energy, and of the angular
momentum (spin) $s$. Thereby the limits of $\bar{z}$ are $\pm\infty$, whereas
those of $\rho$ become unspecified until further notice.

The integrated charge becomes
\begin{equation}
q=\varepsilon_{0}\int\nabla\cdot\bar{\mathbf{E}}\ dV=
2\pi\varepsilon_{0}\int\bigg\{\frac{\partial}{\partial r}(r\int_{-\infty
}^{+\infty}\bar{E}_{r}\ d\bar{z})+
r[\bar{E}_{z}]_{-\infty}^{+\infty}\bigg\}\ dr=0
\end{equation}
because $\bar{E}_{r}$ is antisymmetric and $\bar{E}_{z}$ symmetric with
respect to $\bar{z}$.

The integrated magnetic moment is related to the component
\begin{equation}
j_{\varphi}=\bar{\rho}C_{\varphi}=\varepsilon_{0}(\nabla\cdot\bar{\mathbf{E}%
})c(\cos\alpha)
\end{equation}
of the space-charge current density. Thus%
\begin{multline}
M=\int\int_{-\infty}^{+\infty}\bar{\rho}c(\cos\alpha)\pi r^{2}\ drd\bar{z}=\\
=\pi\varepsilon_{0}c(\cos\alpha)\int\Big\{r\frac{\partial}{\partial r}%
(r\int_{-\infty}^{+\infty}\bar{E}_{r}\ d\bar{z})+
r^{2}\big[\bar{E}_{z}\big]_{-\infty}^{+\infty}\Big\}\ dr=0
\end{multline}
for the same symmetry reasons as those applying to the charge (2.145). It
should be observed that, even if the net magnetic moment vanishes, the local
magnetic field remains nonzero, as well as the local charge density.

An equivalent total mass can be defined from the electromagnetic field energy,
due to the energy relation by Einstein. Consequently we write
\begin{equation}
m=\Big(\frac{1}{c^{2}}\Big)\int w_{f}\ dV
\end{equation}
as given by the field energy density of Eq.(2.25). Applying the condition
(2.131) and using Eqs.(2.134)--(2.139), we then have
\begin{equation}
m\cong2\pi(\varepsilon_{0}/c^{2})\int_{-\infty}^{+\infty}\int r|\bar{E}%
_{r}^{2}|\ drd\bar{z}%
\end{equation}
When adopting the form (2.144) with an antisymmetric field $\bar{E}_{r}$, and
introducing the integral
\begin{equation}
J_{z}=\int_{-\infty}^{+\infty}(\sin k_{0}\bar{z})^{2}e^{-2(\bar{z}/2z_{0}%
)^{2}}d\bar{z}\cong
z_{0}(\pi/2)^{1/2}%
\end{equation}
in the limit $k_{0}z_{0}\gg1$, the total mass finally becomes
\begin{align}
m  &  =a_{0}W_{m}=h\nu_{0}/c^{2}\\
W_{m}  &  =\int\rho R_{5}^{2}\ d\rho\nonumber
\end{align}
where
\begin{equation}
a_{0}=\varepsilon_{0}\pi^{5/2}\sqrt{2}z_{0}(g_{0}/ck_{0}^{2}z_{0})^{2}%
\equiv2a_{0}^{\ast}g_{0}^{2}%
\end{equation}
For this wave packet the energy relations by Planck and Einstein also combine
to
\begin{align}
mc^{2}  &  =h\nu_{0}=(hc/\lambda_{0})(\sin\alpha)\\
\lambda_{0}  &  =2\pi/k_{0}\nonumber
\end{align}
where use has been made of the dispersion relation (2.121) applied to the wave
packet as a whole, i.e.
\begin{align}
\omega_{0}  &  =2\pi\nu_{0}=k_{0}c(\sin\alpha)\\
\nu_{0}  &  =(c/\lambda_{0})(\sin\alpha)\nonumber
\end{align}

Here the slightly reduced phase and group velocity of Eq.(2.121) can be
considered as being associated with a very small nonzero rest mass
\begin{equation}
m_{0}=m[1-(v/c)^{2}]^{1/2}=m(\cos\alpha)
\end{equation}

Turning to the momentum balance governed by Eqs.(2.20)--(2.22), all integrated
components of the forces (2.21) in the $r$ and $\varphi$ directions will
vanish on account of the axial symmetry, and since the $\varphi$ component of
$\bar{\mathbf{F}}_{e}+\bar{\mathbf{F}}_{m}$ will vanish on account of
Eq.(2.137). The only components which remain are
\begin{align}
\bar{F}_{ez}  &  =\int\bar{\rho}\bar{E}_{z}\ dV\\
\bar{F}_{mz}  &  =-\int\bar{\rho}c(\cos\alpha)\bar{B}_{r}\ dV\nonumber
\end{align}
where
\begin{equation}
\bar{\rho}/\varepsilon_{0}=\frac{1}{r}\frac{\partial}{\partial r}(e\bar{E}%
_{r})+\frac{\partial}{\partial z}\bar{E}_{z}%
\end{equation}
Here $\bar{\rho}$ becomes symmetric and $(\bar{E}_{z},\bar{B}_{r})$
antisymmetric with respect to $\bar{z}$. The integrated forces (2.156) will
for this reason also vanish, and the balance equation then acquires its
conventional form. Consequently, the conventional expressions\cite{Mors53}
\begin{equation}
\mathbf{s}=\mathbf{r}\times\mathbf{S}/c^{2}\qquad\mathbf{S}=\bar{\mathbf{E}%
}\times\bar{\mathbf{B}}/\mu_{0}%
\end{equation}
apply for the density $\mathbf{s}$ of the angular momentum, where $\mathbf{r}$
is the radius vector from the origin and $\mathbf{S}$ is the Poynting vector.
The density of angular momentum now becomes
\begin{equation}
s_{z}=\varepsilon_{0}r(\bar{E}_{z}\bar{B}_{r}-\bar{E}_{r}\bar{B}_{z}%
)\cong-\varepsilon_{0}r\bar{E}_{r}\bar{B}_{z}%
\end{equation}
when condition (2.131) applies and use is made of Eqs.(2.134), (2.136),
(2.137), and (2.139). The total angular momentum is given by
\begin{equation}
s=\int s_{z}\ dV=
-2\pi\varepsilon_{0}\int_{-\infty}^{+\infty}\int r^{2}\bar{E}_{r}\bar{B}%
_{z}\ drd\bar{z}%
\end{equation}
and it reduces to the final form
\begin{align}
s  &  =a_{0}r_{0}c(\cos\alpha)W_{s}=h/2\pi\\
W_{s}  &  =-\int\rho^{2}R_{5}R_{7}\ d\rho\nonumber
\end{align}
for the model of a photon having the angular momentum of a boson.

The disappearance of the total integrated volume forces is also supported by
the fact that the solutions (2.125)--(2.130) and (2.134)--(2.139) result in
field strengths $\mathbf{E}+\mathbf{C}\times\mathbf{B}$ and $\bar{\mathbf{E}%
}+\mathbf{C}\times\bar{\mathbf{B}}$ which are readily seen to vanish in lowest
order when $|\cos\alpha|\ll1$.

Here it is noticed that, with the inclusion of a dimensionless profile factor,
relations (2.161) and (2.151) indicate that the form of the angular momentum
agrees with the simple picture of a mass which rotates at the velocity
$c(\cos\alpha)$ around the axis at the distance $r_{0}$.

The result (2.161) and Eq.(2.155) show that a nonzero angular momentum
requires a nonzero rest mass to exist, as represented by $\cos\alpha\neq0$.
Moreover, the factor $\cos\alpha$ in the expression (2.161) and Eq.(2.107) for
the velocity vector $\mathbf{C}$ indicate that the angular momentum, as well
as the nonzero rest mass, are associated with the component $C_{\varphi}$ of
circulation around the axis of symmetry.

First a generating function will now be considered which is finite at the axis
$\rho=0$, and which tends to zero at large $\rho$. These requirements are
fulfilled by
\begin{equation}
R(\rho)=\rho^{\gamma}e^{-\rho}%
\end{equation}
In principle, the factor in front of the exponential part would in a general
case have to be replaced by series of positive or negative powers of $\rho$,
but since we will here proceed to the limit of large $\gamma$, only one term
becomes sufficient. Moreover, the exponential factor in Eq.(2.162) has been
included to secure the general convergence of any moment with $R$ at large
$\rho$. In any case, the forthcoming final results will be found not to depend
explicitly on this exponential factor.

In the evaluation of expressions (2.151) and (2.161) for $W_{m}$ and $W_{s}$
the Euler integral
\begin{equation}
J_{2\gamma-2}=\int_{0}^{\infty}\rho^{2\gamma-2}e^{-2\rho}\ d\rho=
2^{-(2\gamma+3)}\Gamma(2\gamma+1)
\end{equation}
appears in terms of the gamma function $\Gamma(2\gamma+1)$. For $\gamma\gg1$
only the dominant terms of the functions $R_{5}$ and $R_{7}$ of Eqs.(2.142)
and (2.143) will prevail, and the result becomes
\begin{equation}
W_{m}=2^{-(2\gamma+4)}\gamma^{6}(2\gamma-1)\cdot
\Gamma(2\gamma+1)=W_{s}/\gamma
\end{equation}
The function (2.162) has a maximum at the radius
\begin{equation}
\hat{r}=\gamma r_{0}%
\end{equation}
which becomes sharply defined at large $\gamma$, in analogy with an earlier
obtained result\cite{Lehn01}. Combination of Eqs.(2.151), (2.153), (2.161),
(2.164), and (2.165) leads to an effective photon diameter
\begin{equation}
2\hat{r}=\frac{\lambda_{0}}{\pi(\cos\alpha)}%
\end{equation}
This result applies not only to an individual photon of the effective radius
$\widehat{r,}$ but can also stand for the corresponding radius of a radially
polarized dense multiphoton beam.

Turning then to the alternative of a radial part which \emph{diverges} at the
axis, the form
\begin{equation}
R(\rho)=\rho^{-\gamma}e^{-\rho}\qquad\gamma>0
\end{equation}
is taken into consideration, being identical with the form (2.60) for the
electron model. When the radial variable increases monotonically, the function
(2.167) decreases from large values, down to $R=e^{-1}$ at $\rho=1$, and
further to very small values when $\rho$ becomes substantially larger than 1.
Thus $r=r_{0}$ can be taken as a characteristic radial dimension of the configuration.

To obtain \emph{finite} values of the integrated total mass $m$ and angular
momentum $s$ of Eqs.(2.151) and (2.161), a special procedure similar to that
applied to the electron model is being applied. The lower limits of the
integrals (2.151) and (2.161) are specified by
\begin{align}
W_{m}  &  =\int_{\rho_{m}}^{\infty}\rho R_{5}^{2}\ d\rho\\
W_{s}  &  =-\int_{\rho_{s}}^{\infty}\rho^{2}R_{5}R_{7}\ d\rho\nonumber
\end{align}
where $\rho_{m}\ll1$ and $\rho_{s}\ll1$. Making the choice $\gamma\gg1$, these
integrals reduce to
\begin{equation}
W_{m}=\frac{1}{2}\gamma^{5}\rho_{m}^{-2\gamma}\qquad W_{s}=\frac{1}{2}%
\gamma^{5}\rho_{s}^{-2\gamma+1}%
\end{equation}

To secure finite values of $m$ and $s$ we must now permit the characteristic
radius $r_{0}$ and the factor $g_{0}$ in Eqs.(2.132), (2.151), (2.152), and
(2.161) to \textquotedblleft shrink\textquotedblright\ to very small but
nonzero values, as the lower limits $\rho_{m}$ and $\rho_{s}$ approach zero.
This is attained by introducing the relations
\begin{align}
r_{0}  &  =c_{r}\cdot\varepsilon\qquad c_{r}>0\\
g_{0}  &  =c_{g}\cdot\varepsilon^{\beta}\qquad c_{g}>0
\end{align}
where $0<\varepsilon\ll1$, $\beta>0$, $c_{r}$ is a constant with the dimension
of length, and $c_{g}$ one with the dimension of electrostatic potential.
Eqs.(2.151), (2.152), (2.153), (2.161), and (2.169) now combine to
\begin{equation}
m=a_{0}^{\ast}\gamma^{5}c_{g}^{2}(\varepsilon^{2\beta}/\rho_{m}^{2\gamma
})\cong h/\lambda_{0}c
\end{equation}%
\begin{equation}
s=a_{0}^{\ast}\gamma^{5}c_{g}^{2}c_{r}c(\cos\alpha)\cdot
(\varepsilon^{2\beta+1}/\rho_{s}^{2\gamma-1})=h/2\pi
\end{equation}

To obtain finite values of both $m$ and $s$, it is then necessary to satisfy
the conditions
\begin{equation}
\rho_{m}=\varepsilon^{\beta/\gamma}\qquad\rho_{s}=\varepsilon^{(2\beta
+1)/(2\gamma-1)}%
\end{equation}
We are here free to choose $\beta=\gamma\gg1$ by which
\begin{equation}
\rho_{s}\cong\rho_{m}=\varepsilon
\end{equation}
with good approximation. The lower limits of the integrals (2.168) and (2.169)
then decrease linearly with $\varepsilon$ and the radius $r_{0}$. This forms a
\textquotedblleft similar\textquotedblright\ set of geometrical configurations
which thus have a shape being independent of $\rho_{m}$, $\rho_{s}$, and
$\varepsilon$ in the range of small $\varepsilon$.

The ratio of expressions (2.172) and (2.173) finally yields an effective
photon diameter
\begin{equation}
2r_{0}=\frac{\varepsilon\lambda_{0}}{\pi(\cos\alpha)}%
\end{equation}
It should be observed that the wave packet diameters of Eqs.(2.176) and
(2.166) both become independent of the particular values of the parameters
$\gamma$ and $\beta$. As compared to the relatively large photon diameter
(2.166) obtained for a convergent generating function, the diameter (2.176)
based on a divergent such function can shrink to very small dimensions. This
is the case even when $\cos\alpha\ll1$ provided that $\varepsilon\ll\cos
\alpha$. Then the photon model becomes strongly \textquotedblleft
needle-shaped\textquotedblright\ in its transverse directions.

\subsubsection{Features of Present Photon Model}

The possible existence of a nonzero photon rest mass was first called
attention to by Einstein\cite{Eins05}, Bass and Schr\"{o}dinger\cite{Bass55},
and de Broglie and Vigier\cite{Brog72}. It includes such fundamental points as
its relation to the Michelson-Morley experiment, and its so far undetermined
absolute value.

The phase and group velocities of the present non-dispersive wave packets
become slightly smaller than the velocity of light, as expressed by condition
(2.131) and Eq.(2.121). Thereby the velocity constant $c$ can be regarded as
an asymptotic limit at infinite photon energy. A small deviation from this
limit still permits the theory to be compatible with the Michelson-Morley
experiments. With condition (2.131) and Eq.(2.121) the deviation of the phase
and group velocity from $c$ becomes
\begin{equation}
v/c\cong1-\frac{1}{2}(\cos\alpha)^{2}%
\end{equation}
For $\cos\alpha\leq10^{-4}$ corresponding to a photon rest mass $m_{0}%
<0.74\times10^{-39}$ kg, a change in the eight decimal of the recorded
velocity would hardly become detectable.

The Lorentz invariance is satisfied even if there is an axial velocity of
propagation being slightly smaller than $c$. This is due to the velocity
vector $\mathbf{C}$ which forms helical orbits, also having a small component
which circulates around the axis.

The Poynting vector (2.158) is also found from Eqs.(2.134)--(2.139), combined
with the conditions $\gamma\gg1$ and (2.131), to become parallel with the
velocity vector $\mathbf{C}$ in the first approximation.

The previous analysis shows that the physics in presence even of a very small
rest mass becomes fundamentally different from that being based on a rest mass
which is exactly equal to zero. In the latter case we are back from an EMS
mode to a conventional axisymmetric EM mode with its divergent and physically
unacceptable properties. Since the results of the EMS mode hold only for a
nonzero rest mass, the quantum conditions (2.153) for the total energy and
(2.161) for the angular momentum become satisfied by a whole class of small
values of $\cos\alpha$ and $m_{0}$.

As being pointed out by de Broglie and Vigier, such an indeterminateness of
the photon rest mass appears at a first sight to be a serious objection to the
underlying theory. The problem is that the derivations depend simply on the
existence of the nonzero rest mass, but not on its magnitude. To this concept
de Broglie and Vigier add other analogous examples that have been considered
in theoretical physics. Thus, the uncertainty in the absolute value of the
nonzero photon rest mass does not necessarily imply that the corresponding
theory is questionable. On the contrary, as has been seen from the present
analysis, such an unspecified rest mass rather becomes a strength of the
theory. It namely makes variable ranges possible of the effective diameters
for both axisymmetric photon models and corresponding light beams.

As shown by Eq.(2.161) a nonzero angular momentum only becomes possible when
there is a nonzero rest mass (2.155) with an associated factor $\cos\alpha$.
In its turn, this factor is related to a nonzero axial magnetic field
component of the helical field configuration. For a vanishing rest mass we
would thus be back to the conventional case where there is no axial magnetic field.

Turning to the effective photon diameter of these wave packet models, we first
notice that a convergent generating function leads to a diameter (2.166) which
in some cases becomes rather limited, but still does not become small enough
to match atomic dimensions. As an example, an average wave length $\lambda
_{0}=3\times10^{-7}$m and a factor $\cos\alpha=10^{-4}$ which makes the
velocity $v$ deviate from $c$ by $5\times10^{-9}$ only, results in a diameter
$2\hat{r}\cong10^{-3}$m.

A divergent generating function can on the other hand result in a very small
photon diameter (2.176) for a sufficiently small $\varepsilon$, then having
the size of atomic dimensions such as the Bohr radius. This becomes
reconcilable with the ability of the photon to knock out an electron from an
atom in the photoelectric effect. This is consistent with the proposed
\textquotedblleft needle radiation\textquotedblright\ of energy quanta with a
directed momentum as obtained by Einstein\cite{Eins17}.

The obtained wave packet solutions are in some respects similar to the earlier
wave-particle duality by de Broglie. Thus the total energy $h\nu=mc^{2}$ in
the laboratory frame could be regarded to consist of the fraction
$(m-m_{0})c^{2}$ of a ``free'' pilot wave propagating along the axis, plus the
fraction $m_{0}c^{2}$ of a ``bound'' particle-like state of radiation which
circulates around the same axis. The rest mass then merely represents an
integrating part of the total energy. Such a subdivision into a particle and
an associated pilot wave is therefore not necessary in the present case where
the wave packet behaves as an entity, having both particle and wave properties
at the same time.

In the two-slit experiments\cite{Tsuc85} the present needle-shaped wave packet
model has an advantage over a plane wave model. The particle feature in the
form of a very narrow needle-shaped transverse diameter makes the wave packet
model reconcilable with the observed dot-shaped marks. At the same time the
wave nature of the axisymmetric packet makes it possible for interference
phenomena to occur, in the same way as for a plane wave. This dualism and
ability of interference becomes particularly obvious in the zero line width
limit $1/k_{0}z_{0}\longrightarrow0$ where the spectral distribution of
Eq.(2.133) reduces to that of the elementary normal mode of Section E.1.

The analysis performed so far may debouch into the idea that there exist
different quantum states (modes) of the photon representing different
solutions of the same basic equations. In a way this would become analogous to
but not identical with the Copenhagen interpretation. Almost instantaneous
transitions between these states also become imaginable. A similar situation
has already been observed for neutrino oscillations which are associated with
a nonzero neutrino rest mass. Consequently, it is here proposed that analogous
``photon oscillations'' can exist, by which rapid transitions between the
various plane and axisymmetric photon modes take place, under the constraints
of total energy and angular momentum conservation. In this way the photon
could behave differently in different physical situations. Such a preliminary
proposal has, however, to be further investigated.

The photon is the field quantum being responsible for the electromagnetic
interaction. Likewise the weak field interaction acquires quanta, in the form
of the W$^{+}$, W$^{-}$, and Z$^{0}$ bosons\cite{Ryde96}. This raises the
question whether an analogous Proca-type equation applied to the weak field
case could result in axisymmetric solutions being similar to those deduced
here for the electromagnetic field. This may provide the weak-field bosons
with a nonzero rest mass, thereby arriving at a possible alternative to the
Higgs particle concept.

Finally it should be observed that the results of Section E.3 for an
individual radially polarized photon will also apply to a corresponding dense
beam of $N$ photons per unit length and circular cross-section, provided that
the individual photon fields overlap each other. Thereby the right-hand
members of Eqs.(2.151) and (2.161) for the total mass and angular momentum
will include the same factor $N$. As a result, Eqs.(2.166) and (2.176) then
apply also to the beam diameters.

\subsection{Summary of Obtained Main Results in Section II}

The main results which are specific to the present theoretical approach can
now be summarized. These are expected to contribute to an increasing
understanding of a number of fundamental physical phenomena, and they also
predict new features of the electromagnetic field to exist. It should be
noticed here that a recently presented theory by Tauber\cite{Taub03} also has
some parts in common with these features.

Maxwell's equations in the vacuum have been used as a guideline and basis in
the development of quantum electrodynamics (QED), which is therefore expected
to become subject to the typical shortcomings of conventional electromagnetic
theory. The present revised electromagnetics, with its nonzero electric field
divergence, provides \ a way of eliminating these shortcomings, in a first
step of extended quantum electrodynamics (\textquotedblleft
EQED\textquotedblright).

The present theory leads to some results of a general character, of which the
following should be mentioned:

\begin{itemize}
\item The extended equations make it possible for electromagnetic steady
states to exist in the vacuum.

\item New types of wave modes arise, such as a longitudinal purely electric
space-charge (S) wave, and a nontransverse electromagnetic space-charge (EMS) wave.
\end{itemize}

The electrically charged steady states result in a model of charged leptons
such as the electron:

\begin{itemize}
\item For a particle-shaped configuration to possess a nonzero net electric
charge, its characteristic radius has to shrink to that of a point-charge-like
geometry, in agreement with experiments.

\item Despite of the success of the conventional renormalization procedure, a
more satisfactory way from the physical point of view is needed in respect to
the infinite self-energy problem of a point charge. This problem is removed by
the present theory which provides a more surveyable alternative, by
compensating the divergence of the generating function through a shrinking
characteristic radius and thereby leading to finite integrated field quantities.

\item The Lorentz invariance of the electron radius is formally satisfied in
the present theory, by allowing the same radius to shrink to that of a point
charge. The obtained solutions can on the other hand also be applied to the
physically relevant case of a very small but nonzero radius of a configuration
having an internal structure.

\item A steady equilibrium of the average radial forces can under certain
conditions be established, such as to prevent the electron from ``exploding''
under the action of its electric self-charge. This becomes possible for the
point-charge-like model, at parameter values of the plateau region which
correspond to $q^{*}\cong0.99$. The remaining degrees of freedom being
available in the parameter ranges of the plateau have then been used up by the
condition of a radially balanced equilibrium.

\item That there is only a small deviation of $q^{*}$ from unity can, in
itself, be interpreted as an experimental support of the present theory. This
is particularly the case as the deduced result has been obtained from two
independent aspects, namely the minimization of the charge by a variational
analysis, and the determination of the charge from the requirement of a radial
balance of forces.

\item The deduced charge thus deviates only by about one percent from the
measured elementary charge. To remove this small deviation, some additional
quantum mechanical corrections have been proposed. Provided that such
corrections become relevant, the elementary charge would no longer remain as
an independent constant of nature, but can be derived from the present theory
in terms of the velocity of light, Planck's constant, and the permittivity of vacuum.
\end{itemize}

Conventional theory includes a vanishing electric field divergence, and it has
no electric and magnetic field components in the direction of propagation,
i.e. in the case of plane as well as of axisymmetric and screw-shaped wave
modes \cite{Lehn02a}. The Poynting vector then possesses a component only in
the direction of propagation, and there is no angular momentum(spin) with
respect to the same direction. This feature also prevails in a fully quantized
analysis. The electromagnetic field strengths appearing in the Poynting vector
can then be expressed in terms of corresponding quantum mechanical plane wave
expressions. The deduced axisymmetric wave modes lead on the other hand to
models of the individual photon, in the form of EMS wave-packets with a narrow
line width and having the following features:

\begin{itemize}
\item For the photon to possess a nonzero angular momentum (spin) in the
capacity of a boson particle, the axisymmetric field geometry has to become
helical, as well as the Poynting vector.

\item This helical structure is associated with a nonzero rest mass, and it
has field components also in the longitudinal (axial) direction of propagation.

\item The nonzero rest mass can become small enough not to get into conflict
with experiments of the Michelson-Morley type, thereby still preserving the
conditions for a helical field structure.

\item There are wave-packet solutions with a very small characteristic
transverse radius. These have the form of ``needle radiation'' which becomes a
necessary concept for the explanation of the photoelectric effect where a
photon interacts with a single electron in an atom, and of the dot-shaped
marks observed at a screen in double-slit experiments.

\item The axisymmetric wave-packet models behave as one single entity, having
particle and wave properties at the same time.
\end{itemize}

\bigskip

\begin{center}
{\LARGE III. Unified Theory of Bivacuum, Particles Duality, Fields \& Time}

{\Large by A. Kaivarainen}

\bigskip

\textbf{III. A. The new concept of Bivacuum\smallskip}

\emph{A.1. Properties of Bivacuum dipoles - Bivacuum fermions and Bivacuum
bosons\medskip}
\end{center}

\setcounter{section}{3}Our Unified Theory (UT) is a result of long-term
efforts for unification of vacuum, matter and fields from few ground
postulates [21-23; 37, 57]. The Bivacuum concept is a result of new
interpretation and development of Dirac theory [16], pointing to equal
probability of positive and negative energy in Nature.

The Bivacuum is introduced, as a dynamic superfluid matrix of the Universe,
composed from non-mixing \emph{subquantum particles} of opposite polarization,
separated by an energy gap. The hypothetical \emph{microscopic} subquantum
particles and antiparticles have a dimensions less than the Plank length
(10$^{-33}$ cm), zero mass and charge. They form the infinite number of
\emph{mesoscopic} paired vortices - Bivacuum dipoles of three generations with
Compton radii, corresponding to electrons ($e$), muons ($\mu)$ and tauons
($\tau).$ Only such \emph{mesoscopic} collective excitations of subquantum
particles in form of pairs of rotating fast \emph{torus and antitorus} are
quantized. In turn, these Bivacuum 'molecules' compose the \emph{macroscopic}
superfluid ideal liquid, representing the infinitive Bivacuum matrix.

Each of two strongly correlated 'donuts' of Bivacuum dipoles acquire the
opposite mass charge and magnetic moments, compensating each other in the
absence of symmetry shift between them. The latter condition is valid only for
symmetric \emph{primordial} Bivacuum, where the influence of matter and fields
on Bivacuum is negligible. The sub-elementary fermion and antifermion
origination is a result of the Bivacuum dipole symmetry shift toward the torus
or antitorus, correspondingly. The correlation between paired vortical
structures in a liquid medium was theoretically proved by Kiehn [58].

The infinite number of paired vortical structures: [torus ($\mathbf{V}^{+})$ +
antitorus ($\mathbf{V}^{-})$] with the in-phase clockwise or anticlockwise
rotation are named Bivacuum fermions ($\mathbf{BVF}^{\uparrow}%
\,\mathbf{=\ \mathbf{V}^{+}\upuparrows V}^{-}$)$^{i}$ and Bivacuum
antifermions ($\mathbf{BVF}^{\downarrow}\,\mathbf{=V}^{+}%
\mathbf{\downdownarrows V}^{-}$)$^{i},$ correspondingly. Their intermediate -
transition states are named Bivacuum bosons of two possible polarizations:
($\mathbf{BVB}^{+}=\mathbf{V}^{+}\mathbf{\uparrow\downarrow V}^{-}$)$^{i}$ and
($\mathbf{BVB}^{-}=\mathbf{V}^{+}\mathbf{\downarrow\uparrow V}^{-}$)$^{i}$ The
\emph{positive and negative energies of torus and antitorus }($\pm
\mathbf{E}_{\mathbf{V}^{\pm}})$\emph{ }of\emph{\ }three lepton generations
$(i=e,\mu,\tau$), interrelated with their radiuses ($\mathbf{L}_{\mathbf{V}%
^{\pm}}^{n}),$ are quantized as quantum harmonic oscillators of opposite
energies:
\begin{equation}
\lbrack\mathbf{E}_{\mathbf{V}^{\pm}}^{n}=\mathbf{\pm\,m}_{0}\mathbf{c}%
^{2}(\frac{1}{2}+\mathbf{n)=}\tag{3.1}
\mathbf{=\pm\,\hbar\omega}_{0}(\frac{1}{2}+\mathbf{n)}]^{i}%
\;\;\;\;\;\mathbf{n=\,}0,1,2,3..\mathbf{.}^{^{i}}\nonumber
\end{equation}%
\begin{align}
or:  &  \left[  \text{ }\mathbf{E}_{\mathbf{V}^{\pm}}^{n}=\frac{\pm
\hbar\mathbf{c}}{\mathbf{L}_{\mathbf{V}^{\pm}}^{n}}\right]  ^{^{i}%
}\;\;\;\;\;\;\tag{3.1a}\\
where:  &  \left[  \mathbf{L}_{\mathbf{V}^{\pm}}^{n}=\frac{\pm\hbar}%
{\pm\mathbf{m}_{0}\mathbf{c}(\frac{1}{2}+\mathbf{n)}}=\frac{\mathbf{L}_{0}%
}{\frac{1}{2}+\mathbf{n}}\right] \nonumber
\end{align}

where: $\left[  \mathbf{L}_{0}=\hbar/\mathbf{m}_{0}\mathbf{c}\right]
^{^{e,\mu,\tau}}$ is a Compton radii of the electron of corresponding lepton
generation ($i=e,\mu,\tau)$ and $\mathbf{L}_{0}^{e}>>\mathbf{L}_{0}^{\mu
}>\mathbf{L}_{0}^{\tau}.$ The Bivacuum fermions $\left(  \mathbf{BVF}%
^{\updownarrow}\right)  ^{\mu,\tau}$ with smaller Compton radiuses can be
located inside the bigger ones $\left(  \mathbf{BVF}^{\updownarrow}\right)
^{e}$.

The absolute values of increments of torus and antitorus energies
($\Delta\mathbf{E}_{\mathbf{V}^{\pm}}^{i})$, interrelated with increments of
their radii ($\Delta\mathbf{L}_{\mathbf{V}^{\pm}}^{i})$ in primordial Bivacuum
(i.e. in the absence of matter and field influence), resulting from in-phase
symmetric fluctuations are equal:
\begin{gather}
\Delta\mathbf{E}_{\mathbf{V}^{\pm}}^{i}=-\frac{\hbar c}{\left(  \mathbf{L}%
_{^{\mathbf{V}^{\pm}}}^{i}\right)  ^{2}}\Delta\mathbf{L}_{\mathbf{V}^{\pm}%
}^{i}=-\mathbf{E}_{\mathbf{V}^{\pm}}^{i}\frac{\Delta\mathbf{L}_{\mathbf{V}^{\pm}%
}^{i}}{\mathbf{L}_{\mathbf{V}^{\pm}}^{i}}\;\;\;\;or:\;\;\;\nonumber\tag{3.2}\\
\;-\Delta\mathbf{L}_{\mathbf{V}^{\pm}}^{i}=\frac{\pi\left(  \mathbf{L}%
_{^{\mathbf{V}^{\pm}}}^{i}\right)  ^{2}}{\pi\hbar c}\Delta\mathbf{E}%
_{\mathbf{V}^{\pm}}^{i}=\frac{\mathbf{S}_{\mathbf{BVF}^{\pm}}^{i}}{2h\mathbf{c}}\Delta\mathbf{E}%
_{\mathbf{V}^{\pm}}^{i}=\mathbf{L}_{\mathbf{V}^{\pm}}^{i}\frac{\Delta
\mathbf{E}_{\mathbf{V}^{\pm}}^{i}}{\mathbf{E}_{\mathbf{V}^{\pm}}^{i}}\tag{3.2a}\nonumber
\end{gather}

where: $\mathbf{S}_{\mathbf{BVF}^{\pm}}^{i}=\pi\left(  \mathbf{L}%
_{^{\mathbf{V}^{\pm}}}^{i}\right)  ^{2}$ is a square of the cross-section of
torus and antitorus, forming Bivacuum fermions ($\mathbf{BVF}^{\updownarrow})$
and Bivacuum bosons ($\mathbf{BVB}^{\pm})$.

The virtual \emph{mass}, \emph{charge} and \emph{magnetic moments} of torus
and antitorus of $\mathbf{BVF}^{\updownarrow}$ and $\mathbf{BVB}^{\pm}$ are
opposite and in symmetric\emph{ primordial} Bivacuum compensate each other in
their basic $\mathbf{(n=0}$) and excited $\mathbf{(n=1,2,3...}$) states.

The Bivacuum 'atoms': $\mathbf{BVF}^{\updownarrow}\,\mathbf{=[V}%
^{+}\mathbf{\Updownarrow V}^{-}]^{i}$ and $\mathbf{BVB}^{\pm}=[\mathbf{V}%
^{+}\mathbf{\uparrow\downarrow V}^{-}]^{i}$ represent dipoles of three
different poles - the mass ($\mathbf{m}_{V}^{+}$ $=$ $\left\vert
\mathbf{m}_{V}^{-}\right\vert =\mathbf{m}_{0})^{i}$, electric ($e_{+}$ and
$e_{-})$ and magnetic ($\mathbf{\mu}_{+}$ and $\mathbf{\mu}_{-})$ dipoles.

The torus and antitorus (\textbf{V}$^{+}\Updownarrow$ \textbf{V}$^{-})^{i}$ of
Bivacuum fermions of opposite spins $\mathbf{BVF}^{\uparrow}$ and
$\mathbf{BVF}^{\downarrow}$ are both rotating in the same direction: clockwise
or anticlockwise. This determines the positive and negative spins
($\mathbf{S=\pm1/2\hbar})$ of Bivacuum fermions$.$ Their opposite spins may
compensate each other, forming virtual Cooper pairs: [$\mathbf{BVF}^{\uparrow
}$ $\bowtie$ $\mathbf{BVF}^{\downarrow}]$ with neutral boson properties. The
rotation of adjacent $\mathbf{BVF}^{\uparrow}$ and $\mathbf{BVF}^{\downarrow}$
in Cooper pairs is \emph{side- by- side} in opposite directions, providing
zero resulting spin of such pairs and ability to virtual Bose condensation.
The torus and antitorus of Bivacuum bosons $\mathbf{BVB}^{\pm}=[\mathbf{V}%
^{+}\mathbf{\uparrow\downarrow V}^{-}]^{i}$ with spin, equal to zero, are
rotating in opposite directions.

The \emph{energy gap} between the torus and antitorus of symmetric $\left(
\mathbf{BVF}^{\updownarrow}\right)  ^{i}$ or $\left(  \mathbf{BVB}^{\pm
}\right)  ^{i}$ is:
\begin{equation}
\lbrack\mathbf{A}_{BVF}=\mathbf{E}_{\mathbf{V}^{+}}-(-\mathbf{E}%
_{\mathbf{V}^{-}})=
\mathbf{\hbar\omega}_{0}(1+2\mathbf{n)]}^{i}
=\mathbf{m}_{0}^{i}\mathbf{c}^{2}(1+2\mathbf{n)}=\frac{h\mathbf{c}%
}{\text{[\textbf{d}}_{\mathbf{V}^{+}\Updownarrow\mathbf{V}^{-}}\text{]}%
_{n}^{i}}\nonumber\tag{3.3}
\end{equation}

where the characteristic distance between torus ($\mathbf{V}^{+})^{i}$ and
antitorus ($\mathbf{V}^{-})^{i}$ of Bivacuum dipoles \emph{(gap dimension}) is
a quantized parameter:%
\begin{equation}
\text{\lbrack\textbf{d}}_{\mathbf{V}^{+}\Updownarrow\mathbf{V}^{-}}%
\text{]}_{n}^{i}=\frac{h}{\mathbf{m}_{0}^{i}\mathbf{c(}1+2\mathbf{n)}}
\tag{3.4}%
\end{equation}

>From (3.2) and (3.2a) we can see, that at $\mathbf{n\rightarrow0}$, the
energy gap $\mathbf{A}_{BVF}^{i}$ is decreasing till $\mathbf{\hbar\omega}%
_{0}=\mathbf{m}_{0}^{i}\mathbf{c}^{2}$ and the spatial gap dimension
[\textbf{d}$_{\mathbf{V}^{+}\Updownarrow\mathbf{V}^{-}}$]$_{n}^{i}$ is
increasing up to the Compton length $\;\mathbf{\lambda}_{0}^{i}\mathbf{=h}%
/\mathbf{m}_{0}^{i}\mathbf{c}$. On the contrary, the infinitive symmetric
excitation of torus and antitorus is followed by tending the spatial gap
between them to zero: [\textbf{d}$_{\mathbf{V}^{+}\Updownarrow\mathbf{V}^{-}}%
$]$_{n}^{i}\rightarrow0$ at $\mathbf{n\rightarrow\infty.}$ This means that the
quantization of space and energy of Bivacuum are interrelated and discreet.

The Planck aether hypothesis of Winterberg [59], like Quon vacuum model of
Aspden [24] has some similarity with our Bivacuum model. For example,
Winterberg supposed that space is filled by an equal number of positive and
negative Planck mass particles, each occupied the Planck volume and forming
superfluid medium. By the analogy with conventional theory of superfluid
helium theory, Winterberg conjectured the phonon-roton like spectrum of Planck
aether. The rotons are considered as a small vortex rings of Planck radius.
However, this model is too 'rigid'. It do not take into account the
relativistic effects and can not explain the mass and charge origination of
elementary particles, their duality, fields and a lot of other important
phenomena, following from Bivacuum model, as it will be demonstrated in this
section III of our paper. \bigskip

\begin{center}
\emph{A.2 \ The basic (carrying)Virtual Pressure Waves (VPW}$^{\pm}$\emph{)
and }

\emph{Virtual spin waves (VirSW}$^{\pm1/2})$\emph{ of Bivacuum}\smallskip
\end{center}

The emission and absorption of Virtual clouds ($\mathbf{VC}_{j,k}^{+})^{i}$
and anticlouds ($\mathbf{VC}_{j,k}^{-})^{i}$ in primordial Bivacuum, i.e. in
the absence of matter and fields or where their influence on symmetry of
Bivacuum is negligible, are the result of correlated transitions between
different excitation states ($j,k)$ of torus ($\mathbf{V}_{j,k}^{+})^{i}$ and
antitoruses ($\mathbf{V}_{j,k}^{-})^{i},$ forming symmetric $\left[
\mathbf{BVF}^{\updownarrow}\mathbf{]}^{i}\,\ \text{and }\,\mathbf{[BVB}^{\pm
}\right]  ^{i},\,$corresponding to three lepton generations ($i=e,\mu,\tau
):$\thinspace%
\begin{align}
&  \left(  \mathbf{VC}_{j,k}^{+}\right)  ^{i}\equiv\left[  \mathbf{V}_{j}%
^{+}-\mathbf{V}_{k}^{+}\right]  ^{i}\;-\;virtual\;cloud\tag{3.5}\\
&  \left(  \mathbf{VC}_{j,k}^{-}\right)  ^{i}\equiv\left[  \mathbf{V}_{j}%
^{-}-\mathbf{V}_{k}^{-}\right]  ^{i}\;-\;virtual\;anticloud \tag{3.5a}%
\end{align}

where: $j>k$ are the integer quantum numbers of torus and antitorus excitation states.

The virtual clouds: ($\mathbf{VC}_{j,k}^{+})^{i}$ and ($\mathbf{VC}_{j,k}%
^{-})^{i}$ exist in form of collective excitation of \emph{subquantum}
particles and antiparticles of opposite energies, correspondingly. They can be
considered as 'drops' of virtual Bose condensation of subquantum particles of
positive and negative energy.

The process of [$emission\rightleftharpoons absorption]$ of virtual clouds
should be accompanied by oscillation of \textit{virtual pressure
(}$\mathbf{VirP}^{\pm}$\textit{) and excitation of positive and negative
virtual pressure waves (}$\mathbf{VPW}^{+}$\textit{ and }$\mathbf{VPW}%
^{-})_{j,k}$\textit{. }In primordial Bivacuum the virtual pressure waves of
opposite energies totally compensate each other. However, in asymmetric
secondary Bivacuum, in presence of matter and fields, the total compensation
is absent and the resulting virtual pressure is nonzero [22, 23]: $\left(
\Delta\mathbf{VirP}^{\pm}=\left\vert \mathbf{VirP}^{+}\right\vert -\left\vert
\mathbf{VirP}^{-}\right\vert \right)  >0.$

In accordance with our model of Bivacuum, virtual particles and antiparticles
represent the asymmetric Bivacuum dipoles (BVF$^{\updownarrow}$)$^{as}$ and
(BVB$^{\pm})^{as}\;$of three electron generations ($i=e,\mu,\tau)$ in unstable
state, not corresponding to Golden mean conditions (see subsection III.C).
Virtual particles and antiparticles are the result of correlated and opposite
Bivacuum dipole symmetry fluctuations. Virtual particles, like the real
sub-elementary particles, may exist in Corpuscular and Wave phases (see
subsection III.E). The Corpuscular [C]- phase, represents strongly correlated
pairs of asymmetric torus (V$^{+})$ and antitorus (V$^{-})$ of two different
by absolute values energies. The Wave [W]- phase, results from quantum beats
between these states, which are accompanied by emission or absorption of
Cumulative Virtual Cloud (CVC$^{+}$ or CVC$^{-}),$ formed by subquantum particles.

Virtual particles have a mass, charge, spin, etc., but they differs from real
sub-elementary ones by their lower stability (short life-time) and inability
for fusion to stable triplets (see section 3). They are a singlets or very
unstable triplets or other clusters of Bivacuum dipoles ($\mathbf{BVF}%
^{\updownarrow}$)$^{as}$ in contrast to real fermions-triplets.

For Virtual Clouds ($\mathbf{VC}^{\pm})$ and virtual pressure waves
($\mathbf{VPW}^{\pm})$ excited by them$,$ the relativistic mechanics is not
valid. \textit{Consequently, the causality principle also does not work in a
system (interference pattern) of }$\mathbf{VPW}^{\pm}.$\emph{ }

The quantized energies of positive and negative $\mathbf{VPW}^{+}$ and
$\mathbf{VPW}^{-}$ and corresponding virtual clouds and anticlouds, emitted
$\rightleftharpoons$ absorbed by $\left(  \mathbf{BVF}\right)  ^{i}$ and
$\left(  \mathbf{BVB}\right)  ^{i},$ as a result of their transitions between
$\mathbf{j}$ and $\mathbf{k}$ states can be presented as:
\begin{align}
\mathbf{E}_{\mathbf{VPW}_{j,k}^{+}}^{i}\,\,  &  \mathbf{=\hbar\omega}_{0}%
^{i}\mathbf{(j-k)=m}_{0}^{i}\mathbf{c}^{2}\mathbf{(j-k)}\tag{3.6}\\
\mathbf{E}_{\mathbf{VPW}_{j,k}^{-}}^{i}\,  &  \mathbf{=-\hbar\omega}_{0}%
^{i}\mathbf{(j-k)=-m}_{0}^{i}\mathbf{c}^{2}\mathbf{(j-k)} \tag{3.6a}%
\end{align}

The quantized fundamental Compton frequency of \textbf{VPW}$^{\pm}$:
\begin{equation}
\mathbf{q\,\omega}_{0}^{i}=\mathbf{q\,m}_{0}^{i}\mathbf{c}^{2}/\hbar\tag{3.7}%
\end{equation}
$\mathbf{\,}$where: $\mathbf{q=(j-k)=1,2,3..}$ is the quantization number of
$\mathbf{VPW}_{j,k}^{\pm}$ energy.

In symmetric primordial Bivacuum the total compensation of positive and
negative Virtual Pressure Waves is possible:
\begin{equation}
\mathbf{qE}_{\mathbf{VPW}_{j,k}^{+}}^{i}=-\mathbf{qE}_{\mathbf{VPW}_{j,k}^{-}%
}^{i}=\mathbf{\,q\,\hbar\omega}_{0}^{i} \tag{3.8}%
\end{equation}

The density oscillation of $\mathbf{VC}_{j,k}^{+}\,$ and $\mathbf{VC}%
_{j,k}^{-}$ and virtual particles and antiparticles represent \textit{positive
and negative basic virtual pressure waves }$\mathbf{(VPW}_{j,k}^{+}%
$\textit{\ }and\textit{\ }$\mathbf{VPW}_{j,k}^{-}).$\ \ 

The correlated \emph{virtual Cooper pairs} of adjacent Bivacuum fermions
($\mathbf{BVF}_{S=\pm1/2}^{\updownarrow}$), rotating in opposite direction
with resulting spin, equal to zero and Boson properties, can be presented
as$:$%
\begin{equation}
\mathbf{[BVF}_{S=1/2}^{\uparrow}~\mathbf{\bowtie BVF}_{S=-1/2}^{\downarrow
}\mathbf{]}_{S=0}\,\mathbf{\equiv}\tag{3.9}
\mathbf{[(V}^{+}\mathbf{\upuparrows V}^{-}\mathbf{)\bowtie(V}%
^{+}\mathbf{\downdownarrows V}^{-}\mathbf{)]}_{S=0}\nonumber
\end{equation}

Such a pairs, as well as Bivacuum bosons ($\mathbf{BVB}^{\pm}$) in conditions
of ideal equilibrium, like the \emph{Goldstone bosons,} have zero mass and
spin: $S=0$. The virtual clouds ($\mathbf{VC}_{j,k}^{\pm})$, emitted and
absorbed in a course of correlated transitions of $\mathbf{[BVF}^{\uparrow
}\mathbf{\bowtie BVF}^{\downarrow}\mathbf{]}_{S=0}^{j,k}$ between (j) and (k)
sublevels, excite the virtual pressure waves $\mathbf{VPW}^{+}$ and
$\mathbf{VPW}^{-}.$ They compensate the energy of each other totally in
primordial Bivacuum and partly in \emph{secondary Bivacuum }- in presence of
matter and fields.\medskip

\emph{The nonlocal virtual spin waves }$\mathbf{(VirSW}_{j,k}^{\pm
1/2}\mathbf{),}$ with properties of massless collective Nambu-Goldstone modes,
represent oscillation of equilibrium of Bivacuum fermions with opposite spins,
accompanied by origination of intermediate states - Bivacuum bosons
($\mathbf{BVB}^{\pm})$:
\begin{equation}
\mathbf{VirSW}_{j,k}^{\pm1/2}\mathbf{\,\symbol{126}\,}\Big[\mathbf{BVF}%
^{\uparrow}\mathbf{(V}^{+}\mathbf{\upuparrows V}^{-}%
\mathbf{)}\tag{3.10}
\mathbf{\rightleftharpoons BVB}^{\pm}(\mathbf{V}^{+}\Updownarrow\mathbf{V}%
^{-}\mathbf{)}
\rightleftharpoons\mathbf{BVF}^{\downarrow}\mathbf{(V}^{+}%
\mathbf{\downdownarrows V}^{-}\mathbf{)\Big]}\nonumber
\end{equation}

The $\mathbf{VirSW}_{j,k}^{+1/2}$ and $\mathbf{VirSW}_{j,k}^{-1/2}$ are
excited by $\left(  \mathbf{VC}_{j,k}^{\pm}\right)  _{S=1/2}%
^{\circlearrowright}$ and $\left(  \mathbf{VC}_{j,k}^{\pm}\right)
_{S=-1/2}^{\circlearrowleft}$ $\ $of opposite angular momentums: $S_{\pm
1/2}=\pm\frac{1}{2}\hbar=\pm\frac{1}{2}\mathbf{L}_{0}\mathbf{m}_{0}\mathbf{c}$
and frequency, equal to $\mathbf{VPW}_{j,k}^{\pm}$ (3.7):
\begin{equation}
\mathbf{q\omega}_{\mathbf{VirSW}_{j,k}^{\pm1/2}}^{i}=\mathbf{q\omega
}_{\mathbf{VPW}_{j,k}^{\pm}}^{i}\tag{3.10a}
=\mathbf{qm}_{0}^{i}\mathbf{c}^{2}/\mathbf{\hbar}=\mathbf{\,q\,\omega}_{0}%
^{i}\nonumber
\end{equation}

The most probable basic virtual pressure waves $\mathbf{VPW}_{0}^{\pm}$ and
virtual spin waves $\mathbf{VirSW}_{0}^{\pm1/2}$ correspond to minimum quantum
number $\mathbf{q=(j-k)=1.}$

The $\mathbf{VirSW}_{j,k}^{\pm1/2},$ like so-called torsion field, can serve
as a carrier of the phase/spin (angular momentum) and information -
\emph{qubits}, but not the energy.

The Bivacuum bosons ($\mathbf{BVB}^{\pm}),$ may have two polarizations ($\pm
)$, determined by spin state of their actual torus ($\mathbf{V}^{+})$:
\begin{gather}
\mathbf{BVB}^{+}=\mathbf{(V}^{+}\uparrow\downarrow\mathbf{V}^{-}%
\mathbf{),\;\;}\tag{3.11}\qquad
when\mathbf{\;BVF}^{\uparrow}\rightarrow\mathbf{BVF}^{\downarrow}\nonumber\\
\mathbf{BVB}^{-}=\mathbf{(V}^{+}\downarrow\uparrow\mathbf{V}^{-}%
\mathbf{),\;}\tag{3.11a}\qquad
when\mathbf{\;BVF}^{\downarrow}\rightarrow\mathbf{BVF}^{\uparrow}\nonumber
\end{gather}

The \ Bose-Einstein statistics of energy distribution, valid for system of
weakly interacting bosons (ideal gas), do not work for Bivacuum due to strong
coupling of pairs $[\mathbf{BVF}^{\uparrow}\mathbf{\bowtie BVF}^{\downarrow
}]_{S=0}$ and ($\mathbf{BVB}^{\pm}),$ forming virtual Bose condensate
($\mathbf{VirBC}$) with nonlocal properties. The Bivacuum nonlocal properties
can be proved, using the Virial theorem [22,23]. \smallskip

\begin{center}
\emph{A.3 \ Virtual Bose condensation (VirBC), as a base of Bivacuum
nonlocality}\textbf{\smallskip}
\end{center}

It follows from our model of Bivacuum, that the infinite number of Cooper
pairs of Bivacuum fermions $\mathbf{[BVF}^{\uparrow}\mathbf{\bowtie
BVF}^{\downarrow}\mathbf{]}_{S=0}^{i}$ and their intermediate states -
Bivacuum bosons ($\mathbf{BVB}^{\pm})^{i},$ as elements of Bivacuum, have zero
or very small (in presence of fields and matter) translational momentum:
$\mathbf{p}_{\mathbf{BVF}^{\uparrow}\mathbf{\bowtie BVF}^{\downarrow}}%
^{i}=\mathbf{p}_{\mathbf{BVB}}^{i}\;\rightarrow0$ and corresponding de Broglie
wave length tending to infinity: $\ \mathbf{\lambda}_{\mathbf{VirBC}}%
^{i}\,\mathbf{=h}/\mathbf{p}_{\mathbf{BVF}^{\uparrow}\mathbf{\bowtie
BVF}^{\downarrow},\,\mathbf{BVB}}^{i}\;\mathbf{\rightarrow\infty.}$ It leads
to origination of 3D net of virtual adjacent pairs of virtual microtubules
from Cooper pairs $[\mathbf{BVF}^{\uparrow}\mathbf{\bowtie BVF}^{\downarrow
}]_{S=0},$ rotating in opposite direction and ($\mathbf{BVB}^{\pm})_{S=0}$,
which may form single microtubules, with resulting angular momentum, equal to
zero$.$ These twin and single microtubules, termed Virtual Guides
$\mathbf{(VirG}^{\mathbf{BVF}^{\uparrow}\mathbf{\bowtie BVF}^{\downarrow}}$
and $\mathbf{VirG}^{\mathbf{BVB}^{\pm}}\mathbf{),}$ represent a quasi
one-dimensional Bose condensate with nonlocal properties close to that of
'wormholes' (see section 9). Their radiuses are determined by the Compton
radiuses of the electrons, muons and tauons. Their length is limited only by
decoherence effects. In symmetric Bivacuum, unperturbed by matter and fields,
the length of $\mathbf{VirG}$ may have the order of stars and galactics
separation.\medskip

\textbf{Nonlocality,} as the independence of potential on the distance from
its source in the volume or filaments of virtual or real Bose condensate,
follows from application of the Virial theorem to systems of Cooper pairs of
Bivacuum fermions $\mathbf{[BVF}^{\uparrow}\mathbf{\bowtie BVF}^{\downarrow
}\mathbf{]}_{S=0}$ and Bivacuum bosons $\mathbf{(BVB}^{\pm}\mathbf{)}$ [22,23].

The Virial theorem in general form is correct not only for classical, but also
for quantum systems. It relates the averaged kinetic $\overline{\mathbf{T}%
}_{k}\mathbf{(\vec{v})}=\underset{i}{\sum}\overline{\mathbf{m}_{i}%
\mathbf{v}_{i}^{2}/2}$ \ and potential $\overline{\mathbf{V}}\mathbf{(r)}$
energies of particles, composing these systems:%

\begin{equation}
\mathbf{2}\overline{\mathbf{T}}_{k}\mathbf{(\vec{v}})=\sum_{i}\overline
{\mathbf{m}_{i}\mathbf{v}_{i}^{2}}=\sum_{i}\mathbf{\vec{r}}_{i}%
\mathbf{\partial}\overline{\mathbf{V}}\mathbf{/\partial\vec{r}}_{i} \tag{3.12}%
\end{equation}

If the potential energy $\overline{\mathbf{V}}\mathbf{(r)}$ is a
homogeneous\textbf{ }$\mathbf{\varkappa}-order$ function like:%

\begin{equation}
\overline{\mathbf{V}}\mathbf{(r})\sim\mathbf{r}^{\mathbf{\varkappa}%
},\;\;\text{then \ }\mathbf{n=}\frac{2\overline{\mathbf{T}_{k}}}%
{\overline{\mathbf{V}}\mathbf{(r})} \tag{3.12a}%
\end{equation}

For example, for a harmonic oscillator, when $\overline{\mathbf{T}}%
_{k}\,\mathbf{=}\overline{\mathbf{V}},$we have $\,\mathbf{\varkappa=2}$. \ For
Coulomb interaction: $\mathbf{\varkappa=-1}$ and $\mathbf{\bar{T}=-\bar{V}/2}$.

The important consequence of the Virial theorem is that, if the average
kinetic energy and momentum ($\overline{\mathbf{p}})$ of particles in a
certain volume of a Bose condensate (BC) tends to zero:
\begin{equation}
\overline{\mathbf{T}}_{k}\,\mathbf{=\,}\overline{\mathbf{p}}^{2}%
\mathbf{/2m\rightarrow0} \tag{3.13}%
\end{equation}
the interaction between particles in the volume of BC, characterized by the
radius: $\mathbf{L}_{BC}\,\mathbf{=\,}\left(  \hbar/\overline{\mathbf{p}%
}\right)  \rightarrow0,$ becomes nonlocal, as independent on distance between
them:%
\begin{align}
\overline{\mathbf{V}}\mathbf{(r)}  &  \mathbf{\sim r}^{\mathbf{\varkappa}%
}\mathbf{=1=const\;\ \ }\tag{3.14}\\
\mathbf{at\;\ \ \ \varkappa}  &  \mathbf{=2}\overline{\mathbf{T}}%
_{k}\mathbf{/}\overline{\mathbf{V}}\mathbf{(r)=0}\nonumber
\end{align}

Consequently, it is shown, that nonlocality, as independence of potential on
the distance from potential source, is the inherent property of macroscopic
Bose condensate. The individual\ particles (real, virtual or subquantum) in a
state of Bose condensation are spatially indistinguishable due to the
uncertainty principle. The Bivacuum dipoles $[\mathbf{BVF}^{\uparrow
}\mathbf{\bowtie BVF}^{\downarrow}]_{S=0}$ and ($\mathbf{BVB}^{\pm})_{S=0}$
due to quasi one-dimensional Bose condensation are tending to self-assembly by
'head-to-tail' principle. They compose very long virtual microtubules -
Virtual Guides with wormhole properties. The 3D net of these two kind of
Virtual Guides ($\mathbf{VirG}^{\mathbf{BVF}^{\uparrow}\mathbf{\bowtie
BVF}^{\downarrow}}$ and $\mathbf{VirG}^{\mathbf{BVB}^{\pm}}$) represent the
nonlocal fraction of superfluid Bivacuum.\medskip

\begin{center}
\textbf{III.B Basic postulates of Unified Theory and their consequences}%
\emph{\medskip}
\end{center}

There are three basic postulates in our theory, interrelated with each other
[23]:\emph{\ }

\textbf{I. } The absolute values of internal rotational kinetic energies of
torus and antitorus are equal to each other and to the half of the rest mass
energy of the electrons of corresponding lepton generation, independently on
the external group velocity ($\mathbf{v)}$, turning the symmetric Bivacuum
fermions ($\mathbf{BVF}^{\updownarrow})\,$ to asymmetric ones:%
\begin{equation}
\lbrack\mathbf{I]:\;\;\;}\Big(\frac{1}{2}\mathbf{m}_{V}^{+}(\mathbf{v}%
_{gr}^{in})^{2}\,\tag{3.15}
\mathbf{=~}\frac{1}{2}\left\vert -\mathbf{m}_{V}^{-}\right\vert (\mathbf{v}%
_{ph}^{in})^{2}\,~\mathbf{=~}\frac{1}{2}\mathbf{m}_{0}\mathbf{c}%
^{2}\,\mathbf{=const}\Big)_{in}^{i}\nonumber
\end{equation}

where: $\mathbf{m}_{V}^{+}$ \ and $\mathbf{m}_{V}^{-}$ are the 'actual' -
inertial and 'complementary' - inertialess masses of torus ($\mathbf{V}^{+})$
and antitorus ($\mathbf{V}^{-})$; \ the $\mathbf{v}_{gr}^{in}$ and
$\mathbf{v}_{ph}^{in}$ are the \emph{internal} angular group and phase
velocities of subquantum particles and antiparticles, forming torus and
antitorus, correspondingly. In symmetric conditions of \emph{primordial}
Bivacuum and its virtual dipoles, when the influence of matter and fields is
absent: $\mathbf{v}_{gr}^{in}~\mathbf{=v}_{ph}^{in}\mathbf{=c}$ \ and
$\mathbf{m}_{V}^{+}~\mathbf{=m}_{V}^{-}~\mathbf{=m}_{0}.$

It will be proved in section (G.1) of this section of paper, that above
condition means the infinitive life-time of torus and antitorus of
$\mathbf{BVF}^{\updownarrow}$ and $\mathbf{BVB}^{\pm}$. $\medskip$

\textbf{II. } The internal magnetic moments of torus ($\mathbf{V}^{+})$ and
antitorus ($\mathbf{V}^{-})$ of asymmetric Bivacuum fermions $\mathbf{BVF}%
_{as}^{\uparrow}\,\mathbf{=[V}^{+}\mathbf{\uparrow\uparrow V}^{-}]$ and
antifermions: $\mathbf{BVF}_{as}^{\downarrow}\,\mathbf{=[V}^{+}%
\mathbf{\downarrow\downarrow V}^{-}$], when $\mathbf{v}_{gr}^{in}$ $\neq$
$\mathbf{v}_{ph}^{in},$ $\ \mathbf{m}_{V}^{+}$ $\neq$ $\left\vert
-\mathbf{m}_{V}^{-}\right\vert $ and $\left\vert \mathbf{e}_{+}\right\vert
\neq\left\vert \mathbf{e}_{-}\right\vert ,$ are equal to each other and to
that of Bohr magneton: $\left[  \mathbf{\mu}_{B}=\mathbf{\mu}_{0}%
\,\mathbf{\equiv}\frac{1}{2}\left\vert \mathbf{e}_{0}\right\vert \frac{\hbar
}{\mathbf{m}_{0}\mathbf{c}}\right]  $, independently on their external
translational velocity $(\mathbf{v>0)}$ and symmetry shift. In contrast to
permanent magnetic moments of $\mathbf{V}^{+}$ and $\mathbf{V}^{-},$ their
actual and complementary masses $\mathbf{m}_{V}^{+}$ \ and $\left\vert
-\mathbf{m}_{V}^{-}\right\vert $, internal angular velocities ($\mathbf{v}%
_{gr}^{in}$ and $\mathbf{v}_{ph}^{in})$ and electric charges $\left\vert
\mathbf{e}_{+}\right\vert $ and $\left\vert \mathbf{e}_{-}\right\vert $, are
dependent on ($\mathbf{v)}$, however, they compensate each other variations:
\begin{multline}
\lbrack\mathbf{II]:\;\;\;}\Bigg(\left\vert \mathbf{\pm\mu}_{+}\right\vert
\,\mathbf{\equiv}\frac{1}{2}\left\vert \mathbf{e}_{+}\right\vert
\frac{\left\vert \pm\hbar\right\vert }{\left\vert \mathbf{m}_{V}%
^{+}\right\vert \left(  \mathbf{v}_{gr}^{in}\right)  _{rot}}\,\mathbf{=\,}%
\tag{3.16}
\left\vert \mathbf{\pm\mu}_{-}\right\vert \,\mathbf{\equiv}\frac
{1}{2}\left\vert \mathbf{-e}_{-}\right\vert \frac{\left\vert \pm
\hbar\right\vert }{\left\vert \mathbf{-m}_{V}^{-}\right\vert \left(
\mathbf{v}_{ph}^{in}\right)  _{rot}}\mathbf{=}\nonumber\\
=\mathbf{\mu}_{0}\,\mathbf{\equiv}\frac{1}{2}\left\vert \mathbf{e}%
_{0}\right\vert \frac{\hbar}{\mathbf{m}_{0}\mathbf{c}}\mathbf{=const\Bigg)}%
^{i}\nonumber
\end{multline}

This postulate reflects the condition of the invariance of the spin value,
with respect to the external velocity of Bivacuum fermions.

\textbf{III. }The equality of Coulomb interaction between torus and antitorus
$\mathbf{V}^{+}\Updownarrow\mathbf{V}^{-}$ of primordial Bivacuum dipoles of
all three generations $i=e,$ $\mu,$ $\tau$ (electrons, muons and tauons),
providing uniform electric energy density distribution in Bivacuum:
\begin{equation}
\lbrack\mathbf{III]:\ F}_{0}^{i}=\left(  \frac{\mathbf{e}_{0}^{2}}%
{[\mathbf{d}_{\mathbf{V}^{+}\Updownarrow\mathbf{V}^{-}}^{2}]_{n}}\right)
^{e}=\tag{3.17}
\left(  \frac{\mathbf{e}_{0}^{2}}{[\mathbf{d}_{\mathbf{V}^{+}\Updownarrow
\mathbf{V}^{-}}^{2}]_{n}}\right)  ^{\mu}=\nonumber
\left(  \frac{\mathbf{e}_{0}^{2}}{[\mathbf{d}_{\mathbf{V}^{+}\Updownarrow
\mathbf{V}^{-}}^{2}]_{n}}\right)  ^{\tau}\mathbf{\ \ \ }\nonumber
\end{equation}

where: [$\mathbf{d}_{\mathbf{V}^{+}\Updownarrow\mathbf{V}^{-}}$]$_{n}%
^{i}=\frac{h}{\mathbf{m}_{0}^{i}\mathbf{c(}1+2\mathbf{n)}}$ is the separation
between torus and antitorus of Bivacuum three pole dipoles (3.4) at the same
state of excitation ($n$). A similar condition is valid as well for opposite
magnetic poles interaction; $\left\vert \mathbf{e}_{+}\right\vert ~\left\vert
\mathbf{e}_{-}\right\vert =\mathbf{e}_{0}^{2}.$

The important consequences of postulate \textbf{III} are the following
equalities:
\begin{equation}
\left(  \mathbf{e}_{0}\mathbf{m}_{0}\right)  ^{e}=\left(  \mathbf{e}%
_{0}\mathbf{m}_{0}\right)  ^{\mu}=\left(  \mathbf{e}_{0}\mathbf{m}_{0}\right)
^{\tau}=\tag{3.18}
\sqrt{\left\vert \mathbf{e}_{+}\mathbf{e}_{-}\right\vert \left\vert
\mathbf{m}_{V}^{+}\mathbf{m}_{V}^{-}\right\vert }=const\nonumber
\end{equation}

It means that the toruses and antitoruses of symmetric Bivacuum dipoles of
generations with bigger mass: $\mathbf{m}_{0}^{\mu}=206,7$ $\mathbf{m}_{0}%
^{e};$ $\ \mathbf{m}_{0}^{\tau}=3487,28~\mathbf{m}_{0}^{e}$ \ have
correspondingly smaller charges:%
\begin{equation}
\mathbf{e}_{0}^{\mu}=\mathbf{e}_{0}^{e}(\mathbf{m}_{0}^{e}/\mathbf{m}_{0}%
^{\mu});\ \ \ \ \mathbf{e}_{0}^{\tau}=\mathbf{e}_{0}^{e}(\mathbf{m}_{0}%
^{e}/\mathbf{m}_{0}^{\tau}) \tag{3.19}%
\end{equation}
\ \ 

As is shown in the next section, just these conditions provide \emph{the}
\emph{same charge symmetry shift} of Bivacuum fermions of three generations
($i=e,$ $\mu)$ at the different mass symmetry shift between corresponding
torus and antitorus, determined by Golden mean.

It follows from second postulate, that the resulting magnetic moment of
sub-elementary fermion or antifermion ($\mathbf{\mu}^{\pm})$, equal to the
Bohr's magneton, is interrelated with the actual spin of Bivacuum fermion or
antifermion as:
\begin{equation}
\mathbf{\mu}^{\pm}=\left(  \left\vert \mathbf{\pm\mu}_{+}\right\vert
\left\vert \mathbf{\pm\mu}_{-}\right\vert \right)  ^{1/2}=\mathbf{\mu}%
_{B}=\tag{3.20}
\pm\frac{1}{2}\mathbf{\hbar}\frac{\mathbf{e}_{0}}{\mathbf{m}_{0}\mathbf{c}%
}=\mathbf{S}\frac{\mathbf{e}_{0}}{\mathbf{m}_{0}\mathbf{c}}\nonumber
\end{equation}

where: $\mathbf{e}_{0}/\mathbf{m}_{0}\mathbf{c}$ is gyromagnetic ratio of
Bivacuum fermion, equal to that of the electron.

One may see from (3.20), that the spin of the actual torus, equal to that of
the resulting spin of Bivacuum fermion (symmetric or asymmetric), is:%
\begin{equation}
\mathbf{S=}\pm\frac{1}{2}\mathbf{\hbar} \tag{3.21}%
\end{equation}

Consequently, the permanent absolute value of spin of torus and antitorus is a
consequence of 2nd postulate.

The dependence of the \emph{actual inertial} mass ($\mathbf{m}_{V}^{+})$ of
torus \textbf{V}$^{+\text{ }}$of asymmetric Bivacuum fermions ($\mathbf{BVF}%
_{as}^{\uparrow}\,\mathbf{=V}^{+}\mathbf{\upuparrows V}^{-}$) on the external
translational group velocity (\textbf{v)} follows relativistic mechanics:%
\begin{equation}
\mathbf{m}_{V}^{+}=\mathbf{m}_{0}/\sqrt{1-\left(  \mathbf{v/c}\right)  ^{2}%
}=\mathbf{m} \tag{3.22}%
\end{equation}
while the \emph{complementary inertialess} mass ($\mathbf{m}_{V}^{-})$ of
antitorus \textbf{V}$^{-}$ has the reverse velocity dependence:
\begin{equation}
-\mathbf{m}_{V}^{-}=i^{2}\mathbf{m}_{V}^{-}=-\mathbf{m}_{0}\sqrt{1-\left(
\mathbf{v/c}\right)  ^{2}} \tag{3.23}%
\end{equation}

where $i^{2}=-1;$ $\ i=\sqrt{-1}$ and complementary mass in terms of
mathematics is the imaginary parameter.

For Bivacuum antifermions $\mathbf{BVF}_{as}^{\downarrow}\,\mathbf{=V}%
^{+}\mathbf{\downdownarrows V}^{-}$ the relativistic dependences of positive
and negative mass are opposite to those described by (3.22) and (3.23) and the
notions of actual and complementary parameters change place.

The product of actual (inertial) and complementary (inertialess) mass is a
constant, equal to the rest mass of particle squared and reflect the
\emph{mass compensation principle}. It means, that increasing of mass/energy
of the torus is compensated by in-phase decreasing of absolute values of these
parameters for antitorus and vice versa:
\begin{equation}
\left\vert \mathbf{m}_{V}^{+}\right\vert ~\left\vert -\mathbf{m}_{V}%
^{-}\right\vert \,\mathbf{=m}_{0}^{2} \tag{3.24}%
\end{equation}

Taking (3.24) and (3.15) into account, we get for the product of the
\emph{internal} group and phase velocities of positive and negative subquantum
particles, forming torus and antitorus, correspondingly:%
\begin{equation}
\mathbf{v}_{gr}^{in}~\mathbf{v}_{ph}^{in}=\mathbf{c}^{2} \tag{3.24a}%
\end{equation}

A similar symmetric relation is reflecting the \emph{charge compensation
principle}:%
\begin{equation}
\left\vert \mathbf{e}_{+}\right\vert ~\left\vert \mathbf{e}_{-}\right\vert
=\mathbf{e}_{0}^{2} \tag{3.24b}%
\end{equation}

For Bivacuum antifermions $\mathbf{BVF}_{as}^{\downarrow}\,\mathbf{=V}%
^{+}\mathbf{\downdownarrows V}^{-}$ the relativistic dependences of positive
and negative charge, like the positive and negative masses of torus and
antitorus are opposite to that of Bivacuum fermions $\mathbf{BVF}%
_{as}^{\uparrow}\,\mathbf{=V}^{+}\mathbf{\upuparrows V}^{-}$ (see eqs.3.29 and
3.29a). The symmetry of Bivacuum bosons $\left(  \mathbf{BVB}^{\pm}%
\mathbf{=V}^{+}\mathbf{\uparrow\downarrow V}^{-}\right)  ^{i}$ of each
electron's generation ($i=e,\mu,\tau)$ can be ideal and independent on
external velocity, due to opposite relativistic effects of their torus and
antitorus, compensating each other.

The product of actual (inertial) and complementary (inertialess) mass is a
constant, equal to the rest mass of particle squared and reflect the
\emph{mass compensation principle}. It means, that increasing of mass/energy
of the torus is compensated by in-phase decreasing of absolute values of these
parameters for antitorus and vice versa.

The difference between absolute values of total actual and the complementary
energies (3.22 and 3.23) is equal to doubled kinetic energy of asymmetric
\textbf{BVF}$_{as}^{\updownarrow}:$%
\begin{equation}
\left\vert \mathbf{m}_{V}^{+}\right\vert \mathbf{c}^{2}-\left\vert
-\mathbf{m}_{V}^{-}\right\vert \mathbf{c}^{2}=(\mathbf{m}_{V}^{+}%
-\mathbf{m}_{V}^{-})\mathbf{c}^{2}=\tag{3.25}
\mathbf{m}_{V}^{+}\mathbf{v}^{2}=2\mathbf{T}_{k}=\frac{\mathbf{m}%
_{0}\mathbf{v}^{2}}{\sqrt{1-\left(  \mathbf{v/c}\right)  ^{2}}}\nonumber
\end{equation}

The fundamental Einstein equation for total energy of particle can be reformed
and extended, using eqs. 3.25 and 3.24:%
\begin{align}
\mathbf{E}_{tot}  &  =\mathbf{m}_{V}^{+}\mathbf{c}^{2}=\mathbf{mc}%
^{2}=\mathbf{m}_{V}^{-}\mathbf{c}^{2}+\mathbf{m}_{V}^{+}\mathbf{v}%
^{2}\tag{3.26}\\
or  &  :\ \mathbf{E}_{tot}=\mathbf{m}_{V}^{+}\mathbf{c}^{2}=\frac
{\mathbf{m}_{0}^{2}}{\mathbf{m}_{V}^{+}}\mathbf{c}^{2}+\mathbf{m}_{V}%
^{+}\mathbf{v}^{2}\tag{3.26a}\\
or  &  :\ \mathbf{E}_{tot}=\mathbf{m}_{V}^{+}\mathbf{c}^{2}
  =\sqrt{1-(\mathbf{v/c)}^{2}}\mathbf{m}_{0}\mathbf{c}^{2}+2\mathbf{T}_{k}
\tag{3.26b}%
\end{align}

The ratio of (3.23) to (3.22), taking into account (3.24), is:%

\begin{equation}
\frac{\left\vert -\mathbf{m}_{V}^{-}\right\vert }{\mathbf{m}_{V}^{+}}%
=\frac{\mathbf{m}_{0}^{2}}{\left(  \mathbf{m}_{V}^{+}\right)  ^{2}}=1-\left(
\frac{\mathbf{v}}{\mathbf{c}}\right)  ^{2} \tag{3.27}%
\end{equation}

It can easily be transformed to the important formula (3.25).\medskip

\begin{center}
\emph{B.1 The relation between the external and internal parameters }

\emph{of Bivacuum fermions \& quantum roots of Golden mean. }

\emph{The rest mass and charge origination}\textbf{\smallskip}
\end{center}

\ The important formula, unifying a lot of internal and external
(translational-rotational) parameters of $\mathbf{BVF}_{as}^{\updownarrow}$,
taking into account that the product of internal and external phase and group
velocities is equal to light velocity squared:%
\begin{equation}
\mathbf{v}_{ph}^{in}\mathbf{v}_{gr}^{in}=\mathbf{v}_{ph}^{ext}\mathbf{v}%
_{gr}^{ext}=\mathbf{c}^{2} \tag{3.28}%
\end{equation}
can be derived from eqs. (3.15 - 3.27):%

\begin{align}
\left(  \frac{\mathbf{m}_{V}^{+}}{\mathbf{m}_{V}^{-}}\right)  ^{1/2}  &
=\frac{\mathbf{m}_{V}^{+}}{\mathbf{m}_{0}}=\nonumber
  \,\frac{\mathbf{v}_{ph}^{in}}{\mathbf{v}_{gr}^{in}}=\left(  \frac
{\mathbf{c}}{\mathbf{v}_{gr}^{in}}\right)  ^{2}=\tag{3.29}\\
&  =\frac{\mathbf{L}^{-}}{\mathbf{L}^{+}}=\frac{\left\vert \mathbf{e}%
_{+}\right\vert }{\left\vert \mathbf{e}_{-}\right\vert }=\left(
\frac{\mathbf{e}_{+}}{\mathbf{e}_{0}}\right)  ^{2}=\nonumber
 \frac{1}{[1-\left(  \mathbf{v}^{2}\mathbf{/c}^{2}\right)  ^{ext}]^{1/2}}
\tag{3.29a}%
\end{align}

where:
\begin{align}
\mathbf{L}_{V}^{+}\,  &  \mathbf{=\hbar/(m}_{V}^{+}\mathbf{v}_{gr}%
^{in}\mathbf{)}\qquad and\tag{3.30}\qquad
\;\;\;\;\;\mathbf{L}_{V}^{-}    =\hbar/(\mathbf{m}_{V}^{-}\mathbf{v}%
_{ph}^{in})\nonumber\\
\mathbf{L}_{0}\,  &  \mathbf{=\,}\left(  \mathbf{L}_{V}^{+}\mathbf{L}_{V}%
^{-}\right)  ^{1/2}=\tag{3.30a}
\hbar/\mathbf{m}_{0}\mathbf{c}\;-\;Compton\;radius\nonumber
\end{align}
\ are the radii of torus ($\mathbf{V}^{+})$, antitorus ($\mathbf{V}^{-})$ and
the resulting radius of $\mathbf{BVF}_{as}^{\updownarrow}\,\mathbf{=[V}%
^{+}\mathbf{\Updownarrow V}^{-}]$, equal to Compton radius, correspondingly.

The relativistic dependence of the actual charge $\mathbf{e}_{+}$ on velocity
of Bivacuum dipole, following from (3.29a), is:%
\begin{equation}
\mathbf{e}_{+}=\frac{\mathbf{e}_{0}}{[1-\left(  \mathbf{v}^{2}\mathbf{/c}%
^{2}\right)  ^{ext}]^{1/4}} \tag{3.31}%
\end{equation}

The influence of relativistic dependence of \emph{real} particles charge on
the resulting charge and electric field density of Bivacuum, which is known to
be electrically quasi neutral vacuum/bivacuum, is negligible for two reasons:

1. Densities of positive and negative real charges (i.e. particles and
antiparticles) are very small and approximately equal, as it follows from Bo
Lehnert approach (see Chapter II). This quasi-equilibrium of opposite charges
is Lorentz invariant;

2. The remnant uncompensated by real antiparticles charges density at any
velocities can be compensated totally by virtual antiparticles and asymmetric
Bivacuum fermions (BVF) of opposite charges.

The ratio of the actual charge to the actual inertial mass, as it follows from
(3.29 and 3.29a), has also the relativistic dependence:
\begin{equation}
\frac{\mathbf{e}_{+}}{\mathbf{m}_{V}^{+}}=\frac{\mathbf{e}_{0}}{\mathbf{m}%
_{0}}[1-\left(  \mathbf{v}^{2}\mathbf{/c}^{2}\right)  ^{ext}]^{1/4} \tag{3.32}%
\end{equation}

The decreasing of this ratio with velocity increasing is weaker, than what
follows from the generally accepted statement, that charge has no relativistic
dependence in contrast to the actual mass $\mathbf{m}_{V}^{+}$. The direct
experimental study of relativistic dependence of this ratio on the external
velocity (\textbf{v) }may confirm the validity of our formula (3.32) and
general approach.

>From eqs. (3.25; 3.29 and 3.29a) we find for mass and charge symmetry shift:%
\begin{align}
\Delta\mathbf{m}_{\pm}  &  =\mathbf{m}_{V}^{+}-\mathbf{m}_{V}^{-}%
=\mathbf{m}_{V}^{+}\left(  \frac{\mathbf{v}}{\mathbf{c}}\right)
^{2}\tag{3.33}\\
\Delta\mathbf{e}_{\pm}  &  =\mathbf{e}_{+}-\mathbf{e}_{-}=\frac{\mathbf{e}%
_{+}^{2}}{\mathbf{e}_{+}+\mathbf{e}_{-}}\left(  \frac{\mathbf{v}}{\mathbf{c}%
}\right)  ^{2} \tag{3.33a}%
\end{align}

These mass and charge symmetry shifts determines the relativistic dependence
of the \emph{effective} mass and charge of the fermions. In direct experiments
only the actual mass ($\mathbf{m}_{V}^{+})$ and charge ($\mathbf{e}_{\pm})$
can be registered. It means that the complementary mass ($\mathbf{m}_{V}^{-})$
and charge are \emph{hidden} for observation.

\medskip The ratio of charge to mass symmetry shifts is:%
\begin{equation}
\frac{\Delta\mathbf{e}_{\pm}}{\Delta\mathbf{m}_{\pm}}=\frac{\mathbf{e}_{+}%
^{2}}{\mathbf{m}_{V}^{+}~\left(  \mathbf{e}_{+}+\mathbf{e}_{-}\right)  }
\tag{3.34}%
\end{equation}

The mass symmetry shift can be expressed via the squared charges symmetry
shift also in the following manner:%
\begin{equation}
\Delta\mathbf{m}_{\pm}=\mathbf{m}_{V}^{+}-\mathbf{m}_{V}^{-}=\mathbf{m}%
_{V}^{+}\frac{\mathbf{e}_{+}^{2}-\mathbf{e}_{-}^{2}}{\mathbf{e}_{+}^{2}}
\tag{3.35}%
\end{equation}

or using (3.27) this formula turns to:%
\begin{equation}
\frac{\mathbf{e}_{+}^{2}-\mathbf{e}_{-}^{2}}{\mathbf{e}_{+}^{2}}%
=\frac{\mathbf{v}^{2}}{\mathbf{c}^{2}} \tag{3.36}%
\end{equation}

When the mass and charge symmetry shifts of Bivacuum dipoles are small and
$\left\vert \mathbf{e}_{+}\right\vert +\left\vert \mathbf{e}_{-}\right\vert
\simeq2\mathbf{e}_{+}\simeq2\mathbf{e}_{0}$, we get from 3.33a for variation
of charge shift:
\begin{equation}
\Delta\mathbf{e}_{\pm}=\mathbf{e}_{+}-\mathbf{e}_{-}=\frac{1}{2}\mathbf{e}%
_{0}\frac{\mathbf{v}^{2}}{\mathbf{c}^{2}} \tag{3.37}%
\end{equation}

The formula, unifying the \emph{internal} and \emph{external} group and phase
velocities of asymmetric Bivacuum fermions ($\mathbf{BVF}_{as}^{\updownarrow
})$, derived from (3.29) and (3.29a), is$:$%

\begin{equation}
\left(  \frac{\mathbf{v}_{gr}^{in}}{\mathbf{c}}\right)  ^{4}=1-\left(
\frac{\mathbf{v}}{\mathbf{c}}\right)  ^{2} \tag{3.38}%
\end{equation}

where: $\left(  \mathbf{v}_{gr}^{ext}\right)  \equiv\mathbf{v}$ is the
external translational-rotational group velocity of $\mathbf{BVF}%
_{as}^{\updownarrow}.$

At the conditions of \textquotedblright Hidden Harmony\textquotedblright,
meaning the equality of the internal and external rotational group and phase
velocities of asymmetric Bivacuum fermions $\mathbf{BVF}_{as}^{\updownarrow}%
$:
\begin{align}
\left(  \mathbf{v}_{gr}^{in}\right)  _{\mathbf{V}^{+}}^{rot}  &  =\left(
\mathbf{v}_{gr}^{ext}\right)  ^{tr}\equiv\mathbf{v}\tag{3.39}\\
\left(  \mathbf{v}_{ph}^{in}\right)  _{\mathbf{V}^{-}}^{rot}  &  =\left(
\mathbf{v}_{ph}^{ext}\right)  ^{tr} \tag{3.39a}%
\end{align}
and introducing the notation:%
\begin{equation}
\left(  \frac{\mathbf{v}_{gr}^{in}}{\mathbf{c}}\right)  ^{2}=\left(
\frac{\mathbf{v}}{\mathbf{c}}\right)  ^{2}=\left(  \frac{\mathbf{v}_{gr}^{in}%
}{\mathbf{v}_{ph}^{in}}\right)  =\left(  \frac{\mathbf{v}_{gr}^{ext}%
}{\mathbf{v}_{ph}^{ext}}\right)  \equiv\phi\tag{3.40}%
\end{equation}
\ formula (3.38) turns to a simple quadratic equation:%
\begin{gather}
\phi^{2}+\phi-1=0,\;\;\tag{3.41}\\
\text{which has a few modes}:\;\;\nonumber\\
\phi=\frac{1}{\phi}-1\;\;\;or:\;\frac{\phi}{(1-\phi)^{1/2}}=1\tag{3.41a}\\
or:\;\;\frac{1}{(1-\phi)^{1/2}}=\frac{1}{\phi} \tag{3.41b}%
\end{gather}
The solution of (3.41), is equal to \textbf{Golden mean}: $(\mathbf{v/c)}%
^{2}=\phi=0.618$. \ \emph{It is remarkable, that the Golden Mean, which plays
so important role on different Hierarchic levels of matter organization: from
elementary particles to galactics and even in our perception of beauty (i.e.
our mentality), has so deep physical roots, corresponding to Hidden Harmony
conditions} (3.39 and 3.39a). Our theory is the first one, elucidating these
roots [21-23]. This important fact points, that we are on the right track.

The overall shape of asymmetric $\left(  \mathbf{BVF}_{as}^{\updownarrow
}=\mathbf{[V}^{+}\Updownarrow\mathbf{V}^{-}]\right)  ^{i}$ is a
\emph{truncated cone }(Fig.3.1) with plane, parallel to the base with radiuses
of torus ($L^{+})$ and antitorus ($L^{-}),$ defined by eq. (3.30).

Using Golden Mean equation in the form (2.22b), we can see, that all the
ratios (3.29 and 3.29a) at Golden Mean conditions turn to:%
\begin{equation}
\Bigg[\left(  \frac{\mathbf{m}_{V}^{+}}{\mathbf{m}_{V}^{-}}\right)
^{1/2}=\frac{\mathbf{m}_{V}^{+}}{\mathbf{m}_{0}}=\,\frac{\mathbf{v}_{ph}^{in}%
}{\mathbf{v}_{gr}^{in}}\tag{3.42}
=\frac{\mathbf{L}^{-}}{\mathbf{L}^{+}}=\frac{\left\vert \mathbf{e}%
_{+}\right\vert }{\left\vert \mathbf{e}_{-}\right\vert }=\left(
\frac{\mathbf{e}_{+}}{\mathbf{e}_{0}}\right)  ^{2}\Bigg]^{\phi}=\frac{1}{\phi
}\nonumber
\end{equation}

where the actual ($e_{+})$ and complementary ($e_{-})$ charges and
corresponding mass at GM conditions are:
\begin{align}
\mathbf{e}_{+}^{\phi}  &  \mathbf{=\,}\mathbf{e}_{0}\mathbf{/\phi}%
^{1/2}\mathbf{;\;\;\;\;\;e}_{-}^{\phi}\,\mathbf{=e}_{0}\mathbf{\phi}%
^{1/2}\;\;\tag{3.43}\\
\left(  \mathbf{m}_{V}^{+}\right)  ^{\phi}\,  &  \mathbf{=\,}\mathbf{m}%
_{0}\mathbf{/\phi;\;\;\;\;\;\;}\left(  \mathbf{m}_{V}^{-}\right)  ^{\phi
\,}\mathbf{=m}_{0}\mathbf{\,\phi} \tag{3.44}%
\end{align}

using (3.44 and 3.41a) it is easy to see, that the difference between the
actual and complementary mass at GM conditions is equal to the rest mass:%
\begin{equation}
~\big[\left\vert \mathbf{\Delta m}_{V}\right\vert ^{\phi}=\mathbf{m}_{V}%
^{+}-\mathbf{m}_{V}^{-}\tag{3.45}
=\mathbf{m}_{0}(1/\mathbf{\phi-\phi)=m}_{0}\big]^{e,\mu,\tau}\nonumber
\end{equation}

\emph{This is an important result, pointing that just a symmetry shift,
determined by the Golden mean conditions, is responsible for origination of
the rest mass of sub-elementary particles of each of three generation
(}$\emph{i=~}e,\mu,\tau)$\emph{.}

\emph{The same is true for charge origination.} The GM difference between
actual and complementary charges, using relation $\mathbf{\phi=(}%
1/\mathbf{\phi}-1),$ determines corresponding minimum charge of sub-elementary
fermions or antifermions (at $\mathbf{v}_{tr}^{ext}\rightarrow\mathbf{0})$: \
\begin{align}
\mathbf{\phi}^{3/2}\mathbf{e}_{0}  &  =\left\vert \Delta\mathbf{e}_{\pm
}\right\vert ^{\phi}=\left\vert \mathbf{e}_{+}\,\mathbf{-\,e}_{-}\right\vert
^{\phi}\equiv\left\vert \mathbf{e}\right\vert ^{\phi}\text{ \ \ }\tag{3.46}\\
\text{where}  &  \text{:}\;\;\left(  \left\vert \mathbf{e}_{+}\right\vert
\left\vert \mathbf{e}_{-}\right\vert \right)  =\mathbf{e}_{0}^{2}\;\;
\tag{3.46a}%
\end{align}

The absolute values of charge symmetry shifts for electron, muon and tauon at
GM conditions are the same. This result determines the equal absolute values
of empirical rest charges of the electron, positron, proton and antiproton.
However, the mass symmetry shifts at GM conditions, equal to the rest mass of
electrons, muons and tauons are very different.

The ratio of charge to mass symmetry shifts at Golden mean (GM) conditions
($\mathbf{v}_{tr}^{ext}=0)$ is a permanent value for all three electron
generations ($e,$ $\mu,$ $\tau).$ The different values of their rest mass are
taken into account by postulate III and it consequences of their rest mass and
charge relations: $\mathbf{e}_{0}^{\mu}=\mathbf{e}_{0}^{e}(\mathbf{m}_{0}%
^{e}/\mathbf{m}_{0}^{\mu});\ \ \mathbf{e}_{0}^{\tau}=\mathbf{e}_{0}%
^{e}(\mathbf{m}_{0}^{e}/\mathbf{m}_{0}^{\tau})$ (see 3.19) :%
\begin{equation}
\Bigg[\frac{\left\vert \mathbf{\Delta e}_{\pm}\right\vert ^{\phi}}{\left\vert
\mathbf{\Delta m}_{V}\right\vert ^{\phi}}=\frac{\left\vert \mathbf{e}%
^{i}\right\vert ^{\phi}}{\mathbf{m}_{0}^{e}}=\frac{\left\vert \mathbf{e}%
_{+}\right\vert ^{\phi}\phi}{\left\vert \mathbf{m}_{V}^{+}\right\vert ^{\phi}%
}\tag{3.47}
=\frac{\mathbf{e}_{0}\mathbf{\phi}^{3/2}}{\mathbf{m}_{0}}=\frac{\mathbf{e}%
_{0}\mathbf{\phi}^{3/2}}{\mathbf{m}_{0}^{\mu,\tau}}\Bigg]^{e,\mu,\tau
}\nonumber
\end{equation}

where: $\left(  \mathbf{m}_{V}^{+}\right)  ^{\phi}$\textbf{\ }$=\mathbf{m}%
_{0}/\mathbf{\phi}$ \ is the actual mass of unpaired sub-elementary fermion in
[C] phase at Golden mean conditions (see next section); \ $\mathbf{e}%
_{0}\equiv\mathbf{e}_{0}^{e}$; \ $\mathbf{m}_{0}^{e}\equiv\mathbf{m}_{0}.$

Formula (3.47) can be considered as a background of permanent value of
gyromagnetic ratio, equal to ratio of magnetic moment of particle to its
angular momentum (spin). For the electron it is:%
\begin{equation}
\mathbf{\Gamma}~\mathbf{=}\frac{\mathbf{e}_{0}}{2\mathbf{m}_{e}\mathbf{c}}
\tag{3.48}%
\end{equation}

When searching for elementary magnetic charges ($g^{-}$ and $g^{+}$) being
symmetric in respect to the electric ones ($e^{-}$ and $e^{+}$) and named
\emph{monopoles}, the Dirac theory leads to the relation $\mathbf{ge=}\frac
{1}{2}\mathbf{nhc}$ \ between the magnetic monopole and the electric charge of
the same signs, where $\mathbf{n}=1,2,3$ is an integer. It follows from this
definition that the minimal magnetic charge (at $\mathbf{n=1}$) is as big as
$g=67.7e$. The mass of the monopole should then be huge, about $10^{16}%
\operatorname{GeV}%
$. All numerous attempts to reveal such particles experimentally have failed.
But the postulate II of our theory (eq.3.16) explains the absence of magnetic
monopoles, through the absence of a symmetry shift between the internal
opposite actual quantities $|\mathbf{\mu}_{V}^{+}|$ and the complementary
$|\mathbf{\mu}_{V}^{-}|$ \ magnetic moments of the virtual torus and
antitorus, which form Bivacuum fermions. This is in contrast to the mass and
charge symmetry shifts.

\emph{The absence of magnetic monopole - spatially localized magnetic charge,
is one of the important consequences of our model of elementary particles, as
far:}
\begin{equation}
\mathbf{\Delta\mu}^{\pm}\,\mathbf{=\mu}_{V}^{+}\,\mathbf{-\,\mu}_{V}%
^{-}\,\mathbf{=0} \tag{3.49}%
\end{equation}
i.e. the magnetic moments of torus $\mathbf{(V}^{+}\mathbf{)}$ and antitorus
$\mathbf{(V}^{-}\mathbf{)}$ symmetry shift are\ always zero, independently on
the external group velocity of elementary particles. This consequence of our
theory is in line with a lot of unsuccessful attempts to reveal
monopole\textbf{\ }experimentally.\textbf{\medskip}

\begin{center}
III. \textbf{C. \ Fusion of triplets of elementary particles from
sub-elementary fermions }

\textbf{and antifermions at Golden mean conditions \medskip}
\end{center}

The fusion of asymmetric sub-elementary fermions and antifermions of $\mu$ and
$\tau$ generations \ to triplets $<[\mathbf{F}_{\uparrow}^{+}\bowtie
\mathbf{F}_{\downarrow}^{-}]_{x,y}+\mathbf{F}_{\updownarrow}^{\pm}>_{z}%
^{e,p},$ corresponding to electrons/positrons, protons/antiprotons and
neutrons/antineutrons becomes possible also at the Golden mean (GM) conditions
(Fig.3.1). It is accompanied by energy release and electronic and hadronic
$e,h-$gluons origination, equal in sum to the mass defect. It was demonstrated
theoretically, that the vortical structures at certain conditions
self-organizes into vortex crystals [60].

\begin{center}%
\begin{center}
\includegraphics[width=0.8\textwidth]%
{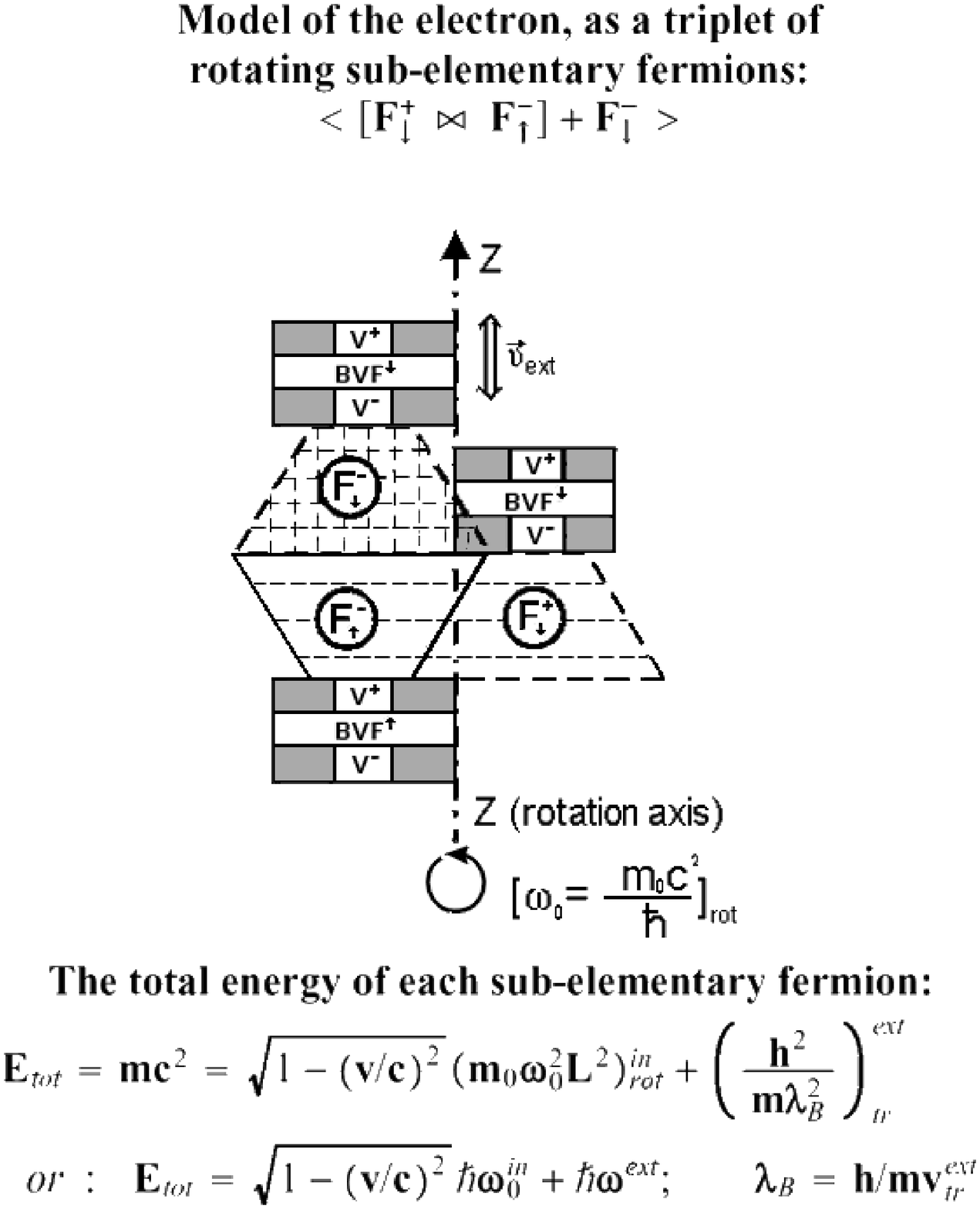}%
\end{center}

\end{center}

\begin{enumerate}
\item \textbf{Fig.3.1 }\ Model of the electron, as a triplets $<[\mathbf{F}%
_{\uparrow}^{+}\bowtie\mathbf{F}_{\downarrow}^{-}]+\mathbf{F}_{\updownarrow
}^{-}>^{e}$, resulting from fusion of unpaired sub-elementary antifermion
$\mathbf{F}_{\updownarrow}^{-}>$ and rotating pair $[\mathbf{F}_{\downarrow
}^{+}\bowtie\mathbf{F}_{\uparrow}^{-}]$ of $\mathbf{\mu}-$generation. The
velocity of rotation of unpaired $\mathbf{F}_{\updownarrow}^{-}>$ around the
same axis of common rotation axis of pair provide the similar mass and charge
symmetry shifts: $\left\vert \mathbf{\Delta m}_{V}\right\vert ^{\phi
}=\mathbf{m}_{0}$\ and $\left\vert \mathbf{\Delta e}^{\pm}\right\vert ^{\phi
}\mathbf{/\phi}^{3/2}~\mathbf{=e}_{0}$, as have the paired sub-elementary
fermions $[\mathbf{F}_{\uparrow}^{+}$ and $\mathbf{F}_{\downarrow}^{-}].$
Three effective anchor $\left(  \mathbf{BVF}^{\updownarrow}\,\mathbf{=[V}%
^{+}\mathbf{\Updownarrow V}^{-}]\right)  _{anc}$ in the vicinity of the
sub-elementary particle base, participate in recoil effects, accompanied
$\mathbf{[C\rightleftharpoons W]}$ pulsation and modulation of basic Bivacuum
pressure waves ($\mathbf{VPW}_{0}^{\pm}).$ The relativistic mass change of
triplets is determined only by symmetry shift of the anchor $\left(
\mathbf{BVF}^{\updownarrow}\,\right)  _{anc}$ of the unpaired sub-elementary
fermion $\mathbf{F}_{\updownarrow}^{\pm}>$ (see section 3.5). \bigskip
\end{enumerate}

\emph{The asymmetry of rotation velocity of torus and antitorus} of
($\mathbf{BVF}_{as}^{\updownarrow}\,\mathbf{=V}^{+}\mathbf{\Updownarrow V}%
^{-}$), is a result of participation of pairs of $\mathbf{BVF}_{as}%
^{\updownarrow}$ of opposite spins $\mathbf{[BVF}^{\uparrow}\mathbf{\bowtie
BVF}^{\downarrow}\mathbf{]}_{S=0}^{i}$ in Bivacuum vorticity. This motion can
be described, as a rolling of pairs of $\mathbf{BVF}_{as}^{\uparrow
}\,\mathbf{=[V}^{+}\mathbf{\upuparrows V}^{-}]$ and $\mathbf{BVF}%
_{as}^{\downarrow}\,\mathbf{=[V}^{+}\mathbf{\downdownarrows V}^{-}]$ with
their \emph{internal} radiuses:
\begin{equation}
\mathbf{L}_{\mathbf{BVF}^{\updownarrow}}^{in}\,\mathbf{=\hbar}/\left\vert
\mathbf{m}_{V}^{+}+\mathbf{m}_{V}^{-}\right\vert _{\mathbf{BVF}^{\updownarrow
}}\mathbf{c} \tag{3.50}%
\end{equation}
around the inside of a larger \emph{external} circle with radius of vorticity:%
\begin{equation}
\mathbf{L}_{\mathbf{BVF}_{as}^{\uparrow}\,\mathbf{\bowtie\,BVF}_{as}%
{}^{\downarrow}}^{ext}\,\tag{3.50a}
\mathbf{=}\frac{\mathbf{\hbar}}{\left\vert \mathbf{m}_{V}^{+}\,\mathbf{-\,m}%
_{V}^{-}\right\vert _{\mathbf{BVF}_{as}^{\uparrow}\,\mathbf{\bowtie\,BVF}%
_{as}{}^{\downarrow}}\cdot\mathbf{c}}\,\nonumber
\mathbf{=\,}\frac{\mathbf{\hbar c}}{\mathbf{m}_{V}^{+}\mathbf{v}%
_{\mathbf{BVF}_{as}^{\uparrow}\,\mathbf{\bowtie\,BVF}_{as}{}^{\downarrow}}%
^{2}}\nonumber
\end{equation}

The increasing of velocity of vorticity $\mathbf{v}_{\mathbf{vor}}$ decreases
both dimensions: $\mathbf{L}_{in}$ and $\mathbf{L}_{\mathbf{ext}}$ till
minimum vorticity radius, including pair of adjacent Bivacuum fermions
$[\mathbf{BVF}_{as}^{\uparrow}\,\mathbf{\bowtie\,BVF}_{a}{}^{\downarrow
}]_{S=0}^{i}\,$with shape of \emph{two identical truncated cones} of the
opposite orientation of planes with common rotation axis (see Fig.3.1). The
corresponding asymmetry of torus $\mathbf{V}^{+}$\textbf{\ }and\textbf{\ }%
$\mathbf{V}^{-}$ is responsible for resulting mass and charge of
$\mathbf{BVF}_{as}^{\updownarrow}$. The trajectory of a fixed point on each of
two $\mathbf{BVF}_{as}^{\updownarrow},$ participating in such dual rotation,
is \textbf{hypocycloid.}\emph{ }

The ratio of internal and external radii: (3.1) and (3.1a) is equal to the
ratio of potential $\mathbf{V}_{p}$ and kinetic $\mathbf{T}_{k}$ energy of the
de Broglie wave, related with its phase $\mathbf{v}_{ph}$ and group
$\mathbf{v}_{gr}$ velocities, as demonstrated in next section 3.4:
\begin{equation}
\frac{\mathbf{L}_{\mathbf{BVF}^{\updownarrow}}^{in}}{\mathbf{L}_{\mathbf{BVF}%
_{as}^{\uparrow}\,\mathbf{\bowtie\,BVF}_{as}{}^{\downarrow}}^{ext}}%
\tag{3.51}
=\frac{\left\vert \mathbf{m}_{V}^{+}+\mathbf{m}_{V}^{-}\right\vert
_{\mathbf{BVF}^{\updownarrow}}}{\left\vert \mathbf{m}_{V}^{+}\,\mathbf{-\,m}%
_{V}^{-}\right\vert _{\mathbf{BVF}_{as}^{\uparrow}\,\mathbf{\bowtie\,BVF}%
_{as}{}^{\downarrow}}}\nonumber
=\frac{\mathbf{V}_{p}}{\mathbf{T}_{k}}=2\frac{\mathbf{v}_{ph}}{\mathbf{v}%
_{gr}}-1\nonumber
\end{equation}
\emph{\smallskip}

At Golden mean conditions, when $\mathbf{v\rightarrow\phi}^{1/2},$ $\left\vert
\mathbf{m}_{V}^{+}\,\mathbf{-\,m}_{V}^{-}\right\vert _{\mathbf{BVF}%
_{as}^{\uparrow}\,\mathbf{\bowtie\,BVF}_{as}{}^{\downarrow}}\rightarrow
\mathbf{m}_{0};$ $\ \left(  \mathbf{m}_{V}^{+}\right)  ^{\phi}=\mathbf{m}%
_{0}/\phi$ \ and \ $\left(  \mathbf{m}_{V}^{-}\right)  ^{\phi}=\mathbf{m}%
_{0}\phi,$ we get for internal (3.50) and external (3.50a) radiuses:%
\begin{align}
\left(  \mathbf{L}_{\mathbf{BVF}^{\updownarrow}}^{in}\right)  ^{\mathbf{\phi}%
}\,  &  \mathbf{=}\frac{\hbar}{\mathbf{m}_{0}(1+2\mathbf{\phi)c}}%
=\frac{\mathbf{L}_{0}}{2.23}\tag{3.52}\\
\mathbf{L}_{\mathbf{BVF}_{as}^{\uparrow}\,\mathbf{\bowtie\,BVF}_{as}%
{}^{\downarrow}}^{ext}\,  &  \mathbf{=}\frac{\mathbf{\hbar}}{\mathbf{m}%
_{0}\mathbf{c}}\,=\mathbf{L}_{0} \tag{3.52a}%
\end{align}

The performed analysis can be considered, as a strong evidence, pointing to
important role of Golden mean in the process of elementary particle fusion
from sub-elementary constituents.\medskip

\textbf{The model of a photon (boson)}, resulting from fusion (annihilation)
of pairs of fermions triplets (electron + positron), are presented by Fig.3.2.
Such a structure can originate also, as a result of six Bivacuum dipoles
opposite symmetry shift, accompanied by the transitions of the excited atoms
and molecules, i.e. systems: [electrons + nuclears] to the ground state.

There are \emph{two possible ways} to make the rotation of adjacent
sub-elementary fermion and sub-elementary antifermion compatible.

One of them is interaction 'side-by-side', like in the 1st and 3d pairs on
Fig.3.2. In such a case they are rotating in opposite directions and their
angular momenta (spins) compensate each other and the resulting spin of such a
pair is zero. The resulting energy of such a pair of sub-elementary particle
and antiparticle is also zero, because their asymmetry with respect to
Bivacuum is exactly opposite, compensating each other.

The other way of compatibility is interaction 'head-to-tail', like in a
central pair of sub-elementary fermions on Fig.3.2. In this configuration they
rotate in the \emph{same direction} and the sum of their spins is:
$\mathbf{s=\pm1\hbar}$. The energy of this pair is a sum of the \emph{absolute
values} of the energies of sub-elementary fermion and antifermion, as far
their resulting symmetry shift is a sum of the symmetry shifts of each of
them. From formula (3.70) of the next section we can see, that if the velocity
of de Broglie wave is equal to the light velocity or very close to it, the
contribution of the rest mass to its total energy is zero or almost zero. In
such a case, pertinent for photon, its total energy is interrelated with
photon frequency ($\mathbf{\nu}_{ph}$) and length like:
\begin{equation}
\mathbf{E}_{ph}\mathbf{=m}_{ph}\mathbf{c}^{2}=\mathbf{h\nu}_{ph}%
~\mathbf{=h}^{2}\mathbf{/(m}_{ph}\mathbf{\lambda}_{ph}^{2}\mathbf{)}
\tag{3.53}%
\end{equation}

where: $\mathbf{m}_{ph}$ is the effective mass of photon and $\mathbf{\lambda
}_{ph}=\mathbf{c/\nu}_{ph}$ is the photon wave length.

It follows from our model, that the minimum value of the photon effective mass
and energy is equal to the sum of\emph{ absolute values} of minimum
mass/energy of central sub-elementary fermion and antifermion: $\mathbf{E}%
_{ph}^{\min}~\mathbf{=~\mathbf{m}_{ph}^{\min}c}^{2}~=2\mathbf{m}_{0}%
\mathbf{c}^{2}$, i.e. the sum of the rest energy of an electron - positron
pair. This consequence of our model is in accordance with available
experimental data.

The law of energy conservation for elementary particles can be reformulated,
as a \emph{law of resulting Bivacuum symmetry conservation}. The additivity of
different forms of energy, as a consequence of the energy conservation law,
means the additivity of Bivacuum dipoles torus and antitorus energy difference
(i.e. forms of kinetic energy) and sum of their absolute values (forms of
potential energy). These energy conservation roots are illustrated for one
particle case by eqs.(3.60 and 3.60a).

\begin{center}%
\begin{center}
\includegraphics[width=0.8\textwidth]%
{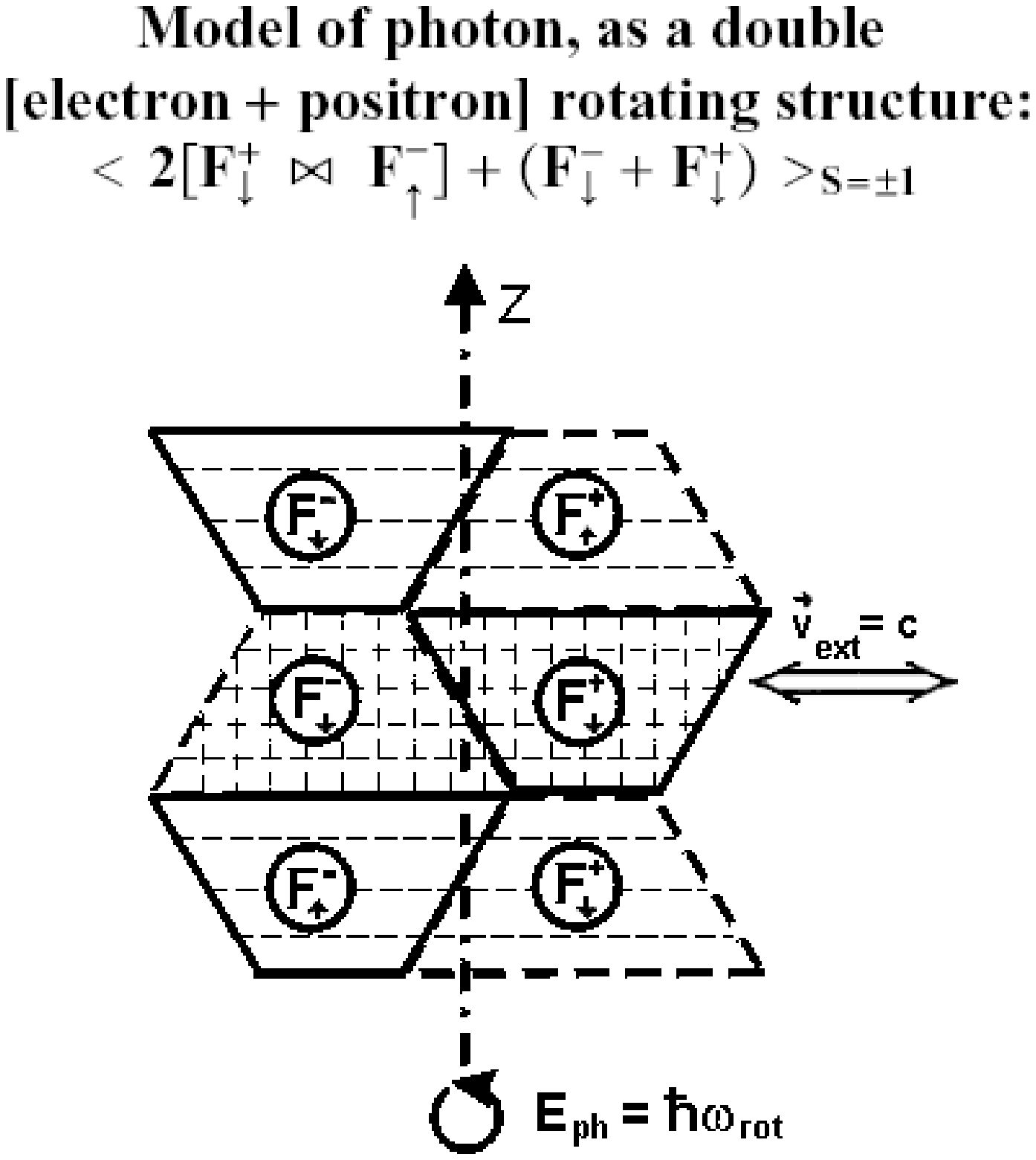}%
\end{center}

\end{center}

\begin{quotation}
\textbf{Fig.3.2} \ Model of photon $\mathbf{<2[F_{\uparrow}^{-}\Join
F_{\downarrow}^{+}]}_{S=0}\mathbf{+\,(F_{\updownarrow}^{-}+F_{\updownarrow
}^{-})}_{S=\pm1}\mathbf{>},$ as result of fusion of electron and positron-like
triplets $<[\mathbf{F}_{\uparrow}^{+}\bowtie\mathbf{F}_{\downarrow}%
^{-}]+\mathbf{F}_{\updownarrow}^{\pm}>$ of sub-elementary fermions , presented
on Fig.3.1. The resulting symmetry shift of such structure is equal to zero,
providing the absence or very close to zero rest mass of photon and its
propagation in primordial Bivacuum with light velocity or very close to it in
the asymmetric secondary Bivacuum. \smallskip
\end{quotation}

We may see, that it has axially symmetric configurations in respect to the
directions of rotation and propagation, which are normal to each other. These
configurations periodically change in the process of sub-elementary fermions
and antifermions correlated $\left[  Corpuscle\leftrightharpoons Wave\right]
$ pulsations in composition of photon (Fig.3.2). The clockwise and counter
clockwise rotation of photons around the z-axes stands for two possible
polarizations of photon. \medskip

\emph{C.1 \ Correlation between new model of hadrons and conventional quark
model of }

\begin{center}
\emph{protons, neutrons and mesons \medskip}
\end{center}

The \emph{proton }($Z=+1;\;S=\pm1/2)$\emph{ is constructed }by the same
principle as the electron (Fig.3.1). It is a result of fusion of pair of
sub-elementary fermion and antifermion $\mathbf{<[F_{\uparrow}^{-}\Join
F_{\downarrow}^{+}]}_{S=0}^{\tau}$ and one unpaired $\left[  \mathbf{\tau
}\,\right]  \,$sub-elementary fermion $\left(  \mathbf{F_{\updownarrow}^{+}%
}\right)  _{S=\pm1/2}^{\tau}$ - \emph{tauons}. These three components of
proton have some similarity with quarks: $\left(  \mathbf{F_{\updownarrow}%
^{+}}\right)  _{S=\pm1/2}^{\tau}\,\symbol{126}\,\,\mathbf{q}^{+}%
\symbol{126}\,\,\mathbf{\tau}^{+}$ and antiquarks $\left(
\mathbf{F_{\updownarrow}^{-}}\right)  _{S=\pm1/2}^{\tau}\,\symbol{126}%
\,\mathbf{q}^{-}\symbol{126}\,\,\mathbf{\tau}^{-}$.

The difference with conventional quark model of protons and neutrons is that
we do not need to use the notion of fractional charge in our proton model:
\begin{align}
\mathbf{p\;}  &  \mathbf{\equiv\,<[F_{\uparrow}^{-}\Join F_{\downarrow}^{+}%
]}_{S=0}\mathbf{+\,}\left(  \mathbf{F_{\uparrow}^{+}}\right)  _{S=\pm
1/2}\mathbf{>}^{\tau}\tag{3.54}\\
or  &  :~\mathbf{p\,}\,\symbol{126}\,\left\langle [\mathbf{q}^{-}\mathbf{\Join
q}^{+}]_{S=0}\,+\left(  \mathbf{q}^{+}\right)  _{S=\pm1/2}\right\rangle
\,\tag{3.54a}\\
or  &  :\;\mathbf{p\,\symbol{126}\,\,}\left\langle [\mathbf{\tau}%
^{-}\mathbf{\Join\tau}^{+}]_{S=0}\,+\left(  \mathbf{\tau}^{+}\right)
_{S=\pm1/2}\right\rangle \tag{3.54b}%
\end{align}
The charges, spins and mass/energy of sub-elementary particles and
antiparticles in pairs $\mathbf{[F_{\uparrow}^{-}\Join F_{\downarrow}^{+}%
]}^{\tau}$ compensate each other. The resulting properties of protons
(\textbf{p}) are determined by unpaired/uncompensated sub-elementary particle
$\mathbf{F_{\uparrow}^{+}>}_{S=\pm1/2}^{\tau}$ \ of heavy $\tau-$electrons
generation (tauons).

\ The \emph{neutron} ($Z=0;\;S=\pm1/2)$ can be presented as:
\begin{align}
\mathbf{n\,\;}  &  \mathbf{\equiv\,<[F_{\uparrow}^{-}\Join F_{\downarrow}%
^{+}]}_{S=0}^{\tau}\,\mathbf{+\,}\tag{3.55}
  \left[  \mathbf{(F_{\uparrow}^{+})}^{\tau}\mathbf{\bowtie
(F_{\downarrow}^{-})}^{e}\right]  _{S=\pm1/2}\mathbf{>}\nonumber\\
or  &  :\mathbf{n}\,\symbol{126}\,\;\left[  \mathbf{q}^{+}\bowtie
\mathbf{q}^{-}\right]  _{S=0}^{\mathbf{\tau}}+\ \left(  \mathbf{q}_{\uparrow
}^{0}\right)  _{S=\pm1/2}^{\mathbf{\tau e}}\tag{3.55a}\\
or  &  :\mathbf{n}\,\symbol{126}\,\;\left[  \mathbf{\tau}^{+}\bowtie
\mathbf{\tau}^{-}\right]  _{S=0}^{\mathbf{\tau}}+\left(  \left[  \mathbf{\tau
}_{\uparrow}^{+}\right]  ^{\mathbf{\tau}}\bowtie\lbrack\mathbf{F}%
_{\updownarrow}^{-}]^{e}\right)  \tag{3.55b}%
\end{align}

where: the neutral quark $\left(  \mathbf{q}_{\uparrow}^{0}\right)
_{S=\pm1/2}^{\mathbf{\tau e}}$ is introduced, as a metastable complex of
positive sub-elementary $\tau-$fermion $\left(  \mathbf{F_{\updownarrow}^{+}%
}\right)  ^{\tau}$ with negative sub-elementary fermion of opposite charge
$[\mathbf{F}_{\updownarrow}^{-}]^{e}$:
\begin{equation}
\left(  \mathbf{q}_{\uparrow}^{0}\right)  _{S=\pm1/2}^{\mathbf{\tau e}%
}=\left(  \left[  \mathbf{q}_{\uparrow}^{+}\right]  \bowtie\lbrack
\mathbf{F}_{\updownarrow}^{-}]^{e}\right)  \tag{3.56}%
\end{equation}

This means that the positive charge of unpaired heavy sub-elementary particle
$\mathbf{(F_{\uparrow}^{+})}^{\tau}$ in neutron ($\mathbf{n}$) is compensated
by the charge of the light sub-elementary fermion $\mathbf{(F_{\downarrow}%
^{-})}^{e}$. In contrast to charge, the spin of unpaired $\mathbf{(F_{\uparrow
}^{+})}^{\tau}$ is not compensated (totally) by spin of
$\mathbf{(F_{\downarrow}^{-})}^{e}$ in neutrons, because of strong mass and
angular momentum difference in conditions of the $\mathbf{(F_{\downarrow}%
^{-})}^{e}$\ confinement. The mass of $\tau$- electron, equal to that of
$\tau$-positron is: $\mathbf{m}_{\tau^{\pm}}=1782(3)$ MeV, the mass of the
regular electron is: $\mathbf{m}_{e^{\pm}}=0,511003(1)$ MeV and the mass of
$\mu-$ electron is: $\mathbf{m}_{\mu^{\pm}}=105,6595(2)\,MeV.$

On the other hand, the mass of proton and neutron are correspondingly:
$\mathbf{m}_{\mathbf{p}}=938,280(3)$ MeV and $\mathbf{m}_{\mathbf{n}}=$
939,573 (3) MeV. They are about two times less than the mass of $\tau$-
electron and equal in accordance to our model to the mass of its unpaired
sub-elementary fermion $\mathbf{(F_{\uparrow}^{+})}^{\tau}.$ This difference
characterizes the energy of neutral massless \emph{gluons} (exchange bosons),
stabilizing the triplets of protons and neutrons. In the case of neutrons this
difference is a bit less (taking into account the mass of $[\mathbf{F}%
_{\updownarrow}^{-}]^{e})$, providing much shorter life-time of isolated
neutrons (918 sec.) than that of protons (%
$>$%
10$^{31}$ years).

In accordance with our \emph{hadron (baryon)} models, each of three quarks
(sub-elementary fermions of $\tau-$ generation) in \textbf{protons} and
\textbf{neutrons} can exist in 3 states (\emph{red}, \emph{green} and
\emph{blue}), but not simultaneously:

1. The \emph{red }state of \textbf{quark/antiquark} means that it is in
corpuscular [C] phase;

2. The \emph{green} state of \textbf{quark/antiquark} means that it is in wave
[W] phase;

3. The \emph{blue }state means that \textbf{quark/antiquark }%
$\mathbf{(F_{\updownarrow}^{\pm})}^{\tau}$ is in the transition
[C]$\Longleftrightarrow$[W] state.

The 8 different combinations of the above defined states of 3 quarks of
protons and neutrons correspond to \emph{8 gluons colors}, stabilizing these
\emph{hadrons }[61]. The gluons with boson properties are represented by pairs
of Cumulative Virtual Clouds ($\mathbf{CVC}^{+}\mathbf{\bowtie CVC}^{-})$,
emitted $\rightleftharpoons$ absorbed\emph{ }in the process of the in-phase
and counterphase $[\mathbf{C\rightleftharpoons W}]$ pulsation of paired quarks
+ antiquark or tauon + antitauon. These 8 gluons, responsible for strong
interaction, can be presented as a different combinations
[C$\leftrightharpoons$W] transition states of $\mathbf{q}^{-}$\textbf{ }and
$\mathbf{q}^{+},$ corresponding to two spin states of proton ($S=\pm
1/2\,\hbar)$, equal to that of unpaired quark - tauon [23].\medskip

\emph{The known experimental values of life-times of}\textbf{ }$\mathbf{\mu}$
and $\mathbf{\tau}$ electrons, corresponding in accordance to our model, to
monomeric asymmetric sub-elementary fermions $\left(  \mathbf{BVF}%
_{as}^{\updownarrow}\right)  ^{\mu,\tau},$ are equal only to $2.19\times
10^{-6}s$ and $3.4\times10^{-13}s$, respectively. We assume here, that
stability of monomeric sub-elementary particles/antiparticles of
$\mathbf{e,\,\mu}$ and $\mathbf{\tau}$ generations, strongly increases as a
result of their fusion in triplets, which became possible at Golden mean conditions.

The well known example of weak interaction, like $\beta-decay$ of the neutron
to proton, electron and $\mathbf{e-}$antineutrino:%
\begin{align}
\mathbf{n\,}  &  \mathbf{\rightarrow p+e}^{-}+\widetilde{\mathbf{\nu}}%
_{e}\tag{3.57}\\
or  &  :\left(  \left[  \mathbf{q}^{+}\bowtie\widetilde{\mathbf{q}}%
^{-}\right]  +\left(  \mathbf{q}_{\uparrow}^{0}\right)  _{S=\pm1/2}%
^{\mathbf{\tau e}}\right)  \rightarrow\tag{3.57a}
  \left(  \left[  \mathbf{q}^{+}\bowtie\widetilde{\mathbf{q}}%
^{-}\right]  +\mathbf{q}^{+}\right)  +\mathbf{e}^{-}+\widetilde{\mathbf{\nu}%
}_{e}\nonumber
\end{align}

is in accordance with our model of elementary particles, neutrino and
antineutrino, are a packet of pairs of interrelated positive and negative
virtual pressure waves $\mathbf{[VPW}^{+}\mathbf{\bowtie VPW}^{-}]$ of two
possible polarization (see section F3).

The reduced sub-elementary fermion of $\tau-$ generation in composition of a
proton or neutron can be considered, as a quark\ and the sub-elementary
antifermion, as an antiquark:
\begin{equation}
\left(  \mathbf{F_{\updownarrow}^{+}}\right)  ^{\tau}\symbol{126}%
\mathbf{\;q}^{+}\;\;\;and\;\;\;\;\left(  \mathbf{F_{\updownarrow}^{-}}\right)
^{\tau}\symbol{126}\mathbf{\;}\widetilde{\mathbf{q}}^{-} \tag{3.58}%
\end{equation}

In the process of $\beta-$decay of the neutron (3.57; 3.57a) the unpaired
negative sub-elementary fermion $[\mathbf{F}_{\updownarrow}^{-}]^{e}$ in the
neutral quark $\left(  \mathbf{q}_{\uparrow}^{0}\right)  _{S=\pm
1/2}^{\mathbf{\tau e}}$ (see 3.56) fuses with asymmetric virtual pair
$\mathbf{[F_{\uparrow}^{-}\Join F_{\downarrow}^{+}]}_{S=0}^{e},$ which emerged
from the vicinal to neutron polarized Cooper pair of Bivacuum fermion and
antifermion. The result of this fusion is a release of the real electron and
electronic antineutrino:
\begin{equation}
\lbrack\mathbf{F}_{\updownarrow}^{-}]^{e}+\mathbf{[F_{\uparrow}^{-}\Join
F_{\downarrow}^{+}]}_{S=0}^{e}\rightarrow\mathbf{e}^{-}+\widetilde
{\mathbf{\nu}}_{e} \tag{3.58a}%
\end{equation}

The antineutrino $\widetilde{\mathbf{\nu}}_{e}$ is a consequence of inelastic
recoil effect in Bivacuum matrix, accompanied the electron fusion and its
separation from hadron. The neutrino and antineutrino of three generation can
be considered as the asymmetric superposition of positive and negative virtual
pressure waves: $\mathbf{[VPW}^{+}\mathbf{\Join~VPW}^{-}]_{e,\mu,\tau}$ with
clockwise or counterclockwise rotation of their plane, which determines
spirality/spin of neutrino ($\pm1/2\hbar)$.

The energy of 8 gluons, corresponding to different superposition of
[$\mathbf{CVC}^{+}\bowtie\mathbf{CVC}^{-}]_{S=0,1}$, emitted and absorbed with
the in-phase $[\mathbf{C\,\rightleftharpoons W}]$ pulsation of pair [quark +
antiquark] in the baryons triplets:
\begin{equation}
\lbrack\mathbf{F}_{\uparrow}^{+}\bowtie\mathbf{F}_{\downarrow}^{-}%
]_{S=0,1}^{\tau}=[\mathbf{q}^{+}\mathbf{+}\widetilde{\mathbf{q}}^{-}]_{S=0,1}
\tag{3.59}%
\end{equation}
is about 50\% of energy/mass of quarks. This explains, why the mass of
isolated unstable tauon and antitauon is about 2 times bigger, than their mass
in composition of proton or neutron.

It looks, that our model of elementary particles is compatible with existing
data and avoid the strong assumption of fractional charge. We anticipate that
future experiments, like deep inelastic scattering inside the hadrons, will be
able to choose between models of fractal and integer charge of the
quarks.\medskip

One of the versions of elementary particle fusion have some similarity with
thermonuclear\emph{ fusion} and can be as follows. The rest mass of
\emph{isolated} sub-elementary fermion/antifermion \emph{before} fusion of the
electron or proton, is equal to the rest mass of unstable muon or tauon,
correspondingly. The 200 times decrease of muons mass is a result of mass
defect, equal to the binding energy of triplets: electrons or positrons. It is
provided by origination of electronic \emph{e-gluons} and release of the huge
amount of excessive kinetic (thermal) energy, for example in form of high
energy photons or \emph{e-neutrino} beams.

In protons, as a result of fusion of three $\tau-$electrons/positrons, the
contribution of hadron \emph{h-gluon} energy to mass defect is only about 50\%
of their mass. However, the absolute hadron fusion energy yield is higher,
than that of the electrons and positrons.

\emph{Our hypothesis of stable electron/positron and hadron fusion from
short-living }$\mu$\emph{ and }$\tau$\emph{ - electrons, as a precursor of
electronic and hadronic quarks, correspondingly, can be verified using special
collider} [22, 23].

In accordance to our Unified Theory, there are two different mechanisms of
stabilization of the electron and proton structures in form of triplets of
sub-elementary fermions/antifermions of the reduced $\mu$ and $\tau$
generations, correspondingly, preventing them from exploding under the action
of self-charge:

1. Each of sub-elementary fermion/antifermion, representing asymmetric pair of
torus ($\mathbf{V}^{+})$ and antitorus ($\mathbf{V}^{-}),$ as a charge,
magnetic and mass dipole, is stabilized by the Coulomb, magnetic and
gravitational attraction between torus and antitorus;

2. The stability of triplet, as a whole, is provided by the exchange of
Cumulative Virtual Clouds (CVC$^{+}$ and CVC$^{-})$ between three
sub-elementary fermions/antifermions in the process of their
$\mathbf{[C\rightleftharpoons W]}$ pulsation. In the case of proton and
neutron, the 8 transition states corresponds to 8 \emph{h-gluons} of hadrons,
responsible for strong interaction. In the case of the electron or positron,
the stabilization of triplets is realized by 8 lighter \emph{e-gluons }[23].
The process of internal exchange of pairs $\mathbf{[F_{\uparrow}^{-}\Join
F_{\downarrow}^{+}]}_{S=0,1}^{e,p}$ with unpaired sub-elementary fermion in
triplets is accompanied by the external energy exchange with Bivacuum dipoles,
modulating positive and negative virtual pressure waves [$\mathbf{VPW}%
^{+}\mathbf{\bowtie VPW}^{-}].$ The feedback reaction between Bivacuum dipoles
and elementary particles is also existing [23]. \bigskip

\begin{center}
\textbf{D. \ Total, potential and kinetic energies of elementary de Broglie
waves\medskip}
\end{center}

The total energy of sub-elementary particles of triplets of the electrons or
protons $\mathbf{<[F_{\uparrow}^{-}\Join F_{\downarrow}^{+}]}_{S=0}%
\mathbf{+\,}\left(  \mathbf{F_{\updownarrow}^{\pm}}\right)  _{S=\pm
1/2}\mathbf{>}^{e,p}$ we can present in three modes, as a sum of total
potential $\mathbf{V}_{tot}$ and kinetic $\mathbf{T}_{tot}$ energies,
including the internal and external contributions$:$%
\begin{align}
\mathbf{E}_{tot}  &  =\mathbf{V}_{tot}+\mathbf{T}_{tot}\,\tag{3.60}
 =\frac{1}{2}(\mathbf{m}_{V}^{+}+\,\mathbf{m}_{V}^{-})\mathbf{c}^{2}%
+\frac{1}{2}(\mathbf{m}_{V}^{+}-\,\mathbf{m}_{V}^{-})\mathbf{c}^{2}\nonumber\\
\mathbf{E}_{tot}  &  =\mathbf{m}_{V}^{+}\mathbf{c}^{2}\tag{3.60a}
 =\frac{1}{2}\mathbf{m}_{V}^{+}(2\mathbf{c}^{2}-\mathbf{v}^{2})+\frac{1}%
{2}\mathbf{m}_{V}^{+}\mathbf{v}^{2}\;\nonumber\\
\mathbf{E}_{tot}  &  =\mathbf{2T}_{k}(\mathbf{v/c)}^{2}=\tag{3.60b}
  \frac{1}{2}\mathbf{m}_{V}^{+}\mathbf{c}^{2}[1+\mathbf{R}^{2}]+\frac{1}%
{2}\mathbf{m}_{V}^{+}\mathbf{v}^{2}\nonumber
\end{align}

where: $\mathbf{R\,}=\mathbf{m}_{0}/\mathbf{m}_{V}^{+}=\,\sqrt
{1-(\mathbf{v/c)}^{2}}$ \ is the dimensionless relativistic factor;
$\mathbf{v}$ is external translational - rotational velocity of particle.
\ One may see, that $\mathbf{E}_{tot}\rightarrow\mathbf{m}_{0}\mathbf{c}%
^{2}\;\;\;$at$\;\ \;\mathbf{v\rightarrow0\;\;}$and$\mathbf{\ \;m}_{V}%
^{+}\rightarrow\mathbf{m}_{0}.$

Taking into account that $\frac{1}{2}(\mathbf{m}_{V}^{+}-\,\mathbf{m}_{V}%
^{-})\mathbf{c}^{2}=\frac{1}{2}\mathbf{m}_{V}^{+}\mathbf{v}^{2}$\ and
$\mathbf{c}^{2}=\mathbf{v}_{gr}\mathbf{v}_{ph},$ where $\mathbf{v}_{gr}%
\equiv\mathbf{v,}$ then dividing the left and right parts of (3.60 and 3.60a)
by $\frac{1}{2}\mathbf{m}_{V}^{+}\mathbf{v}^{2},$ we get:%
\begin{equation}
2\frac{\mathbf{c}^{2}}{\mathbf{v}^{2}}-1=2\frac{\mathbf{v}_{ph}}%
{\mathbf{v}_{gr}}-1\tag{3.61}
=\frac{(\mathbf{m}_{V}^{+}+\,\mathbf{m}_{V}^{-})\mathbf{c}^{2}}{\mathbf{m}%
_{V}^{+}\mathbf{v}^{2}}=\frac{\mathbf{m}_{V}^{+}+\,\mathbf{m}_{V}^{-}%
}{\mathbf{m}_{V}^{+}-\,\mathbf{m}_{V}^{-}}\ \ \ \ \nonumber
\end{equation}

Comparing formula (3.61) with known relation for relativistic de Broglie wave
for ratio of its potential and kinetic energy (Grawford, 1973), we get the
confirmation of our definitions of potential and kinetic energies of
elementary particle in (3.60):
\begin{equation}
2\frac{\mathbf{v}_{ph}}{\mathbf{v}_{gr}}-1=\frac{\mathbf{V}_{tot}}%
{\mathbf{T}_{tot}}=\frac{\mathbf{m}_{V}^{+}+\,\mathbf{m}_{V}^{-}}%
{\mathbf{m}_{V}^{+}-\,\mathbf{m}_{V}^{-}}\ \tag{3.62}%
\end{equation}
\newline In Golden mean conditions, necessary for triplet fusion, the ratio
$(\mathbf{V}_{tot}/\mathbf{T}_{tot})^{\phi}=(1/\phi+\phi)=2.236.$

Consequently, the total potential ($\mathbf{V}_{tot}\mathbf{)}$ and kinetic
($\mathbf{T}_{tot})$ energies of sub-elementary fermions and their increments
can be presented as:
\begin{align}
\mathbf{V}_{tot}\,  &  \mathbf{=}\frac{1}{2}\mathbf{(\mathbf{m}_{V}%
^{+}+\mathbf{m}_{V}^{-})\mathbf{c}^{2}}\tag{3.63}
  \mathbf{=}\frac{1}{2}\mathbf{m}_{V}^{+}~(2\mathbf{c}^{2}-\mathbf{v}%
^{2})=\frac{1}{2}\frac{\mathbf{\hbar c}}{\mathbf{L}_{\mathbf{V}_{tot}}%
}\eqslantgtr\mathbf{V}_{tot}^{\phi};\ \ \ \ \ \nonumber\\
\Delta\mathbf{V}_{tot}  &  =\Delta\mathbf{m}_{V}^{+}\mathbf{c}^{2}%
-\mathbf{\Delta T}_{tot}\tag{3.63a}
 =-\frac{1}{2}\frac{\hbar\mathbf{c}}{\mathbf{L}_{\mathbf{V}_{tot}}}%
\frac{\Delta\mathbf{L}_{\mathbf{V}_{tot}}}{\mathbf{L}_{\mathbf{V}_{tot}}%
}=-\mathbf{V}_{p}\,\frac{\Delta\mathbf{L}_{\mathbf{V}_{tot}}}{\mathbf{L}%
_{\mathbf{V}_{tot}}}\nonumber
\end{align}

where: the characteristic velocity of potential energy, squared, is related to
the group velocity of particle ($\mathbf{v})$, as $\mathbf{v}_{p}%
^{2}=\mathbf{c}^{2}(2-\mathbf{v}^{2}/\mathbf{c}^{2})$ and\ the characteristic
\emph{curvature of potential energy} of elementary particles is:%
\begin{equation}
\mathbf{L}_{\mathbf{V}_{tot}}    =\frac{\hbar}{\mathbf{(\mathbf{m}_{V}%
^{+}+\mathbf{m}_{V}^{-})\mathbf{c}}}\eqslantless\mathbf{L}_{0}^{\phi}\text{
\ \ }\tag{3.64}\qquad
 \text{at }\left(  \frac{\mathbf{v}_{tot}}{\mathbf{c}}\right)
^{2}\eqslantgtr\phi\nonumber
\end{equation}

The total kinetic energy of unpaired sub-elementary fermion of triplets
includes the internal vortical dynamics and external translational one, which
determines their de Broglie wave length, ($\mathbf{\lambda}_{B}=2\pi
\mathbf{L}_{\mathbf{T}_{ext}}):$
\begin{align}
\mathbf{T}_{tot}  &  =\frac{1}{2}\left\vert \mathbf{m}_{V}^{+}\,\mathbf{-\,m}%
_{V}^{-}\right\vert \mathbf{c}^{2}\,\mathbf{=}\frac{1}{2}\mathbf{m}_{V}%
^{+}\mathbf{v}^{2}\tag{3.65}
  =\frac{1}{2}\frac{\mathbf{\hbar c}}{\mathbf{L}_{\mathbf{T}_{tot}}%
}\eqslantgtr\mathbf{T}_{tot}^{\phi};\ \ \ \ \ \ \ \ \ \ \ \ \ \ \nonumber\\
\mathbf{\Delta T}_{tot}  &  =\mathbf{T}_{tot}\frac{1+\mathbf{R}^{2}%
}{\mathbf{R}^{2}}\frac{\Delta\mathbf{v}}{\mathbf{v}}\tag{3.65a}
  =-\mathbf{T}_{k}\,\frac{\Delta\mathbf{L}_{\mathbf{T}_{tot}}}{\mathbf{L}%
_{\mathbf{T}_{tot}}}\nonumber
\end{align}

where the characteristic \emph{curvature of kinetic energy} of sub-elementary
particles in triplets is:%
\begin{equation}
\mathbf{L}_{\mathbf{T}_{tot}}    =\frac{\hbar}{\mathbf{(\mathbf{m}_{V}%
^{+}-\mathbf{m}_{V}^{-})\mathbf{c}}}\eqslantless\mathbf{L}_{0}^{\phi}\text{
~\ \ \ }\tag{3.65b}
  \text{at }\left(  \frac{\mathbf{v}_{tot}}{\mathbf{c}}\right)
^{2}\eqslantgtr\phi\nonumber
\end{equation}

It is important to note, that in compositions of triplets
$\mathbf{<[F_{\uparrow}^{-}\Join F_{\downarrow}^{+}]}_{S=0}\mathbf{+\,}\left(
\mathbf{F_{\updownarrow}^{\pm}}\right)  _{S=\pm1/2}\mathbf{>}^{e,p}$ the
\emph{minimum} values of \emph{total} potential and kinetic energies and the
\emph{maximum} values of their characteristic curvatures correspond to that,
determined by Golden mean conditions (see eqs. 3.52 and 3.52a). In our
formulas above it is reflected by corresponding inequalities. In accordance to
our theory, the Golden mean conditions determine a threshold for triplets
fusion from sub-elementary fermions.

The increment of total energy of elementary particle is a sum of total
potential and kinetic energies increments:%
\begin{equation}
\mathbf{\Delta E}_{tot}=\mathbf{\Delta V}_{tot}+\mathbf{\Delta T}%
_{tot}=\tag{3.66}
-\mathbf{V}_{tot}\,\frac{\Delta\mathbf{L}_{\mathbf{V}_{tot}}}{\mathbf{L}%
_{\mathbf{V}_{tot}}}-\mathbf{T}_{tot}\,\frac{\Delta\mathbf{L}_{\mathbf{T}%
_{tot}}}{\mathbf{L}_{\mathbf{T}_{tot}}}\nonumber
\end{equation}

The well known Dirac equation for energy of a free relativistic particle,
following also from Einstein relativistic formula (3.22), can be easily
derived from (3.60a), multiplying its left and right part on $\mathbf{m}%
_{V}^{+}\mathbf{c}^{2}$ and using introduced mass compensation principle
(3.24):
\begin{equation}
\mathbf{E}_{tot}^{2}=(\mathbf{m}_{V}^{+}\mathbf{c}^{2})^{2}=\left(
\mathbf{m}_{0}\mathbf{c}^{2}\right)  ^{2}+\left(  \mathbf{m}_{V}^{+}\right)
^{2}\mathbf{v}^{2}\mathbf{c}^{2} \tag{3.67}%
\end{equation}

where: $\mathbf{m}_{0}^{2}=\left\vert \mathbf{m}_{V}^{+}~\mathbf{m}_{V}%
^{-}\right\vert $ \ and the actual inertial mass of torus of unpaired
sub-elementary fermion in triplets is equal to regular mass of particle:
$\mathbf{m}_{V}^{+}=\mathbf{m}_{0}.$

Dividing the left and right parts of (3.67) to $\mathbf{m}_{V}^{+}%
\mathbf{c}^{2},$ we may present the total energy of an elementary de Broglie
wave, as a sum of \emph{internal and external} energy contributions, in
contrast to previous sum of \emph{total potential and kinetic} energies (3.60):%

\begin{align}
\mathbf{E}_{tot}  &  =\mathbf{m}_{V}^{+}\mathbf{c}^{2}=\mathbf{E}%
^{in}+\mathbf{E}^{ext}=\tag{3.68}\\
&  =\frac{\mathbf{m}_{0}}{\mathbf{m}_{V}^{+}}\left(  \mathbf{m}_{0}%
\mathbf{c}^{2}\right)  _{rot}^{in}+\left(  \mathbf{m}_{V}^{+}\mathbf{v}%
^{2}\right)  _{tr}^{ext}\tag{3.68a}\\
\mathbf{E}_{tot}  &  =\mathbf{R}\left(  \mathbf{m}_{0}\mathbf{c}^{2}\right)
_{rot}^{in}+\left(  \mathbf{m}_{V}^{+}\mathbf{v}^{2}\right)  _{tr}%
^{ext}\tag{3.69}\\
or  &  :\;\;\mathbf{E}_{tot}=\mathbf{m}_{V}^{+}\mathbf{c}^{2}=(\mathbf{m}%
_{V}^{+}-\,\mathbf{m}_{V}^{-})\mathbf{c}^{4}/\mathbf{v}^{2}\tag{3.69a}\\
\mathbf{E}_{tot}  &  =\mathbf{h\nu}_{C\rightleftharpoons W}=\mathbf{R}\left(
\mathbf{m}_{0}\mathbf{c}^{2}\right)  _{rot}^{in}+\tag{3.70}
\left[  \frac{\mathbf{h}^{2}}{\mathbf{m}_{V}^{+}\lambda_{\mathbf{B}}^{2}%
}\right]  _{tr}^{ext}\nonumber
\end{align}

where: $\mathbf{R}\equiv\sqrt{1-\left(  \mathbf{v/c}\right)  ^{2}}$ is
relativistic factor, dependent on the \emph{external} translational velocity
$(\mathbf{v)}$ of particle; $\mathbf{m}_{V}^{+}=\mathbf{m}_{0}/\mathbf{R}$
$=\mathbf{m\;}$is the actual inertial mass of sub-elementary fermion; the
external translational de Broglie wave length is: $\lambda_{\mathbf{B}}%
=\frac{h}{\mathbf{m}_{V}^{+}\,\mathbf{v}}$ and $\mathbf{\nu}%
_{C\rightleftharpoons W}$ is the resulting frequency of corpuscle - wave
pulsation (see next section).

We can easily transform formula (3.69) to a mode, including the internal
rotational parameters of sub-elementary fermion, necessary for the rest mass
and charge origination:
\begin{equation}
\mathbf{E}_{tot}=\mathbf{R}\,\left(  \mathbf{m}_{0}\mathbf{\omega}_{0}%
^{2}\mathbf{L}_{0}^{2}\right)  _{rot}^{in}\,+\left[  (\mathbf{m}_{V}%
^{+}-\,\mathbf{m}_{V}^{-})\mathbf{c}^{2}\right]  _{tr}^{ext} \tag{3.71}%
\end{equation}

where:\ $\mathbf{L}_{0}=\hbar/\mathbf{m}_{0}\mathbf{c}$ is the Compton radius
of sub-elementary particle; $\mathbf{\omega}_{0}=\mathbf{m}_{0}\mathbf{c}%
^{2}/\hbar$ is the angular Compton frequency of sub-elementary fermion
rotation around the common axis in a triplet (Fig.3.1).

For potential energy of a sub-elementary fermion, we get from (3.69),
assuming, that $\left(  \mathbf{m}_{V}^{+}\mathbf{v}^{2}\right)
=2\mathbf{T}_{tot}$ and $\mathbf{E}_{tot}=\mathbf{V}_{tot}+\mathbf{T}_{tot}:$
\begin{equation}
\mathbf{V}_{tot}=\mathbf{E}_{tot}-\frac{1}{2}\left(  \mathbf{m}_{V}%
^{+}\mathbf{v}^{2}\right)  \tag{3.72}
=\mathbf{R}^{tot}\left(  \mathbf{m}_{0}\mathbf{c}^{2}\right)  _{rot}%
^{in}+\frac{1}{2}\left(  \mathbf{m}_{V}^{+}\mathbf{v}^{2}\right) \nonumber
\end{equation}

The difference between potential and kinetic energies, as analog of
Lagrangian, from (3.71) is:%
\begin{equation}
\mathbf{%
\mathcal{L}%
=V}_{p}-\frac{1}{2}\left(  \mathbf{m}_{V}^{+}\mathbf{v}^{2}\right)
_{tr}^{ext}=\mathbf{R}\left(  \mathbf{m}_{0}\mathbf{c}^{2}\right)
_{\mathbf{rot}}^{in} \tag{3.73}%
\end{equation}

It follows from (3.71 - 3.73), that at $\mathbf{v}\rightarrow\mathbf{c,}$ the
\emph{total} relativistic factor, involving both the external and internal
translational - rotational dynamics of sub-elementary fermions in triplets:
$\mathbf{R}^{tot}\equiv\sqrt{1-\left(  \mathbf{v}_{tot}\mathbf{/c}\right)
^{2}}\;\rightarrow0$ \ and the rest mass contribution to total energy of
sub-elementary particle also tends to zero: $\mathbf{R}^{tot}\left(
\mathbf{m}_{0}\mathbf{c}^{2}\right)  _{rot}^{in}\rightarrow0.$ Consequently,
the total potential and kinetic energies tend to equality $\mathbf{V}%
_{tot}\rightarrow\mathbf{T}_{tot},$ and the Lagrangian to zero. \emph{This is
a conditions for harmonic oscillations of the photon, propagating in
unperturbed Bivacuum. \medskip}

The important formula for doubled external kinetic energy can be derived from
(3.69), taking into account that the relativistic relation between the actual
and rest mass is $\mathbf{m}_{V}^{+}=\mathbf{m}_{0}/\mathbf{R:}$%
\begin{align}
2\mathbf{T}_{k}  &  =\mathbf{m}_{V}^{+}\mathbf{v}^{2}=\mathbf{m}_{V}%
^{+}\mathbf{c}^{2}-\mathbf{R\,m}_{0}\mathbf{c}^{2}\tag{3.74}
  =\frac{\mathbf{m}_{0}\mathbf{c}^{2}}{\mathbf{R}}(1^{2}-\mathbf{R}%
^{2})\ \ or:\nonumber\\
2\mathbf{T}_{k}  &  =\frac{\mathbf{m}_{0}\mathbf{c}^{2}}{\mathbf{R}%
}(\mathbf{1}-\mathbf{R)(1+R)}\tag{3.74a}
  \mathbf{=(1+R)}\left[  \mathbf{m}_{V}^{+}\mathbf{c}^{2}-\mathbf{\,m}%
_{0}\mathbf{c}^{2}\right] \nonumber
\end{align}

This formula is a background of the introduced in section (III: 7) notion of
\emph{Tuning energy} of Bivacuum Virtual Pressure Waves ($\mathbf{VPW}^{\pm})$.

\ Our expressions (3.60 - 3.74a) are more general, than the well known (3.67),
as far they take into account the properties of both poles (actual and
complementary) of Bivacuum dipoles and subdivide the total energy of particle
to the internal and external or to kinetic and potential ones.\medskip

\begin{center}
\textbf{E.\ The dynamic mechanism of corpuscle-wave duality\medskip}
\end{center}

It is generally accepted, that the manifestation of corpuscle - wave duality
of a particle is dependent on the way in which the observer interacts with a
system. However, the mechanism of duality, as a background of quantum physics,
is still obscure.

\ It follows from our theory, that the [corpuscle (C) $\rightleftharpoons$
wave (W)] duality represents modulation of the internal (hidden) quantum beats
frequency between the asymmetric 'actual' (torus) and 'complementary'
(antitorus) states of sub-elementary fermions or antifermions by the external
- empirical de Broglie wave frequency of these particles [23]. The [C] phase
of each sub-elementary fermions of triplets $<[\mathbf{F}_{\uparrow}%
^{+}\bowtie\mathbf{F}_{\downarrow}^{-}]+\mathbf{F}_{\updownarrow}^{\pm}>^{i}$
of elementary particles, like electrons and protons, exists as a mass and an
electric and magnetic asymmetric dipole.

The [$\mathbf{C\rightarrow W}$] transition is a result of two stages superposition:

\emph{The 1st stage} of transition is a reversible dissociation of charged
sub-elementary fermion in [C] phase $\left(  \mathbf{F}_{\updownarrow}^{\pm
}\right)  _{\mathbf{C}}^{\mathbf{e}^{\pm}}$ to charged Cumulative Virtual
Cloud $(\mathbf{CVC}^{\pm})_{\mathbf{F}_{\updownarrow}^{\pm}}^{\mathbf{e}%
^{\pm}-\mathbf{e}_{anc}^{\pm}}$ of subquantum particles and the
\emph{'anchor'} Bivacuum fermion in \textbf{C }phase $\left(  \mathbf{BVF}%
_{anc}^{\updownarrow}\right)  _{\mathbf{C}}^{\mathbf{e}_{anc}^{\pm}}$:
\begin{equation}
\mathbf{(I)}\emph{:}\text{ \ }\Bigg[\left(  \mathbf{F}_{\updownarrow}^{\pm
}\right)  _{\mathbf{C}}^{\mathbf{e}^{\pm}}\overset
{\text{\textbf{Recoil/Antirecoil}}}{\underset{}{<==========>}}\tag{3.75}
\left(  \mathbf{BVF}_{anc}^{\updownarrow}\right)  _{\mathbf{C}}^{\mathbf{e}%
_{anc}^{\pm}}+(\mathbf{CVC}^{\pm})_{\mathbf{F}_{\updownarrow}^{\pm}%
}^{\mathbf{e}^{\pm}-\mathbf{e}_{anc}^{\pm}}\Bigg]^{i}\nonumber
\end{equation}

where notations $\mathbf{e}^{\pm},$ $\mathbf{e}_{anc}^{\pm}$ and
$\mathbf{e}^{\pm}-\mathbf{e}_{anc}^{\pm}$ mean, correspondingly, the total
charge, the anchor charge and their difference, pertinent to $\mathbf{CVC}%
^{\pm}.$

\emph{The 2nd stage} of [$\mathbf{C\rightarrow W]}$ transition is a reversible
dissociation of the anchor Bivacuum fermion $(\mathbf{BVF}_{anc}%
^{\updownarrow})^{i}=\mathbf{[V}^{+}\Updownarrow\mathbf{V}^{-}]_{anc}^{i}$ to
symmetric $(\mathbf{BVF}^{\updownarrow})^{i}$ and the anchor cumulative
virtual cloud ($\mathbf{CVC}^{\pm})_{\mathbf{BVF}_{anc}^{\updownarrow}},$ with
frequency, equal to the empirical frequency of de Broglie wave of particle$:$
\begin{equation}
\mathbf{(II):}\text{ \ }\left(  \mathbf{BVF}_{anc}^{\updownarrow}\right)
_{\mathbf{C}}^{\mathbf{e}_{anc}^{\pm}}\tag{3.76}
<\overset{\text{\textbf{Recoil/Antirecoil}}}{\underset{}{===========}%
}>\nonumber
\left[  \left(  \mathbf{BVF}^{\updownarrow}\right)  ^{0}+(\mathbf{CVC}^{\pm
})_{\mathbf{BVF}_{anc}^{\updownarrow}}^{\mathbf{e}_{anc}^{\pm}}\right]
_{W}^{i}\nonumber
\end{equation}

The 2nd stage takes a place if $(\mathbf{BVF}_{anc}^{\updownarrow})^{i}$ is
asymmetric, i.e. in the case of nonzero external translational - rotational
velocity of particle, when the magnetic field originates. The beat frequency
of $(\mathbf{BVF}_{anc}^{\updownarrow})^{e,p}$ is equal to that of the
empirical de Broglie wave frequency: $\mathbf{\omega}_{B}=\hbar/(\mathbf{m}%
_{V}^{+}\mathbf{L}_{B}^{2})$. The higher is the external kinetic energy of
fermion, the higher is frequency $\mathbf{\omega}_{B}$ and stronger magnetic
field, generated by the anchor Bivacuum fermion. The frequency of the stage
(II) oscillations modulates the internal frequency of
$\mathbf{[C\rightleftharpoons W]}$ pulsation: $\left[  \mathbf{R~\omega}%
_{0}\,\mathbf{=R~m}_{0}\mathbf{c}^{2}\mathbf{/\hbar}\right]  ^{e,p},$ related
to contribution of the rest mass energy to the total energy of the de Broglie
wave [23].

The $[\mathbf{C\;}\rightleftharpoons\;\mathbf{W}]\;$pulsations of unpaired
sub-elementary fermion $\mathbf{F}_{\updownarrow}^{\pm}>$, of triplets of the
electrons or protons $<[\mathbf{F}_{\uparrow}^{+}\bowtie\mathbf{F}%
_{\downarrow}^{-}]+\mathbf{F}_{\updownarrow}^{\pm}>^{e,p}$ are in counterphase
with the in-phase pulsation of paired sub-elementary fermion and antifermion,
modulating Bivacuum virtual pressure waves ($\mathbf{VPW}^{\pm}):$
\begin{equation}
\text{\ }[\mathbf{F}_{\uparrow}^{+}\bowtie\mathbf{F}_{\downarrow}^{-}%
]_{W}^{e,p}<\overset{\mathbf{CVC}^{+}+\mathbf{CVC}^{-}}{\underset{}{=======}%
}>[\mathbf{F}_{\uparrow}^{+}\bowtie\mathbf{F}_{\downarrow}^{-}]_{C}^{e,p}
\tag{3.77}%
\end{equation}

The basic frequency of $\mathbf{[C\rightleftharpoons W]}$ pulsation,
corresponding to Golden mean conditions, $\mathbf{(v/c)}^{2}%
\mathbf{=0,618=\phi,}$ is equal to that of the 1st stage frequency (5.1) at
zero external translational velocity (\textbf{v}$_{tr}^{ext}=0;$
$\mathbf{R=1}).$ This frequency is the same as the basic Bivacuum virtual
pressure waves ($\mathbf{VPW}_{q=1}^{\pm})$ and virtual spin waves
($\mathbf{VirSW}_{q=1}^{S=\pm1/2})$ frequency (3.10)$:$ $\left[
\mathbf{\omega}_{q=1}\,\mathbf{=m}_{0}\mathbf{c}^{2}\mathbf{/\hbar}\right]
^{i}.$

The empirical parameters of de Broglie wave of elementary particle are
determined by asymmetry of the torus and antitorus of the \emph{anchor}
Bivacuum fermion $(\mathbf{BVF}_{anc}^{\updownarrow})^{e,p}=\mathbf{[V}%
^{+}\Updownarrow\mathbf{V}^{-}]_{anc}^{e,p}$ (Fig.3.1) and the frequency of
its reversible dissociation to symmetric $(\mathbf{BVF}^{\updownarrow})^{i}$
and the anchor cumulative virtual cloud ($\mathbf{CVC}_{anc}^{\pm})-$ stage
(\textbf{II}) of duality mechanism (3.76).

The total energy, charge and spin of triplets - fermions, moving in space with
velocity (\textbf{v)} is determined by the unpaired sub-elementary fermion
$\left(  \mathbf{F}_{\updownarrow}^{\pm}\right)  _{z},$ as far the paired ones
in $[\mathbf{F}_{\uparrow}^{+}\bowtie\mathbf{F}_{\downarrow}^{-}]_{x,y}$ of
triplets compensate each other. From (3.69 and 3.70) it is easy to get:%
\begin{align}
\mathbf{E}_{tot}  &  =\mathbf{m}_{V}^{+}\mathbf{c}^{2}=\hbar\mathbf{\omega
}_{\mathbf{C\rightleftharpoons W}}\tag{3.78}
 =\mathbf{R}\left(  \mathbf{\hbar\omega}_{0}\right)  _{rot}^{in}+\left(
\mathbf{\hbar\omega}_{B}^{ext}\right)  _{tr}=\nonumber
 \mathbf{R}\left(  \mathbf{m}_{0}\mathbf{c}^{2}\right)  _{rot}%
^{in}\,+(\mathbf{m}_{V}^{+}\mathbf{v}^{2})_{tr}^{ext}\nonumber\\
\mathbf{E}_{tot}  &  =\mathbf{m}_{V}^{+}\mathbf{c}^{2}=\mathbf{E}_{rot}%
^{in}+\mathbf{E}_{tr}^{ext}\tag{3.78a}
 =\mathbf{R}\left(  \mathbf{m}_{0}\mathbf{\omega}_{0}^{2}\mathbf{L}_{0}%
^{2}\right)  _{\mathbf{rot}}^{in}+\left(  \frac{\mathbf{h}^{2}}{\mathbf{m}%
_{V}^{+}\mathbf{\lambda}_{B}^{2}}\right)  _{\mathbf{tr}}^{ext}\nonumber
\end{align}

where: $\mathbf{R}=\sqrt{1-(\mathbf{v/c)}^{2}}$ is the relativistic factor;
$\mathbf{v}\equiv\mathbf{v}_{tr}^{ext}$\textbf{\ }is the external
translational group velocity; $\mathbf{\lambda}_{B}=h/\mathbf{m}_{V}%
^{+}\mathbf{v=2\pi L}_{B}$ is the external translational de Broglie wave
length; the actual inertial mass is $\mathbf{m}_{V}^{+}=\mathbf{m}%
=\mathbf{m}_{0}/\mathbf{R;}$ $\ \mathbf{L}_{0}^{i}=\hbar/\mathbf{m}_{0}%
^{i}\mathbf{c}$ $\ $is a Compton radius of the elementary particle.

It follows from our approach, that the fundamental phenomenon of
$\mathbf{corpuscle-wave}$ \textbf{duality} (Fig.3.3) is a result of modulation
of the primary - carrying frequency of the internal
$\mathbf{[C\rightleftharpoons W]}^{in}$ pulsation of individual sub-elementary
fermions (\emph{1st stage}): $\left(  \omega^{in}\right)  ^{i}=\mathbf{R}%
\omega_{0}^{i}=\mathbf{Rm}_{0}^{i}\mathbf{c}^{2}/\hbar=\mathbf{Rc/L}_{0}^{i}$,
by the frequency of the external empirical de Broglie wave of triplet:
$\omega_{B}^{ext}=\mathbf{m}_{V}^{+}\mathbf{v}_{ext}^{2}/\hbar=2\pi
\mathbf{v}_{ext}\mathbf{/L}_{B}$, equal to angular frequency of
$\mathbf{[C\rightleftharpoons W]}_{anc}$ pulsation of the anchor Bivacuum
fermion $(\mathbf{BVF}_{anc}^{\updownarrow})^{i}$ (\emph{2nd stage)}$.$

The contribution of this external dynamics modulation of the internal one,
determined by asymmetry of the \emph{anchor} $(\mathbf{BVF}_{anc}%
^{\updownarrow})^{i}=\mathbf{[V}^{+}\Updownarrow\mathbf{V}^{-}]_{anc}^{i}$ to
the total energy of particle, is determined by second terms in (3.78) and
(3.78a):
\begin{align}
\mathbf{E}_{B}^{ext}  &  =\left(  \mathbf{\hbar\omega}_{B}^{ext}\right)
_{tr}=\left(  \frac{\mathbf{h}^{2}}{\mathbf{m}_{V}^{+}\mathbf{\lambda}_{B}%
^{2}}\right)  _{\mathbf{tr}}=\tag{3.79}
\left[  (\mathbf{m}_{V}^{+}-\,\mathbf{m}_{V}^{-})\mathbf{c}^{2}\right]
_{tr}^{ext}\nonumber\\
&  =\left(  \mathbf{m}_{V}^{+}\mathbf{v}^{2}\right)  _{tr}^{ext}=\left(
\mathbf{m}_{V}^{+}\mathbf{\omega}_{B}^{2}\mathbf{L}_{B}^{2}\right)
_{\mathbf{rot}}^{ext} \tag{3.79a}%
\end{align}

This \emph{external} contribution is increasing with particle acceleration and
tending to light velocity. At $\mathbf{v\rightarrow c,}$ $\left(
\mathbf{m}_{V}^{+}\mathbf{v}^{2}\right)  _{tr}^{ext}\rightarrow\mathbf{m}%
_{V}^{+}\mathbf{c}^{2}=\mathbf{E}_{tot}.$

In contrast to\emph{ external} translational contribution of triplets, the
\emph{internal} rotational contribution of individual unpaired sub-elementary
fermions \emph{inside the triplets} is tending to zero at the same conditions:%
\begin{equation}
\mathbf{E}_{rot}^{in}=\mathbf{R}\left(  \mathbf{\hbar\omega}_{0}\right)
_{rot}^{in}=\mathbf{R}\left(  \mathbf{m}_{0}\mathbf{\omega}_{0}\mathbf{L}%
_{0}^{2}\right)  _{\mathbf{rot}}^{in}=\tag{3.80}
\mathbf{R\,}\left(  \mathbf{m}_{0}\mathbf{c}^{2}\right)  _{\mathbf{rot}}%
^{in}\rightarrow0\text{ \ \ at \ }\mathbf{v\rightarrow c}\nonumber
\end{equation}

as far at $\mathbf{v\rightarrow c}$, the $\mathbf{R}=\sqrt{1-(\mathbf{v/c)}%
^{2}}$ $\rightarrow0.$

For a regular nonrelativistic electron the carrier frequency is $\omega
^{in}=R\omega_{0}^{e}\,\symbol{126}\,10^{21}s^{-1}>>\omega_{B}^{ext}.$
\ However, for relativistic case at $\mathbf{v\rightarrow c,}$ the situation
is opposite: $\omega_{B}^{ext}>>\omega^{in}$ at $\omega^{in}\rightarrow0.$

\begin{center}%
\begin{center}
\includegraphics[width=0.8\textwidth]%
{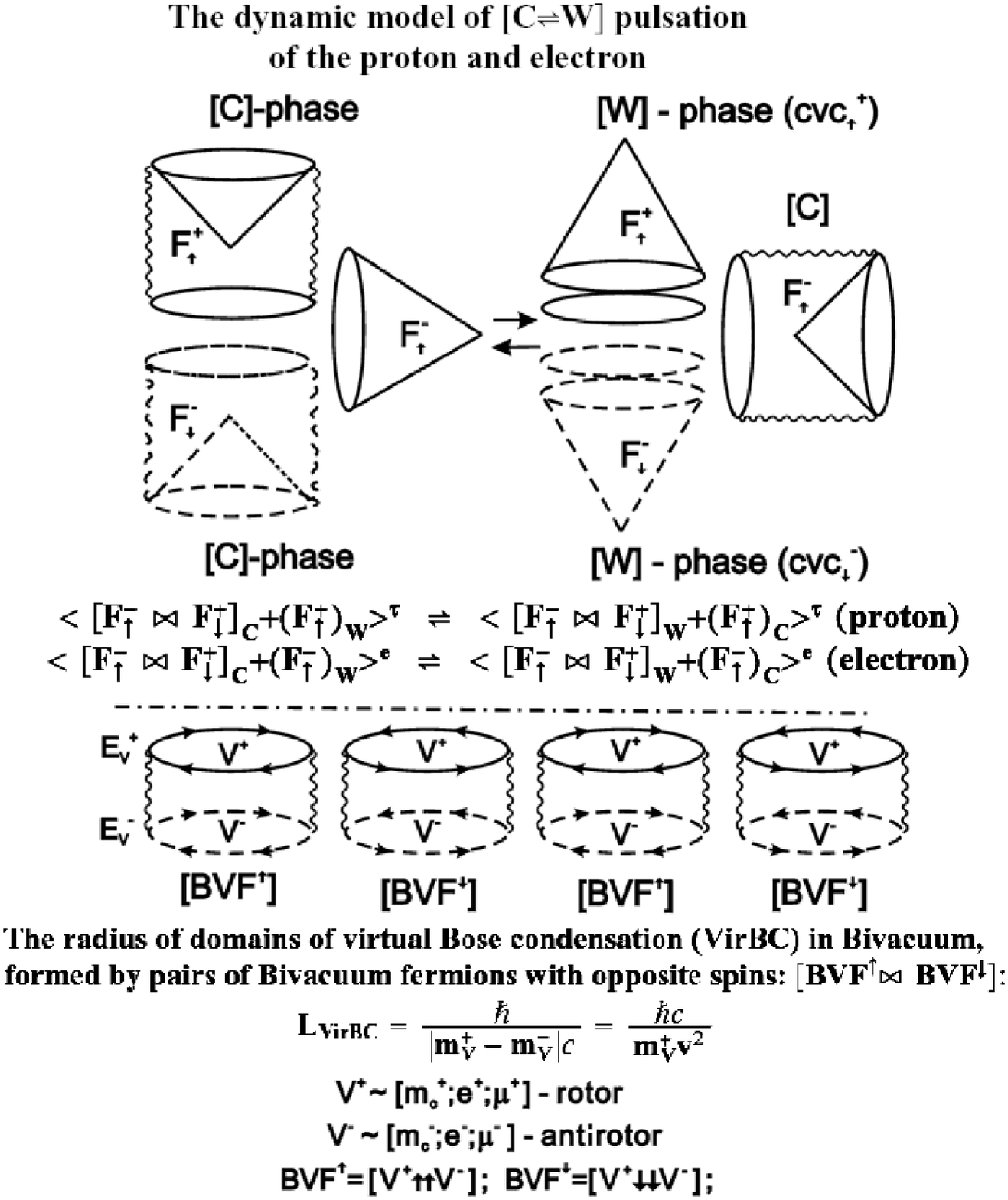}%
\end{center}

\end{center}

\begin{quotation}
\textbf{Fig. 3.3.} \ Dynamic model of \textbf{[}$\mathbf{C\rightleftharpoons
W}]$ pulsation of triplets of sub-elementary fermions/antifermions (the
reduced by fusion to triplets $\mu$ and $\tau$ electrons$)$ composing,
correspondingly, electron and proton $<[\mathbf{F}_{\uparrow}^{+}%
\bowtie\mathbf{F}_{\downarrow}^{-}]+\mathbf{F}_{\updownarrow}^{\pm}>^{e,p}$.
The pulsation of the pair $\mathbf{[F}_{\uparrow}^{-}\mathbf{\Join
F}_{\downarrow}^{+}\mathbf{],}$ modulating virtual pressure waves of Bivacuum
(VPW$^{+}$ and VPW$^{-}),$ is counterphase to pulsation of unpaired
sub-elementary fermion/antifermion $\mathbf{F}_{\updownarrow}^{\pm}>$.\bigskip
\end{quotation}

The properties of the \emph{anchor} Bivacuum fermion $\mathbf{BVF}%
_{anc}^{\updownarrow}$ where analyzed [22], at three conditions:

1. The external translational velocity ($\mathbf{v)}$ is zero;

2. The external translational velocity corresponds to Golden mean
($\mathbf{v=c\phi}^{1/2});$

3. The relativistic case, when $\mathbf{v\sim c.}$

Under nonrelativistic conditions ($\mathbf{v<<c),}$ the de Broglie wave
(modulation) frequency is low: 2$\pi\left(  \mathbf{\nu}_{B}\right)
_{tr}<<(\mathbf{\omega}^{in}=\mathbf{R\omega}_{0})$. However, in relativistic
case ($\mathbf{v\;\symbol{126}\,c)}$, the modulation frequency of the 'anchor'
($\mathbf{BVF}_{anc}^{\updownarrow}$), equal to that of de Broglie wave, can
be higher, than the internal one : 2$\pi\left(  \mathbf{\nu}_{B}\right)
_{tr}\geq\mathbf{\omega}^{in}.$

The paired sub-elementary fermion and antifermion of $\mathbf{[F_{\uparrow
}^{-}\Join F_{\downarrow}^{+}]}_{S=0}$ also have the 'anchor' Bivacuum fermion
and antifermion ($\mathbf{BVF}_{anc}^{\updownarrow}$), similar to that of
unpaired. However, the opposite energies of their $\mathbf{[C}$\textbf{\ }%
$\rightleftharpoons\mathbf{W}]$ pulsation compensate each other in accordance
with proposed model. \bigskip

\begin{center}
\textbf{ F. The nature of electrostatic, magnetic and gravitational
potentials, }

\textbf{based on Unified theory\ \ \medskip}

\emph{F.1 Electromagnetic dipole radiation as a consequence of }

\emph{charge oscillation}
\end{center}

The [$emission\,$\thinspace$\rightleftharpoons absorption]$ of
photons\emph{\ }in a course of elementary fermions - triplets
$\mathbf{<[F_{\uparrow}^{-}\Join F_{\downarrow}^{+}]}_{S=0}\mathbf{+\,}\left(
\mathbf{F_{\uparrow}^{+}}\right)  _{S=\pm1/2}\mathbf{>}^{e,\tau}$ vibrations
can be described by known mechanism of the electric dipole radiation
($\mathbf{\varepsilon}_{EH})$, induced by charge acceleration ($a$), following
from Maxwell equations [68]:
\begin{equation}
\mathbf{\varepsilon}_{EH}=\frac{2e^{2}}{3c^{3}}a^{2} \tag{3.81}%
\end{equation}

The resulting frequency of $[C\rightleftharpoons W]$ pulsation of each of
three sub-elementary fermions in triplets is a sum of internal frequency
contribution ($\mathbf{R\,\omega}_{0}^{in})$ and the external frequency
($\mathbf{\omega}_{B})$ of de Broglie wave from (3.78):
\begin{equation}
\left[  \mathbf{\omega}_{C\rightleftharpoons W}\,\mathbf{=}\,\mathbf{R\,\omega
}_{0}^{in}\mathbf{\,+\omega}_{B}\right]  ^{i} \tag{3.82}%
\end{equation}

where: $\mathbf{R}=\sqrt{1-(\mathbf{v/c)}^{2}}$ is relativistic
factor$\mathbf{.}$

The acceleration can be related only with external translational dynamics
which determines the empirical de Broglie wave parameters of particles.
Acceleration is a result of alternating change of the charge deviation from
the position of equilibrium: $\mathbf{\Delta\lambda}_{B}\mathbf{(t)}=\left(
\mathbf{\lambda}_{B}^{t}\mathbf{-\lambda}_{0}\right)  \sin$\textbf{
}$\mathbf{\omega}_{B}\mathbf{t}$ \ with de Broglie wave frequency of triplets:
$\omega_{B}=\hbar/(m_{V}^{+}L_{B}^{2}),$ where $L_{B}=\hbar/m_{V}%
^{+}\mathbf{v}$. It is accompanied by oscillation of the instant de Broglie
wave length ($\mathbf{\lambda}_{B}^{t})$.

The acceleration of charge in the process of $\mathbf{C\leftrightharpoons W}$
pulsation of the anchor $\mathbf{BVF}_{anc}^{\updownarrow}$ can be expressed
as:%
\begin{equation}
\mathbf{a=\omega}_{B}^{2}\mathbf{\Delta\lambda}_{B}\mathbf{(t)} \tag{3.83}%
\end{equation}
\begin{equation}
\mathbf{a=\omega}_{B}^{2}\left(  \mathbf{\lambda}_{B}\mathbf{-\lambda}%
_{0}\right)  \sin\mathbf{\ \omega}_{B}\mathbf{t} \tag{3.84}%
\end{equation}

where: $\mathbf{\lambda}_{B}^{t}\mathbf{=2\pi L}_{B}$ is the instant de
Broglie wave length of the particle and $\mathbf{\lambda}_{0}\mathbf{=h/m}%
_{0}\mathbf{c}$ is the Compton length of triplet.

The intensity of dipole radiation of pulsing $\mathbf{BVF}_{anc}%
^{\updownarrow}$ from 3.82 and 3.83 is:%

\begin{equation}
\mathbf{\varepsilon}_{EM}\,\mathbf{=\,}\frac{2}{3c^{3}}\mathbf{\omega}_{B}%
^{4}~\left(  \mathbf{d}_{\mathbf{E}}^{t}\right)  ^{2} \tag{3.85}%
\end{equation}

where the oscillating electric dipole moment is: $\mathbf{d}_{\mathbf{E}}%
^{t}=\mathbf{e}\left(  \mathbf{\lambda}_{B}\mathbf{-\lambda}_{0}\right)  .$

Consequently, in accordance with our model of duality, the EM dipole radiation
is due to modulation of the anchor Bivacuum fermion BVF$_{anc}^{\updownarrow}$
frequency of $\mathbf{C\rightleftharpoons W}$ pulsation of unpaired
sub-elementary fermion by thermal vibrations of triplets. The electromagnetic
field, is a result of Corpuscle - Wave pulsation of photon (Fig.3.2) and its
fast rotation with angle velocity ($\omega_{rot}$), equal to pulsation
frequency. The clockwise or anticlockwise direction of photon rotation as
respect to direction of its propagation, corresponds to its two
polarization.$\medskip$

\begin{center}
\emph{F.2 }\ \emph{The different kind of Bivacuum dipoles perturbation,
induced by dynamics of elementary particles \medskip}
\end{center}

In the process of [$\mathbf{C\leftrightharpoons W}$] pulsation of
sub-elementary particles in triplets $<[\mathbf{F}_{\uparrow}^{+}%
\bowtie\mathbf{F}_{\downarrow}^{-}]+\mathbf{F}_{\updownarrow}^{\pm}>^{e,p}$
the [local (internal)$~\Leftrightarrow\,$distant (external)] compensation
effects stand for the energy conservation law. The \emph{local} effects are
pertinent for the [C] phase of particles. They are confined in the volume of
sub-elementary fermions and are determined by gravitational interaction
between opposite actual and complementary mass, Coulomb attraction between
opposite electric charges and magnetic attraction between magnetic moments of
asymmetric torus and antitorus of sub - elementary fermions. These potential
attraction forces are balanced by the energy gap between torus (V$^{+})$ and
antitorus (V$^{-})$ and centripetal force of the axial rotation of triplets.
The axis of triplet rotation is strictly related, in accordance with our
model, with the direction of its translational propagation. It is supposed,
that like magnetic field force lines, this rotation follows the \emph{right
hand screw rule }and is responsible for \emph{magnetic field} origination. The
total energy of triplet, the angular frequency of its rotation and the
velocity of its translational propagation are interrelated.

The \emph{distant} (external) effects, related with certain polarization and
alignment of Bivacuum dipoles (circular and linear) are related with
[$\mathbf{C\rightarrow W]}$ transition of unpaired $\mathbf{F}_{\updownarrow
}^{\pm}>^{e,p}$. It is accompanied by the recoil effect and and the distant
elastic deformation of Bivacuum superfluid matrix, shifting correspondingly a
symmetry of Bivacuum dipoles. The reverse$\ [\mathbf{W}\rightarrow\mathbf{C]}$
transition represents the \emph{antirecoil effect. }The latter is accompanied
by giving back the delocalized recoil energy and restoration of the unpaired
sub-elementary fermion and the whole triplet properties.

The recoil - antirecoil and polarization effects on surrounding Bivacuum
dipoles (BVF$^{\uparrow}$ and BVF$^{\downarrow})$ in form of spherical elastic
waves, excited by $[\mathbf{C\rightleftharpoons W]}$ pulsations of
\emph{unpaired} sub-elementary fermions $\mathbf{F}_{\updownarrow}^{+}>$ and
antifermions\emph{ }$\mathbf{F}_{\updownarrow}^{-}>$ of triplets
$<[\mathbf{F}_{\uparrow}^{+}\bowtie\mathbf{F}_{\downarrow}^{-}]+\mathbf{F}%
_{\updownarrow}^{\pm}>^{e,p},$ are \emph{opposite}. This determines the
opposite effects of positive and negative charged $\mathbf{F}_{\updownarrow
}^{-}>$ on symmetry of Bivacuum torus (\textbf{V}$^{+})$ and antitorus
(\textbf{V}$^{-})$ of Bivacuum dipoles $\mathbf{BVF}^{\updownarrow},$
accompanied by their mass and charge polarization and acquiring the external
momentum. Corresponding energy shift between torus and antitorus is dependent
on distance $(R)$ from charge as ($\overrightarrow{r}/R$\textbf{).} The
induced by such mechanism linear alignment of Bivacuum bosons (\textbf{BVB}%
$^{\pm})$ or Cooper pairs $\mathbf{[BVF}^{\uparrow}\mathbf{\bowtie
BVF}^{\downarrow}]$ with opposite direction of $\mathbf{BVF}^{\uparrow}%
$\textbf{ }and\textbf{ }$\mathbf{BVF}^{\downarrow}$ rotation between remote
$\mathbf{F}_{\updownarrow}^{+}>$ and $\mathbf{F}_{\updownarrow}^{-}>$ of
different triplets, like bundles of virtual microtubules stands for
\emph{electrostatic field and its 'force lines} origination.

On the other hand, pulsations of each of paired sub-elementary fermions\emph{
}$[\mathbf{F}_{\uparrow}^{+}\bowtie\mathbf{F}_{\downarrow}^{-}]$\emph{ }have
similar and symmetric effect on excitation of (\textbf{V}$^{+})$ and
(\textbf{V}$^{-})$ of $\mathbf{BVF}^{\updownarrow}$ $\mathbf{=[V}%
^{+}\mathbf{\Updownarrow V}^{-}\mathbf{],}$ independently on the charge of
unpaired fermion and direction of triplets motion. Corresponding excitation of
positive and negative virtual pressure waves (VPW$^{+}$ and VPW$^{-})$ is a
background of \emph{gravitational} attraction between triplets, like between
pulsing spheres in liquid medium. The influence of the in-phase recoil effect
of pulsing $[\mathbf{F}_{\uparrow}^{+}\bowtie\mathbf{F}_{\downarrow}^{-}]$ on
the external momentum and energy of torus (\textbf{V}$^{+})$ and antitorus
(\textbf{V}$^{-})$ of Bivacuum dipoles $\mathbf{BVF}^{\updownarrow}=\left[
\mathbf{V}^{+}\Updownarrow\mathbf{V}^{-}\right]  $ should be symmetric and
equal by absolute value to increment of energy of unpaired $\mathbf{F}%
_{\updownarrow}^{\pm}>$. \emph{\medskip}

The fast rotation of pairs of sub-elementary fermion and antifermion
$[\mathbf{F}_{\uparrow}^{+}\bowtie\mathbf{F}_{\downarrow}^{-}]$ of triplets
with opposite charges and actual magnetic moments should induce a polarization
of Bivacuum dipoles around the vector of triplets propagation, shifting the
spin equilibrium $\mathbf{[BVF}^{\uparrow}\mathbf{\leftrightharpoons BVB}%
^{\pm}\rightleftharpoons\mathbf{BVF}^{\downarrow}]$ to the left or right and
stimulate their 'head-to-tail' circular structures assembly in form of closed
virtual microtubules. \emph{Our conjecture is that corresponding
circular/axial 'polymerization' of polarized Bivacuum dipoles around direction
of current, coinciding with that of triplets propagation, stands for curled
magnetic field origination.} The fast chaotic thermal motion of conducting
electrons in metals or charged particles in plasma became more ordered in
electric current, increasing correspondingly the magnetic cumulative effects
due to triplet fast rotation in the same plane. \medskip

\textbf{It is possible to present the above explanation of field nature and
origination in more formal way}. The total energies of [$\mathbf{C\rightarrow
W}$] and [$\mathbf{W\rightarrow C}$] transitions of particles we present using
general formula (3.60). However, here we take into account the corresponding
$recoil~\leftrightharpoons~antirecoil$ effects and the reversible conversion
of\emph{ }the internal - local (Loc) gravitational, Coulomb and magnetic
potentials to the external - distant (Dis)\ Bivacuum matrix
perturbation,\ stimulated by these transitions. The [$\mathbf{C\rightarrow W}%
$] transition, accompanied by three kinds of \emph{recoil} effects, can be
described as:
\begin{align}
\mathbf{E}^{C\rightarrow W}  &  =\mathbf{m}_{V}^{+}\mathbf{c}^{2}%
=\mathbf{V}_{tot}+\tag{3.86}
\left[  (\mathbf{E}_{G})_{\operatorname{Re}c}^{Loc}-(\mathbf{E}%
_{G})_{\operatorname{Re}c}^{Dist}\right]  +\nonumber\\
&  +\mathbf{T}_{tot}+\left[  (\mathbf{E}_{E})_{\operatorname{Re}c}%
^{Loc}-(\mathbf{E}_{E})_{\operatorname{Re}c}^{Dist}\right]  _{tr}%
+\tag{3.86a}\\
&  +\left[  (\mathbf{E}_{H})_{\operatorname{Re}c}^{Loc}-(\mathbf{E}%
_{H})_{\operatorname{Re}c}^{Dist}\right]  _{rot} \tag{3.86b}%
\end{align}

In the process of the reverse [$\mathbf{W\rightarrow C}$] transition the
unpaired sub-elementary fermion $\mathbf{F}_{\updownarrow}^{\pm}>$ of triplet
$\mathbf{<[F_{\uparrow}^{-}\Join F_{\downarrow}^{+}]}_{S=0}\mathbf{+\,}\left(
\mathbf{F_{\updownarrow}^{\pm}}\right)  _{S=\pm1/2}\mathbf{>}$ gets back the
\emph{antirecoil} (\emph{ARec}) energy in form of relaxation of Bivacuum
matrix, transforming back the Bivacuum matrix elastic deformation and
VPW$^{\pm}$ excitation:%

\begin{align}
&  \mathbf{E}^{W\rightarrow C}=\mathbf{m}_{V}^{+}\mathbf{c}^{2}=\nonumber
\mathbf{V}_{tot}+\left[  -(\mathbf{E}_{G})_{A\operatorname{Re}c}%
^{Loc}+(\mathbf{E}_{G})_{A\operatorname{Re}c}^{Dist}\right]  +\tag{3.87}\\
&  +\mathbf{T}_{tot}+\left[  -(\mathbf{E}_{E})_{A\operatorname{Re}c}%
^{Loc}+(\mathbf{E}_{E})_{A\operatorname{Re}c}^{Dist}\right]  _{tr}%
+\tag{3.87a}\\
&  +\left[  -(\mathbf{E}_{H})_{A\operatorname{Re}c}^{Loc}+(\mathbf{E}%
_{H})_{A\operatorname{Re}c}^{Dist}\right]  _{rot}+ \tag{3.87b}%
\end{align}

where:\ $\mathbf{V}_{tot}\,\mathbf{=}\frac{1}{2}\mathbf{(\mathbf{m}_{V}%
^{+}+\mathbf{m}_{V}^{-})\mathbf{c}^{2}=}\frac{1}{2}\mathbf{\mathbf{m}_{V}%
^{+}(2c}^{2}-\mathbf{v}^{2})$ is a total potential energy of triplet (3.63),
equal to that of unpaired sub-elementary fermion, and $\mathbf{T}%
_{tot}\,\mathbf{=}\frac{1}{2}\mathbf{(\mathbf{m}_{V}^{+}-\mathbf{m}_{V}%
^{-})\mathbf{c}^{2}}$ is its total kinetic energy (3.65).

The reversible conversions of the\emph{ local gravitational potential }
$\pm(\mathbf{V}_{G})_{\operatorname{Re}c,A\operatorname{Re}c}^{Loc}$ to the
distant one $\pm(\mathbf{V}_{G})_{\operatorname{Re}c,A\operatorname{Re}%
c}^{Dist},$ accompanied the [$\mathbf{C\leftrightharpoons W}]$ pulsation of
unpaired sub-elementary fermions of triplets, interrelated with paired ones
$[\mathbf{F}_{\uparrow}^{+}\bowtie\mathbf{F}_{\downarrow}^{-}]$. The latter
excite the positive VPW$^{+}$ and negative VPW$^{-}$ spherical virtual
pressure waves, propagating in space with light velocity. They are a
consequence of transitions of torus \textbf{V}$^{+}$ and antitorus
\textbf{V}$^{-}$ of surrounding Bivacuum dipoles $\mathbf{BVF}^{\updownarrow
}=\left[  \mathbf{V}^{+}\Updownarrow\mathbf{V}^{-}\right]  $ between the
excited and ground states.

The $\pm(\mathbf{E}_{E})_{\operatorname{Re}c,~Arec}^{Loc}$ and \ $\pm
(\mathbf{E}_{E})_{\operatorname{Re}c,~A\operatorname{Re}c}^{Dist}$ are the
local and distant electrostatic potential oscillations; the $\pm
(\mathbf{E}_{H})_{\operatorname{Re}c,Arec}^{Loc}$ and \ $\pm(\mathbf{E}%
_{H})_{\operatorname{Re}c,A\operatorname{Re}c}^{Dist}$ are the local and
distant magnetic potentials oscillations, accompanied
$[\mathbf{C\leftrightharpoons W}]$ pulsations and rotation of triplets.

Let us consider in more detail the interconversions of the \emph{internal -
local} and the \emph{external - distant} gravitational, Coulomb and magnetic
potentials of charged elementary particles, like electron or proton. \medskip

\begin{center}
\emph{F3. The new approach to quantum gravity and antigravity. }

\emph{The mechanism of relativistic mass increasing and }

\emph{a reason for gravitational and inertial mass equality. }\medskip
\end{center}

The unified right parts of eqs. (3.86) and (3.87), describing the excitation
of \emph{gravitational waves}, which represent superposition of positive and
negative virtual pressure waves ($\mathbf{VPW}^{+}$ and $\mathbf{VPW}^{-}),$
are:%
\begin{align}
&  \overline{\mathbf{V}}_{tot}^{\mathbf{C\rightleftharpoons W}}=\mathbf{V}%
_{tot}\pm\left[  (\mathbf{E}_{G})_{\operatorname{Re}c}^{Loc}-(\mathbf{E}%
_{G})_{\operatorname{Re}c}^{Dist}\right]  ~\ \symbol{126}~\tag{3.88}\\
&  ~\symbol{126}~\mathbf{VPW}^{\pm}\pm\left[  \mathbf{\Delta VPW}%
^{+}\mathbf{+~\Delta VPW}^{-}\right]  \tag{3.88a}%
\end{align}
This formula reflects the fluctuations of \emph{potential energy} of triplets,
induced by $\mathbf{[C\rightleftharpoons W]}$ pulsation of unpaired
sub-elementary fermions (3.75 and 3.76), interrelated with and paired ones
(5.3), which are responsible for gravitational field.

The more detailed presentation of (3.88) is:%
\begin{multline}
\overline{\mathbf{V}}_{tot}^{\mathbf{C\rightleftharpoons W}}=\frac{1}%
{2}\mathbf{(\mathbf{m}_{V}^{+}+\mathbf{m}_{V}^{-})\mathbf{c}^{2}\pm}%
\tag{3.89}\\
\pm\Bigg\{\left[  \mathbf{G}\frac{(\mathbf{m}_{V}^{+}\mathbf{m}_{V}^{-}%
)}{\mathbf{L}_{G}}\right]  ^{Loc}-\nonumber
\left[  \left(  \frac{\mathbf{m}_{0}}{\mathbf{M}_{Pl}}\right)  ^{2}%
\mathbf{(\mathbf{m}_{V}^{+}+\mathbf{m}_{V}^{-})\mathbf{c}^{2}}\right]
^{Dis}\Bigg\}\nonumber
\end{multline}

The local \textit{internal} gravitational interaction between the opposite
mass poles of the mass-dipoles of unpaired sub-elementary fermions
(antifermions) $\left(  \mathbf{F_{\updownarrow}^{\pm}}\right)  _{S=\pm1/2}$
turns reversibly to the \emph{external} distant one: \
\begin{equation}
\left[  \mathbf{G}\frac{\left\vert \mathbf{m}_{V}^{+}\mathbf{m}_{V}%
^{-}\right\vert }{\mathbf{L}_{G}}\right]  ^{Loc}\overset{\mathbf{C\rightarrow
W}}{\underset{\mathbf{W\rightarrow C}}{\rightleftarrows}}\tag{3.90}
\bigg[\mathbf{\beta}^{i}\mathbf{(\mathbf{m}_{V}^{+}+\mathbf{m}_{V}%
^{-})\mathbf{c}^{2}}\nonumber
\mathbf{=\mathbf{\beta}^{i}\mathbf{m}_{V}^{+}(2c}^{2}-\mathbf{v}%
^{2})\bigg]^{Dis}\nonumber
\end{equation}

where: $\mathbf{L}_{G}=\hbar/\mathbf{(\mathbf{m}_{V}^{+}+\mathbf{m}_{V}%
^{-})\mathbf{c}}$ is a characteristic curvature of potential energy (3.64);
$\mathbf{M}_{Pl}^{2}=\mathbf{\hbar c/G}$ is a Plank mass; $\mathbf{m}_{0}%
^{2}=\left\vert \mathbf{m}_{V}^{+}\mathbf{m}_{V}^{-}\right\vert $ is a rest
mass squared; $\mathbf{\beta}^{i}=\left(  \frac{\mathbf{m}_{0}^{i}}%
{\mathbf{M}_{Pl}}\right)  ^{2}$ is the introduced earlier dimensionless
gravitational fine structure constant [21-23]. For the electron $\beta
^{e}=1.739\times10^{-45}$ and $\sqrt{\beta^{e}}=\frac{\mathbf{m}_{0}^{e}%
}{\mathbf{M}_{Pl}}=0.41\times10^{-22}$

The excitation of the \emph{external -} distant spherical virtual pressure
waves of positive and negative energy: \textbf{VPW}$^{+}$ and \textbf{VPW}%
$^{-}$ is a result of torus and antitorus energy fluctuations, accompanied
$\mathbf{[C\leftrightharpoons W]}$ pulsation of paired sub-elementary fermions
$\mathbf{[F_{\uparrow}^{-}\Join F_{\downarrow}^{+}]}_{S=0}$, strictly equal by
absolute values to fluctuation energy of the unpaired $\left(
\mathbf{F_{\updownarrow}^{\pm}}\right)  _{S=\pm1/2}$ one. It is important to
note, that the introduced gravitational field does not depend on charge of
triplet, determined by unpaired sub-elementary fermion $\mathbf{<[F_{\uparrow
}^{-}\Join F_{\downarrow}^{+}]}_{S=0}\mathbf{+\,}\left(
\mathbf{F_{\updownarrow}^{\pm}}\right)  _{S=\pm1/2}\mathbf{>}$, in contrast to
electrostatic and magnetic field. \medskip

>From the proposed mechanism of gravitation and similar values of
$\mathbf{(\mathbf{m}_{V}^{+})}$ in the left and right parts of eq. 3.90 it
directly follows \emph{the equality of gravitational and inertial mass}. The
\emph{inertia} itself is a consequence of particles system symmetry and
\emph{effective }mass relativistic increasing with acceleration. Consequently,
inertia is proportional to the number of particles in this system and (i.e.
total mass of the system) and a generalized Le Chatelier's Principle, as a
reluctance/metastability of system to additional symmetry shift. \medskip

In accordance to our hypothesis [21-23], \textbf{the mechanism of
gravitational attraction} is similar to Bjerknes attraction between pulsing
spheres in liquid medium - Bivacuum. The dependence of Bjerknes force on the
distance between centers of pulsing objects is quadratic: $\mathbf{F}%
_{Bj}~\ \symbol{126}~\mathbf{1/r}^{2}$:%
\begin{equation}
\mathbf{F}_{G}=\mathbf{F}_{Bj}=\frac{1}{\mathbf{r}^{2}}\mathbf{\pi
\mathbf{\rho}_{G}R}_{1}^{2}\mathbf{R}_{2}^{2}\mathbf{v}^{2}\cos\beta\tag{3.91}%
\end{equation}

where $\mathbf{\rho}_{G}$ is density of liquid, i.e. virtual density of
secondary Bivacuum. It is determined by Bivacuum dipoles (BVF$^{\updownarrow}$
and BVB$^{\pm})$ symmetry shift; $\mathbf{R}_{1}$ and $\mathbf{R}_{2}$
radiuses of pulsing/gravitating spheres; $\mathbf{v}$ is velocity of spheres
surface oscillation (i.e. velocity of \textbf{VPW}$^{\pm},$ excited by
$\mathbf{[C\rightleftharpoons W]}$ pulsation of elementary particles, which
can be assumed to be equal to light velocity: $\mathbf{v=c}$); $\beta$ is a
phase shift between pulsation of spheres.

It is important to note, that on the big enough distances the \emph{attraction
}may\emph{ }turn to\emph{ repulsion}. The latter effect, depending on the
phase shift of $\mathbf{[C\rightleftharpoons W]}$ pulsation of interacting
remote triplets ($\beta)$, can explain the revealed acceleration of the
Universe expansion. The possibility of artificial phase shift of
$\mathbf{[C\rightleftharpoons W]}$ pulsation of coherent elementary particles
of any object may (for example by magnetic field) may change its gravitational
attraction to repulsion. The volume and radius of pulsing spheres ($R_{1}$ and
$R_{2})$ in such approach is determined by sum of volume of hadrons, composing
gravitating systems in solid, liquid, gas or plasma state. The gravitational
attraction or repulsion is a result of increasing or decreasing of virtual
pressure of subquantum particles between interacting systems as respect to its
value outside them. This model can serve as a background for new quantum
gravity theory.

The effective radiuses of gravitating objects $\mathbf{R}_{1}$ and
$\mathbf{R}_{2}$ can be calculated from the effective volumes of the objects:%
\begin{equation}
\mathbf{V}_{1,2}=\frac{4}{3}\mathbf{\pi R}_{1,2}^{3}=\mathbf{N}_{1,2}\frac
{4}{3}\mathbf{\pi L}_{p,n}^{3} \tag{3.91a}%
\end{equation}

where: $\mathbf{N}_{1,2}=\mathbf{M}_{1,2}/\mathbf{m}_{p,n}$ is the number of
protons and neutrons in gravitating bodies with mass $\mathbf{M}_{1,2};$
$\mathbf{m}_{p,n}$ is the mass of proton and neutron; $\mathbf{L}_{p,n}%
=\hbar/\mathbf{m}_{p,n}\mathbf{c}$ is the Compton radius of proton and neutron.

>From (3.91a) we get for effective radiuses:
\[
\mathbf{R}_{1,2}=\left(  \frac{\mathbf{M}_{1,2}}{\mathbf{m}_{p,n}}\right)
^{1/3}\mathbf{L}_{p,n}=\left(  \frac{\mathbf{M}_{1,2}}{\mathbf{m}_{p,n}%
}\right)  ^{1/3}\frac{\hbar}{\mathbf{m}_{p,n}\mathbf{c}}%
\]

Putting this to (3.91) we get for the force of gravitational interaction
between two macroscopic objects, each of them formed by atoms with coherently
pulsing protons and neutrons:%
\begin{equation}
\mathbf{F}_{G}=\frac{1}{\mathbf{r}^{2}}\pi\mathbf{\rho}_{Bv}\frac{\left(
\mathbf{M}_{1}\mathbf{M}_{2}\right)  ^{2/3}}{\mathbf{m}_{p,n}^{4/3}}\left(
\frac{\hbar}{\mathbf{m}_{p,n}}\right)  ^{4}\frac{1}{\mathbf{c}^{2}}
\tag{3.91b}%
\end{equation}

Equalizing this formula with Newton's one: $\mathbf{F}_{G}^{N}=\frac
{1}{\mathbf{r}^{2}}\mathbf{G}\left(  \mathbf{M}_{1}\mathbf{M}_{2}\right)  ,$
we get the expression for gravitational constant:
\begin{equation}
\mathbf{G=\pi}\frac{\mathbf{\rho}_{G}}{\sqrt[3]{\mathbf{M}_{1}\mathbf{M}_{2}}%
}\frac{\mathbf{\hbar}^{2}\mathbf{/c}^{2}}{\sqrt[3]{\mathbf{m}_{p,n}^{16}}}
\tag{3.91c}%
\end{equation}

The condition of gravitational constant permanency from (3.91c), is the
anticipated from our theory interrelation between the mass of gravitating
bodies $\sqrt[3]{\mathbf{M}_{1}\mathbf{M}_{2}}$ and the virtual density
$\mathbf{\rho}_{G}$ of secondary Bivacuum, determined by Bivacuum fermions
symmetry shift and excitation in gravitational field:
\begin{equation}
\mathbf{G=const,\ \ \ \ \ }if\ \ \ \ \ \ \ \ \frac{\mathbf{\rho}_{G}}%
{\sqrt[3]{\mathbf{M}_{1}\mathbf{M}_{2}}}=\mathbf{const} \tag{3.91d}%
\end{equation}

where, taking into account (3.90):
\begin{equation*}
\sqrt[3]{\mathbf{M}_{1}\mathbf{M}_{2}}~\ \symbol{126}~\mathbf{\rho}_{G}~\ 
=\frac{\frac{1}{2}\mathbf{\left(  \frac{\mathbf{m}_{0}}{\mathbf{M}_{Pl}%
}\right)  ^{2}(\mathbf{m}_{V}^{+}+\mathbf{m}_{V}^{-})}}{\frac{3}{4}\mathbf{\pi
L}_{G}^{3}}\mathbf{~}
\mathbf{=}\frac{2}{3}\frac{\mathbf{\left(  \frac{\mathbf{m}_{0}}%
{\mathbf{M}_{Pl}}\right)  ^{2}\mathbf{m}_{V}^{+}(2}-\mathbf{v}^{2}%
/\mathbf{c}^{2})}{\mathbf{\pi L}_{G}^{3}}%
\end{equation*}

assuming, that the radius of characteristic volume of asymmetry of Bivacuum
fermion, responsible for gravitation is:
\[
\mathbf{L}_{G}=\frac{\hbar}{\mathbf{(\mathbf{m}_{V}^{+}-\mathbf{m}_{V}^{-})c}}%
\]

we get for reduced gravitational density:%
\begin{equation}
\mathbf{\rho}_{G}=\frac{2}{3}\frac{1}{\mathbf{\pi}\hbar^{3}}\mathbf{\left(
\frac{\mathbf{m}_{0}}{\mathbf{M}_{Pl}}\right)  }^{2}\cdot\tag{3.91e}
\mathbf{(\mathbf{m}_{V}^{+}+\mathbf{m}_{V}^{-})(\mathbf{m}_{V}%
^{+}-\mathbf{m}_{V}^{-})}^{3}\mathbf{c}^{3}\nonumber
\end{equation}

we may see that at conditions of ideal symmetry of primordial Bivacuum, i.e.
in the absence of matter and fields the virtual density of Bivacuum and
gravitational interaction is equal to zero ($\mathbf{\rho}_{G}=0)$, as far
$\mathbf{(\mathbf{m}_{V}^{+}-\mathbf{m}_{V}^{-})=0.}$ \medskip

\begin{center}
\emph{F.4.\ Possible nature of neutrino and antineutrino\medskip\ }
\end{center}

We put forward a conjecture, that neutrinos or antineutrinos represent a
\emph{stable non elastic} Bivacuum excitations in form of packet of positive
and negative virtual pressure waves, slightly uncompensated each other:
\[
\mathbf{\nu,}\widetilde{\mathbf{\nu}}~\ \symbol{126}~\mathbf{[VPW}%
^{+}\mathbf{\bowtie VPW}^{-}\mathbf{]~\symbol{126}~VirP}^{+}-\mathbf{VirP}^{-}%
\]
However, their nonlocal asymmetry is compensating the local asymmetry in
Bivacuum, accompanied origination of positrons or electrons. This asymmetry
can be described as a small difference in energy of sub-elementary fermion and
antifermion $\mathbf{[F_{\uparrow}^{-}\Join F_{\downarrow}^{+}]}_{S=0}$,
composing pair in triplets $\mathbf{<[F_{\uparrow}^{-}\Join F_{\downarrow}%
^{+}]+\,}\left(  \mathbf{F_{\updownarrow}^{\pm}}\right)  \mathbf{>}$. It can
be positive for electrons and negative for positrons and expressed as energy
increment of elementary de Broglie wave:%
\begin{align*}
\mathbf{E}_{\nu}^{i}  &  \mathbf{=}\pm\left[  \left(  \mathbf{m}_{V}%
^{+}\mathbf{c}^{2}\right)  _{\mathbf{[F_{\uparrow}^{-}\Join F_{\downarrow}%
^{+}]}}^{\mathbf{F_{\downarrow}^{+}}}~\mathbf{-~}\left(  \mathbf{m}_{V}%
^{-}\mathbf{c}^{2}\right)  _{\mathbf{[F_{\uparrow}^{-}\Join F_{\downarrow}%
^{+}]}}^{\mathbf{F_{\downarrow}^{-}}}\right]  ^{i}\\
or  &  :\mathbf{\ }\left[  \mathbf{\Delta m}_{V}^{\pm}\mathbf{c}%
^{2}~\mathbf{=\Delta(Rm}_{0}\mathbf{c}^{2}\mathbf{)+\Delta m}_{V}^{\pm
}\mathbf{v}^{2}\right]  ^{i}%
\end{align*}

We suppose that this increment can be defined by gravitational fine structure
constant $\mathbf{\beta}^{i}=\left(  \frac{\mathbf{m}_{0}^{i}}{\mathbf{M}%
_{Pl}}\right)  ^{2}$ different for each lepton generation:
\[
\mathbf{E}_{\nu}^{i}=\mathbf{\beta}^{i}\mathbf{m}_{V}^{\pm}\mathbf{c}%
^{2~}\mathbf{=\beta}^{i}\left[  \mathbf{(Rm}_{0}\mathbf{c}^{2}\mathbf{)+m}%
_{V}^{\pm}\mathbf{v}^{2}\right]
\]

The compensation of the local symmetry shift, related to particles/matter
dynamics and transitions by spatially delocalized symmetry shifts in huge
number of Bivacuum dipoles around these particles in form of different fields
potentials reaction - looks to be a general phenomena. This idea of
interconversion between local and non-local effects can be formulated as\emph{
symmetry shifts compensation principle or: }
\[
\text{\textbf{The total\ sum of symmetry\ shifts, related\ to}}%
\]%
\[
\text{\textbf{Matter and Bivacuum transitions\ in\ the\ Universe\ is\ zero}%
}\mathbf{:}\text{{}}%
\]

\begin{equation}
\frac{1}{Z}\left[  \overset{\infty}{\sum}\mathbf{P}_{k}\Delta\mathbf{(m}%
_{V}^{+}~\mathbf{-~m}_{V}^{-}\mathbf{)}_{k}\mathbf{c}^{2}~\mathbf{+~}%
\overset{\infty}{\sum}\mathbf{P}_{j}\Delta\mathbf{(m}_{V}^{+}\mathbf{v}%
^{2}\mathbf{)}_{j}\right]  =\mathbf{~}\mathbf{0} \tag{3.91f}%
\end{equation}

where: $\mathbf{Z=}\overset{\infty}{\sum}\mathbf{P}_{k}+$ $\overset{\infty
}{\sum}\mathbf{P}_{j}\ $is the total partition function, i.e. sum of
probabilities of all transitions, conversions, fluctuations of energy in the
Universe, including fields: $\Delta\mathbf{(m}_{V}^{+}~\mathbf{-~m}_{V}%
^{-}\mathbf{)}_{k}\mathbf{c}^{2}$ and matter: $\Delta\mathbf{(m}_{V}%
^{+}\mathbf{v}^{2}\mathbf{)}_{j}$; $\ \mathbf{m}_{V}^{+}$\textbf{ }and\textbf{
}$\mathbf{m}_{V}^{-}$ are the actual and complementary mass of torus
(\textbf{V}$^{+})$ and antitorus (\textbf{V}$^{-})$ of each Bivacuum dipoles
and elementary particle in the Universe. The right part of 3.91f in the
bracket reflects the changes of energy of matter and the left part - the
interrelated changes of energy of secondary Bivacuum. This \emph{Bivacuum
symmetry shifts compensation principle} can be considered, as a background for
the \emph{energy conservation law.} \medskip

\begin{center}
\emph{F.5 The Bivacuum dipoles symmetry shift and linear and circular
ordering, }

\emph{as a background of electrostatic and magnetic fields origination\medskip
}
\end{center}

\emph{Let us consider now the origination of electrostatic and magnetic
fields}, as a consequence of proposed models of elementary particles and their
duality. The unified \emph{right} parts of eqs. (3.86) and (3.86a) can be
subdivided to translational \emph{(electrostatic)} and rotational
(\emph{magnetic)} contributions, determined by corresponding degrees of
freedom of Cumulative Virtual Cloud ($\mathbf{CVC}_{tr,rot}^{\pm})$:
\begin{align}
\overline{\mathbf{T}}_{tot}^{\mathbf{C\rightleftharpoons W}}  &
=\mathbf{T}_{tot}\pm\left[  (\mathbf{E}_{E})_{\operatorname{Re}c}%
^{Loc}-(\mathbf{E}_{E})_{\operatorname{Re}c}^{Dist}\right]  _{tr}\pm
\tag{3.92}\\
&  \pm\left[  (\mathbf{E}_{H})_{\operatorname{Re}c}^{Loc}-(\mathbf{E}%
_{H})_{\operatorname{Re}c}^{Dist}\right]  _{rot} \tag{3.92a}%
\end{align}

where the most probable total kinetic energy of particle can be expressed only
via its actual inertial mass ($\mathbf{\mathbf{m}_{V}^{+}}$) and external
velocity ($\mathbf{v)}$:%
\begin{equation}
\mathbf{T}_{tot}=\frac{1}{2}\mathbf{(\mathbf{m}_{V}^{+}-\mathbf{m}_{V}%
^{-})\mathbf{c}^{2}=}\frac{1}{2}\mathbf{\mathbf{m}_{V}^{+}v^{2}} \tag{3.93}%
\end{equation}

\emph{The comparison of (3.93) with (3.90) prove that the inertial and
gravitational mass are equal.}

Formula (3.92-3.92a) reflects the fluctuations of the most probable total
kinetic energy, accompanied $\mathbf{[C\rightleftharpoons W]}$ pulsation of
unpaired sub-elementary fermion, responsible for linear - electrostatic and
curled - magnetic fields origination. The more detailed form of (3.92) is:%
\begin{align}
&  \overline{\mathbf{T}}_{tot}^{\mathbf{C\rightleftharpoons W}}=\frac{1}%
{2}\mathbf{(\mathbf{m}_{V}^{+}-\mathbf{m}_{V}^{-})\mathbf{c}^{2}\pm}%
\tag{3.94}
\left\{  \left[  \frac{\left\vert \mathbf{e}_{+}\mathbf{e}%
_{-}\right\vert }{\mathbf{L}_{T}}\right]  ^{Loc}-\left[  \frac{\mathbf{e}^{2}%
}{\hbar\mathbf{c}}\mathbf{(\mathbf{m}_{V}^{+}-\mathbf{m}_{V}^{-}%
)\mathbf{c}^{2}}\right]  _{tr}^{Dis}\right\}  \pm\nonumber\\
&  \pm\Bigg\{\left[  \mathbf{K}_{\mathbf{HE}}\frac{\left\vert \mathbf{\mu}%
_{+}\mathbf{\mu}_{-}\right\vert }{\mathbf{L}_{T}}\right]  ^{Loc}-\tag{3.94a}
\left[  \mathbf{K}_{\mathbf{HE}}\frac{\mathbf{\mu}_{0}^{2}}{\hbar
\mathbf{c}}\mathbf{(\mathbf{m}_{V}^{+}-\mathbf{m}_{V}^{-})\mathbf{c}^{2}%
}\right]  _{rot}^{Dis}\Bigg\}\nonumber
\end{align}

The oscillation of electrostatic translational contribution, taking into
account the obtained relation between mass and charge symmetry shifts (3.35):
$\mathbf{m}_{V}^{+}-\mathbf{m}_{V}^{-}=\mathbf{m}_{V}^{+}\frac{\mathbf{e}%
_{+}^{2}-\mathbf{e}_{-}^{2}}{\mathbf{e}_{+}^{2}}$ can be expressed as:%

\begin{equation}
\left[  \frac{\left\vert \mathbf{e}_{+}\mathbf{e}_{-}\right\vert }%
{\mathbf{L}_{T}}\right]  ^{Loc}\overset{\mathbf{C\rightarrow W}}%
{\underset{\mathbf{W\rightarrow C}}{\rightleftarrows}}\left[  \mathbf{\alpha
}\left(  \mathbf{m}_{V}^{+}\mathbf{\mathbf{c}^{2}}\frac{\mathbf{e}_{+}%
^{2}-\mathbf{e}_{-}^{2}}{\mathbf{e}_{+}^{2}}\right)  \right]  _{tr}^{Dis}
\tag{3.95}%
\end{equation}

where: $\mathbf{L}_{T}=\hbar/\mathbf{(\mathbf{m}_{V}^{+}-\mathbf{m}_{V}%
^{-})\mathbf{c}}$ is a characteristic curvature of kinetic energy (3.65b);
$\left\vert \mathbf{e}_{+}\mathbf{e}_{-}\right\vert =\mathbf{e}_{0}^{2}$ is a
rest charge squared; $\alpha=\mathbf{e}^{2}/\hbar\mathbf{c}$ is the well known
dimensionless electromagnetic fine structure constant.

The right part of (3.95) taking into account that: $\mathbf{e}_{+}%
^{2}-\mathbf{e}_{-}^{2}=\left(  \mathbf{e}_{+}-\mathbf{e}_{-}\right)  \left(
\mathbf{e}_{+}+\mathbf{e}_{-}\right)  $ characterizes the electric dipole
moment of triplet, equal to that of unpaired sub-elementary fermion $\left(
\mathbf{F_{\updownarrow}^{\pm}}\right)  .$

Like in the case of gravitational potential, it is assumed, that the local
\textit{internal} Coulomb potential between opposite charges of torus and
antitorus of unpaired sub-elementary fermions (antifermions) $\left(
\mathbf{F_{\updownarrow}^{\pm}}\right)  _{S=\pm1/2}$ turn reversibly to the
\emph{external} distant one due to elastic recoil$\leftrightharpoons
$antirecoil effects, induced by $\mathbf{C\rightleftharpoons W}$ pulsation
of$\left(  \mathbf{F_{\updownarrow}^{\pm}}\right)  _{S=\pm1/2}$. \medskip

\begin{center}
\emph{The factors, responsible for Coulomb interaction between elementary
particles\medskip}
\end{center}

There are three factors, which determines the attraction or repulsion between
opposite or similar elementary charges, correspondingly. They are provided by
the recoil $\rightleftharpoons$ antirecoil effects, induced by
[$\mathbf{C\rightleftharpoons W}]$ pulsation and emission $\rightleftharpoons$
absorption of positive or negative cumulative virtual clouds CVC$^{+}$ or
CVC$^{-}$ of the unpaired sub-elementary fermion $\left(
\mathbf{F_{\updownarrow}^{\pm}}\right)  _{S=\pm1/2}$ of triplets:

1. The opposite (attraction) or similar (repulsion) Bivacuum dipoles symmetry
shift and polarization;

2. Polymerization or depolymerization of Bivacuum dipoles of opposite or
similar symmetry/polarization shifts;

3. The different kind of interference of virtual pressure waves of the
opposite or similar by sign energy:
\begin{equation}
\left[  \mathbf{VPW}^{\pm}\mathbf{~+~VPW}^{\mp}\right]  \text{ or }\left[
\mathbf{VPW}^{\pm}\mathbf{~+~VPW}^{\pm}\right]  \tag{3.96}%
\end{equation}

excited by corresponding cumulative virtual clouds - opposite or similar:%

\begin{equation}
\left[  \mathbf{CVC}^{\pm}\mathbf{~+~CVC}^{\mp}\right]  \text{ or }\left[
\mathbf{CVC}^{\pm}\mathbf{~+~CVC}^{\pm}\right]  \tag{3.96a}%
\end{equation}

These virtual clouds from subquantum particles, in accordance to our model of
duality, are emitted $\rightleftharpoons$ absorbed by the unpaired
sub-elementary fermions or/and sub-elementary antifermions in triplets of
elementary particles $\mathbf{<[F_{\uparrow}^{-}\Join F_{\downarrow}^{+}%
]}_{S=0}\mathbf{+\,}\left(  \mathbf{F_{\updownarrow}^{\pm}}\right)
_{S=\pm1/2}\mathbf{>}$.

\emph{The 1st factor} is a basic one. The attraction between opposite charges
is a consequence of tendency of Bivacuum to decrease the resulting symmetry
shift, decreasing the separation between charges. The repulsion between
similar charges also is a consequence of decreasing of resulting symmetry
shift, being possible in this case only because of distance increasing between
charges. Both of these processes are the consequence of energy conservation
law, formulated as eq. 3.91f, i.e. tendency of the Bivacuum symmetry
increments to zero.

\emph{The 2nd factor} is a consequence of the 1st one. The electrostatic
Coulomb attraction between opposite electric charges can be a result of
Bivacuum bosons $\mathbf{BVB}^{\pm}=$ $[\mathbf{V}^{+}\uparrow\downarrow
\mathbf{V}^{-}]$ or Cooper pairs of Bivacuum dipoles:%

\begin{equation}
\left\{  \left(  \mathbf{BVF}^{\uparrow}=[\mathbf{V}^{+}\upuparrows
\mathbf{V}^{-}]\right)  \Join\left(  \mathbf{BVF}^{\downarrow}=[\mathbf{V}%
^{+}\downdownarrows\mathbf{V}^{-}]\right)  \right\}  \tag{3.96b}%
\end{equation}
polarized by the unpaired sub-elementary fermions of triplets due to charge
symmetry shift: This induces the linear assembly of Bivacuum dipoles, their
"head to tail" polymerization between $\left(  \mathbf{F_{\updownarrow}^{\pm}%
}\right)  _{S=\pm1/2}\mathbf{>}$ of two opposite charges (i.e. electron and proton).

These quasi one-dimensional single $\sum\mathbf{BVB}^{\pm}$ and twin
$\sum(\mathbf{BVF}^{\uparrow}\Join\left(  \mathbf{BVF}^{\downarrow}\right)  $
virtual microtubules, like 'springs', are responsible for the 'force lines'
origination, connecting the opposite charges. The energy of attraction has the
following dependence on the length ($R)$ of virtual 'springs':
$\overrightarrow{r}/R,$ where $\overrightarrow{r}$ is the unitary
radius-vector. On the other hand, the repulsion between similar charges by
sign is a result of the opposite phenomena - chaotization of Bivacuum dipoles
orientation between such charges. So, the repulsion between similar charges
can be explained also in terms of Bivacuum entropy increasing.

\emph{The 3d factor} is determined by interaction of positive and negative
subquantum particles density oscillation, representing virtual pressure waves:
$\mathbf{VPW}^{+}\mathbf{~}$and$\mathbf{~VPW}^{-}.$ Its effect on attraction
or repulsion of charges also can be explained in terms of tending of system:
$\left[  \text{Charges + Bivacuum}\right]  $ to minimum symmetry shift in
space between charges in accordance to energy conservation law in form of eq.
3.91f.\medskip

\emph{The oscillation of magnetic dipole radiation contribution} in the
process of [$\mathbf{C\leftrightharpoons W}]$ pulsations of sub-elementary
fermions between local and distant modes is not accompanied by magnetic
moments symmetry shift, but only by the oscillation of separation between
torus and antitorus of BVF$^{\updownarrow}.$ It can be described as:%
\begin{gather}
\left[  \mathbf{K}_{HE}^{i}\frac{\left\vert \mathbf{\mu}_{+}\mathbf{\mu}%
_{-}\right\vert }{\mathbf{L}_{T}}\right]  ^{Loc}~\overset{\mathbf{C\rightarrow
W}}{\underset{\mathbf{W\rightarrow C}}{\rightleftarrows}}\tag{3.97}
~\left[  \mathbf{K}_{HE}^{i}\frac{\mathbf{\mu}_{0}^{2}}{\hbar\mathbf{c}%
}\mathbf{(\mathbf{m}_{V}^{+}-\mathbf{m}_{V}^{-})\mathbf{c}^{2}}\right]
_{rot}^{Dis}\nonumber\\
or:~\left[  \mathbf{K}_{HE}^{i}\frac{\left\vert \mathbf{\mu}_{+}\mathbf{\mu
}_{-}\right\vert }{\mathbf{L}_{T}}\right]  ^{Loc}\overset{\mathbf{C\rightarrow
W}}{\underset{\mathbf{W\rightarrow C}}{\rightleftarrows}}\tag{3.97a}
\left[  \mathbf{K}_{HE}^{i}\frac{\mathbf{\mu}_{0}^{2}}{\hbar\mathbf{c}%
}~\mathbf{\mathbf{m}_{V}^{+}\omega_{T}^{2}L}_{T}^{2}\right]  _{rot}%
^{Dis}\nonumber
\end{gather}

where: $\frac{\mathbf{\mu}_{0}^{2}}{\hbar\mathbf{c}}=\gamma$ is the magnetic
fine structure constant, introduced in our theory. The magneto - electric
conversion coefficient $\mathbf{K}_{HE}$ we find from the equality of the
electrostatic and magnetic contributions:
\begin{equation}
\mathbf{E}_{H}=\mathbf{E}_{E}=\mathbf{T}_{rec}=\frac{1}{2}\frac{\mathbf{e}%
^{2}}{\hbar\mathbf{c}}\mathbf{(\mathbf{m}_{V}^{+}-\mathbf{m}_{V}%
^{-})\mathbf{c}^{2}}\tag{3.98}
\mathbf{=}\frac{1}{2}\mathbf{K}_{HE}^{i}\frac{\mathbf{\mu}_{0}^{2}}%
{\hbar\mathbf{c}}~\mathbf{\mathbf{m}_{V}^{+}\omega_{T}^{2}L}_{T}^{2}\nonumber
\end{equation}
\ 

These equality is a consequence of equiprobable energy distribution between
translational (electrostatic) and rotational (magnetic) independent degrees of
freedom of an unpaired sub-elementary fermion and its cumulative virtual cloud
(CVC$^{\pm})$. This becomes evident for the limiting case of photon in vacuum.
The sum of these two contributions is equal to%
\begin{equation}
\mathbf{E}_{H}+\mathbf{E}_{E}=\alpha~\mathbf{\mathbf{m}_{V}^{+}v_{res}^{2}}
\tag{3.98a}%
\end{equation}

where $\mathbf{v_{res}}$ is a resulting translational - rotational recoil velocity.

>From the above conditions it follows, that:%
\begin{equation}
\mathbf{K}_{HE}\frac{\mathbf{\mu}_{0}^{2}}{\hbar\mathbf{c}}=\mathbf{K}%
_{HE}\frac{\hbar\mathbf{e}_{0}^{2}}{4\mathbf{m}_{0}^{2}\mathbf{c}^{3}}%
=\frac{\mathbf{e}_{0}^{2}}{\hbar\mathbf{c}} \tag{3.99}%
\end{equation}

where $\mathbf{\mu}_{0}^{2}=\left\vert \mathbf{\mu}_{+}\mathbf{\mu}%
_{-}\right\vert =\left(  \frac{1}{2}\mathbf{e}_{0}\frac{\hbar}{\mathbf{m}%
_{0}\mathbf{c}}\right)  ^{2}$ is the Bohr magneton.

Consequently, the introduced magneto-electric conversion coefficient is%
\begin{equation}
\mathbf{K}_{HE}^{e,p}=\left(  \frac{\mathbf{m}_{0}^{e,p}\mathbf{c}}{\hbar
/2}\right)  ^{2}=\left(  \frac{2}{\mathbf{L}_{0}^{e,p}}\right)  ^{2}
\tag{3.100}%
\end{equation}

where $\mathbf{L}_{0}^{e,p}=\hbar/\mathbf{m}_{0}^{e,p}\mathbf{c}$ is the
Compton radius of the electron or proton.\medskip

\emph{Origination of magnetic field} is assumed to be a result of dynamic
equilibrium%
\begin{equation}
\lbrack\mathbf{BVF}_{S=+1/2}^{\uparrow}~\rightleftarrows\mathbf{BVB}^{\pm
}\rightleftarrows~\mathbf{BVF}_{S=-1/2}^{\downarrow}] \tag{3.100b}%
\end{equation}
shift between Bivacuum fermions and Bivacuum antifermions to the left or
right, accompanied by corresponding shift of equilibrium between Bivacuum
bosons of zero spin and magnetic moment of opposite polarization:%
\begin{equation}
\left\langle \mathbf{BVB}^{+}\equiv\lbrack\mathbf{V}^{+}\mathbf{\uparrow
\downarrow V}^{-}]\right\rangle \rightleftharpoons\left\langle \lbrack
\mathbf{V}^{+}\mathbf{\downarrow\uparrow V}^{-}]\equiv\mathbf{BVB}%
^{-}\right\rangle \tag{3.100c}%
\end{equation}
and circulation in the plane, normal to direction of charged particles
propagation in the current. It should be related to asymmetric properties of
the 'anchor' $\mathbf{BVF}_{anc}^{\updownarrow}$ of unpaired sub-elementary
fermion $\left(  \mathbf{F_{\updownarrow}^{\pm}}\right)  _{S=\pm1/2}%
\mathbf{>}$ of triplets $\mathbf{<[F_{\uparrow}^{-}\Join F_{\downarrow}^{+}%
]}_{S=0}\mathbf{+\,}\left(  \mathbf{F_{\updownarrow}^{\pm}}\right)
_{S=\pm1/2}\mathbf{>}$ and fast rotation of \emph{pairs} of charge and
magnetic dipoles $\mathbf{[F_{\uparrow}^{-}\Join F_{\downarrow}^{+}]}_{S=0}$
in plane, normal to \emph{directed} motion of triplets, i.e. current. This
statement is a consequence of the empirical fact, that the magnetic field can
be exited only by the electric current: $\overrightarrow{\mathbf{j}%
}\mathbf{=n}$ $\mathbf{e}\overrightarrow{\mathbf{v}_{j}},$ i.e. \emph{directed
motion }of the charged particles. The cumulative effect of rotation of many of
the electrons of current in plane, normal to current and direction of rotation
is determined by the \emph{hand screw rule} and induce the circular structure
formation around $\overrightarrow{\mathbf{j}}$ in Bivacuum$.$ These
axisymmetric structures are the result of 'head-to-tail' vortical assembly of
Bivacuum dipoles, i.e. by the same principle as between opposite charges. The
vortex - like motion of these axial circled structures along the force lines
of the magnetic field occur due to a small symmetry shift between mass and
charge of torus and antitorus of $\mathbf{BVF}^{\updownarrow}$ in accordance
with (2.11) and (2.13a). This asymmetry is produced by perturbation of
Bivacuum dipoles, induced by the unpaired sub-elementary fermions of triplets.
The unpaired sub-elementary fermions and antifermions have the opposite
influence on symmetry shift between torus and antitorus, interrelated with
their opposite influence on the direction of the $[\mathbf{BVF}_{S=+1/2}%
^{\uparrow}~\rightleftarrows~\mathbf{BVB}^{\pm}\rightleftarrows~\mathbf{BVF}%
_{S=-1/2}^{\downarrow}]$ equilibrium shift.

The equilibrium constant between Bivacuum fermions of opposite spins,
characterizing their uncompensated magnetic moment, we introduce, using
(3.27), as function of the external translational velocity of $\mathbf{BVF}%
^{\updownarrow}$:%
\begin{multline}
\mathbf{K}_{BVF^{\uparrow}\rightleftharpoons BVF^{\downarrow}}=\frac
{\mathbf{BVF}_{S=+1/2}^{\uparrow}}{\mathbf{BVF}_{S=-1/2}^{\downarrow}%
}\tag{3.101}
=\exp\left[  -\frac{\mathbf{\alpha(m}_{V}^{+}~\mathbf{-~m}_{V}^{-}%
)}{\mathbf{m}_{V}^{+}}\right]  =\nonumber\\
=\exp\left[  -\mathbf{\alpha}\frac{\mathbf{v}^{2}}{\mathbf{c}^{2}}\right]
=\exp\left[  -\frac{\mathbf{\omega}_{T}^{2}\mathbf{L}_{T}^{2}}{\mathbf{c}^{2}%
}\right] \nonumber
\end{multline}

The similar values of $\mathbf{K}_{BVF^{\uparrow}\rightleftharpoons
BVF^{\downarrow}}$ have the axial distribution with respect to the current
vector ($\mathbf{j)}$ of charges$\mathbf{.}$ The conversion of Bivacuum
fermions or Bivacuum antifermions to Bivacuum bosons ($\mathbf{BVB}^{\pm
}=\mathbf{V}^{+}\Updownarrow\mathbf{V}^{-})$ with different probability, also
may provide an increasing or decreasing of the equilibrium constant
$\mathbf{K}_{BVF^{\uparrow}\rightleftharpoons BVF^{\downarrow}}.$ The
corresponding probability difference is dependent on the direction of current
related, in-turn, with direction of unpaired sub-elementary fermion
circulation, affecting in opposite way on the angular momentum and stability
of $\mathbf{BVF}_{S=+1/2}^{\uparrow}$ and $\mathbf{BVF}_{S=-1/2}^{\downarrow
}.$

The magnetic field tension can be presented as a gradient of the constant of
equilibrium:
\begin{equation}
\mathbf{H=grad}\left(  \mathbf{K}_{BVF^{\uparrow}\rightleftharpoons
BVF^{\downarrow}}\right)  \tag{3.102}
=(\overrightarrow{r}/R)\mathbf{K}_{BVF^{\uparrow}\rightleftharpoons
BVF^{\downarrow}}\nonumber
\end{equation}

The chaotic thermal velocity of the 'free' conductivity electrons in metals
and ions at room temperature is very high even in the absence of current, as
determined by Maxwell-Boltzmann distribution:
\begin{equation}
\mathbf{v}_{T}~\mathbf{=}\sqrt{\frac{\mathbf{kT}}{\mathbf{m}_{\mathbf{V}}^{+}%
}}\symbol{126}10^{7}~cm/s \tag{3.103}%
\end{equation}

It proves, that not the acceleration, but the ordering of the electron
translational and rotational dynamics in space, provided by current, is a main
reason of the curled magnetic field excitation. In contrast to conventional
view, the electric current itself is not a \emph{primary}, but only a
\emph{secondary} reason of magnetic field origination, as the charges
translational and rotational dynamics 'vectorization factor'.\medskip

\begin{center}
\emph{F6. Interpretation of the Maxwell displacement current,}

\emph{ based on Bivacuum model\medskip}
\end{center}

The magnetic field origination in Bivacuum can be analyzed also from more
conventional point of view.

Let us analyze the 1st Maxwell equation, interrelating the circulation of
vector of magnetic field tension $\mathbf{H}$ along the closed contour
$\mathbf{L}$ with the conduction current (\textbf{j)} and \emph{displacement
current } $\mathbf{j}_{d}\,\mathbf{=}\frac{1}{4\pi}\frac{\partial\mathbf{E}%
}{\partial t}$ through the surface, limited by $\mathbf{L}:$%

\begin{equation}
\oint_{\mathbf{L}}\mathbf{H\,}dl=\frac{4\pi}{c}\int_{\mathbf{S}}\left(
\mathbf{j}+\frac{1}{4\pi}\frac{\partial\mathbf{E}}{\partial t}\right)
d\mathbf{s} \tag{3.104}%
\end{equation}

where ($\mathbf{s)}$ is the element of surface, limited with contour
($l\mathbf{).}$

The existence of the displacement current: $\mathbf{j}_{d}=\frac{1}{4\pi}%
\frac{\partial\mathbf{E}}{\partial t}$ is in accordance with our model of
Bivacuum, as the oscillating virtual dipoles (BVF$^{\updownarrow}$ and
BVB$^{\pm})$ continuum.

In condition of \emph{primordial} Bivacuum of the ideal virtual dipoles
symmetry (i.e. in the absence of matter and fields) these charges totally
compensate each other.

However, even in primordial symmetric Bivacuum the oscillations of distance
and energy gap between torus and antitorus of Bivacuum dipoles is responsible
for \emph{displacement current. }This alternating current generates
corresponding\emph{ displacement magnetic field. }

Corresponding virtual dipole oscillations are the consequence of the in-phase
transitions of $\mathbf{V}^{+}$ and $\mathbf{V}^{-}$ between the excited and
ground states, compensating each other. These transitions are accompanied by
spontaneous emission and absorption of positive and negative virtual pressure
waves: $\mathbf{VPW}^{+}$\emph{ }and\emph{ }$\mathbf{VPW}^{-}.$

The primordial displacement current and corresponding displacement magnetic
field \ can be enhanced by presence of matter. It is a consequence of
fluctuations of potential energy of sub-elementary fermions of triplets,
accompanied by excitation of positive and negative virtual pressure waves,
which in accordance to our approach, is interrelated with gravitational field
of particles (see 3.88 - 3.90). \medskip

\begin{center}
\emph{F7. New kind of current in secondary Bivacuum, additional to
displacement one\medskip}
\end{center}

\emph{\ }This additional current is a consequence of vibrations of
BVF$^{\updownarrow}$, induced by recoil-antirecoil effects, accompanied
[$\mathbf{C\leftrightharpoons W}]$ transitions of unpaired sub-elementary
fermion of triplets $\mathbf{<[F_{\uparrow}^{-}\Join F_{\downarrow}^{+}%
]}_{S=0}\mathbf{+\,}\left(  \mathbf{F_{\uparrow}^{+}}\right)  _{S=\pm
1/2}\mathbf{>}^{e,\tau}.$

The corresponding elastic deformations of Bivacuum fermions $\mathbf{(BVF}%
^{\updownarrow}\mathbf{)}\equiv\lbrack\mathbf{V}^{+}\Updownarrow\mathbf{V}%
^{-}]$ are followed by small charge-dipole symmetry zero-point oscillations
($\mathbf{v}^{ext}=0)$ with amplitude, determined by the most probable
resulting translational - rotational recoil velocity ($\mathbf{v}_{rec}).$ At
conditions $\mathbf{e}_{+}\simeq\mathbf{e}_{-}\simeq\mathbf{e}_{0}$ and
$\left\vert \mathbf{e}_{+}-\mathbf{e}_{-}\right\vert <<\mathbf{e}_{0},$\ i.e.
at small perturbations of torus and antitorus: $\mathbf{V}^{+}$ and
$\mathbf{V}^{-}$ we have for the charge symmetry shift oscillation amplitude:%

\begin{equation}
\Delta\mathbf{e}_{\pm}=\mathbf{e}_{+}-\mathbf{e}_{-}=\frac{1}{2}\mathbf{e}%
_{0}\frac{\mathbf{v}_{rec}^{2}}{\mathbf{c}^{2}} \tag{3.105}%
\end{equation}

The resulting most probable recoil velocity using (3.98) can be defined from
two expressions of recoil momentum, standing for electromagnetism:
\begin{align}
\mathbf{T}_{rec}  &  =\frac{1}{2}\alpha\mathbf{(\mathbf{m}_{V}^{+}%
-\mathbf{m}_{V}^{-})\mathbf{c}}^{2}=\frac{1}{2}\alpha~\mathbf{\mathbf{m}%
_{V}^{+}v_{res}^{2}}\tag{3.106}\\
\mathbf{v}_{rec}^{2}  &  =\alpha~\mathbf{v_{res}^{2}} \tag{3.106a}%
\end{align}

The minimum value of recoil velocity, corresponding to zero \emph{external}
translational velocity of triplets and internal velocity, determined by Golden
mean conditions $(\mathbf{v}_{\mathbf{res}}\mathbf{/c)}^{2}\mathbf{=~}%
\phi=0.61803398,$ can be considered as the \emph{velocity of zero-point
oscillations}:
\begin{equation}
\left(  \mathbf{v}_{rec}^{2}\right)  ^{\min}\equiv\left(  \mathbf{v}_{0}%
^{2}\right)  _{HE}^{\min}=\mathbf{\alpha\phi}~\mathbf{c}^{2} \tag{3.107}%
\end{equation}

where: $\alpha=e^{2}/\hbar c=0,0072973506;$\ \ $\mathbf{\alpha\phi=}\left(
\mathbf{v}_{rec}^{2}\right)  ^{\min}/\mathbf{c}^{2}=4.51\cdot10^{-3}.$

The alternating \emph{recoil current (j}$_{rec}^{EH})$, additional to that of
Maxwell \emph{displacement current (j}$_{d}$,) existing in presence of charged
particles even in the absence of conducting current ($\mathbf{j=0)}$ is equal
to product of (3.105) and square root of (3.107). At Golden mean conditions
(\textbf{v/c)}$^{2}=\phi$ this \emph{recoil current} is:
\begin{equation}
\left(  \mathbf{j}_{rec}^{\phi}\right)  ^{EH}=\left(  \Delta\mathbf{e}_{\pm
}\right)  ^{\phi}\left(  \mathbf{v}_{rec}\right)  ^{\min}=\frac{1}%
{2}\mathbf{\alpha}^{1/2}\mathbf{\phi}^{3/2}~\mathbf{e}_{0}\mathbf{c}
\tag{3.108}%
\end{equation}

Corresponding gravitational contribution of recoil velocity, related to
vibration of potential energy of particle (3.90) is much smaller, as far
$\beta<<\alpha$:
\begin{equation}
\mathbf{V}_{rec}=\frac{1}{2}\beta\mathbf{(\mathbf{m}_{V}^{+}+\mathbf{m}%
_{V}^{-})\mathbf{c}}^{2}\tag{3.109}
=\frac{1}{2}\beta~\mathbf{\mathbf{m}_{V}^{+}c^{2}(2-v}^{2}/\mathbf{c^{2}%
})\nonumber
\end{equation}

The zero-point characteristic potential recoil velocity squared at GM
conditions ($\mathbf{v}^{2}/\mathbf{c^{2})}^{\phi}\mathbf{=\phi}$ is:
\begin{align}
\left(  \mathbf{v}_{0}^{2}\right)  _{G}  &  =\beta\mathbf{c^{2}(2-\phi
);\ }\nonumber\qquad
\mathbf{\ \ }\left(  \mathbf{v}_{0}^{2}\right)  _{G}/\mathbf{c}^{2}  
=\beta\mathbf{(2-\phi)}\nonumber\\
\left(  \mathbf{v}_{0}\right)  _{G}  &  =\mathbf{c}\beta^{1/2}\mathbf{(2-\phi
)}^{1/2}\tag{3.109a}
  =1,446\cdot10^{-12}\ cm/s\nonumber
\end{align}

Consequently, the Maxwell equation (3.104) can be modified, taking into
account the EH recoil current, as%

\begin{equation}
\oint_{\mathbf{L}}\mathbf{H\,}dl\tag{3.110}
=\frac{4\pi}{c}\int_{\mathbf{S}}\left(  \mathbf{j}+\frac{1}{4\pi}%
\frac{\partial\mathbf{E}}{\partial t}+\mathbf{j}_{rec}^{EH}\right)
d\mathbf{s}\nonumber
\mathbf{=I}_{tot}\nonumber
\end{equation}

where: $\mathbf{I}_{tot}$ is total current via surface ($\mathbf{S).}$

We have to note, that $\mathbf{j}_{rec}^{EH}$ is nonzero not only in the
vicinity of particles, but as well in any remote space regions of Bivacuum,
perturbed by electric and magnetic potentials. This consequence of our theory
coincides with the extended electromagnetic theory of Bo Lehnert (section II
of this paper), also considering current in vacuum, additional to displacement one.

In accordance with the known Helmholtz theorem, each kind of vector field
(\textbf{F),} tending to zero at infinity, can be presented, as a sum of the
gradient of some scalar potential ($\phi)$ and a rotor of vector potential
(\textbf{A):}
\begin{equation}
\mathbf{F=grad\,}\varphi+\mathbf{rot\,A} \tag{3.111}%
\end{equation}

The scalar and vector potentials are convenient to use for description of
electromagnetic field, i.e. photon properties. They are characterized by the
interrelated translational and rotational degrees of freedom, indeed (Fig.3.2).

To explain the \emph{ability of secondary Bivacuum to keep the average
(macroscopic) mass and charge equal to zero,} we have to postulate, that the
mass and charge symmetry shifts oscillations of Bivacuum fermions and
antifermions, forming virtual Cooper pairs:
\begin{equation}
\mathbf{(BVF}^{\uparrow}\mathbf{)}_{S=+1/2}^{\pm}\tag{3.112}
\equiv\lbrack\mathbf{V}^{+}\uparrow\uparrow\mathbf{V}^{-}]^{\pm}%
\ \bowtie\ [\mathbf{V}^{+}\downarrow\downarrow\mathbf{V}^{-}]^{\mp}%
\ \equiv\nonumber
\ \mathbf{(BVF}^{\downarrow}\mathbf{)}_{S=-1/2}^{\mp}\nonumber
\end{equation}
are opposite by sign, but equal by the absolute value. Consequently, the
polarized secondary Bivacuum (i.e. perturbed by matter and field) can be
considered, as a \emph{plasma of the in-phase oscillating virtual dipoles
(BVF)} of opposite resulting charge and mass/energy. \medskip

\begin{center}
\emph{F8. \ The mechanism,} \emph{increasing the refraction index of Bivacuum
\medskip}
\end{center}

By definition, the \emph{torus }is a figure, formed by rotation of a circle
with maximum radius, corresponding to minimum quantum number $\mathbf{n=0}$
(see 3.1a) $\mathbf{L}_{\mathbf{V}^{\pm}}^{i}=\frac{2\mathbf{\hbar}%
}{\mathbf{m}_{0}^{i}\mathbf{c}}$, around the axis, shifted from the center of
the circle at the distance $\mathbf{\pm\Delta L}_{EH,G}\mathbf{.}$ The
electromagnetic ($EH)$ and gravitational ($G)$ vibrations of positions
$(\mathbf{\pm\Delta L}_{EH,G}\mathbf{)}_{V^{\pm}}$ of the big number of
recoiled \textbf{BVF}$_{rec},$ induced by the elastic
recoil$\leftrightharpoons$antirecoil deformations of Bivacuum matrix, are
accompanied by vibrations of square and volume of torus ($\mathbf{V}^{+})$ and
antitorus ($\mathbf{V}^{-})$ of perturbed Bivacuum dipoles: $(\mathbf{BVF}%
_{rec}^{\updownarrow})^{i}=\mathbf{[V}^{+}\Updownarrow\mathbf{V}^{-}%
]_{rec}^{i}$. The electromagnetic and gravitational increments of square
($\mathbf{\Delta S}_{\mathbf{V}^{\pm}}^{E,G}$) and volume ($\mathbf{\Delta
V}_{\mathbf{V}^{\pm}}^{E,G})$ of toruses and antitoruses of $(\mathbf{BVF}%
_{rec}^{\updownarrow})^{i},$ as a consequence of their center vibrations can
be presented, correspondingly, as:%
\begin{align}
\mathbf{\Delta S}_{\mathbf{V}^{\pm}}^{EH,G}  &  =4\pi^{2}\left\vert
\mathbf{\Delta L}_{EH,G}\right\vert _{V^{\pm}}^{EH,G}\cdot\mathbf{L}%
_{\mathbf{V}^{\pm}}\tag{3.113}\\
\mathbf{\Delta V}_{\mathbf{V}^{\pm}}^{EH,G}  &  =4\pi^{2}\left\vert
\mathbf{\Delta L}_{EH,G}\right\vert _{V^{\pm}}^{EH,G}\cdot\mathbf{L}%
_{\mathbf{V}^{\pm}}^{2} \tag{3.113a}%
\end{align}

At conditions of zero-point oscillations, corresponding to Golden Mean (GM),
when the ratio ($\mathbf{v}_{0}\mathbf{/c)}^{2}\mathbf{=\phi}$ and external
translational velocity ($\mathbf{v)}$ is zero, the maximum shifts of center of
secondary Bivacuum dipoles \emph{in vicinity of pulsing elementary particles}
due to electromagnetic and gravitational recoil-antirecoil (zero-point)
vibrations are, correspondingly:
\begin{align}
(\mathbf{\Delta L_{EH}^{i})}_{V^{\pm}}^{\phi}  &  =\left(  \mathbf{\tau
}_{C\rightleftharpoons W}^{\phi}\;\mathbf{v}_{EH}^{\phi}\right)
^{i}=\tag{3.114}
\frac{\mathbf{\hbar}}{\mathbf{m}_{0}^{i}\mathbf{c}}\,\mathbf{(\alpha\phi
)}^{1/2}=0,067\,(\mathbf{L}_{\mathbf{V}^{\pm}}^{i})\nonumber\\
(\mathbf{\Delta L_{G}^{i})}_{V^{\pm}}^{\phi}  &  =\left(  \mathbf{\tau
}_{C\rightleftharpoons W}^{\phi}\;\mathbf{v}_{G}^{\phi}\right)  ^{i}%
=\tag{3.114a}
  \frac{\mathbf{\hbar}}{\mathbf{m}_{0}^{i}\mathbf{c}}\,\mathbf{\beta}%
^{1/2}\mathbf{\mathbf{(2-\phi})}^{1/2}=\nonumber
3,27\cdot10^{-23}\,(\mathbf{L}_{\mathbf{V}^{\pm}}^{i})\nonumber
\end{align}

where: the recoil$~\leftrightharpoons~$antirecoil oscillation period is
$\left[  \mathbf{\tau}_{C\rightleftharpoons W}^{\phi}=\mathbf{1/\omega
}_{C\rightleftharpoons W}^{\phi}=\hbar/\mathbf{m}_{0}^{i}\mathbf{c}%
^{2}\right]  ^{i};$\ the recoil$\leftrightharpoons$antirecoil most probable
velocity of zero-point oscillations, which determines the electrostatic and
magnetic fields is: $\mathbf{v}_{EH}^{\phi}=\mathbf{c(\alpha\phi)}%
^{1/2}=0.201330447\times10^{8}%
\operatorname{m}%
\operatorname{s}%
^{-1}$ and $\mathbf{(\alpha\phi)}^{1/2}=0,067$ $\ $the corresponding
zero-point velocity, which determines gravitational field is: $\mathbf{v}%
_{G}^{\phi}=\mathbf{c\,\beta}_{e}^{1/2}\mathbf{\mathbf{(2-\phi})}%
^{1/2}=1,446\cdot10^{-12}\,%
\operatorname{m}%
\operatorname{s}%
^{-1}\;$and~ $\mathbf{\beta}_{e}^{1/2}\mathbf{\mathbf{(2-\phi})}%
^{1/2}=0,48\cdot10^{-22}.$

The dielectric permittivity of Bivacuum and corresponding refraction index,
using our theory of refraction index of matter [63, 64], can be presented as a
ratio of volume of Bivacuum fermions and bosons in symmetric \emph{primordial
}Bivacuum\emph{ }(\textbf{V}$_{\mathbf{pr}})$ to their volume in
\emph{secondary} Bivacuum: \textbf{V}$_{\mathbf{\sec}}=\mathbf{V}%
_{BVF}~-~(\mathbf{r/}r)\mathbf{\Delta V}_{\mathbf{BVF}_{rec}}^{E,G}$,
perturbed by matter and fields. The secondary Bivacuum is optically more
dense, if we assume that the volume, occupied by Bivacuum fermion torus and
antitorus, is excluded for photons. The Coulomb and gravitational potentials
and the related excluded volumes of perturbed Bivacuum fermions/antifermions
decline with distance ($r)$ as:%
\begin{equation}
(\overset{\longrightarrow}{\mathbf{r}}/r)\mathbf{\Delta V}_{\mathbf{BVF}%
_{rec}}^{EH}\text{ \ \ \ and \ \ \ \ }(\overset{\longrightarrow}{\mathbf{r}%
}/r)\mathbf{\Delta V}_{\mathbf{BVF}_{rec}}^{G} \tag{3.115}%
\end{equation}

where: ($r$) is a distance from the charged and/or gravitating particle and
$\overset{\longrightarrow}{\mathbf{r}}$ is the unitary radius vector. Taking
all this into account, we get for permittivity of secondary Bivacuum:%

\begin{multline}%
[c]{c}%
\mathbf{\varepsilon}=\mathbf{n}^{2}=\left(  \frac{\mathbf{c}}{\mathbf{v}%
_{EH,G}}\right)  ^{2}=\frac{N~\mathbf{V}_{\mathbf{pr}}}{N~\mathbf{V}%
_{\mathbf{\sec}}}
=\frac{\mathbf{V}_{BVF}}{\mathbf{V}_{BVF}~-~(\mathbf{r/}r)\mathbf{\Delta
V}_{\mathbf{BVF}_{rec}}^{EH,G}}=\\
=\frac{1}{(1-\mathbf{r/}r)\mathbf{\Delta V}_{\mathbf{BVF}_{rec}}%
^{EH,G}/\mathbf{V}_{BVF}}%
\tag{3.116}%
\end{multline}

\begin{equation}
\mathbf{n}^{2}=\frac{\mathbf{1}}{\mathbf{1}-(\mathbf{r/}r)~3\pi\left\vert
\mathbf{\Delta L}\right\vert _{V^{\pm}}^{EH,G}\cdot\mathbf{L}_{\mathbf{V}%
^{\pm}}} \tag{3.116a}%
\end{equation}

where: the velocity of light propagation in asymmetric secondary Bivacuum of
higher virtual density, than in primordial one, is notated as: $\mathbf{v}%
_{EH,G}=\mathbf{c}_{EH,G};$ \ the volume of primordial Bivacuum fermion
is\ $\mathbf{V}_{BVF}=(4/3)\pi\mathbf{L}_{V^{\pm}}^{3}$ and its increment in
secondary Bivacuum:\ $\mathbf{\Delta V}_{\mathbf{BVF}_{rec}}^{E,G}%
=\mathbf{\Delta V}_{\mathbf{V}^{\pm}}^{E,G}$ (3.113a).

$(\mathbf{r/}r)$ is a ratio of unitary radius-vector to distance between the
source of $\mathbf{[C\leftrightharpoons W]}$ pulsations (elementary particle)
and perturbed by the electrostatic, magnetic and gravitational potential
$\mathbf{BVF}_{rec}^{EH,G}.$

Putting (3.114) into formula (3.116a) we get for the refraction index of
Bivacuum and relativistic factor $\mathbf{(R}_{E}\mathbf{)}$ in the vicinity
of charged elementary particle (electron, positron or proton, antiproton) the
following expression:%
\begin{equation}
\left[  \varepsilon=\mathbf{n}^{2}=\left(  \frac{\mathbf{c}}{\mathbf{c}_{EH}%
}\right)  ^{2}\right]  _{E}\tag{3.117}
=\frac{1}{1-(\mathbf{r/}r)~3\pi\mathbf{(\alpha\phi)}^{1/2}}\lesssim
2.71\nonumber
\end{equation}

where: $\ \ 1\lesssim\mathbf{n}^{2}\lesssim2,71\ $is tending to 1 at
$r\rightarrow\infty.$

The Coulomb relativistic factor:%
\begin{equation}
\mathbf{R}_{EH}=\sqrt{1-\frac{\left(  \mathbf{c}_{EH}\right)  ^{2}}%
{\mathbf{c}^{2}}}=\tag{3.118}
\sqrt{(\mathbf{r/}r)~0,631}\lesssim(\mathbf{r/}r)^{1/2}~0.794\nonumber
\end{equation}

$\ \ 0\lesssim\mathbf{R}_{E}\lesssim0,794$ is tending to zero at
$r\rightarrow\infty.$

In similar way, using (3.114a) and (3.116a), for the refraction index of
Bivacuum and the corresponding relativistic factor $\mathbf{(R}_{G}\mathbf{)}$
of gravitational vibrations of Bivacuum fermions ($\mathbf{BVF}^{\updownarrow
})$ in the vicinity of pulsing elementary particles at zero-point conditions,
we get:
\begin{equation}
\left[  \varepsilon=\mathbf{n}^{2}=\left(  \frac{\mathbf{c}_{G}}{\mathbf{c}%
}\right)  ^{2}\right]  _{G}=\tag{3.119}
\frac{1}{1-(\mathbf{r/}r)3\pi\mathbf{(\beta}^{e}\mathbf{)}^{1/2}%
(2\mathbf{-\phi})^{1/2}}\curlyeqsucc1\nonumber
\end{equation}

where $\mathbf{(\beta}^{e}\mathbf{)}^{1/2}(\mathbf{2-\phi})^{1/2}%
=0.48\times10^{-22}.$

The gravitational relativistic factor:%

\begin{equation}
\mathbf{R}_{G}=\sqrt{1-\left(  \frac{\mathbf{c}_{G}}{\mathbf{c}}\right)  ^{2}%
}=\tag{3.120}
\sqrt{(\mathbf{r/}r)~0,48\cdot10^{-22}}\lesssim(\mathbf{r/}r)^{1/2}%
~0,69\cdot10^{-11}\nonumber
\end{equation}

Like in previous case, the Bivacuum refraction index, increased by
gravitational potential, is tending to its minimum value: $\mathbf{n}%
^{2}\rightarrow1$ at the increasing distance from the source: $r\rightarrow
\infty$.

The charge - induced refraction index increasing of secondary Bivacuum, in
contrast to the mass - induced one, is independent of lepton generations of
Bivacuum dipoles ($e,\mu,\tau).$

The formulas (3.117) and (3.119) for Bivacuum dielectric permittivity and
refraction index near elementary particles, perturbed by their Coulomb and
gravitational potentials, point out that bending and scattering probability of
photons on charged particles is much higher, than that on neutral particles
with similar mass.

We have to point out, that the \emph{light velocity} in conditions: $\left[
\mathbf{n}_{EH,G}^{2}=\mathbf{c/v}_{EH,G}=\mathbf{c/c}_{EH,G}\right]  >1$ is
not longer a scalar, but a vector, determined by the gradient of Bivacuum
fermion symmetry shift:%
\begin{equation}
grad~\Delta\left\vert \mathbf{m}_{V}^{+}\,\mathbf{-\,m}_{V}^{-}\right\vert
_{EH,G}~\mathbf{c}^{2}=grad~\Delta\left(  \mathbf{m}_{V}^{+}\mathbf{v}%
^{2}\right)  \tag{3.121}%
\end{equation}
and corresponding gradient of torus and antitorus equilibrium constant
increment: $\mathbf{\Delta K}_{\mathbf{V}^{+}\updownarrow\mathbf{V}^{-}%
}=\mathbf{1-m}_{V}^{-}/\mathbf{m}_{V}^{+}=\left(  \mathbf{c}_{EH,G}%
/\mathbf{v}\right)  ^{2}$:%
\begin{align}
&  grad[\mathbf{\Delta K}_{\mathbf{V}^{+}\updownarrow\mathbf{V}^{-}%
}=\mathbf{1-m}_{V}^{-}/\mathbf{m}_{V}^{+}]=\tag{3.122}\\
&  =grad\left(  \frac{\mathbf{c}_{EH,G}}{\mathbf{c}}\right)  ^{2}=grad\frac
{1}{\mathbf{n}^{2}} \tag{3.122a}%
\end{align}
\medskip

The other important consequence of: $\left[  \mathbf{n}^{2}\right]  _{E,G}>1$
is that \emph{the contributions of the rest mass energy of photons and
neutrino to their total energy is not zero}, as far the electromagnetic and
gravitational relativistic factors (\textbf{R}$_{EH,G}$\textbf{)} are greater
than zero. It follows from the basic formula for the total energy of de
Broglie wave (the photon in our case):
\begin{equation}
\mathbf{E}_{tot}=\mathbf{m}_{V}^{+}\mathbf{c}^{2}=\hbar\mathbf{\omega
}_{\mathbf{C\rightleftharpoons W}}\tag{3.123}
=\mathbf{R}\left(  \mathbf{m}_{0}\mathbf{c}^{2}\right)  _{rot}^{in}\,+\left(
\mathbf{\hbar\omega}_{B}^{ext}\right)  _{tr}\nonumber
\end{equation}

where the gravitational relativistic factor of electrically neutral objects:
$\mathbf{R}_{G}=\sqrt{(\mathbf{r/}r)3,08\cdot10^{-22}}\lesssim(\mathbf{r/}%
r)^{1/2}~1.75\times10^{-11}.$

This consequence is also consistent with the models of the photon and
neutrino, developed by Bo Lehnert (see section II of this paper).

We can see, that in conditions of \emph{primordial }Bivacuum, when
$r\rightarrow\infty,$ the $\mathbf{n}_{EH,G}\rightarrow1$, $\mathbf{R}%
_{EH,G}\rightarrow0$ \ and the contribution of the rest mass energy
$\mathbf{R}\left(  \mathbf{m}_{0}\mathbf{c}^{2}\right)  _{rot}^{in}\,$\ tends
to zero. At these limiting conditions the frequency of photon
$[Corpuscle\leftrightharpoons Wave]$ pulsation is equal to the frequency of
the photon as a wave: \
\begin{equation}
\mathbf{E}_{ph}=\hbar\mathbf{\omega}_{\mathbf{C\rightleftharpoons W}%
}=\mathbf{\hbar\omega}_{ph}=h\frac{\mathbf{c}}{\lambda_{ph}} \tag{3.124}%
\end{equation}

The results of our analysis explain the bending of light beams, under the
influence of strong gravitational potential in another way, than by Einstein's
general theory of relativity. A similar idea of polarizable vacuum and it
permittivity variations has been developed by Dicke [65], Fock [32] and
Puthoff [31], as a background of 'vacuum engineering'.

For the spherically symmetric star or planet it was shown using Dicke model,
that the dielectric constant $\mathbf{K}$\ of polarizable vacuum is given by
the exponential form:
\begin{equation}
\mathbf{K=exp(2GM/rc}^{2}\mathbf{)} \tag{3.125}%
\end{equation}

where $\mathbf{G}$ is the gravitational constant, $\mathbf{M}$ is the mass,
and $\mathbf{r}$ is the distance from the mass center.

For comparison with expressions derived by conventional General Relativity
techniques, the following approximation of the formula above is sufficient
[31]:
\begin{equation}
\mathbf{K\approx1+}\frac{2GM}{rc^{2}}\mathbf{+}\frac{1}{2}\left(  \frac
{2GM}{rc^{2}}\right)  ^{2} \tag{3.126}%
\end{equation}

Our approach propose the concrete mechanism of Bivacuum optical density
increasing near charged and gravitating particles, inducing light beams
bending.\medskip

\begin{center}
\textbf{G.} \textbf{The Principle of least action, as a consequence of
Bivacuum}

\textbf{basic} \textbf{Virtual Pressure Waves (VPW}$_{q=1}^{\pm})$\textbf{
resonance interaction with particles.}

\textbf{ Possible source of energy for overunity devices.\smallskip\ \ }
\end{center}

Let us analyze the Maupertuis-Lagrange formula for \emph{action}:
\begin{equation}
\mathbf{S}_{ext}\mathbf{=}\overset{t_{1}}{\underset{t_{0}}{\int}}%
\mathbf{2T}_{k}^{ext}\cdot d\mathbf{t} \tag{3.127}%
\end{equation}

The \emph{principle of Least action}, choosing one of few possible
trajectories of system changes from one configuration to another at the
permanent total energy of a system of elementary particles: $\mathbf{E}%
_{tot}=\mathbf{m}_{V}^{+}\mathbf{c}^{2}=const$ has a form:%
\begin{equation}
\Delta\mathbf{S}_{ext}=0 \tag{3.128}%
\end{equation}

This means, that the optimal trajectory of particle corresponds to minimum
variations of the external energy of it wave B.

The time interval: $\mathbf{t=t}_{1}-\mathbf{t}_{2}\mathbf{=nt}_{B}$ is equal
or bigger than the period of the external (modulation) frequency of wave B
($\mathbf{t}_{B}=1/\mathbf{\nu}_{B})$:
\begin{equation}
\mathbf{t=t}_{1}-\mathbf{t}_{2}\,\mathbf{=nt}_{B}=\mathbf{n/\nu}_{B}
\tag{3.129}%
\end{equation}
\ Using eqs.(3.127 and 3.74; 3.74a), we get for the action:%

\begin{align}
\mathbf{S}_{ext}\,  &  \mathbf{=\,}2\mathbf{T}_{k}^{ext}\,\cdot
\mathbf{t=\mathbf{m}_{V}^{+}\mathbf{v}^{2}\cdot t}\tag{3.130}
 \mathbf{=\,}\mathbf{(1+R)[m}_{V}^{+}\mathbf{c}^{2}\,\mathbf{-\,\,m}%
_{0}\mathbf{c}^{2}]^{i}\mathbf{\cdot t\,}\mathbf{=}\nonumber\\
or  &  :\;\;\mathbf{S}_{ext}=\mathbf{\mathbf{m}_{V}^{+}\mathbf{v}^{2}\cdot
t=(1+R)}\,\mathbf{TE\cdot t\,} \tag{3.130a}%
\end{align}

where relativistic factor: $\;\mathbf{R}=\sqrt{1-\left(  \mathbf{v/c}\right)
^{2}}\;$and $\ \left(  \mathbf{v}\right)  \mathbf{\;}$is the resulting
external translational velocity.

We introduce here the important notions of \emph{Tuning energy} of Bivacuum
and corresponding frequency difference, as:%
\begin{equation}
\mathbf{TE=E}_{tot}-\mathbf{E}_{rot}^{in}\,\mathbf{=[m}_{V}^{+}\mathbf{c}%
^{2}\,\mathbf{-\,q\,m}_{0}\mathbf{c}^{2}]^{i}\tag{3.131}
=\mathbf{\mathbf{\hbar\omega}_{TE}=\hbar}[\mathbf{\omega}_{\mathbf{C\ }%
\rightleftharpoons\mathbf{W}}-\mathbf{q\,\omega}_{\mathbf{0}}]^{i}%
=\frac{\mathbf{\mathbf{m}_{V}^{+}\mathbf{v}^{2}}}{\mathbf{1+R}}\nonumber
\end{equation}

where: $\mathbf{q=j-k}$ \ is a quantum number characterizing excitation of
Bivacuum virtual pressure ($\mathbf{VPW}_{q}^{\pm})$ and spin ($\mathbf{VirSW}%
_{q})$ waves (see section 3.1.2).

The frequency ($\mathbf{\omega}_{TE}^{i})$ of Bivacuum Tuning energy is
defined as:
\begin{equation}
\mathbf{\omega}_{TE}=(\mathbf{m}_{V}^{+}\,\mathbf{-\,q\,m}_{0})^{i}%
\mathbf{c}^{2}/\hbar=[\mathbf{\omega}_{\mathbf{C\ }\rightleftharpoons
\mathbf{W}}-\mathbf{q\,\omega}_{\mathbf{0}}^{i}] \tag{3.132}%
\end{equation}

The influence of Tuning energy of Bivacuum on matter is a result of
\emph{forced combinational resonance} between virtual pressure waves
($\mathbf{VPW}_{q}^{\pm}$) of Bivacuum with quantized frequency (3.10a):%
\begin{equation}
\mathbf{\omega}_{\mathbf{VPW}_{q}^{\pm}}^{i}\mathbf{=q\mathbf{\omega
}_{\mathbf{0}}^{i}=q\,m}_{0}^{i}\mathbf{c}^{2}\mathbf{/\hbar} \tag{3.133}%
\end{equation}
and [$\mathbf{C\rightleftharpoons W}$] pulsation of elementary particles
$\mathbf{\omega}_{\mathbf{C\ }\rightleftharpoons\mathbf{W}}$.

The condition of combinational resonance is:
\begin{align}
&  \mathbf{\omega}_{\mathbf{VPW}_{q}}^{i}=q\mathbf{\,\omega}_{0}%
^{i}=\mathbf{\,\omega}_{\mathbf{C\rightleftharpoons W}}\tag{3.134}\\
or  &  :\;\mathbf{E}_{VPW}=\;\mathbf{q\,m}_{0}^{i}\mathbf{c}^{2}%
=\mathbf{\,m}_{\mathbf{V}}^{+}\mathbf{c}^{2} \tag{3.134a}%
\end{align}

The energy exchange between $\mathbf{VPW}_{q}^{+}$ $+$\ $\mathbf{VPW}_{q}%
^{-}\,$and real particles in the process of $\mathbf{[C\rightleftharpoons W}]$
pulsation of pairs $[\mathbf{F}_{\uparrow}^{+}\bowtie\mathbf{F}_{\downarrow
}^{-}]_{x,y}$ of real fermions - triplets $<[\mathbf{F}_{\uparrow}^{+}%
\bowtie\mathbf{F}_{\downarrow}^{-}]_{x,y}+\mathbf{F}_{\updownarrow}^{\pm}%
>_{z}^{i}$ at \emph{pull-in -range} state accelerate them, if
$\mathbf{q=2,3,4,}$ i.e. bigger than $\mathbf{1,}$ driving to resonant
conditions (3.134)$\mathbf{.}$ On the other hand, at the minimum energy of
$\mathbf{VPW}_{q=1}^{\pm}$, when $\mathbf{q=j-k=1}$, their interaction with
triplets slow the translational velocity of particles down and the
deceleration (cooling) effect takes a place.

\ In accordance to rules of combinational resonance of Bivacuum virtual
pressure waves with elementary particles, we have the following relation
between quantized energy and frequency of $\mathbf{VPW}_{q}^{\pm}$ and
energy/frequency of triplets $\mathbf{C\rightleftharpoons W}$ pulsation in
resonance conditions:%

\begin{align}
\mathbf{E}_{\mathbf{VPW}^{\pm}}\,  &  \mathbf{=\hbar\omega}_{\mathbf{VPW}%
^{\pm}}^{i}=\mathbf{q\hbar\omega}_{0}^{i}\tag{3.135}
 =\mathbf{\hbar\omega}_{\mathbf{C\rightleftharpoons W}}^{i}=\mathbf{R}%
\,\hbar\mathbf{\omega}_{0}\,\mathbf{+\,\hbar\omega}_{B}\nonumber\\
\mathbf{E}_{\mathbf{VPW}^{\pm}}\,  &  \mathbf{=q\,m}_{0}^{i}\mathbf{c}%
^{2}=\mathbf{R\,m}_{0}^{i}\mathbf{c}^{2}+\mathbf{m}_{V}^{+}\mathbf{v}%
^{2}\nonumber\\
or  &  :\mathbf{E}_{\mathbf{VPW}^{\pm}}=\mathbf{R\,m}_{0}^{i}\mathbf{c}%
^{2}\tag{3.136}
  +\frac{\mathbf{m}_{0}^{i}\mathbf{c}^{2}(\mathbf{v/c)}^{2}}{\mathbf{R}%
}\nonumber\\
or  &  :\;\;\mathbf{q=R+}\frac{(\mathbf{v/c)}^{2}}{\mathbf{R}} \tag{3.136a}%
\end{align}

where the \textbf{ VPW}$_{q}^{\pm}\;$quantum number: $\mathbf{q=j-k}%
=1,2,3...($integer$\;$numbers)

The angle frequency of de Broglie waves of particles $\left(  \mathbf{\omega
}_{B}\right)  _{1,2,3}$, is dependent on the external translational velocity
($\mathbf{v)}_{1,2,3}$ :%
\begin{equation}
\left(  \mathbf{\omega}_{B}=\mathbf{\hbar/2m}_{V}^{+}\mathbf{L}_{B}%
^{2}=\mathbf{m}_{V}^{+}\mathbf{v}^{2}/2\hbar\right)  _{1,2,3} \tag{3.137}%
\end{equation}

The important relation between the translational most probable velocity of
particle (v) and quantization number ($\mathbf{q),}$ corresponding to resonant
interaction of Bivacuum $\mathbf{VPW}_{q}^{\pm}$ with pulsing particles,
derived from (3.136) is:%
\begin{equation}
\mathbf{v=c}\left(  \frac{\mathbf{q}^{2}\mathbf{-1}}{\mathbf{q}^{2}}\right)
^{1/2} \tag{3.138}%
\end{equation}

At the conditions of triplets fusion, when $\mathbf{q=1,}$ the resonant
conditions correspond to \emph{translational} velocity of particle equal to
zero: $\mathbf{v}_{n=1}\,\mathbf{=0.}$

When the quantized energy of \ $\mathbf{E}_{\mathbf{VPW}_{n}^{\pm}%
}=\mathbf{q\,m}_{0}^{i}\mathbf{c}^{2}$ corresponds to $\mathbf{q=2,}$ the
resonant translational velocity of particle from (3.138) should be:
$\mathbf{v}_{q=2}\,\mathbf{=c\times0.866=2,6\times10}^{10}$cm/s.

At $\mathbf{q=3,}$we have from (3.138) for resonant velocity: $\mathbf{v}%
_{q=3}\,\mathbf{=c\times0.942=2.83\times10}^{10}$cm/s.

It is natural to assume, that if the velocity of particles ($\mathbf{v)}$,
corresponds to $\mathbf{q<1.5,}$ the interaction of these pulsating particles
with basic \textbf{VPW}$_{n=1}^{\pm}$ should slow down their velocity, driving
it to resonant conditions: $\mathbf{q=1},\;\mathbf{v\rightarrow0.}$ \emph{The
2nd and 3d laws of thermodynamics for the closed systems, reflecting the
'spontaneous' cooling of matter and tending the entropy to zero [66], can be a
consequence of just this condition. }

On the other hand, if velocity of particles is high enough and corresponds to
$\mathbf{q>1,5}$ in (3.138), their pull-in range interaction with excited
$\mathbf{VPW}_{n=2}^{\pm}$ can accelerate them up to conditions:
$\mathbf{q=2,\,}$ $\mathbf{v\rightarrow2.6\times10}^{10}$cm/s. If the starting
particles velocity corresponds to $\mathbf{q>2.5}$, their forced resonance
with even more excited $\mathbf{VPW}_{n=3}^{\pm}$ should accelerate them up to
conditions: $\mathbf{q=3,\,}$corresponding to $\mathbf{v=2,83\times10}^{10}%
$cm/s. The described mechanism of Bivacuum - Matter interaction, can be a
general physical background of all kinds of \emph{overunity devices} [21, 67].
The coherent electrons and protons of hot enough plasma in stars and in
artificial conditions also may get the additional energy from high-frequency
virtual pressure waves of Bivacuum $\mathbf{VPW}_{n=2,3..}^{\pm},$ excited by
strong gravitational and/or magnetic fields.

We can see from the formulas above, that the action of Bivacuum Tuning energy
due to interaction of Bivacuum low frequency $\mathbf{VPW}_{n=1}^{\pm}$ with
particles, is responsible for realization of fundamental principle of Least
action: $\Delta\mathbf{S}_{ext}\,\mathbf{\rightarrow0}$ \ at
$\mathbf{TE\rightarrow0}$, corresponding to minimization of translational
kinetic energy of particles $\mathbf{v}_{tr}^{ext}\rightarrow0$ at
$\mathbf{q\rightarrow1}$.

The action of $\mathbf{TE}$ on virtual Cooper pairs of sub-elementary fermions
is opposite to that on real particles. It increases the velocity and kinetic
energy of virtual particles and finally may turn them to real ones. Such a
mechanism of real particles (like photon) origination may work in the process
of photons emission by excited atoms, molecules and accelerated charges.

\emph{The second law of thermodynamics}, formulated as a spontaneous
irreversible transferring of the heat energy from the warm body to the cooler
body or surrounding medium, also means slowing down the kinetic energy of
particles, composing this body. Consequently, the 2nd law of thermodynamics,
as well as Principle of least action, can be a consequence of Tuning energy
($\mathbf{TE}$) minimization, slowing down particles thermal
\emph{translational} dynamics at pull-in range conditions:
\begin{equation}
  \mathbf{TE\,}\mathbf{=\hbar(\mathbf{\omega}_{\mathbf{C\ }\rightleftharpoons
\mathbf{W}}\rightarrow\mathbf{\omega}_{\mathbf{0}})}\overset{\mathbf{v\,}%
\rightarrow0}{\mathbf{\rightarrow}}\mathbf{0})\tag{3.139}\qquad
  at\qquad \left(  \mathbf{1,5}<\mathbf{q}\right)  \;\overset{\mathbf{v\,}%
\rightarrow0}{\rightarrow}\,\;(\mathbf{q\,\,}\mathbf{=\,1)}\nonumber
\end{equation}

\emph{The third law of thermodynamics} states, that the entropy of equilibrium
system is tending to zero at the absolute temperature close to zero. Again,
this may be a consequence of forced combinational resonance between basic
$\mathbf{VPW}_{n=1}^{\pm}$ and particles $\mathbf{[C\rightleftharpoons W}]$
pulsation, when \emph{translational} velocity of particles \textbf{v\thinspace
}$\rightarrow0$ and $\mathbf{TE=\hbar(\mathbf{\omega}_{\mathbf{C\ }%
\rightleftharpoons\mathbf{W}}\rightarrow\mathbf{\omega}_{\mathbf{0}}%
)\overset{\mathbf{v\,}\rightarrow0}{\rightarrow}0}$ \ at$\;(\mathbf{q<1,5)}%
\overset{\mathbf{v\,}\rightarrow0}{\rightarrow}\,\;(\mathbf{q\,\,}%
\mathbf{=\,1)}.$ At these conditions in accordance with our Hierarchic theory
of condensed matter [63, 64] the de Broglie wave length of atoms is tending to
infinity and state of macroscopic Bose condensation of ultimate coherence and
order, i.e. minimum entropy.

This result of our Unified theory could explain the energy conservation law,
independently of the Universe cooling. Decreasing of thermal kinetic energy in
the process of cooling is compensated by increasing of potential energy of the
matter particle-particle interaction, accompanied Bose condensation. The
energy of resonant interaction of matter with Bivacuum virtual pressure waves
of basic energy: $\mathbf{E}_{VPW}=n\,\mathbf{\hbar\omega}_{0}=n$%
\textbf{\thinspace}$\mathbf{m}_{0}\mathbf{c}^{2}$ at $n=1$ is also increasing
at $\mathbf{v}\rightarrow0$, keeping the total energy of the Universe
permanent. \emph{ }\medskip

\begin{center}
\emph{G.1 The new approach to problem of Time}\textbf{\medskip}
\end{center}

It follows from the section above, that the Principle of least action is a
consequence of basic Tuning energy ($\mathbf{TE}_{q=1}$) of Bivacuum influence
on particles, driving the properties of matter on all hierarchical levels to
Golden mean condition. It is shown, using the formula for action (3.130a) that
the introduced dimensionless internal \emph{pace of time} for any closed
coherent system is determined by the pace of its kinetic energy change
(anisotropic in general case), related to changes of Tuning energy (eq.3.131):%
\begin{equation}
\lbrack\mathbf{dt/t}    \mathbf{=d}\ln\mathbf{t=-\,d}\ln\mathbf{T}%
_{k}]_{x,,y,z}=\tag{3.140}
\mathbf{\mathbf{-d\,}\ln[\mathbf{(1+R)TE}}]_{x,y,z}\nonumber
\end{equation}
Using relations (3.140 and 3.131), the pace of the internal time and time
itself for closed system of particles can be presented via their acceleration
and velocity:%

\begin{equation}
\left[  \frac{\mathbf{dt}}{\mathbf{t}}=\mathbf{d}\ln\mathbf{t}=-\frac
{\mathbf{d}\overrightarrow{\mathbf{v}}}{\overrightarrow{\mathbf{v}}}%
\frac{2-(\mathbf{v/c)}^{2}}{1-(\mathbf{v/c)}^{2}}\right]  _{x,y,z} \tag{3.141}%
\end{equation}%
\begin{equation}
\Bigg[\mathbf{t}   \mathbf{=-}\frac{\overrightarrow{\mathbf{v}}}%
{\mathbf{d}\overrightarrow{\mathbf{v}}\mathbf{/dt}}\frac{1-(\mathbf{v/c)}^{2}%
}{2-(\mathbf{v/c)}^{2}}\tag{3.142}
  =\mathbf{-}\frac{\mathbf{[\mathbf{(1+R)TE]}}}%
{\mathbf{d\,[\mathbf{(1+R)TE]/dt}}}\Bigg]_{x,y,z}\nonumber
\end{equation}

The pace of time and time itself are positive ($\mathbf{t>0)}$, if the
particle motion is slowing down ($\mathbf{d}\overrightarrow{\mathbf{v}%
}\mathbf{/dt\,<0\;}$and$\mathbf{\;d}\overrightarrow{\mathbf{v}}\mathbf{<0)}$
and negative, if particles are accelerating. For example, at temperature
decreasing the time and its pace are positive. At temperature increasing they
are negative. Oscillations of atoms and molecules in condensed matter, like
pendulums, are accompanied by alternation the sign of acceleration and,
consequently, sign of time. In the absence of acceleration ($\mathbf{d}%
\overrightarrow{\mathbf{v}}\mathbf{/dt=0\;\;}$and$\mathbf{\;\ dv=0)}$, the
time is infinitive and its pace zero:
\begin{gather*}
\mathbf{t}\mathbf{\rightarrow\infty
\;\;\;\;\;\;\;and\;\;\;\;\;\;\;\;\;\;\;\;\;}\frac{\mathbf{dt}}{\mathbf{t}%
}\rightarrow0\;\;\;\\
at\;\;\;\mathbf{d}\overrightarrow{\mathbf{v}}\mathbf{/dt}\mathbf{\rightarrow
0\;\;\;and\;\;\;\;v=const}%
\end{gather*}

The postulated principle [I] of conservation of internal kinetic energy of
torus (\textbf{V}$^{+})$ and antitorus (\textbf{V}$^{-})$ of symmetric and
asymmetric Bivacuum fermions/antifermions: $\left(  \mathbf{BVF}%
_{as}^{\updownarrow}\right)  ^{\phi}$ $\equiv\mathbf{F}_{\updownarrow}^{\pm}$
(3.15) in fact reflects the condition of infinitive life-time of Bivacuum
dipoles in symmetric and asymmetric states. The latter means a stability of
sub-elementary fermions and elementary particles, formed by them.

In scale of the Universe, the decreasing of all-pervading Tuning energy
($\mathbf{\mathbf{TE)}}$ in (3.142):%
\begin{equation}
\left\{  \mathbf{d\,[\mathbf{(1+R)TE]/dt}}\right\}  <0\;\;\;\;(\mathbf{t>0)}
\tag{3.143}%
\end{equation}
means positive direction of \thinspace\textquotedblright
TIME\ ARROW\textquotedblright\ \ $\mathbf{t>0.}$ This corresponds to tending
of the actual energy of particles to the energy of rest mass: $\mathbf{m}%
_{V}^{+}\mathbf{c}^{2}\,\mathbf{\rightarrow\,q\,m}_{0}\mathbf{c}^{2}$
\ $\mathbf{at\;v\rightarrow0}$, meaning cooling of the Universe, if
quantization number of virtual pressure waves (VPW$^{\pm}$) is minimum:
$\mathbf{q=1}$.

The presented approach to the TIME problem differs from the conventional one,
following from relativistic theory, however, they do not contradict each
other. For example, from (3.142) we can see, that in relativistic conditions,
when as a result of acceleration $(\mathbf{d}\overrightarrow{\mathbf{v}%
}\mathbf{/dt)>0}$\textbf{ }and\textbf{ } $\mathbf{v\rightarrow c}$ the
negative internal time of closed system is tending to zero, like it follows
from Lorentz relations for the \emph{'own'} time of system.

The internal time (3.142) turns to infinity and its pace (3.141) to zero in
the absence of acceleration and deceleration in a closed system $(\mathbf{d}%
\overrightarrow{\mathbf{v}}\mathbf{/dt)=0}$. The permanent
($\mathbf{t\rightarrow\infty)}$ collective motion of the electrons and atoms
of $^{4}\mathbf{He}$ in superconductors and superfluid liquids,
correspondingly, with constant velocity ($\mathbf{v=const)}$ in the absence of
collisions and accelerations are good examples.

In Bivacuum with superfluid properties the existence of stable excitations,
like quantized toruses/antitoruses of Bivacuum dipoles and vortical filaments
- virtual guides (VirG), described in section 3.9, are also in-line with out
time theory.

Each closed real system of elementary particles and macroscopic objects,
rotating around common center on stable orbits, like in atoms, planetary
systems, galactics, etc. is characterized by different centripetal
acceleration ($-\mathbf{d}\overrightarrow{\mathbf{v}}\mathbf{/dt=}%
\overrightarrow{\mathbf{v}}^{2}/\mathbf{R=\omega}^{2}\mathbf{R)}$ and velocity
($\overrightarrow{\mathbf{v}}\mathbf{)}$. Corresponding characteristic times
of such systems is equal to periods of their cycles: $\mathbf{t=T=2\pi/\omega
}$. The closed systems of virtual particles and virtual waves in Bivacuum do
not follow the causality principle and the notion of time for such a systems
is uncertain. In conditions of primordial symmetric Bivacuum, when the real
and virtual particles are absent, the time is absent also.

At conditions: [$\mathbf{v=c=const],}$ valid\ for the case of photons, we get
from (3.142) the uncertainty for time, like: $\mathbf{t=0/0.\;}$The similar
result we get for state of virtual Bose condensate ($\mathbf{VirBC}$) in
Bivacuum, when the \emph{external translational} velocity of Bivacuum fermions
($\mathbf{BVF}^{\updownarrow})$ and Bivacuum bosons ($\mathbf{BVB}^{\pm})$ is
equal to zero ($\mathbf{v=0=const)}$. The latter condition corresponds to
totally symmetric $\mathbf{BVF}^{\updownarrow}$ and $\mathbf{BVB}^{\pm}$, of
primordial Bivacuum, when their torus (V$^{+})$ and antitorus (V$^{-})$ mass,
charge and magnetic moments are equal to each other (see eqs.3.16 and 3.29).
It follows from (3.142) that the internal time for each selected closed system
of particles is a parameter, characterizing the average velocity and
acceleration of these particles, i.e. this system internal dynamics. \medskip

\begin{center}
\textbf{H. Virtual Replicas (VR) of material objects in Bivacuum\smallskip}
\end{center}

The basically new concept of Virtual replica ($\mathbf{VR}$) or virtual
hologram of any material object in Bivacuum, is introduced in our Unified
theory [21-23]. The \textbf{VR }is result of interference of \emph{primary}
all-pervading quantized Virtual Pressure Waves ($\mathbf{VPW}_{q}^{+}$ and
$\mathbf{VPW}_{q}^{-})$ and Virtual Spin waves ($\mathbf{VirSW}_{\mathbf{q}%
}^{\mathbf{S=\pm1/2}}$) of Bivacuum, working as the \emph{\textquotedblright
reference waves\textquotedblright} in hologram formation and the same waves,
modulated by de Broglie waves of atoms and molecules\emph{, }representing the
\emph{\textquotedblright object waves\textquotedblright} $\mathbf{VPW}%
_{\mathbf{m}}^{\pm}$ \ and $\mathbf{VirSW}_{\mathbf{m}}^{\pm1/2}.$ The
frequencies of \emph{basic} \emph{reference }virtual pressure waves
($\mathbf{VPW}_{q=1}^{\pm}\equiv\mathbf{VPW}_{0}^{\pm})$ and virtual spin
waves ($\mathbf{VirSW}_{q=1}^{\pm1/2}\equiv\mathbf{VirSW}_{0}^{\pm1/2})$ of
Bivacuum are equal to Compton frequencies, like the carrying frequencies of
$\mathbf{[C\rightleftharpoons W]}$ pulsation of sub-elementary fermions of
triplets at zero external translational velocity of particles $\mathbf{(v=0)}$
(Fig.3.1):%
\begin{equation*}
\lbrack\mathbf{\omega}_{VPW_{0}}\,\mathbf{=\omega}_{VirSW_{0}}=\mathbf{\omega
}_{0}
=\mathbf{\omega}_{\mathbf{C\rightleftharpoons W}}^{\mathbf{v=0}}%
=\mathbf{qm}_{0}\mathbf{c}^{2}/\hbar]\,^{i}%
\end{equation*}

\emph{Two kinds of primary wave modulation (}$\mathbf{VPW}_{\mathbf{m}}^{\pm}$
\ and $\mathbf{VirSW}_{\mathbf{m}}^{\pm1/2})$ are realized by cumulative
virtual clouds ($\mathbf{CVC}^{\pm}$), emitted/absorbed in the process of
$\mathbf{[C\rightleftharpoons W]}$ pulsation of pairs: $[\mathbf{F}_{\uparrow
}^{+}\bowtie\mathbf{F}_{\downarrow}^{-}]_{C}\rightleftharpoons\lbrack
\mathbf{F}_{\uparrow}^{+}\bowtie\mathbf{F}_{\downarrow}^{-}]_{W}$ \ of
elementary triplets (electrons, protons, neutrons) $<[\mathbf{F}_{\uparrow
}^{+}\bowtie\mathbf{F}_{\downarrow}^{-}]+\mathbf{F}_{\updownarrow}^{\pm}%
>^{i},$ and the \emph{recoil angular momentum}, generated by $\mathbf{CVC}%
^{\pm1/2}$ of unpaired sub-elementary fermion\textbf{\ }$\mathbf{F}%
_{\updownarrow}^{\pm}>^{i}$, correspondingly$:$%

\begin{equation}
\lbrack\mathbf{(F}_{\uparrow}^{+}\mathbf{\bowtie\,F}_{\downarrow}%
^{-}\mathbf{)}_{C}\,+\,\left(  \mathbf{F}_{\updownarrow}^{\pm}\right)
_{W}]\tag{3.144}
\overset{+\mathbf{CVC}^{\pm}\,\mathbf{-\,}\text{\textbf{Recoil}}}%
{\underset{-\mathbf{CVC}^{\pm}\,\mathbf{+\,}\text{\textbf{Antirecoil}}%
}{<=========>}}\nonumber
\lbrack\mathbf{(F}_{\uparrow}^{+}\mathbf{\bowtie\,F}_{\downarrow}%
^{-}\mathbf{)}_{W}\,+\,\left(  \mathbf{F}_{\updownarrow}^{\pm}\right)
_{C}]\nonumber
\end{equation}

The in-phase $[\mathbf{C\rightleftharpoons W}]$ pulsation of a sub-elementary
fermion $\mathbf{F}_{\downarrow}^{+}$ and antifermion $\mathbf{F}_{\uparrow
}^{-}$ of pair $[\mathbf{F}_{\uparrow}^{-}\bowtie\mathbf{F}_{\downarrow}%
^{+}],$ accompanied by reversible [emission $\rightleftharpoons$ absorption]
of cumulative virtual clouds $\mathbf{CVC}^{+}$ and $\mathbf{CVC}^{-}$ and the
'object' virtual pressure wave ($\mathbf{VPW}_{\mathbf{m}}^{+}$ and
$\mathbf{VPW}_{\mathbf{m}}^{-})$ excitation. However, the recoil energy and
the angular momenta of CVC$^{+}$ and CVC$^{-}$ of $\mathbf{F}_{\uparrow}^{-}$
and $\mathbf{F}_{\downarrow}^{+}$ of pairs compensate each other and the
resulting recoil energy of $[\mathbf{F}_{\uparrow}^{-}\bowtie\mathbf{F}%
_{\downarrow}^{+}]$ is zero.

Superposition of \textbf{VPW}$_{m}^{+\text{ }}$and \textbf{VPW}$_{m}^{-},$
excited by cumulative virtual clouds: \textbf{CVC}$^{+}$ and \textbf{CVC}%
$^{-}$, emitted and absorbed in the process of the in-phase
$\mathbf{[C\rightleftharpoons W}]$ pulsation of pairs $[\mathbf{F}_{\uparrow
}^{+}\bowtie\mathbf{F}_{\downarrow}^{-}]$ of rotating triplets (Fig.3.1),
activate \emph{quantized whirls} in Bivacuum. The stability of VR of object,
as a \emph{hierarchical system of quantized metastable torus-like and vortex
filaments structures formed by }$\mathbf{VPW}_{\mathbf{m}}^{\pm}$\emph{\ and
by }$\mathbf{VirSW}_{\mathbf{m}}^{\pm1/2}$\emph{\ excited by paired and
unpaired sub-elementary fermions, correspondingly, }in superfluid
Bivacuum,\emph{\ }could be responsible for so-called
\textbf{\textquotedblright phantom effect\textquotedblright} of object after
its removing to another distant place. $\medskip$

\begin{center}
\emph{H.1 Bivacuum perturbations, induced by the oscillation of the total
energy of de Broglie waves, accompanied by their thermal vibrations and recoil
}$\rightleftharpoons$ \emph{antirecoil effects\smallskip}
\end{center}

In contrast to the situation with unpaired sub-elementary fermion $\left(
\mathbf{F}_{\updownarrow}^{\pm}\right)  $ in triplets, the recoil/antirecoil
momenta and energy, accompanying the in-phase emission/absorption of
$\mathbf{CVC}_{\mathbf{S=+1/2}}^{+}$ and $\mathbf{CVC}_{\mathbf{S=-1/2}}^{-}$
by $\mathbf{F}_{\uparrow}^{+}$ and $\mathbf{F}_{\downarrow}^{-}$ of pair
$[\mathbf{F}_{\uparrow}^{+}\bowtie\mathbf{F}_{\downarrow}^{-}],$ totally
compensate each other in the process of their $\mathbf{[C\rightleftharpoons
W]}$ pulsation. Such pairs display the properties of neutral particles with
zero spin and zero rest mass:%

\begin{equation}
\lbrack\mathbf{F}_{\uparrow}^{-}\bowtie\mathbf{F}_{\downarrow}^{+}%
]_{C}\;\tag{3.145}
\underset{\mathbf{[E}_{CVC^{+}}\mathbf{\,+\,E}_{CVC^{-}}\mathbf{]-\Delta
VP}^{\mathbf{F}_{\uparrow}^{+}\mathbf{\bowtie\,F}_{\downarrow}^{-}}%
\,}{\overset{\mathbf{[E}_{CVC^{+}}\mathbf{\,+\,E}_{CVC^{-}}\mathbf{]\,+\Delta
VP}^{\mathbf{F}_{\uparrow}^{+}\mathbf{\bowtie\,F}_{\downarrow}^{-}}%
}{<================>}\;}\nonumber
\lbrack\mathbf{F}_{\uparrow}^{-}\bowtie\mathbf{F}_{\downarrow}^{+}%
]_{W}\nonumber
\end{equation}

The total energy increment of elementary particle, equal to that of each of
sub-elementary fermions of triplet, generated in nonequilibrium processes,
accompanied by entropy change, like melting, boiling, etc., can be presented
in a few manners:
\begin{align}
\Delta\mathbf{E}_{tot}  &  =\mathbf{\Delta(m}_{V}^{+}\mathbf{c}^{2}%
)=\tag{3.146}
\mathbf{\Delta}\left(  \frac{\mathbf{m}_{0}\mathbf{c}^{2}}{[1-\left(
\mathbf{v/c}\right)  ^{2}]^{1/2}}\right)  =\nonumber
\frac{\mathbf{m}_{0}\mathbf{v}}{\mathbf{R}^{3}}\mathbf{\Delta v=}%
\frac{\mathbf{p}}{\mathbf{R}^{2}}\mathbf{\Delta v=}\frac{\mathbf{h}%
}{\mathbf{\lambda}_{B}\,\mathbf{R}^{2}}\mathbf{\Delta v}\nonumber\\
or  &  :\;\Delta\mathbf{E}_{tot}=\tag{3.147}
\mathbf{\Delta\lbrack(m}_{V}^{+}-\mathbf{m}_{V}^{-})\mathbf{c}%
^{2}(\mathbf{c/v)}^{2}]=\nonumber
\frac{\mathbf{2T}_{k}}{\mathbf{R}^{2}}\frac{\Delta\mathbf{v}}{\mathbf{v}%
}\;\ \nonumber\\
or  &  :\,\Delta\mathbf{E}_{tot}=\frac{\mathbf{2T}_{k}}{\mathbf{R}^{2}}%
\frac{\Delta\mathbf{v}}{\mathbf{v}}=\tag{3.147a}
\mathbf{\Delta}[\mathbf{R}\left(  \mathbf{m}_{0}\mathbf{c}^{2}\right)
_{rot}^{in}]+\mathbf{\Delta}(\mathbf{m}_{V}^{+}\mathbf{v}^{2})_{tr}%
^{ext}\nonumber
\end{align}

where:$\;\mathbf{R=}\sqrt{1-(\mathbf{v/c)}^{2}}$ is the relativistic
factor;$\;\;\Delta\mathbf{v}$\textbf{\ }is\textbf{\ }the increment of
the\textbf{\ }external translational velocity of particle; the actual inertial
mass of sub-elementary particle is: $\mathbf{m}_{V}^{+}=\mathbf{m}%
_{0}/\mathbf{R;}$ \ $\mathbf{p=\mathbf{m}_{V}^{+}v=h/\lambda}_{B}$\ \ is the
external translational momentum of unpaired sub-elementary particle
$\mathbf{F}_{\updownarrow}^{\pm}>^{i}$, equal to that of whole triplet
$<[\mathbf{F}_{\uparrow}^{+}\bowtie\mathbf{F}_{\downarrow}^{-}]+\mathbf{F}%
_{\updownarrow}^{\pm}>^{i};$ \ $\mathbf{\lambda}_{B}=\mathbf{h/p}$ \ is the de
Broglie wave of particle; \ $\mathbf{2T}_{k}=\mathbf{m}_{V}^{+}\mathbf{v}^{2}$
is a doubled kinetic energy$;$ \ $\mathbf{\Delta\ln v=}\Delta\mathbf{v/v}$.

The increments of \emph{internal} rotational and \emph{external} translational
contributions to total energy of the de Broglie wave (see eq. 3.147a) are, correspondingly:%

\begin{align}
\mathbf{\Delta}[\mathbf{R}\left(  \mathbf{m}_{0}\mathbf{c}^{2}\right)
_{rot}^{in}]  &  =\mathbf{-2T}_{k}(\Delta\mathbf{v/v)}\tag{3.148}\\
\mathbf{\Delta}(\mathbf{m}_{V}^{+}\mathbf{v}^{2})_{tr}^{ext}  &
=\mathbf{\Delta(2T}_{k})_{tr}^{ext}=\tag{3.148a}
\mathbf{2T}_{k}\frac{1\mathbf{+R}^{2}}{\mathbf{R}^{2}}\frac{\Delta
\mathbf{v}}{\mathbf{v}}\nonumber
\end{align}

The time derivative of total energy of elementary de Broglie wave, following
from (3.60 - 3.60b) is:%
\begin{equation}
\frac{\mathbf{dE}_{tot}}{\mathbf{dt}}\,\mathbf{=\,}\frac{\mathbf{2T}_{k}%
}{\mathbf{R}^{2}\mathbf{v}}\frac{d\mathbf{v}}{\mathbf{dt}}\,\mathbf{=}%
\frac{\mathbf{2T}_{k}}{\mathbf{R}^{2}}\frac{d\ln\mathbf{v}}{\mathbf{dt}}
\tag{3.149}%
\end{equation}

Between the increments of energy of triplets, equal to that of unpaired
$\Delta\mathbf{E}_{tot}=$ $\Delta\mathbf{E}_{\mathbf{F}_{\updownarrow}^{\pm}}$
\ and increments of modulated $\mathbf{CVC}_{\mathbf{m}}^{+}$ and
$\mathbf{CVC}_{\mathbf{m}}^{-},$ emitted by pair $[\mathbf{F}_{\uparrow}%
^{-}\bowtie\mathbf{F}_{\downarrow}^{+}]$ in the process of
$\mathbf{[C\rightarrow W]}$ transition, the direct correlation is existing. \ 

These cumulative virtual clouds modulated by particle's de Broglie wave
($\mathbf{\lambda}_{B}=\mathbf{h/\mathbf{m}_{V}^{+}v)}$: $\mathbf{CVC}%
_{\mathbf{m}}^{+}$ and $\mathbf{CVC}_{\mathbf{m}}^{-}$ of paired
sub-elementary fermions,\textbf{\ }superimposed with basic virtual pressure
waves ($\mathbf{VPW}_{0}^{\pm})$ of Bivacuum, turn them to the \emph{object
waves }$(\mathbf{VPW}_{\mathbf{m}}^{\pm})$, necessary for virtual hologram of
the object formation:
\begin{align}
&  \Delta\mathbf{E}_{\mathbf{F}_{\uparrow}^{+}\;}^{\mathbf{F}_{\uparrow}%
^{-}\bowtie\,\mathbf{F}_{\downarrow}^{+}}\,\,=\frac{\mathbf{h}}%
{\mathbf{\lambda}_{B}\,\mathbf{R}^{2}}\mathbf{\Delta v=}\tag{3.150}
 \mathbf{\frac{\mathbf{2T}_{k}}{\mathbf{R}^{2}}\mathbf{\Delta\ln v}%
\;\;}\,\overset{\mathbf{CVC}_{\mathbf{m}}^{+}}{-\longrightarrow}%
\,\Delta(\mathbf{VPW}_{\mathbf{m}}^{+})\nonumber\\
&  -\Delta\mathbf{E}_{\mathbf{F}_{\downarrow}^{-}\;}^{\mathbf{F}_{\uparrow
}^{-}\bowtie\,\mathbf{F}_{\downarrow}^{+}}\,\,\,\overset{\mathbf{CVC}%
_{\mathbf{m}}^{-}}{-\longrightarrow}\,\Delta(\mathbf{VPW}_{\mathbf{m}}^{-})
\tag{3.150a}%
\end{align}
\ 

The virtual pressure waves represent oscillations of corresponding virtual
pressure ($\mathbf{VirP}_{m}^{\pm}).$

The increment of total energy of fermion or antifermion, equal to increment of
its unpaired sub-elementary fermion can be presented via increments of paired
sub-elementary fermions (3.148 and 3.148a), like:%
\begin{multline}
\Delta\mathbf{E}_{tot}=\Delta\mathbf{E}_{\mathbf{F}_{\updownarrow}^{+}%
}=\tag{3.151}
\frac{1}{2}\left(  \Delta\mathbf{E}_{\mathbf{F}_{\uparrow}^{+}\;}%
^{\mathbf{F}_{\uparrow}^{+}\mathbf{\bowtie\,F}_{\downarrow}^{-}}%
-\Delta\mathbf{E}_{\mathbf{F}_{\downarrow}^{-}\;}^{\mathbf{F}_{\uparrow}%
^{+}\mathbf{\bowtie\,F}_{\downarrow}^{-}}\right)  +\nonumber
\frac{1}{2}\left(  \Delta\mathbf{E}_{\mathbf{F}_{\uparrow}^{+}\;}%
^{\mathbf{F}_{\uparrow}^{+}\mathbf{\bowtie\,F}_{\downarrow}^{-}}%
+\Delta\mathbf{E}_{\mathbf{F}_{\downarrow}^{-}\;}^{\mathbf{F}_{\uparrow}%
^{+}\mathbf{\bowtie\,F}_{\downarrow}^{-}}\right)  =\nonumber\\
=\Delta\mathbf{T}_{k}^{+}+\Delta\mathbf{V}^{+}\nonumber
\end{multline}

where, the contributions of the kinetic and potential energy increments to the
total energy increment, interrelated with increments of positive and negative
virtual pressures ($\Delta\mathbf{VirP}^{\pm}),$ are, correspondingly:
\begin{align}
&  \Delta\mathbf{T}_{k}=\tag{3.152}
\frac{1}{2}\left(  \Delta\mathbf{E}_{\mathbf{F}_{\uparrow}^{+}%
\;}^{\mathbf{F}_{\uparrow}^{+}\mathbf{\bowtie\,F}_{\downarrow}^{-}}%
-\Delta\mathbf{E}_{\mathbf{F}_{\downarrow}^{-}\;}^{\mathbf{F}_{\uparrow}%
^{+}\mathbf{\bowtie\,F}_{\downarrow}^{-}}\right)  ~\symbol{126}\nonumber
\frac{1}{2}\left(  \Delta\mathbf{VirP}^{+}-\Delta\mathbf{VirP}%
^{-}\right)  \ \symbol{126}\nonumber
~\mathbf{\alpha}\Delta\left(  \mathbf{m}_{V}^{+}\mathbf{v}^{2}\right)
_{\mathbf{F}_{\updownarrow}^{\pm}}\nonumber\\
&  \Delta\mathbf{V~}\tag{3.152a}
 \mathbf{=}\frac{1}{2}\left(  \Delta\mathbf{E}_{\mathbf{F}_{\uparrow}^{+}%
\;}^{\mathbf{F}_{\uparrow}^{+}\mathbf{\bowtie\,F}_{\downarrow}^{-}}%
+\Delta\mathbf{E}_{\mathbf{F}_{\downarrow}^{-}\;}^{\mathbf{F}_{\uparrow}%
^{+}\mathbf{\bowtie\,F}_{\downarrow}^{-}}\right)  ~\symbol{126}\nonumber
~\frac{1}{2}\left(  \Delta\mathbf{VirP}^{+}+\Delta\mathbf{VirP}%
^{-}\right)  \ \nonumber
 \symbol{126}~\mathbf{\beta}\Delta\left(  \mathbf{m}_{V}^{+}~\mathbf{+~m}%
_{V}^{-}\right)  \mathbf{c}_{\mathbf{F}_{\updownarrow}^{\pm}}^{2}\nonumber
\end{align}

\emph{The specific information of any object is imprinted in its Virtual
Replica (VR)}, because cumulative virtual clouds ($\mathbf{CVC}_{\mathbf{m}%
}^{\pm}$) of the object's elementary particles and their superposition with
Bivacuum pressure waves and Virtual spin waves: $\mathbf{VPW}_{\mathbf{m}%
}^{\pm}$ and $\mathbf{VirSW}_{\mathbf{m}}^{\pm1/2}$ \ are modulated by
frequency, phase and amplitude of the thermal de Broglie waves of molecules,
composing this object. Comparing eqs. 3.152a and 3.88a we may see, that the
modulated gravitational virtual pressure waves form a part of VR. \medskip

\begin{center}
\emph{H.2 \ Modulation of the basic Virtual Pressure Waves (}$VPW_{q}^{\pm}%
)$\emph{\ and Virtual Spin Waves (}$VirSW_{q}^{\pm1/2})$\emph{\ }

\emph{of Bivacuum by molecular translations and librations\medskip}
\end{center}

The external translational/librational kinetic energy of particle
($\mathbf{T}_{k})_{tr,lb}$ is directly related to its de Broglie wave length
($\mathbf{\lambda}_{B}),$ the group (\textbf{v}$\mathbf{),}$\textbf{\ }phase
velocity ($\mathbf{v}_{ph})$ and frequency ($\mathbf{\nu}_{B}=\mathbf{\omega
}_{B}/2\pi)$:%
\begin{equation}
\Big(\mathbf{\lambda}_{B}\,\mathbf{=\,}\frac{\mathbf{h}}{\mathbf{m}_{V}%
^{+}\mathbf{v}}\,\mathbf{=\,}\frac{\mathbf{h}}{\mathbf{2m}_{V}^{+}%
\mathbf{T}_{k}}=\tag{3.153}
\frac{\mathbf{v}_{ph}}{\mathbf{\nu}_{B}}=\mathbf{2\pi}\frac{\mathbf{v}_{ph}%
}{\mathbf{\omega}_{B}}\Big)_{tr,lb}\nonumber
\end{equation}

where the de Broglie wave frequency is related to its length and kinetic
energy of particle as: \
\begin{equation}
\left[  \mathbf{\nu}_{B}=\frac{\mathbf{\omega}_{B}}{2\pi}=\frac{h}%
{2\mathbf{m}_{V}^{+}\mathbf{\lambda}_{B}^{2}}=\frac{\mathbf{m}_{V}%
^{+}\mathbf{v}^{2}}{2h}\right]  _{tr,lb} \tag{3.154}%
\end{equation}

It follows from our model, that zero-point frequency of [$C\rightleftharpoons
W]$ pulsation ($\mathbf{\omega}_{0})^{i}$ of sub-elementary fermions and
antifermions, forming triplets of elementary particles $<[\mathbf{F}%
_{\uparrow}^{+}\bowtie\mathbf{F}_{\downarrow}^{-}]+\mathbf{F}_{\updownarrow
}^{\pm}>^{i}$, accompanied by [\emph{emission} $\rightleftharpoons
\,absorbtion]$ of cumulative virtual clouds $\mathbf{CVC}^{\pm}$ has has the
same value, as a basic (reference) frequency $\left(  \mathbf{\omega}%
_{q=1}=\mathbf{\omega}_{0}\mathbf{=m}_{0}\mathbf{c}^{2}\mathbf{/\hbar}\right)
^{i}$ of Bivacuum.

The total energy of de Broglie wave and resulting frequency of pulsation
($\mathbf{\omega}_{\mathbf{C\rightleftharpoons W}})$ (eq. 3.78) is a result of
modulation of the internal frequency, related to the rest mass of particle, by
the most probable frequency of de Broglie wave of the whole particle
($\mathbf{\omega}_{B})$, determined by its most probable external momentum:
$\mathbf{p}=\mathbf{m}_{V}^{+}\mathbf{v.}$

In a composition of condensed matter this value is different for thermal
librations and translation of molecules. The corresponding most probable
modulation frequencies of translational and librational de Broglie waves are
possible to calculate, using our Hierarchic theory of condensed matter and
based on this theory computer program [57, 63, 64].

The interference between the basic $\mathbf{(q=j-k=1)}$ reference virtual spin
waves ($\mathbf{VirSW}_{q=1}^{\pm1/2})$ and virtual pressure\ waves
\textbf{VPW}$_{q=1}^{\pm}$ of Bivacuum with virtual 'object waves', modulated
by the librational and translational de Broglie waves ($\mathbf{\lambda
}_{lb,tr}\,\mathbf{=h/m}_{V}^{+}\mathbf{v}$) of molecules, produce the
holographic-like image or Virtual Replicas (VR) of the object.

The \emph{frequencies} of virtual pressure waves ($\mathbf{VPW}_{\mathbf{m}%
}^{\pm})$ and virtual spin waves ($\mathbf{VirSW}_{\mathbf{m}}^{\pm1/2})$ are
modulated by de Broglie waves of the object particles. Corresponding
modulation frequencies are related to frequencies of librational
($\mathbf{\omega}_{lb})$ and\ translational ($\mathbf{\omega}_{tr})$ de
Broglie waves of molecules of matter (3.154) in accordance to rules of
combinational resonance:%

\begin{align}
\mathbf{\omega}_{\mathbf{VPW}_{\mathbf{m}}^{\pm}}^{i}  &  =\mathbf{R}%
\,\mathbf{\omega}_{0}^{i}\,\,\mathbf{+\,g\,\omega}_{tr}+\tag{3.155}
\mathbf{r\,\omega}_{lb}    \cong\left[  \mathbf{R}\,\mathbf{\omega}_{0}%
^{i}\,\,\mathbf{+\,g\,\omega}_{tr}\right] \nonumber\\
\mathbf{\omega}_{\mathbf{VirSW}_{\mathbf{m}}^{\pm1/2}}^{i}  &  =\mathbf{R}%
\,\mathbf{\omega}_{0}^{i}\,\mathbf{+\,r\,\omega}_{lb}\,+\tag{3.155a}
\mathbf{g\,\omega}_{tr}  \cong\mathbf{R}\,\mathbf{\omega}_{0}%
^{i}\,\mathbf{+\,r\,\omega}_{lb}\nonumber\\
\mathbf{R}  &  =\sqrt{1-\left(  \mathbf{v/c}\right)  ^{2};}\;\nonumber\qquad
\;\;\mathbf{g,r}    =1,2,3...(\text{integer}\;\text{numbers)}\nonumber
\end{align}

Each of 24 collective excitations of condensed matter, introduced in our
Hierarchic theory [22, 63], has his own characteristic frequency and can be
imprinted in Virtual Replica of the object, as a corresponding pattern.

Three kinds of modulations: \ \emph{frequency, amplitude and phase} of the
object ($\mathbf{VPW}_{\mathbf{m}}^{\pm})$ and ($\mathbf{VirSW}_{\mathbf{m}%
}^{\pm1/2})$ by de Broglie waves of the object's molecules may be described by
known relations [68]:

1. \emph{The frequencies} of virtual pressure waves ($\omega_{_{VPW^{\pm}}%
}^{M})$ and spin waves ($\omega_{_{VirSW^{\pm}}}^{M}),$ \emph{modulated} by
translational and librational de Broglie waves of the object's molecules, can
be presented as:
\begin{align}
\mathbf{\omega}_{_{VPW_{m}^{\pm}}}^{M}  &  \mathbf{=\,R}\mathbf{\omega}%
_{0}^{i}+\mathbf{\Delta\omega}_{B}^{tr}\cos\mathbf{\omega}_{B}^{tr}%
\,t\tag{3.156}\\
\mathbf{\omega}_{_{VirSW_{m}^{\pm1/2}}}^{M}  &  \mathbf{=\,R\omega}_{0}%
^{i}+\mathbf{\Delta\omega}_{B}^{lb}\cos\mathbf{\omega}_{B}^{lb}\,t
\tag{3.156a}%
\end{align}

The Compton pulsation frequency of elementary particles is equal to basic
frequency of Bivacuum virtual waves at $\mathbf{q=j-k=1}$: $\ $%
\begin{equation}
\mathbf{\omega}_{0}^{i}=\mathbf{m}_{0}^{i}\mathbf{c}^{2}\mathbf{/\hbar=\omega
}_{VPW_{q=1}^{\pm},ViSW_{q=1}}^{i} \tag{3.157}%
\end{equation}
Such kind of modulation is accompanied by two satellites with frequencies:
$\mathbf{\ }$($\mathbf{\omega}_{0}^{i}+\mathbf{\omega}_{B}^{tr,lb})$ \ and
$\,$($\mathbf{\omega}_{0}^{i}-\mathbf{\omega}_{B}^{tr,lb})=\mathbf{\Delta
\omega}_{tr,lb}^{i}$. The latter is named frequency deviation$.$ In our case:
\ $\mathbf{\omega}_{0}^{e}\,(\symbol{126}\,10^{21}s^{-1})>>\mathbf{\omega}%
_{B}^{tr,lb}\,(\symbol{126}\;10^{12}s^{-1})$ and $\mathbf{\Delta\omega
}_{tr,lb}>>$ $\mathbf{\omega}_{B}^{tr,lb}.$

The temperature of condensed matter and phase transitions may influence the
modulation frequencies of de Broglie waves of its molecules.\medskip

2. \emph{The amplitudes of virtual pressure waves (VPW}$_{m}^{\pm}%
)$\emph{\ and virtual spin waves VirSW}$_{m}^{\pm1/2}$\emph{ (informational
waves) modulated by the object are dependent on translational and librational
de Broglie waves frequencies as:}%
\begin{align}
\mathbf{A}_{VPW_{m}^{\pm}}  &  \mathbf{\approx A}_{0}\mathbf{(}\sin
\mathbf{R\omega}_{0}^{i}\mathbf{t\,}\tag{3.158}
  \mathbf{+\gamma\,\omega}_{B}^{tr}\sin\mathbf{\,t\cdot}\cos\mathbf{\omega
}_{B}^{tr}\mathbf{\,}t)\nonumber\\
\mathbf{I}_{VirSW_{m}^{\pm1/2}}  &  \mathbf{\approx I}_{0}\mathbf{(}%
\sin\mathbf{R\omega}_{0}^{i}\mathbf{t\,}\tag{3.158a}
  \mathbf{+\gamma\,\omega}_{B}^{lb}\sin\mathbf{\,t\cdot}\cos\mathbf{\omega
}_{B}^{lb}\mathbf{\,t)}\nonumber
\end{align}

where: the informational/spin field amplitude is determined by the amplitude
of Bivacuum fermions [$BVF^{\uparrow}\rightleftharpoons BVF^{\downarrow}]$
equilibrium constant oscillation: $\mathbf{I}_{S}\equiv\mathbf{I}%
_{\mathbf{VirSW}^{\pm1/2}}\,\symbol{126}\,\mathbf{K}_{BVF^{\uparrow
}\rightleftharpoons BVF^{\downarrow}}\mathbf{(t)}$

The index of frequency modulation is defined as: $\mathbf{\gamma}=$%
\textbf{\ }$\mathbf{(\Delta\omega}_{tr,lb}/\mathbf{\omega}_{B}^{tr,lb}).$ The
carrying zero-point pulsation frequency of particles is equal to the basic
frequency of Bivacuum virtual waves: $\ \mathbf{\omega}_{VPW_{0}^{\pm
},ViSW_{0}}^{i}\mathbf{=\omega}_{0}^{i}$. Such kind of modulation is
accompanied by two satellites with frequencies: $\mathbf{\ }$($\mathbf{\omega
}_{0}^{i}+\mathbf{\omega}_{B}^{tr,lb})$ and\ ($\mathbf{\omega}_{0}%
^{i}-\mathbf{\omega}_{B}^{tr,lb})=\mathbf{\Delta\omega}_{tr,lb}.$ In our case:
\ $\mathbf{\omega}_{0}^{e}(\symbol{126}\,10^{21}s^{-1})>>\mathbf{\omega}%
_{B}^{tr,lb}(\symbol{126}\;10^{12}s^{-1})$ and $\mathbf{\gamma}>>1.$

The fraction of molecules in state of mesoscopic molecular Bose condensation
(mBC), representing, coherent clusters [22,63] is a factor, influencing the
amplitude ($A_{0}$) and such kind of modulation of Virtual replica of the object.

3. \emph{The} \emph{phase modulated VPW}$_{m}^{\pm}$\textbf{\ \ }\emph{and
VirSW}$_{m}^{\pm1/2}$\textbf{\ \ }by de Broglie waves of molecules, related to
their translations and librations, can be described like:
\begin{align}
\mathbf{A}_{VPW_{m}^{\pm}}^{M}  &  \mathbf{=A}_{0}\sin\mathbf{(R\omega}%
_{0}\mathbf{t+\Delta\varphi}_{tr}\sin\mathbf{\omega}_{B}^{tr}\mathbf{t)}%
\tag{3.159}\\
\mathbf{I}_{VirSW_{m}^{\pm1/2}}^{M}  &  \mathbf{=I}_{0}\sin\mathbf{(R\omega
}_{0}\mathbf{t+\Delta\varphi}_{lb}\sin\mathbf{\omega}_{B}^{lb}\mathbf{t)}
\tag{3.159a}%
\end{align}

The value of phase increment $\mathbf{\Delta\varphi}_{tr,lb}$ of modulated
virtual waves of Bivacuum ($VPW_{m}^{\pm}$ and $VirSW_{m}^{\pm1/2}),$ contains
the information about geometrical properties of the object.

The phase modulation takes place, if the phase increment $\mathbf{\Delta
\varphi}_{tr,lb}$ is independent on the modulation frequency $\mathbf{\omega
}_{B}^{tr,lb}$.

\emph{The virtual holographic image}, resulting from \emph{interference
pattern} of the virtual \emph{object waves}: $\mathbf{VPW}_{m}^{\pm}$ and
$\mathbf{VirSW}_{m}^{\pm1/2}$, modulated (scattered) by translations and
librations of molecules, with similar \emph{reference waves} of Bivacuum
($\mathbf{VPW}_{q}^{\pm}$ and $\mathbf{VirSW}_{q}^{\pm1/2})$ contains full
information about the object's \emph{internal} dynamic and spatial properties
and may be named \emph{''Virtual Replica \ (VR)'' of the object. } \medskip

\begin{center}
\textbf{I. \ Possible mechanism of entanglement between remote }

\textbf{elementary particles via Virtual Guides of spin, momentum and energy
(VirG}$_{\mathbf{S,M,E}})$\textbf{\smallskip}
\end{center}

In accordance to our theory, the instant nonlocal quantum entanglement between
two or more distant \emph{similar} elementary particles (electrons, protons,
\ neutrons, photons), named [Sender (S)] and [Receiver (R)], revealed in a lot
of experiments, started by Aspect and Grangier [69], involves a few stages:

\textbf{1.} Superposition of their nonlocal and distant components of Virtual
replicas ($\mathbf{VR}$) or Virtual hologram, formed by interference of
modulated by de Broglie waves of object Virtual spin waves: $\left[
\mathbf{VirSW}_{\mathbf{m}}^{\pm1/2}(S)<==>\mathbf{VirSW}_{\mathbf{m}}%
^{\pm1/2}(R)\right]  $ and Virtual pressure \ waves: $\left[  \mathbf{VPW}%
_{\mathbf{m}}^{\pm}(S)<==>\mathbf{VPW}_{\mathbf{m}}^{\pm}(R)\right]  $ with
corresponding \emph{reference waves} of Bivacuum $\mathbf{VirSW}_{\mathbf{q}%
}^{\pm1/2}$ and $\mathbf{VPW}_{\mathbf{q}}^{\pm},$ described in Section3.8.2;

\textbf{2.} Tuning (frequency and phase synchronization) of de Broglie waves
of remote interacting identical particles (like electrons, protons) and their
complexes in form of atoms and molecules, as a condition of Virtual Guides of
spin-momentum-energy ($\mathbf{VirG}_{\mathbf{SME}})$ between [S] and [R]
formation (see Fig.3.4).

Superpositions of counterphase ($\mathbf{VirSW}_{\mathbf{m}}^{\pm1/2})$ of [S]
and [R], exited by their unpaired sub-elementary fermions $F_{\updownarrow
}^{\pm}\rangle$ of triplets $<[\mathbf{F}_{\uparrow}^{+}\bowtie\mathbf{F}%
_{\downarrow}^{-}]+\mathbf{F}_{\updownarrow}^{\pm}>^{i}$ of opposite spins in
form of virtual standing waves, may stimulate formation of two kinds of
Virtual Guides:

a) single \emph{nonlocal virtual guides }$\mathbf{VirG}_{SME}^{\left(
\mathbf{BVB}^{\pm}\right)  ^{i}}$ - virtual microtubules from Bivacuum bosons
($\mathbf{BVB}^{\pm})^{i}$. In this case the $\mathbf{VirG}_{SME}^{\left(
\mathbf{BVB}^{\pm}\right)  ^{i}}$ is not rotating as a whole around its axis
and the resulting spin is zero.

b) twin \emph{nonlocal virtual guides }$\mathbf{VirG}_{SME}^{\mathbf{[BVF}%
^{\uparrow}\mathbf{\bowtie BVF}^{\downarrow}]^{i}}$\ from Cooper pairs of
Bivacuum fermions $\mathbf{[BVF}^{\uparrow}\mathbf{\bowtie BVF}^{\downarrow
}]^{i}\mathbf{.}$ In this case each of two adjacent microtubules rotate around
their own axes in opposite directions. The resulting angular moment (spin) of
such pair is also zero.

Two remote coherent triplets - elementary particles, like (electron
-\ electron) or (proton - proton) with similar frequency of
$\mathbf{[C\rightleftharpoons W]}_{e,p}$ pulsation and opposite spins (phase)
can be connected by Virtual guides ($\mathbf{VirG}_{SME}^{i})$ of spin (S),
momentum (M) and energy (E) from Sender to Receiver of both kinds. The spin -
information (qubits), momentum and kinetic energy instant transmission via
such $\mathbf{VirG}_{SME}^{i}$ from [S] and [R] is possible. The same is true
for two synchronized photons (bosons) of opposite spins.

The double $\mathbf{VirG}_{SME}^{\mathbf{[BVF}^{\uparrow}\mathbf{\bowtie
BVF}^{\downarrow}]^{i}}$ can be transformed to single $\mathbf{VirG}%
_{SME}^{\left(  \mathbf{BVB}^{\pm}\right)  ^{i}}$ by conversion of opposite
Bivacuum fermions: $\mathbf{BVF}^{\uparrow}=[\mathbf{V}^{+}\upuparrows
\mathbf{V}^{-}]$ and $\mathbf{BVF}^{\downarrow}=[\mathbf{V}^{+}\downdownarrows
\mathbf{V}^{-}]$ to the pair of Bivacuum bosons of two possible alternatives
of polarization:%
\begin{equation*}
\mathbf{BVB}^{+}    =[\mathbf{V}^{+}\uparrow\downarrow\mathbf{V}^{-}]\;\;\qquad
\;\;or\qquad\mathbf{BVB}^{-}    =[\mathbf{V}^{+}\downarrow\uparrow
\mathbf{V}^{-}]
\end{equation*}

Superposition of two nonlocal virtual spin waves excited by similar elementary
particles (electrons or protons) of Sender $\left(  \mathbf{VirSW}%
_{\mathbf{m}}^{\mathbf{S=+1/2}}\right)  _{\mathbf{S}}$ and Receiver $\left(
\mathbf{VirSW}_{\mathbf{m}}^{\mathbf{S=-1/2}}\right)  _{\mathbf{R}}$ of the
same pulsation frequency and opposite spins, i.e. opposite phase of
$\mathbf{[C\rightleftharpoons W]}$ pulsation, forms \emph{Virtual Guide }of
spin, momentum and energy\emph{ (}$\mathbf{VirG}_{SME})^{i}$
(Fig.3.4).\medskip$:$ \
\begin{gather}
\Bigg[<[\mathbf{F}_{\downarrow}^{+}\bowtie\mathbf{F}_{\uparrow}^{-}%
]_{C}+\left(  \mathbf{F}_{\downarrow}^{-}\right)  _{W}>_{\mathbf{S}%
}\nonumber
\overset{\mathbf{VirSW}_{S}}{\rightarrowtail}\;\overset{\mathbf{BVB}^{+}%
}{\underset{\mathbf{BVB}^{-}}{=\Bumpeq=\Bumpeq=}}\mathbf{\;}\overset
{\mathbf{VirSW}_{R}}{\leftarrowtail}\,\nonumber
<\left(  \mathbf{F}_{\uparrow}^{-}\right)  _{C}+[\mathbf{F}_{\downarrow}%
^{-}\bowtie\mathbf{F}_{\uparrow}^{+}]_{W}>_{\mathbf{R}}\Bigg]^{i}\nonumber\\
or:[\mathbf{n}_{+}\mathbf{BVB}^{+}(\mathbf{V}^{+}\uparrow\downarrow
\mathbf{V}^{-})\,\tag{3.160}
\mathbf{+\,n}_{-}\mathbf{BVB}^{-}(\mathbf{V}^{+}\downarrow\uparrow
\mathbf{V}^{-}]^{i}=\left(  \mathbf{VirG}_{SME}^{ext}\right)  ^{i}%
\;\;\nonumber
\end{gather}

The spin exchange via $\mathbf{VirG}_{SME}^{i}$ is accompanied either by the
instant change of Bivacuum boson polarization: $\left[  \mathbf{BVB}%
^{+}\rightleftharpoons\mathbf{BVB}^{-}\right]  ^{i}$ or by instant spin state
change of both Bivacuum fermions$\mathbf{,}$ forming virtual Cooper pairs in
the double virtual microtubule:%
\begin{equation}
\lbrack\mathbf{BVF}^{\uparrow}\bowtie\mathbf{BVF}^{\downarrow}]^{i}%
\ \overset{(S=+1/2)\rightarrow(S=-1/2)}{\rightleftharpoons}\ \tag{3.161}
\lbrack\mathbf{BVF}^{\downarrow}\bowtie\mathbf{BVF}^{\uparrow}]^{i}\nonumber
\end{equation}

The radius of virtual microtubules of $\mathbf{VirG}_{SME}^{i}$ is dependent
on generation of torus and antitorus ($i=e,\mu,\tau)$, forming them:%
\begin{equation*}
\mathbf{L}_{V}^{e}\,    \mathbf{=}\mathbf{\hbar}/\mathbf{m}_{0}%
^{e}\mathbf{c>>L}_{V}^{\mu}\,\mathbf{=}
\mathbf{\hbar}/\mathbf{m}_{0}^{\mu}\mathbf{c>L}_{V}^{\tau}\,\mathbf{=\hbar
}/\mathbf{m}_{0}^{\mu}\mathbf{c}%
\end{equation*}

The radius of $\mathbf{VirG}_{SME}^{e}$, connecting two remote electrons, is
the biggest one ($\mathbf{L}^{e}).$ The radius of $\mathbf{VirG}_{SME}^{\tau}%
$, connecting two protons or neutrons ($\mathbf{L}^{\tau})$ is about
3.5$\times10^{3}$ times smaller$.$ The entanglement between similar atoms in
pairs, like hydrogen, oxygen, carbon or nitrogen can be realized via complex
virtual guides of atoms ($\mathbf{VirG}_{SME}^{at})$, representing
\emph{multishell constructions. }

The increments of momentum $\Delta\mathbf{p=\Delta(m}_{V}^{+}\mathbf{v)}%
_{tr,lb}$ and kinetic ($\Delta$\textbf{T}$_{k})_{tr,lb}$ energy transmission
from [S] to [R] of \emph{selected generation of elementary particles }is
determined by the translational and librational velocity variation
($\mathbf{\Delta v)}$ of nuclei of (S). This means, that energy/momentum
transition from [S] to [R] is possible, if they are in nonequilibrium state.

The variation of kinetic energy of atomic nuclei under external force
application, induces nonequilibrium in a system $\mathbf{(S+R)}$ and
decoherence of $\mathbf{[C\rightleftharpoons W}]$ pulsation of protons and
neutrons of [S] and [R]. The nonlocal energy transmission from [S] to [R] is
possible, if the decoherence is not big enough for disassembly of the virtual
microtubules and their systems in the case of atoms. The electronic
$\mathbf{VirG}_{SME}^{e},$ as more coherent (not so dependent on thermal
vibrations), can be responsible for stabilization of the complex atomic
Virtual Guides $\sum\mathbf{VirG}_{SME}^{e,p,n}.$

The values of the energy and velocity increments of free elementary particles
are interrelated by (3.147).

The instantaneous energy flux via ($\mathbf{VirG}_{SME}$)$^{i}$, is mediated
by pulsation of energy and radii of torus ($\mathbf{V}^{+})$ and antitorus
($\mathbf{V}^{-})$ of Bivacuum bosons: $\mathbf{BVB}^{+}\mathbf{=[V}%
^{+}\mathbf{\uparrow\downarrow V}^{-}\mathbf{]}$. Corresponding energy
increments of the actual torus and complementary antitorus of $\mathbf{BVB}%
^{\pm},$ forming ($\mathbf{VirG}_{SME}$)$^{i}$\textbf{, }are directly related
to increments of Sender particle external velocity $\left(  \Delta
\mathbf{v}\right)  $:%

\begin{align}
\Delta\mathbf{E}_{V^{+}}\;  &  =+\Delta\mathbf{m}_{V}^{+}c^{2}=\tag{3.162}
\Big(+\frac{\mathbf{p}^{+}}{\mathbf{R}^{2}}\left(  \Delta\mathbf{v}%
\right)  _{\mathbf{F}_{\uparrow}^{+}}^{[\mathbf{F}_{\uparrow}^{+}%
\bowtie\mathbf{F}_{\downarrow}^{-}]}=\nonumber
\mathbf{m}_{V}^{+}\mathbf{c}^{2}\frac{\Delta\mathbf{L}_{V^{+}}}%
{\mathbf{L}_{V^{+}}}\Big)_{N,S}\text{ \ \ \ actual}\nonumber\\
\Delta\mathbf{E}_{V^{-}}  &  =-\Delta\mathbf{m}_{V}^{-}c^{2}=\tag{3.162a}
\Big(-\frac{\mathbf{p}^{-}}{\mathbf{R}^{2}}\left(  \Delta\mathbf{v}%
\right)  _{\mathbf{F}_{\uparrow}^{-}}^{[\mathbf{F}_{\uparrow}^{+}%
\bowtie\mathbf{F}_{\downarrow}^{-}]}=\nonumber
-\mathbf{m}_{V}^{-}\mathbf{c}^{2}\frac{\Delta\mathbf{L}_{V^{-}}%
}{\mathbf{L}_{V^{-}}}\Big)_{N,S}\text{ \ \ complementary}\nonumber
\end{align}

where: $\mathbf{p}^{+}\mathbf{=m}_{V}^{+}\mathbf{v;\;}$\ $\mathbf{p}%
^{-}\mathbf{=m}_{V}^{-}\mathbf{v}$ are the actual and complementary momenta;
$\mathbf{L}_{V^{+}}=\hbar/\mathbf{m}_{V}^{+}\mathbf{c}$ and $\mathbf{L}%
_{V^{-}}=\hbar/\mathbf{m}_{V}^{-}\mathbf{c}$ are the radii of torus and
antitorus of $\mathbf{BVB}^{\pm}=[\mathbf{V}^{+}\Updownarrow\mathbf{V}^{-}]$,
forming $\mathbf{VirG}_{S,M,E}^{in,ext}.$

The nonlocal energy exchange between [S] and [R] is accompanied by the instant
pulsation of radii of tori (V$^{+})$ and antitori (V$^{-})$ of
BVF$^{\updownarrow}$ and BVB$^{\pm}$, accompanied by corresponding pulsation
$\left\vert \Delta\mathbf{L}_{V^{\pm}}/\mathbf{L}_{V^{\pm}}\right\vert $ of
the whole virtual microtubule $\mathbf{VirG}_{SME}$ (Fig.3.4)$.$

Most effectively the proposed mechanism of spin (information), momentum and
energy exchange can work between Sender and Receiver, containing coherent
molecular clusters - mesoscopic Bose condensate (mBC) [21, 22, 63].

The instantaneous angular momentum (spin) exchange between [S] and [R] does
not need the radius pulsation, but only the instantaneous polarization change
of Bivacuum dipoles $\left(  \mathbf{BVB}^{+}\rightleftharpoons\mathbf{BVB}%
^{-}\right)  ^{i}$ or\emph{ Cooper pairs }[$\mathbf{BVF}^{\uparrow}%
\bowtie\mathbf{BVF}^{\downarrow}]~\rightleftharpoons~[\mathbf{BVF}%
^{\downarrow}\bowtie\mathbf{BVF}^{\uparrow}],$ forming $\mathbf{VirG}_{SME}.$

\begin{center}%
\begin{center}
\includegraphics[width=0.8\textwidth]%
{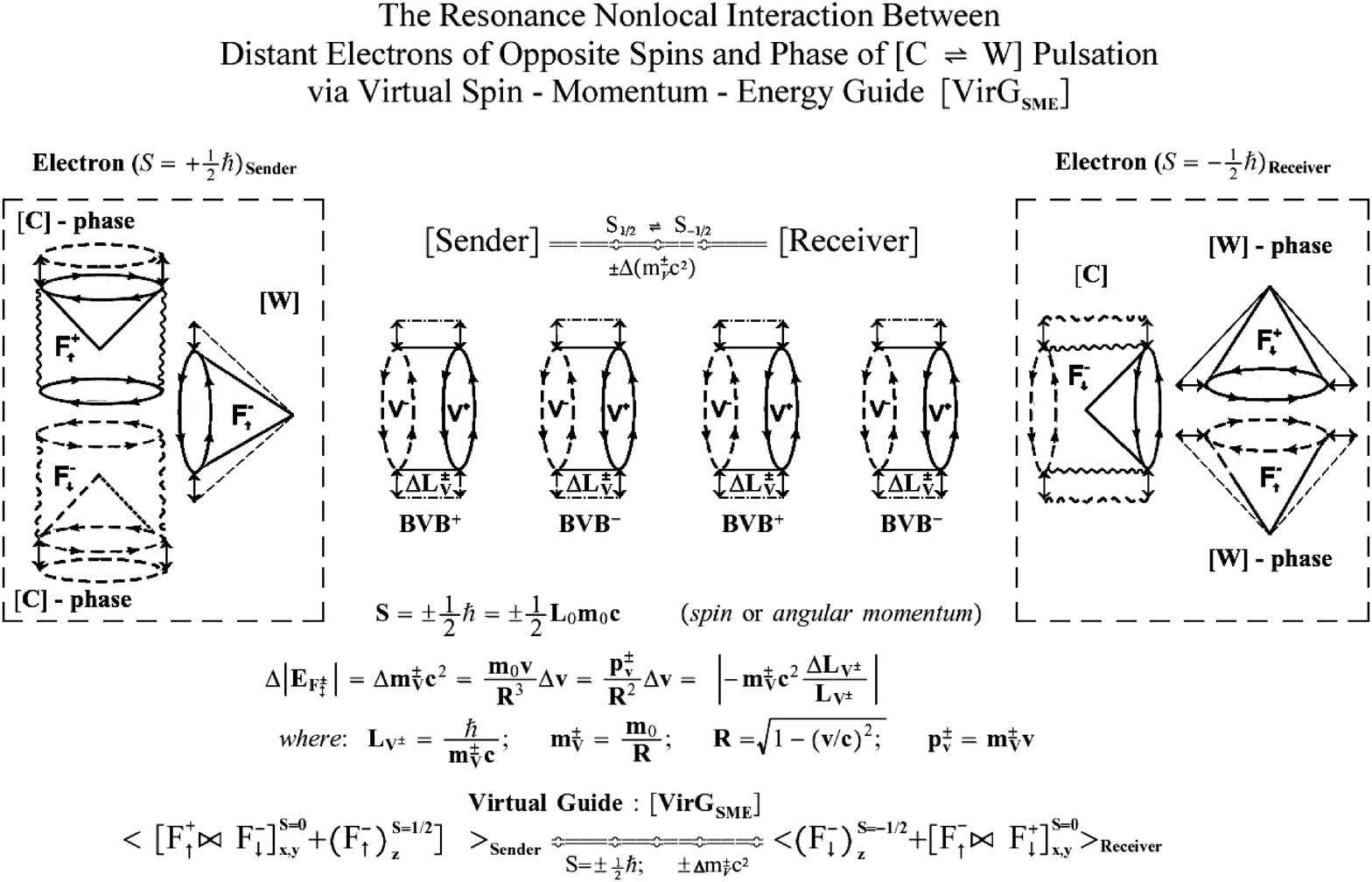}%
\end{center}

\end{center}

\begin{quotation}
\textbf{Fig.3.4.} The mechanism of nonlocal Bivacuum mediated interaction
(entanglement) between two distant unpaired sub-elementary fermions of 'tuned'
elementary triplets (particles) of the opposite spins $<\mathbf{[F}_{\uparrow
}^{+}\mathbf{\bowtie F}_{\downarrow}^{-}\mathbf{]+F}_{\uparrow}^{-}%
>_{\text{\textbf{Sender}}}^{i}$ and $<[\mathbf{F}_{\downarrow}^{+}%
\mathbf{\bowtie F}_{\uparrow}^{-}\mathbf{]+F}_{\downarrow}^{-}%
>_{\text{\textbf{Receiver}}}^{i},$ with close frequency of
[$\mathbf{C\rightleftharpoons W}]$ pulsation and close de Broglie wave length
($\lambda_{\mathbf{B}}\mathbf{\,=h/m}_{V}^{+}\mathbf{v})$ of particles. The
tunnelling of momentum and energy increments: $\mathbf{\Delta}\left\vert
\mathbf{m}_{\mathbf{V}}^{\pm}\mathbf{c}^{2}\right\vert $ $\symbol{126}%
\,\mathbf{\Delta}\left\vert \mathbf{VirP}^{+}\right\vert \pm\mathbf{\Delta
}\left\vert \mathbf{VirP}^{-}\right\vert $\ from Sender to Receiver and
vice-verse via Virtual spin-momentum-energy Guide [\textbf{VirG}%
$_{\mathbf{SME}}^{i}]$ is accompanied by instantaneous pulsation of diameter
($\mathbf{2\Delta L}_{\mathbf{V}}^{\pm}\mathbf{)}$ of this virtual
microtubule, formed by Bivacuum bosons \textbf{BVB}$^{\pm}$ or double
microtubule, formed by Cooper pairs of Bivacuum fermions: [$\mathbf{BVF}%
^{\uparrow}\bowtie\mathbf{BVF}^{\downarrow}].$ The spin state exchange between
[S] and [R] can be realized by the instantaneous change polarization of Cooper
pairs: [$\mathbf{BVF}^{\uparrow}\bowtie\mathbf{BVF}^{\downarrow}%
]\rightleftharpoons\lbrack\mathbf{BVF}^{\downarrow}\bowtie\mathbf{BVF}%
^{\uparrow}]$ and Bivacuum bosons: \textbf{BVB}$^{+}\rightleftharpoons$
\textbf{BVB}$^{-}.$ \medskip
\end{quotation}

In virtual microtubules ($\mathbf{VirG}_{SME}^{i}$)$^{i}$ the time and its
'pace' are uncertain: $\mathbf{t=0/0}$, as far the external velocities
(\textbf{v)} and accelerations (\textbf{dv/dt)} of Bivacuum dipoles, composing
such virtual Bose condensate, are zero (see eqs. 3.141 and 3.142).\emph{\ }%
\medskip

\begin{center}
\emph{I.1 }\ \emph{The role of tuning} \emph{force (}$\mathbf{F}%
_{\mathbf{VPW}^{\pm}})$\emph{ of virtual }

\emph{pressure waves }$\mathbf{VPW}_{q}^{\pm}$\emph{ of Bivacuum in
entanglement\medskip}
\end{center}

The tuning between \textbf{two similar elementary} particles: 'sender (S)' and
'receiver (R)' via $\mathbf{VirG}_{SME}^{i}$ may be qualitatively described,
using well known model of \emph{damped harmonic oscillators,} interacting with
all-pervading virtual pressure waves ($\mathbf{VPW}^{\pm})$ of Bivacuum with
fundamental frequency ($\mathbf{\omega}_{0}=\mathbf{m}_{0}\mathbf{c}^{2}%
/\hbar).$ The criteria of tuning - synchronization of [S] and [R] is the
equality of the amplitude probability of resonant energy exchange of Sender
and Receiver with virtual pressure waves ($\mathbf{VPW}_{0}^{\pm})$:
$\mathbf{A}_{C\rightleftharpoons W}^{S}\,\mathbf{=A}_{C\rightleftharpoons
W}^{R}$, resulting from minimization of frequency difference ($\mathbf{\omega
}_{S}\,\mathbf{-\,\omega}_{0})\rightarrow0$ and $\left(  \mathbf{\omega}%
_{R}\,\mathbf{-\,\omega}_{0}\right)  \rightarrow0$:%
\begin{align}
&  \mathbf{A}_{C\rightleftharpoons W}^{S}\sim\left[  \frac{1}{\left(
\mathbf{m}_{V}^{+}\right)  _{S}}\frac{\mathbf{F}_{\mathbf{VPW}^{\pm}}}{\left(
\mathbf{\omega}_{S}^{2}\,\mathbf{-\,\omega}_{0}^{2}\right)  +\operatorname{Im}%
\mathbf{\,\gamma\omega}_{S}}\right] \tag{3.163}\\
&  \left[  \mathbf{A}_{C\rightleftharpoons W}^{R}\right]  _{x,y,z}\sim\left[
\frac{1}{\left(  \mathbf{m}_{V}^{+}\right)  _{R}}\frac{\mathbf{F}%
_{\mathbf{VPW}^{\pm}}}{\left(  \mathbf{\omega}_{R}^{2}\,\mathbf{-\,\omega}%
_{0}^{2}\right)  +\operatorname{Im}\mathbf{\,\gamma\omega}_{R}}\right]
\tag{3.163a}%
\end{align}

where the frequencies of $\mathbf{C\rightleftharpoons W}$ pulsation of
particles of Sender ($\mathbf{\omega}_{S}$) and Receiver ($\mathbf{\omega}%
_{R})$ are:
\begin{align}
\mathbf{\omega}_{R}  &  =\mathbf{\omega}_{\mathbf{C\ }\rightleftharpoons
\mathbf{W}}=\mathbf{R}\,\mathbf{\omega}_{0}^{in}\,+\left(  \mathbf{\omega}%
_{B}^{ext}\right)  _{R}\tag{3.164}\\
\mathbf{\omega}_{S}  &  =\mathbf{\omega}_{\mathbf{C\ }\rightleftharpoons
\mathbf{W}}=\mathbf{R}\,\mathbf{\omega}_{0}^{in}\,+\left(  \mathbf{\omega}%
_{B}^{ext}\right)  _{S} \tag{3.164a}%
\end{align}
\ \ 

$\mathbf{\gamma}$ is a damping coefficient due to \emph{decoherence effects},
generated by local fluctuations of Bivacuum deteriorating the phase/spin
transmission via $\mathbf{VirG}_{SME}$; \ $\ \left(  \mathbf{m}_{V}%
^{+}\right)  _{S,R}$ are the actual mass of (S) and (R); \ $\left[
\mathbf{F}_{\mathbf{VPW}}\right]  $ is a \emph{tuning} \emph{force of virtual
pressure waves }$\mathbf{VPW}^{\pm}$\emph{ of Bivacuum with tuning energy
}$\mathbf{E}_{VPW}=\mathbf{q\,m}_{0}\mathbf{c}^{2}$\emph{ and wave length
}$\mathbf{L}_{VPW}=\hbar/\mathbf{m}_{0}\mathbf{c}$\textit{\ }
\begin{equation}
\mathbf{F}_{\mathbf{VPW}^{\pm}}=\frac{\mathbf{E}_{VPW}}{\mathbf{L}_{VPW}%
}=\frac{\mathbf{q}}{\mathbf{\hbar}}\mathbf{m}_{0}^{2}\mathbf{c}^{3}
\tag{3.165}%
\end{equation}

The most probable \emph{tuning force} has a minimum energy, corresponding to
$\mathbf{q=j-k=1.}$

The influence of \emph{virtual tuning force} ($\mathbf{F}_{\mathbf{VPW}^{\pm}%
})$ stimulates the synchronization of [S] and [R] pulsations, i.e. $\omega
_{R}$ $\rightarrow\omega_{S}\rightarrow\omega_{0}$. This fundamental frequency
$\mathbf{\omega}_{0}=\mathbf{m}_{0}\mathbf{c}^{2}/\hbar$ is the same in any
space volume, including those of Sender and Receiver.\medskip

The $\mathbf{VirG}_{SME}$ represent quasi \textbf{1D} macroscopic virtual Bose
condensate with a configuration of single microtubules, formed by Bivacuum
bosons ($\mathbf{BVB}^{\pm})$ or with configuration of double microtubules,
composed from Cooper pairs as described in previous section.

The effectiveness of entanglement between two or more similar elementary
particles is dependent on synchronization of their
$[\mathbf{C\rightleftharpoons W}]$ pulsation frequency and 'tuning' the phase
of these pulsations via nonlocal virtual guide ($\mathbf{VirG}_{SME}$)$^{S,R}$
between Sender and Receiver under the action of the virtual pressure waves
\textbf{VPW}$_{q=1}^{\pm}$ and Tuning energy of Bivacuum.

The mechanism proposed may explain the experimentally confirmed nonlocal
interaction between coherent elementary particles [69], atoms and between
remote coherent clusters of molecules.

Our theory predicts that the same mechanism, involving a nonlocal net of
$\mathbf{VirG}_{SME},$ may provide the entanglement even between macroscopic
systems, including biological ones. It is possible, if the frequency and phase
of \thinspace$\lbrack C\rightleftharpoons W]$\ pulsations of clusters of their
particles in state of mesoscopic Bose condensation are 'tuned' to each other
[22]. \medskip

\begin{center}
\textbf{J. \ Experimental data, confirming Unified theory (UT)\smallskip}

\emph{J.1 \ The electromagnetic radiation of nonuniformly accelerating
charges\smallskip}
\end{center}

It follows from our theory, that the charged particle, \emph{nonuniformly}
accelerating in cyclotron, synchrotron or in undulator, can be a source of
photons. The same is true for dipole radiation (section III.F.1).

>From eqs.(3.74, 3.74a and 3.98) we get general expression for electromagnetic
radiation, dependent on the doubled kinetic energy $\mathbf{\Delta(2T}%
_{k}\mathbf{)=\Delta(m}_{V}^{+}\mathbf{v}^{2}\mathbf{)}$ of alternately
accelerated charged particle and related \emph{inelastic} recoil effects,
accompanied its [$\mathbf{C\rightleftharpoons W}]$ pulsation:%
\begin{gather}
\mathbf{\hbar\omega}_{EH}\,\mathbf{+\,\Delta2T}_{k}\,\mathbf{=}\tag{3.166}
\mathbf{\,}\mathbf{\Delta}\left\{  \mathbf{(1+R)[m}_{V}^{+}\mathbf{c}%
^{2}\,\mathbf{-\,m}_{0}\mathbf{c}^{2}]\right\}  =\nonumber
\mathbf{\Delta\lbrack(1+R)TE]=}\nonumber\\
or:\;\;\mathbf{\hbar\omega}_{EH}\,\mathbf{+\,\Delta2T}_{k}\tag{3.166a}
=\mathbf{\Delta}\left[  \left\vert \alpha\mathbf{m}_{V}^{+}\mathbf{v}%
^{2}\right\vert _{rec}+\mathbf{m}_{V}^{+}\mathbf{v}^{2}\right]  ^{ext}%
\nonumber\\
\text{ where }\mathbf{R}=\sqrt{1-\left(  \mathbf{v/c}\right)  ^{2}}\text{ is
relativistic factor}\nonumber
\end{gather}

We can see from this formula, that the alternation of kinetic energy of
charged particle, resulting from its alternating acceleration, can be
accompanied by electromagnetic radiation. This effect occurs, if the jump of
kinetic energy $\mathbf{\Delta2T}_{k}$ and corresponding \emph{inelastic}
recoil energy jump: $\Delta\left\vert \alpha\mathbf{m}_{V}^{+}\mathbf{v}%
^{2}\right\vert _{rec}$ exceed the energetic threshold necessary for photon
origination. The [$\mathbf{C\rightleftharpoons W}]$ pulsations of all three
sub-elementary fermions of triplets of charged elementary particles:$\;\langle
\lbrack\mathbf{F}_{\uparrow}^{+}\bowtie\mathbf{F}_{\downarrow}^{-}%
]_{W}+\left(  \mathbf{F}_{\uparrow}^{-}\right)  _{C}\rangle_{p,e}%
\rightleftharpoons\langle\lbrack\mathbf{F}_{\uparrow}^{+}\bowtie
\mathbf{F}_{\downarrow}^{-}]_{C}+\left(  \mathbf{F}_{\uparrow}^{-}\right)
_{W}\rangle_{p,e}$, modulated by external translational dynamics, participate
in photon creation. In accordance with our model (Fig.3.2) photons are the
result of fusion of the electron and positron like triplets. They can be a
result of correlated asymmetric excitation of three pairs of Bivacuum Cooper
pairs $3[\mathbf{BVF}^{\uparrow}\mathbf{\bowtie BVF}^{\downarrow}\mathbf{]}$
by accelerated electrons or positrons, like in the case of excited atoms and molecules.

It is the Tuning energy ($\mathbf{TE}$) of Bivacuum: $\mathbf{TE=[m}_{V}%
^{+}\mathbf{c}^{2}\,\mathbf{-\,qm}_{0}\mathbf{c}^{2}]$, that induces the
transitions of the excited state of particle to its ground state,
corresponding to the rest mass energy:
\begin{equation}
\mathbf{m}_{V}^{+}\mathbf{c}^{2}\,\overset{TE}{\rightarrow}\mathbf{\,m}%
_{0}\mathbf{c}^{2}\;\;\;\;at\;\;\mathbf{v}_{tr}^{ext}\rightarrow
0,\;\;if\;\;\mathbf{q=1} \tag{3.167}%
\end{equation}

There are huge numbers of experimental data confirming this consequence of our
theory for electromagnetic radiation. The gravitational radiation in form of
Virtual Pressure Waves (VPW$^{\pm})$ in similar conditions is also predictable.

\emph{At the permanent (uniform) acceleration of the charged elementary
particle, }moving along the hyperbolic trajectory, the radiation is absent
because it does not provide the fulfilment of condition of overcoming of
activation barrier, necessary for three Cooper pairs symmetry shift. The
photon radiation by charged particles is possible only on the conditions of
nonuniform and big enough jumps of particles accelerations.

Some similarity is existing between the mechanisms of inelastic phonon
excitation in solids, detected by $\gamma-~$resonance spectroscopy, and photon
excitation in Bivacuum by alternatively accelerated particle.\medskip

The one more consequence of Unified Theory is that the radiation of photons,
induced by accelerations of charged elementary particle, should be strongly
asymmetric and coincide with direction of charged particle propagation in
space [21]. This is also well confirmed result by analysis of synchrotron and
undulator radiation. \smallskip

\begin{center}
\emph{J.2 \ The double turn (720}$^{0}),$\emph{\ as a condition of the
fermions spin state reversibility}\textbf{\smallskip}
\end{center}

It is known fact, that the total rotating cycle for spin of the electrons or
positrons is not 360$^{0},$\ but 720$^{0},\,$i.e. \textit{double turn} by
external magnetic field of special configuration, is necessary to return
elementary fermions to starting state [70]. The correctness of our model of
elementary particles was testified by its ability to explain this nontrivial
fact [21, 22] on example of the electron or proton:%

\begin{equation}
<[F_{\uparrow}^{+}\bowtie\,F_{\downarrow}^{-}]_{x,y}\,+\,\left(
\mathbf{F}_{\uparrow}^{-}\right)  _{z}>^{e,\tau} \tag{3.168}%
\end{equation}

The external rotating \textbf{H }field interact with sub-elementary
fermions/antifermions of triplets in two stage manner ($2\cdot360^{0})$,
changing their spins. The angle of spin rotation of sub-elementary particle
and antiparticles of neutral pairs $[F_{\uparrow}^{+}\bowtie\,F_{\downarrow
}^{-}]$ are the additive parameters. It means that turn of resulting spin of
\emph{pair} on 360$^{0}$ includes reorientation spins of each $F_{\uparrow
}^{+}$ and $\,F_{\downarrow}^{-}$ only on 180$^{0}$. Consequently, the full
spin turn of pair $[F_{\uparrow}^{+}\bowtie\,F_{\downarrow}^{-}]$ resembles
that of M\"{o}bius transformation.

The spin of unpaired sub-elementary fermion $\mathbf{F}_{\uparrow}^{-}>,$ in
contrast to paired ones, makes a \emph{full turn} each 360$^{0},$ i.e. twice
in 720$^{0}$ cycle$:$
\begin{align}
&  <[\left(  F_{\uparrow}^{+}\right)  _{x}\bowtie\,\left(  F_{\downarrow}%
^{-}\right)  _{y}]\,+\,\left(  \mathbf{F}_{\uparrow}^{-}\right)
_{z}>\rightarrow\tag{3.169}\\
\,\overset{360^{0}}{\longrightarrow}\,  &  <[\left(  F_{\downarrow}%
^{+}\right)  _{x}\overset{180^{0}+180^{0}}{\bowtie}\,\left(  F_{\uparrow}%
^{-}\right)  _{y}]\,+\,\left(  \mathbf{F}_{\uparrow}^{-}\right)
_{z}>\,\rightarrow\,\nonumber\\
\overset{360^{0}}{\longrightarrow}\,  &  <[\left(  F_{\uparrow}^{+}\right)
_{x}\bowtie\,\left(  F_{\downarrow}^{-}\right)  _{y}]\,+\,\left(
\mathbf{F}_{\uparrow}^{-}\right)  _{z}>\nonumber
\end{align}

The difference between the intermediate - 2nd stage and the original one in
(3.169) is in opposite spin states of paired sub-elementary particle and
antiparticle:
\begin{equation}
  [\left(  F_{\uparrow}^{+}\right)  _{x}\bowtie\,\left(  F_{\downarrow}%
^{-}\right)  _{y}]\tag{3.170}
  \overset{360^{0}}{\longrightarrow}[\left(  F_{\downarrow}^{+}\right)
_{x}\overset{180^{0}+180^{0}}{\bowtie}\,\left(  F_{\uparrow}^{-}\right)
_{y}]\nonumber
\end{equation}
It follows from our consideration, that the 3D spatial organization of the
electron (positron) is asymmetric, and some difference $\left(  F_{\uparrow
}^{+}\right)  _{x}$ $\neq$\ $[\left(  F_{\downarrow}^{+}\right)  _{x}$ \ and
$\left(  F_{\downarrow}^{-}\right)  _{y}$ $\neq\,\left(  F_{\uparrow}%
^{-}\right)  _{y}$ is existing. The $\mathbf{[C\rightleftharpoons W}]$
pulsation of unpaired sub-elementary fermion $\left(  \mathbf{F}_{\uparrow
}^{-}\right)  _{z}$ is counterphase and spatially compatible in basic state of
the electron triplet with $\left(  F_{\downarrow}^{-}\right)  _{y}$ and in the
intermediate state - with its partner $\left(  F_{\downarrow}^{+}\right)
_{x}$ in triplets (3.169).

One more known \textquotedblright strange\textquotedblright\ experimental
result can be explained by our dynamic model of triplets of elementary
particles. The existence of two paired in-phase pulsating sub-elementary
fermions (3.170) with opposite parameters, exchanging by spin, charge and
energy in the process of their $\mathbf{[C\rightleftharpoons W]}$ pulsation,
can be responsible for the \emph{two times stronger magnetic field}, generated
by electron, as compared with those, generated by rotating sphere with charge
$\left\vert e^{-}\right\vert .$ \medskip

\begin{center}
\emph{J.3 \ Michelson-Morley experiment, as a possible evidence }

\emph{for the Virtual Replica of the Earth existing\medskip}
\end{center}

The experiments, performed in 1887 by Michelson-Morley and similar later
experiments of higher precision, has been based on checking the difference of
light velocity in the direction of Earth motion on its orbit around the Sun
and in the direction normal or opposite to the this one. In the case of fixed
aether (vacuum) with condensed matter properties, \emph{independent of the
Earth motion}, one may anticipate that the difference in these two light
velocities should exist, if the value of refraction index of vacuum/aether is
independent on velocity and direction of the Earth propagation in space. The
absence of any difference was interpreted, as the absence of the Earth -
independent fixed aether. This conclusion was used by Einstein in his
postulate of permanency of light velocity, \emph{but different time} in
different inertial systems, termed 'the relativity principle'.

The conjecture of virtual replica (VR), following from our corpuscle-wave
duality and Bivacuum models, allows the another interpretation of
Michelson-Morley experiments. The VR of the Earth or any other material object
represents a standing Bivacuum virtual pressure waves (VPW$^{+}$ and
VPW$^{-})$, modulated by the object's particles corpuscle - wave pulsation
(see section 8).

The distant component of VR may have at least as big diameter, as the Earth
atmosphere and it moves in space together with planet. It is obvious, that in
such 'virtual shell' of the Earth the light velocity could be the same in any directions.

I propose the experiment, which may confirm the existence of both: the VR and
the Aether/Bivacuum, as a superfluid medium with certain mechanical
properties, like compressibility providing the VPW$^{\pm}$ existing. For this
end we assume that the properties of VR on distance of about few hundred
kilometers from the planet surface differs from that on the surface.

If we perform one series of the Michelson-Morley like experiments on the
satellite, rotating with the same angle velocity as the Earth, i.e. fixed as
respect to the Earth surface and another series of experiments on the surface,
the difference in results will confirm our Virtual Replica theory and the
Bivacuum model with Aether properties. The absence of difference in light
velocity in opposite direction as respect to Earth trajectory can be explained
in two ways:

1. As a result of equality of light velocity in any directions, independently
on direction of Earth translational propagation in space (confirmation of the
Einstein relativity principle and the absence of the Aether);

2. As a result of certain correlation between the translational velocity of
the material object (Earth or satellite) and Bivacuum perturbation (Bivacuum
dipoles symmetry shift) in the same direction, increasing the refraction index
of Bivacuum and light velocity (see section 6.6). This explanation is
compatible with the Aether concept.

Consequently, the absence of difference in light velocity in Michelson-Morley
like experiments, in any case is not a strong evidence for the Aether
absence.\medskip

\begin{center}
\emph{J.4. Interaction of particles with their Virtual Replicas, as a
background }

\emph{of two slit experiments explanation\medskip}
\end{center}

In accordance with our model, the electron and proton (Fig.3.1) are the
triplets $\langle\lbrack\mathbf{F}_{\uparrow}^{-}\bowtie\mathbf{F}%
_{\downarrow}^{+}]+\mathbf{F}_{\downarrow}^{-}\rangle^{e,\tau},$ formed by two
negatively charged sub-elementary fermions of opposite spins ($\mathbf{F}%
_{\uparrow}^{-}$\thinspace\thinspace and \thinspace$\mathbf{F}_{\downarrow
}^{-})$ and one uncompensated sub-elementary antifermion ($\mathbf{F}%
_{\downarrow}^{+})$ of $\mu$ and $\tau$ generation. The symmetric pair of
sub-elementary fermion and antifermion: $\left[  \mathbf{F}_{\uparrow}%
^{-}\bowtie\mathbf{F}_{\downarrow}^{+}\right]  $ are pulsing between
Corpuscular [C] and Wave [W] states in-phase, compensating the influence of
energy, spin and charge of each other.

It follows from our model, that the charge, spin, energy and momentum of the
electron and positron are determined just by uncompensated/unpaired
sub-elementary fermion ($\mathbf{F}_{\updownarrow}^{\pm})$. The parameters of
($\mathbf{F}_{\updownarrow}^{\pm})$ are correlated strictly with similar
parameters of pair $\left[  \mathbf{F}_{\uparrow}^{-}\bowtie\mathbf{F}%
_{\downarrow}^{+}\right]  $ due to conservation of symmetry of properties of
sub-elementary fermions and antifermions in triplets$.$ It means that
energy/momentum and, consequently, de Broglie wave length and frequency of
uncompensated sub-elementary fermion ($\mathbf{F}_{\updownarrow}^{\pm})$
determine the empirical de Broglie wave properties of the whole particle
(electron, positron).

The frequency of de Broglie wave and its length can be expressed from eq. 3.78
as:%
\begin{align}
\mathbf{\nu}_{B}  &  =\frac{(\mathbf{m}_{V}^{+}\mathbf{v}^{2})_{tr}^{ext}}%
{h}=\frac{\mathbf{v}}{\lambda_{B}}=\mathbf{\nu}_{\mathbf{C\rightleftharpoons
W}}-\mathbf{R\nu}_{0}\tag{3.171}\\
or  &  :\;\mathbf{\nu}_{B}=\frac{\mathbf{m}_{V}^{+}\mathbf{c}^{2}}%
{h}-\mathbf{R\nu}_{0} \tag{3.171a}%
\end{align}

where: $\ \mathbf{\nu}_{0}=\mathbf{m}_{0}\mathbf{c}^{2}/h=\omega_{0}/2\pi;$
\ $\lambda_{B}=h/\mathbf{m}_{V}^{+}\mathbf{v}$

In a nonrelativistic case, when $\mathbf{v<<c}$ and the relativistic factor
$\mathbf{R=}\sqrt{1-\left(  \mathbf{v/c}\right)  ^{2}}\simeq1,$ the energy of
de Broglie wave is close to Tuning energy ($\mathbf{TE}$) of Bivacuum
(3.131):
\begin{equation}
\mathbf{E}_{B}=h\mathbf{\nu}_{B}\simeq\mathbf{m}_{V}^{+}\mathbf{c}%
^{2}-\mathbf{m}_{0}\mathbf{c}^{2}=\mathbf{TE} \tag{3.172}%
\end{equation}

The fundamental phenomenon of de Broglie wave is a result of modulation of the
carrying internal frequency of $\mathbf{[C\rightleftharpoons W]}$ pulsation
($\omega_{in}=$\textbf{R}$\omega_{0}=\mathbf{Rm}_{0}\mathbf{c}^{2}/\hbar)$ by
the angular frequency of the de Broglie wave: $\omega_{B}=\mathbf{m}_{V}%
^{+}\mathbf{v}_{tr}^{2}/\hbar=2\pi\mathbf{v/\lambda}_{B},$ equal to the
frequency of beats between the actual and complementary torus and antitorus of
the \emph{anchor} Bivacuum fermion ($\mathbf{BVF}_{anc}^{\updownarrow}$) of
unpaired $\mathbf{F}_{\updownarrow}^{\pm}$. The Broglie wave length
$\mathbf{\lambda}_{B}=h/(\mathbf{m}_{V}^{+}\mathbf{v})$ and mass symmetry
shift of $\mathbf{BVF}_{anc}^{\updownarrow}$ is determined by the external
translational momentum of particle: $\overrightarrow{\mathbf{p}}%
=\mathbf{m}_{V}^{+}\overrightarrow{\mathbf{v}}.$ For nonrelativistic particles
$\omega_{B}<<\omega_{0}.$ For relativistic case, when $\mathbf{v}$ is close to
$\mathbf{c}$ and $\mathbf{R}\simeq0\mathbf{,}$ the de Broglie wave frequency
is close to resulting frequency of $\mathbf{[C\rightleftharpoons W]}$
pulsation: $\omega_{B}\simeq\omega_{\mathbf{C\rightleftharpoons W}}.$

In accordance with our model of duality, the reversible
$[\mathbf{C\rightleftharpoons W}]$\ pulsations of $\mathbf{F}_{\updownarrow
}^{\pm}$ and $\mathbf{BVF}_{anc}^{\updownarrow}$ are accompanied by
\emph{outgoing and incoming} Cumulative Virtual Cloud ($\mathbf{CVC}^{\pm}$),
composed of subquantum particles of opposite energy. On this point, our
understanding of duality and wave properties of particle coincide with that of
Bohm and Hiley [17].

Introduced in our theory notion of \emph{Virtual replica (VR) or virtual
hologram} of any material object in Bivacuum is a result of interference of
basic Virtual Pressure Waves ($\mathbf{VPW}_{q=1}^{\pm})$ and Virtual Spin
Waves ($\mathbf{VirSW}_{q=1}^{\pm1/2})$ of Bivacuum (reference waves), with
virtual \textquotedblright object waves\textquotedblright\ ($\mathbf{VPW}%
_{\mathbf{m}}^{\pm})$ and ($\mathbf{VirSW}_{\mathbf{m}}^{\pm1/2}),$
representing $\mathbf{CVC}^{+}$ and $\mathbf{CVC}^{-}$ of pair $[\mathbf{F}%
_{\uparrow}^{-}\bowtie\mathbf{F}_{\downarrow}^{+}],$ modulated by de Broglie
waves of the whole particles (see section 3.8).

The feedback influence of Bivacuum \emph{Virtual replica} of the triplet on
its original and corresponding momentum exchange may induce the wave - like
behavior of even a singe separated elementary fermion, antifermion
$\langle\lbrack\mathbf{F}_{\uparrow}^{-}\bowtie\mathbf{F}_{\downarrow}%
^{+}]+\mathbf{F}_{\updownarrow}^{\pm}\rangle^{e,p}$ (Fig.3.1) or boson, like
the photon (Fig.3.2)$.$

The reason for periodical character of the electron's trajectory in our model
(self-interference) can also be a result of periodic momentum oscillation,
produced by $[\mathbf{C\rightleftharpoons W}]$\ pulsation of the anchor
\textbf{BVF}$_{anc}$\textbf{ }of unpaired sub-elementary fermion
($\mathbf{F}_{\downarrow}^{-})$ in triplet. In the case of photon, the
momentum oscillation is equal to its frequency ($\nu_{p}=\mathbf{c/\lambda
}_{p})$, as far the relativistic factor $\mathbf{R=}\sqrt{1-\left(
\mathbf{v/c}\right)  ^{2}}$\textbf{ }in (3.171) is zero or very close to zero
near strongly gravitating objects. It is provided by the in-phase
$[\mathbf{C\rightleftharpoons W}]$\ pulsation of the anchor $\mathbf{BVF}%
_{anc}^{\updownarrow}$ of central pair of sub-elementary fermions with similar
spin orientations (Fig.3.2). The momentums of two side pairs of sub-elementary
fermions of photon with opposite spins compensate each other, because their
$[\mathbf{C\rightleftharpoons W}]$\ pulsations are counterphase.

The duality properties of elementary particles can be understood, as far they
can be \emph{simultaneously} in the corpuscle $[\mathbf{C]}$ and wave
$[\mathbf{W}]$ phase. Our theory is able to prove this condition in two
different ways:

1. In each triplet of elementary particle $\langle\lbrack\mathbf{F}_{\uparrow
}^{-}\bowtie\mathbf{F}_{\downarrow}^{+}]+\mathbf{F}_{\updownarrow}^{\pm
}\rangle$ the $[\mathbf{C\rightleftharpoons W}]$\ pulsations of the pair
$[\mathbf{F}_{\uparrow}^{-}\bowtie\mathbf{F}_{\downarrow}^{+}]$ is
counterphase with unpaired sub-elementary fermion $\mathbf{F}_{\updownarrow
}^{\pm}\rangle$. It means, that when $[\mathbf{F}_{\uparrow}^{-}%
\bowtie\mathbf{F}_{\downarrow}^{+}]$ is in the $[\mathbf{C]}$ phase, the
$\mathbf{F}_{\updownarrow}^{\pm}\rangle$ is in the [$\mathbf{W}]$ phase and
vise versa;

2. In the absence of particle accelerations, meaning its permanent kinetic
energy, the internal time of $[\mathbf{C]}$ and $\ [\mathbf{W}]$ phases is the
same. Consequently, in the internal reference system, both phase of unpaired
or paired sub-elementary fermions in different de Broglie wave semiperiods can
be considered as simultaneous.

This statement follows from our notion of time and its pace for closed system
(see section 7.1).

The formula for \emph{pace of time} for any closed coherent system or free
particle is determined by the pace of its kinetic energy: ($\mathbf{m}_{V}%
^{+}\mathbf{v}^{2}/2)_{x,y,z}=\mathbf{m}_{0}\mathbf{v}^{2}/\left(
2\sqrt{1-(\mathbf{v/c)}^{2}}\right)  _{x,y,z}$ change, which is anisotropic in
a general case:%
\begin{equation}
\lbrack\frac{\mathbf{dt}}{\mathbf{t}}\mathbf{=-\,dT}_{k}/\mathbf{T}%
_{k}]_{x,,y,z} \tag{3.173}%
\end{equation}

For the internal time itself it follows from (3.173):%

\begin{equation}
\left[  \mathbf{t=-}\frac{\mathbf{v}}{\mathbf{dv/dt}}\frac{1-(\mathbf{v/c)}%
^{2}}{2-(\mathbf{v/c)}^{2}}\right]  _{x,y,z} \tag{3.174}%
\end{equation}

We can see, that in the absence of acceleration
($\mathbf{dv/dt=0\;\;and\;\ dv=0)}$, i.e. permanent kinetic energy, momentum
and, consequently, permanent de Broglie wave frequency and length, the time
for such particle is infinitive and pace of time is zero:
\begin{equation}
\mathbf{t}\mathbf{\rightarrow\infty\;\;\;and\;\;\;}\frac{\mathbf{dt}%
}{\mathbf{t}}\rightarrow0\; \tag{3.175}%
\end{equation}

These conditions mean that the $[\mathbf{C]}$ and $\ [\mathbf{W}]$ phases of
triplets - elementary particles, moving with permanent velocity, may be
considered, as simultaneous. The oscillation of kinetic energy and velocity of
triplets occur only in transition states $\mathbf{[C\rightarrow W]}$ or
$\mathbf{[W\rightarrow C]},$ accompanied by $recoil~\leftrightharpoons
~antirecoil$ effects.

We can see from the above analysis, that our model does not need the Bohmian
\textquotedblright quantum potential\textquotedblright\ [17] or de Broglie's
\textquotedblright pilot wave\textquotedblright\ for explanation of wave-like
behavior of elementary particles.

Scattering of the photon on a free electron will affect its velocity,
momentum, mass, wave B frequency, length, its virtual replica (VR) and its
feedback influence on the electron, following by change of the interference picture.

Our theory predicts that, applying of the EM field to \emph{singe electrons}
with frequency resonant to their de Broglie frequency, should be accompanied
by alternative acceleration of the electrons, modulation of their internal
time and Virtual Replica with the same frequency, accompanied by 'washing out'
the interference pattern in two-slit experiment. This consequence of our
explanation of two-slit experiment can be easily verified. \medskip

\begin{center}
\textbf{K. The main Conclusions of Unified Theory (section III)\medskip}
\end{center}

1. A new Bivacuum model, as the infinite dynamic superfluid matrix of virtual
dipoles, named Bivacuum fermions (\textbf{BVF}$^{\updownarrow}$)$^{i}$ and
Bivacuum bosons (\textbf{BVB}$^{\pm})^{i}$, formed by correlated torus
(\textbf{V}$^{+})$ and antitorus (\textbf{V}$^{-}),$ as a collective
excitations of subquantum particles and antiparticles of opposite energy,
charge and magnetic moments and separated by energy gap, is developed. In
primordial non polarized Bivacuum, i.e. in the absence of matter and fields,
these parameters of torus and antitorus totally compensate each other. Their
spatial and energetic properties correspond to three generations: electrons,
muons and tauons ($i=e,\mu,\tau)$. The positive and negative Virtual Pressure
Waves (\textbf{VPW}$^{\pm})$ and Virtual Spin Waves (\textbf{VirSW}%
$^{S=\pm1/2})$ are the result of emission and absorption of positive and
negative Virtual Clouds (\textbf{VC}$^{\pm}),$ resulting from transitions of
\textbf{V}$^{+}$ and \textbf{V}$^{-}$ between different state of excitation.

2. It is demonstrated that symmetry shift between \textbf{V}$^{+}$ and
\textbf{V}$^{-}$ parameters to the left or right, opposite for Bivacuum
fermions $\mathbf{BVF}^{\uparrow}$\textbf{ }and antifermions\textbf{
}$\mathbf{BVF}^{\downarrow}$ with relativistic dependence on their external
rotational-translational velocity, is accompanied by sub-elementary fermion
and antifermion formation. The formation of sub-elementary fermions and their
fusion to stable triplets of elementary fermions, like electrons and protons
$\langle\lbrack\mathbf{F}_{\uparrow}^{-}\bowtie\mathbf{F}_{\downarrow}%
^{+}]+\mathbf{F}_{\updownarrow}^{\pm}\rangle^{e,p}$, corresponding to the rest
mass and charge origination, become possible at the certain velocity of
angular rotation of Cooper pairs of $\left[  \mathbf{BVF}^{\uparrow
}\mathbf{\bowtie BVF}^{\downarrow}\right]  $ around a common axis. It is
shown, that this rotational-translational velocity is determined by Golden
Mean condition: (\textbf{v/c)}$^{2}=\phi=0.618.$ The photon is a result of
fusion (annihilation) of two triplets of particle and antiparticle: [electron
+ positron] or [proton + antiproton]. The photon represents a rotating sextet
of sub-elementary fermions and antifermions with axial structural symmetry and
minimum energy $\mathbf{2m}_{0}^{e}\mathbf{c}^{2}.$

3. The fundamental physical roots of Golden Mean condition: (\textbf{v/c)}%
$^{2}=\mathbf{v}_{gr}^{ext}/\mathbf{v}_{ph}^{ext}=\phi$ are revealed as the
equality of internal and external group and phase velocities of torus and
antitorus of sub-elementary fermions, correspondingly: $\mathbf{v}_{gr}^{in}=$
$\mathbf{v}_{gr}^{ext};\ $ $\mathbf{v}_{ph}^{in}=\mathbf{v}_{ph}^{ext}.$ These
equalities are named 'Hidden Harmony Conditions'.

4. The new expressions for total, potential and kinetic energies of de Broglie
waves of elementary particles were obtained. The former represents the
extended basic Einstein and Dirac formula for free particle: $\mathbf{E}%
_{tot}=\mathbf{m}_{V}^{+}\mathbf{c}^{2}=\sqrt{1-(\mathbf{v/c)}^{2}}%
\mathbf{m}_{0}\mathbf{c}^{2}+\mathbf{h}^{2}/\mathbf{m}_{V}^{+}\mathbf{\lambda
}_{B}^{2}.$ The new formulas take into account the contributions of the actual
mass/energy of torus (\textbf{V}$^{+})$ and those of complementary antitorus
(\textbf{V}$^{-}),$ correspondingly, of asymmetric sub-elementary fermions to
the total ones. The shift of symmetry between the mass and other parameters of
torus and antitorus of sub-elementary fermions are dependent on their
\emph{internal} rotational-translational dynamics in triplets and the
\emph{external} translational velocity of the whole triplets.

5. A dynamic mechanism of [corpuscle (\textbf{C}) $\rightleftharpoons$ wave
(\textbf{W})] duality is proposed. It involves the modulation of the internal
(hidden) quantum beats frequency between the asymmetric 'actual' (torus) and
'complementary' (antitorus) states of sub-elementary fermions or antifermions
by the external - empirical de Broglie wave frequency of the whole particles.

6. It is demonstrated, that the elastic deformations and collective
excitations of Bivacuum matrix, providing field origination, are a consequence
of reversible $\left[  recoil\leftrightharpoons antirecoil\right]  $\ effects,
generated by correlated $\left[  Corpuscle\leftrightharpoons Wave\right]  $
pulsation of sub-elementary fermions/antifermions of triplets and their fast
rotation. The linear and circular alignment of Bivacuum dipoles are
responsible for electrostatic and magnetic field origination. The
gravitational waves and field are the result of positive and negative virtual
pressure waves excitation (VPW$^{+}$ and VPW$^{-})$ by the in-phase $\left[
\mathbf{C\leftrightharpoons W}\right]  $ pulsation of pairs $[\mathbf{F}%
_{\uparrow}^{-}\bowtie\mathbf{F}_{\downarrow}^{+}]$ of triplets $\langle
\lbrack\mathbf{F}_{\uparrow}^{-}\bowtie\mathbf{F}_{\downarrow}^{+}%
]+\mathbf{F}_{\updownarrow}^{\pm}\rangle,$ counterphase to that of unpaired
$\mathbf{F}_{\updownarrow}^{\pm}\rangle.$ Such virtual waves provide the
attraction or repulsion between pulsing remote particles, depending on phase
shift of pulsations, as in the case of hydrodynamic Bjerknes force action. The
zero-point vibrations of particle and evaluated zero-point velocity of these
vibrations are also the result of $\left[  recoil\leftrightharpoons
antirecoil\right]  $\ effects, accompanied by $\left[
\mathbf{C\leftrightharpoons W}\right]  $ pulsation of triplets.

7. Maxwell's \emph{displacement current} and the \emph{additional instant
current} are the consequences of Bivacuum virtual dipoles (\textbf{BVF}%
$^{\Updownarrow}$ and \textbf{BVB}$^{\pm}$) excitations and vibrations,
correspondingly. The excitations and vibrations are accompanied by the elastic
deformations of secondary Bivacuum matrix, induced by presence of matter and
fields. The increasing of the excluded for photons volume of tori and
antitori, enhance the refraction index of Bivacuum and decrease the light
velocity near strongly gravitating and charged objects. The nonzero
contribution of the rest mass energy to photons and neutrino energy is a
consequence of the enhanced refraction index of secondary Bivacuum.

8. It is shown that the Principle of least action and realization of 2nd and
3d laws of thermodynamics for closed systems - can be a result of slowing down
the dynamics of particles and their kinetic energy decreasing, under the
influence of the basic - lower frequency Virtual Pressure Waves ($\mathbf{VPW}%
_{q=1}^{\pm})$ with minimum quantum number $q=j-k=1.$ This is a consequence of
induced combinational resonance between [$\mathbf{C\leftrightharpoons W]}$
pulsation of particles and basic $\mathbf{VPW}_{q=1}^{\pm}$ of Bivacuum. The
new notion of Bivacuum Tuning Energy (TE), responsible for forcing of
particles pulsation frequency to resonance conditions with $\mathbf{VPW}%
_{q=1}^{\pm}$, is introduced.

9. It is demonstrated, that the dimensionless 'pace of time'
($\mathbf{dt/t=-dT}_{k}\mathbf{/T}_{k}$) and time itself for each closed
system are determined by the change of this system kinetic energy. They are
positive, if the particles of the system are slowing down under the influence
of $\mathbf{VPW}_{q=1}^{\pm}$ and Tuning energy of Bivacuum. The
$\mathbf{(dt/t)}$ and $\mathbf{(t)}$ are negative in the opposite case. This
new concept of time does not contradict the relativistic theory.

10. The notion of Virtual Replica (VR) or virtual hologram of any material
object is developed. The \textbf{VR }is a result of interference of
all-pervading quantized Virtual Pressure Waves ($\mathbf{VPW}_{q}^{+}$ and
$\mathbf{VPW}_{q}^{-})$ and Virtual Spin waves ($\mathbf{VirSW}_{\mathbf{q}%
}^{\mathbf{S=\pm1/2}}$) of Bivacuum, working as \emph{\textquotedblright
reference waves\textquotedblright} in holograms formation, with modulated by
de Broglie waves of matter atoms and molecules\emph{\ -\ \textquotedblright
object waves\textquotedblright} $\mathbf{VPW}_{\mathbf{m}}^{\pm}$ \ and
$\mathbf{VirSW}_{\mathbf{m}}^{\pm1/2}.$ In our Unified theory the virtual
pressure waves are identified with gravitational waves. The mechanism of
gravitation may have common features with the hydrodynamic Bjerknes force
between pulsating spheres. Their attraction of repulsion is dependent on the
phase shift between pulsating spheres.

11. A possible Mechanism of Quantum entanglement between remote elementary
particles via Virtual Guides of spin, momentum and energy ($\mathbf{VirG}%
_{\mathbf{S,M,E}})$ is proposed. The $\mathbf{VirG}_{\mathbf{S,M,E}},$
connecting similar and coherent electrons and nuclears of atoms of Sender(S)
and Receiver(R). The single $\mathbf{VirG}_{\mathbf{S,M,E}}^{\mathbf{BVB}%
^{\pm}}$ can be assembled from Bivacuum bosons $\mathbf{(BVB}^{\pm}%
\mathbf{)}^{i}$ and the twin $\mathbf{VirG}_{\mathbf{S,M,E}}^{\mathbf{BVF}%
^{\uparrow}\mathbf{\bowtie BVF}^{\downarrow}}$ from adjacent microtubules,
rotating in opposite directions, are formed by Cooper pairs of Bivacuum
fermions [$\mathbf{BVF}^{\uparrow}\mathbf{\bowtie BVF}^{\downarrow}]^{i}$. The
spin/information transmission via Virtual Guides is accompanied by
reorientation of spins of tori and antitori of conjugated Bivacuum dipoles.
The momentum and energy transmission from S to R is realized by the instant
pulsation of diameter of such virtual microtubule. The Virtual Guides of both
kinds represent the quasi 1D virtual Bose condensate with nonlocal properties,
similar to that of 'wormholes'.

12. It is demonstrated that the consequences of Unified theory (UT) of
Bivacuum, matter and fields are in good accordance with known experimental
data. Among them: electromagnetic radiation of nonuniformly accelerating
charges and the double turn ($2\times360=720^{0})$ of the electron's spin in
external magnetic field, necessary for total reversibility of spin state of
fermion. The two slit experiment also get its explanation even in a single
particles case, as a consequence of interference of particle in the wave phase
with its own virtual replica (VR).\bigskip

\section{General Conclusions}

There are several important physical features being common to the two present
theories, thus reinforcing their results. Among the basic properties there are
the similarities with the Dirac theory, the symmetry with respect to the
concepts of matter and antimatter, the non-appearance of magnetic monopoles,
and a space-charge vacuum current in addition to the displacement current.
Further there are the models of the leptons having an internal vortex-like
structure and force balance, the photon models with a small but nonzero rest
mass and a slightly reduced velocity of propagation, and the simultaneous
particle-wave behavior.

There are further results by which the two theories can be considered to
become complementary to each other. Among the characteristic results of
Theory(I), one can thus mention the deduced point-charge-like state of the
electron model, being associated with a very small but nonzero effective
radius. This removes the self-energy problem and it presents a more surveyable
alternative to the renormalization procedure. Further results are the deduced
electronic charge which deviates by only one percent from its experimental
value, the photon wave-packet models having limited spatial extensions
associated with a nonzero angular momentum, and the deduced needle-radiation
under certain conditions. Theory(I) has so far been limited to applications on
leptons, individual photons, and certain types of light beams.

A number of characteristic results from Theory(II) are also of special
interest. This Unified theory includes the following new issues:

- the new model of vacuum, named Bivacuum;

- the extension of special theory of relativity, unifying it with quantum mechanics;

- the dynamic mechanism of corpuscle-wave duality;

- the mechanism of the rest mass and charge origination, as a result of
Bivacuum dipoles symmetry shift, determined by Golden mean conditions;

- the mechanism of elementary particles fusion from sub-elementary particles;

- the new approach to quantum gravity and to electric and magnetic phenomena,
related to specific perturbation of Bivacuum by corpuscle-wave pulsation of particles;

- the new approach to time problem;

- the mechanism of nonlocality, as a result of Bose condensation, the virtual
one in Bivacuum and real in superconductors and superfluids;

- the concept of virtual replica or virtual hologram of any material object,
making possible the direct and feedback reaction of matter with Bivacuum. The
resonance interaction of basic virtual pressure waves (VPW$^{+}$ and
VPW$^{-})$ with pulsing particles, slowing down their dynamics, display itself
as a Principle of least action, 2nd and 3d law of thermodynamics. \bigskip

\section*{Acknowledgements}

The authors are indebted to Miss Anna Forsell for valuable and skillful help
with the manuscript of this paper.

\begin{center}
\bigskip

REFERENCES to Section III
\end{center}

\begin{quotation}
57. Kaivarainen A. (1995). Hierarchic Concept of Matter and Field. Water,
biosystems and elementary particles. New York, NY, pp. 485, ISBN 0-9642557-0-7.

58. Kiehn R.M. (1998). The Falaco Soliton: Cosmic strings in a swimming pool;
Coherent structures in fluids are deformable topological torsion defects. At:
IUTAM-SMFLO conf. at DTU, Denmark, May 25, 1997; URL:
http://www.uh.edu/\symbol{126}rkiehn

59. Winterberg F. (2003). Plank mass plasma vacuum conjecture. Z. Naturforsch,
58a, 231-267.

60. Jin D.Z., Dubin D.H. E. (2000). Characteristics of two-dimensional
turbulence that self-organizes into vortex crystals. Phys. Rev. Lett., 84(7), 1443-1447.

61. Okun' L.B. (1998). Physics of Elementary particles. "Nauka", Moscow, pp. 272.

62. Berstetski V., Lifshhitz M., Pitaevski L. (1989). Course of theoretical
physics, Physics, vol. IV: Quantum electrodynamics. Nauka, Moscow.

63. Kaivarainen A. (2001). New Hierarchic theory of condensed matter and its
computerized application to water and ice. In the Archives of Los-Alamos: http://arXiv.org/abs/physics/0102086.

64. Kaivarainen A. (2001a). Hierarchic theory of matter, general for liquids
and solids: ice, water and phase transitions. American Institute of Physics
(AIP) Conference Proceedings (ed. D.Dubois), vol. 573, 181-200.

65. Dicke R. H. (1957). "Gravitation without a principle of equivalence," Rev.
Mod. Phys. 29, 363-376.

66. Glansdorf P., Prigogine I. (1971). Thermodynamic theory of structure,
stability and fluctuations. Wiley and Sons, N.Y.

67. Naudin J--L. \ website "The Quest For Overunity": http://members.aol.com/jnaudin509/

Okun' L.B. (1998). Physics of Elementary particles. "Nauka", Moscow, pp. 272.

68. Prochorov A.M. Physics. Big Encyclopedic Dictionary. Moscow, 1999.

69. Aspect A. and Grangier P. (1983). Experiments on Einstein-Podolsky-Rosen
type correlations with pairs of visible photons. In: Quantum Theory and
Measurement. Eds. Wheeler J.A., Zurek W.H. Princeton University Press.

70. Davies P. (1985). The search for a grand unified theory of nature. A
Touchstone book, published by Simon \& Schuster, Inc. NY.

\end{quotation}

\end{document}